\documentclass[11pt]{article}
\pdfoutput=1
\usepackage[a4paper,text={16.8cm,22.4cm}]{geometry}
\usepackage{amsmath,amsfonts,braket,slashed,amssymb,tikz,bm,psfrag,graphicx,color,dsfont}
\usepackage{multicol}
\usepackage{ctable}
\usepackage[small,labelfont=bf]{caption}
\usepackage{hyperref}
\usepackage{subcaption}
\usepackage{cleveref}
\usepackage{bbm}

\numberwithin{equation}{section}

\RequirePackage[sort&compress,square,comma,numbers]{natbib}
\allowdisplaybreaks
\newcommand{\spac}{{\hspace{0.3mm}}}

\usepackage{subfiles}

\usepackage{graphicx}
\usepackage{mwe}

\makeatletter
\newsavebox\zb@x
\newcounter{z@@m}
\usepackage{calc}
\newdimen\B@r\newdimen\P@r
\newdimen\@zw\newdimen\@zh\newdimen\@zd

\newcommand{\zoombox}[2][0]{%
  \leavevmode%
  \sbox\zb@x{#2}%
  \setlength\B@r{1pt*\ratio{\wd\zb@x}{\ht\zb@x+\dp\zb@x}}%
  \setlength\P@r{1pt*\ratio{\paperwidth}{\paperheight}}%
  \ifdim\B@r>\P@r\relax%
    \setlength\@zw{\wd\zb@x}\setlength\@zh{\@zw*\ratio{\paperheight}{\paperwidth}}%
    \setlength\@zd{(\@zh-\ht\zb@x-\dp\zb@x)*\real{0.5}+\dp\zb@x}%
    \setlength\@zh{\@zh-\@zd}%
  \else%
    \setlength\@zh{\ht\zb@x+\dp\zb@x}%
    \setlength\@zw{\@zh*\ratio{\paperwidth}{\paperheight}}%
    \setlength\@zh{\ht\zb@x}\setlength\@zd{\dp\zb@x}%
  \fi%
  \makebox[0pt][l]{\makebox[\wd\zb@x][c]{\makebox[\@zw][l]{%
    \pdfdest name {zbfs\thez@@m} fitr
      width  \@zw\space
      height \@zh\space
      depth  \@zd\space
  }}}%
  \pdfdest name {zb\thez@@m} fitr
    width  \wd\zb@x\space
    height \ht\zb@x\space
    depth  \dp\zb@x\space
  \immediate\pdfannot 
    width  \wd\zb@x\space
    height \ht\zb@x\space
    depth  \dp\zb@x\space
  {%
    /Subtype/Link/H/N
    /Border [0 0 #1 [1 2]]
    /A <<
      /S/JavaScript
      /JS (
        if(typeof(zoomed)=='undefined'||!zoomed){
          var lastView=this.viewState;
          if(app.fs.isFullScreen) this.gotoNamedDest('zbfs\thez@@m');
          else this.gotoNamedDest('zb\thez@@m');
          zoomed=true;
        }else{
          this.viewState=lastView;
          zoomed=false;
        }
      )
    >>
  }%
  \usebox{\zb@x}%
  \stepcounter{z@@m}%
} 
\makeatother
\begin{document}

\begin{titlepage}

\begin{flushright}
\normalsize
MITP/21-025\\ 
CERN-TH-2021-148\\
IPPP/21/37

\end{flushright}

\vspace{0.5cm}
\begin{center}
\Large\bf\boldmath
Flavor probes of axion-like particles
\end{center}

\vspace{0.1cm}
\begin{center}
Martin Bauer$^a$, Matthias Neubert$^{b,c}$, Sophie Renner$^{d,e}$,\\[3pt]
Marvin Schnubel$^{b}$ and Andrea Thamm$^{f}$\\
\vspace{0.5cm} 
{\sl ${}^a$Institute for Particle Physics Phenomenology, Department of Physics\\
Durham University, Durham, DH1 3LE, United Kingdom\\[2mm]
${}^b$PRISMA$+$ Cluster of Excellence \& Mainz Institute for Theoretical Physics\\
Johannes Gutenberg University, 55099 Mainz, Germany\\[2mm]
${}^c$Department of Physics \& LEPP, Cornell University, Ithaca, NY 14853, U.S.A.\\[2mm]
${}^d$SISSA International School for Advanced Studies \& INFN, Sezione di Trieste\\
Via Bonomea 265, 34136, Trieste, Italy\\[3mm]
${}^e$Theoretical Physics Department, CERN, 1211 Geneva, Switzerland\\[2mm]
${}^f$School of Physics, The University of Melbourne, Victoria 3010, Australia}\\
\end{center}

\vspace{0.5cm}
\begin{abstract}
Axions and axion-like particles (ALPs) are well-motivated low-energy relics of high-energy extensions of the Standard Model (SM). We investigate the phenomenology of an ALP with flavor-changing couplings, and present a comprehensive analysis of quark and lepton flavor-changing observables within a general ALP effective field theory.  Observables studied include rare meson decays, flavor oscillations of neutral mesons, rare lepton decays, and dipole moments. 
We derive bounds on the general ALP couplings as a function of its mass, consistently taking into account the ALP lifetime and branching ratios. We further calculate quark flavor-changing effects that are unavoidably induced by running and matching between the new physics scale and the scale of the measurements. This allows us to derive bounds on benchmark ALP models in which only a single (flavorless or flavor-universal) ALP coupling to SM particles is present at the new physics scale, and in this context we highlight the complementarity and competitiveness of flavor bounds with constraints derived from collider, beam dump and astrophysical measurements. We find that searches for ALPs produced in meson decays provide some of the strongest constraints in the MeV-GeV mass range, even for the most flavorless of ALP models.
Likewise, we discuss the interplay of flavor-conserving and flavor-violating couplings of the ALP to leptons, finding that constraints from lepton flavor-violating observables generally depend strongly on both.
Additionally, we analyze whether an ALP can provide an explanation for various experimental anomalies including those observed in rare $B$-meson decays, measurements at the ATOMKI and KTeV experiments, and in the anomalous magnetic moments of the muon and the electron. 
\end{abstract}
\end{titlepage}

\tableofcontents
\newpage

%%%%%%%%%%%%%%%%%%%%%%%%%%%%%%%%%%%%%%%%%%%%%%%%%%%%%%%%%%%%%%%%%%%%%%%%%%%%%%%
\section{Introduction}
\label{sec:intro}
Axions and axion-like particles (collectively referred to as ALPs in this work) are pseudo Nambu--Goldstone bosons (PNGBs), which appear in the spontaneous breaking of a global symmetry. Their name derives from the QCD axion, which was introduced by Peccei, Quinn and others to address the strong CP problem \cite{Peccei:1977hh,Peccei:1977ur,Weinberg:1977ma,Wilczek:1977pj}. In this work we study the reach of flavor observables in searches for ALPs, finding that they can set the most stringent constraints in the mass range between an MeV and several GeV, thus filling the gap between tight limits derived from astrophysical and beam dump experiments (for sub-MeV masses) and collider bounds (for multi-GeV masses). ALP candidates in this mass window are motivated by the fact that the typical light and weakly coupled QCD axion faces the so-called ``axion quality problem'' \cite{Georgi:1981pu,Lazarides:1985bj,Kamionkowski:1992mf,Holman:1992us,Barr:1992qq}: Any global symmetry is believed to be broken by the effects of quantum gravity, and in the effective theory for the QCD axion this conjecture implies that higher-dimensional operators introduce corrections to the axion potential, which can destabilize its minimum, thereby reintroducing the strong CP problem. Requiring these corrections to be small favors heavy-axion solutions to the strong CP problem with ALP masses in the MeV, GeV or even TeV region \cite{Rubakov:1997vp,Berezhiani:2000gh,Hook:2014cda,Fukuda:2015ana,Gherghetta:2016fhp,Dimopoulos:2016lvn}. This mass range is further motivated by supersymmetric and composite-Higgs models featuring light PNGBs. For example, the R-axion is the PNGB of the R-symmetry breaking together with supersymmetry at low energies \cite{Bellazzini:2017neg}, while non-minimal coset structures in composite-Higgs models predict pseudo Nambu--Goldstone bosons in addition to the Higgs boson \cite{Gripaios:2009pe,Ferretti:2013kya}.

ALP couplings to gauge bosons are present in most models. In fact, the coupling of the axion to gluons is a required feature of models solving the strong CP problem, while a PNGB from a composite-Higgs model originates from the same coset as the Higgs doublet and can inherit some of the same couplings. While the SM fermions are uncharged under the Peccei--Quinn symmetry in KSVZ models \cite{Kim:1979if, Shifman:1979if}, additional ALP couplings to fermions are present in DFSZ-type QCD axion models proposed in \cite{Dine:1981rt,Zhitnitsky:1980tq} and also occur for PNGBs originating from supersymmetric or composite-Higgs UV completions. Here, we use a model-independent approach and consider the complete set of leading operators describing ALP interactions with the SM. They are suppressed by the characteristic mass scale of the new physics sector, implying that a heavy new sector gives rise to weak ALP couplings. 

If the underlying global symmetry is flavor-dependent, the ALP can acquire flavor-violating couplings to quarks and leptons, and this would provide new sources of flavor and CP violation in addition to the SM Yukawa couplings. Examples include generalized DFSZ models \cite{Davidson:1984ik,Peccei:1986pn,Krauss:1986wx,Geng:1988nc}, in which the Peccei--Quinn charges of the quarks are not flavor-universal \cite{Celis:2014iua,Alves:2017avw,DiLuzio:2017ogq,MartinCamalich:2020dfe}, and axi-flavon models, in which the ALP is a light Froggatt--Nielsen flavon with couplings that can also address the strong CP problem \cite{Ema:2016ops,Calibbi:2016hwq,Alanne:2019xaz}. Even if the underlying global symmetry is flavor-universal, flavor-violating ALP couplings are induced radiatively \cite{Choi:2017gpf,MartinCamalich:2020dfe,Bauer:2020jbp}. This would be the case in the original axion models as well as in DFSZ models, where the Peccei--Quinn charges of the quarks are taken to be generation universal \cite{Dine:1981rt,Zhitnitsky:1980tq}. Models with global lepton flavor symmetries can give rise to lepton flavor-violating ALP couplings~\cite{Gelmini:1982zz, Anselm:1985bp}.

The presence of flavor violation opens up the possibility to discover ALPs in rare, flavor-changing processes. Indeed, lepton and quark flavor transitions provide some of the most sensitive tests of new physics beyond the SM. In these processes, modern flavor experiments can indirectly probe mass scales well beyond the energy reach of the LHC. Exotic flavor-changing decays of mesons or leptons can also produce direct evidence for ALPs, which could provide information about the symmetry structure of a new physics sector otherwise out of reach of direct searches.

In this paper we start from the most general set of dimension-5 operators describing the couplings of an ALP to the SM particles in an effective field theory defined up to a scale $\Lambda$, which can be substantially larger than the electroweak scale. The couplings of the ALP to the SM fields are set by physics at the UV scale, and any structure of the couplings (arising for example from flavor symmetries in the UV theory) is naturally imposed at this scale. The running of the Wilson coefficients from the UV scale to below the electroweak scale and the matching contributions at appropriate thresholds can lead to significant changes in the ALP coupling structure and generate quark flavor-violating ALP couplings \cite{Chala:2020wvs,Bauer:2020jbp}, as summarized in Section~\ref{sec:ALPcouplings}. In the renormalization-group (RG) evolution of the ALP couplings we include the relevant anomalous dimensions at two-loop order in gauge couplings and one-loop order in Yukawa interactions as derived in \cite{Bauer:2020jbp}. At a low scale of order 2\,GeV, and for the case of a very light ALP, we match our effective theory onto an effective chiral Lagrangian describing the ALP interactions with photons and light pseudoscalar mesons. In particular, we study the effects of the consistent treatment of the weak decay processes $K^+\to\pi^+ a$ and $\pi^+\to e^+\nu_e\spac a$ as derived in \cite{Bauer:2021wjo}, including also the corresponding calculation of $K_L\to \pi^0a$. We also consistently match the ALP effective theory to an ALP-nucleon Lagrangian, taking into account the finite ALP mass for the first time.

A comprehensive study of the quark flavor phenomenology of an ALP is presented in Section~\ref{sec:quarks}. We first discuss a large set of relevant observables, including exotic two-body decays such as $K\to\pi a$, $B\to\pi a$ and $B\to K^{(*)}a$, $D\to\pi a$ and $D_{(s)}\to K a$, as well as $\pi^+\to e^+\bar\nu_\ell a$ and radiative $J/\psi$ and $\Upsilon$ decay. We then go on to study virtual ALP effects in the leptonic decays $B_{d,s}\to\mu^+\mu^-$, $B_{d,s}$--$\bar B_{d,s}$ mixing, and the chromomagnetic moment of the top quark. In all cases we present a detailed analysis of current flavor bounds. Quark flavor bounds on ALPs have also been widely studied in the literature, taking various approaches. 
For recent studies of flavor constraints on flavor-diagonal ALP-quark couplings from quark flavor transitions see \cite{Freytsis:2009ct,Batell:2009jf,Dolan:2014ska,Dobrich:2018jyi,Carmona:2021seb,Alonso-Alvarez:2021ett,Guerrera:2021yss,Choi:2021kuy}. 
Bounds on flavor non-universality, flavor off-diagonal quark couplings, the coupling to gluons, and the coupling to $W$ bosons were derived in \cite{Arias-Aragon:2017eww}, \cite{Carmona:2021seb,Bonnefoy:2020llz,MartinCamalich:2020dfe,Ishida:2020oxl,Bjorkeroth:2018dzu,Dobrich:2018jyi,Choi:2017gpf,Bauer:2016rxs}, \cite{Aloni:2018vki,Chakraborty:2021wda,Gori:2020xvq} and \cite{Izaguirre:2016dfi,Gavela:2019wzg,Gori:2020xvq,Sun:2021jpw}, respectively. 

As well as calculating observables in terms of the relevant ALP couplings at the scale of the measurement, we present detailed studies of the ALP phenomenology in a set of eight benchmark models, in which a single ALP couplings is assumed to be non-zero at the UV scale $\Lambda$. These are the three couplings to the different types of gauge fields, and the couplings to the five chiral fermion multiplets of the SM, which for simplicity we assume to be flavor-universal at the UV scale. In these benchmark models all flavor-changing ALP couplings are induced radiatively via RG evolution and matching. The benchmarks thus provide useful estimates of the minimal amount of flavor effects present in any ALP model. 
An advantage of this approach is that it allows the immediate comparison of flavor bounds with other constraints and regions of interest -- for example bounds from collider or astrophysical observables -- within common parameter spaces. In particular, we highlight the complementarity between the flavor constraints and ALP contributions to rare decays of Higgs and $Z$ bosons ($h\to aa$, $h\to Za$ and $Z\to\gamma a$) \cite{Bauer:2017nlg}. Our approach also demonstrates the universal nature of flavor bounds on ALP models; any nonzero ALP coupling in the UV unavoidably generates quark flavor-changing effects at low energies.

Furthermore, we discuss possible ALP effects on observables whose current experimental values deviate from their SM predictions. This includes the apparent violation of lepton universality in rare $B$-meson decays observed by LHCb \cite{Aaij:2017vbb,Aaij:2019wad,Aaij:2021vac}, the anomalies in excited Beryllium and Helium decays measured by the ATOMKI collaboration \cite{Krasznahorkay:2018snd,Krasznahorkay:2019lyl}, and the excess in neutral pion decays $\pi^0\to e^+e^-$ observed by KTeV \cite{Abouzaid:2006kk}.

Section~\ref{sec:leptons} is dedicated to an analysis of flavor-violating ALP couplings to leptons, which have also received a lot of attention in recent years \cite{Bjorkeroth:2018dzu,Cornella:2019uxs,Endo:2020mev,Calibbi:2020jvd,Ishida:2020oxl,Bonnefoy:2020llz,Escribano:2020wua,DiLuzio:2020oa,Ma:2021jkp}, and provides an update to our work of Ref.~\cite{Bauer:2019gfk}. In contrast to the quark sector, lepton flavor-violating (LFV) ALP couplings are not radiatively induced if the ALP has flavor-conserving couplings at the new physics scale. We consider tree-level lepton flavor violation and compute the contributions to the electromagnetic form factors from diagrams with internal ALPs. We discuss the rare muon decays $\mu\to ea$, $\mu\to ea\gamma$, $\mu\to 3e$, $\mu\to e\gamma$, $\mu\to e$ conversion, muonium--antimuonium oscillations, and the ALP contributions to the anomalous magnetic and electric dipole moments of the muon and the electron. Loop-induced lepton flavor transitions with the ALP in the loop can provide the dominant contribution to the decay widths $\mu\to 3e$ and the analogous decays of tau leptons, if the ALP is too heavy to be produced resonantly in $\mu\to ea$ \cite{Bauer:2019gfk,Cornella:2019uxs}. For heavy ALPs, the radiative lepton decay $\mu\to e\gamma$ currently provides the strongest constraint on its LFV coupling to a muon and an electron, highlighting the complementarity of searches for ALPs in resonant and non-resonant lepton decays. We also present current constraints and projections for ALPs from existing and future flavor experiments. 

We discuss three benchmark scenarios with tree-level flavor-violating ALP--lepton couplings at the UV scale. Specifically, we allow for either tau-muon, muon-electron or tau-electron transitions, in addition to flavor-diagonal couplings to leptons. 
In each case, we show the parameter space for which the ALP contribution can address tensions between the measurements and the SM predictions of the anomalous magnetic moment of the muon \cite{Aoyama:2020ynm,Abi:2021gix} and the electron \cite{Hanneke:2010au,Parker:2018vye,Morel:2020dww}. We discuss the impact of the flavor-conserving couplings of the ALP on its phenomenology in LFV observables, and the complementarity of lepton flavor-violating and -conserving constraints in different ALP mass ranges.

This paper is structured as follows: In Section \ref{sec:ALPcouplings} we describe the effective ALP Lagrangian at different scales taking into account running and matching effects from the new physics scale to the scale of the measurement. We discuss all relevant ALP couplings, including those to mesons and nucleons for finite ALP masses, as well as possible decay modes. Section \ref{sec:quarks} is dedicated to a comprehensive analysis of flavor-changing ALP couplings to quarks and a selection of low-energy anomalies. Section \ref{sec:leptons} discusses flavor-changing ALP couplings to leptons including a detailed exploration of the anomalous magnetic moments of the electron and the muon. At the start of Sections \ref{sec:quarks} and \ref{sec:leptons}, we provide short introductions outlining novel aspects of our analyses. We conclude in Section \ref{sec:conclusions}.

\clearpage
%%%%%%%%%%%%%%%%%%%%%%%%%%%%%%%%%%%%%%%%%%%%%%%%%%%%%%%%%%%%%%%%%%%%%%%%%%%%%%%
\section{The effective ALP Lagrangian}
\label{sec:ALPcouplings}

In this section we summarize results derived in \cite{Bauer:2017ris,Bauer:2020jbp,Bauer:2021wjo},  which are relevant for the phenomenology of ALP effects on flavor observables. In particular, we emphasize the important fact that RG evolution effects unavoidably generate ALP couplings to all SM fermions in the effective theory at and below the electroweak scale, irrespective of whether such couplings exist at the UV scale of global symmetry breaking. Also, the ALP couplings to left-handed down-type quarks necessarily contain flavor off-diagonal entries at and below the electroweak scale. This has significant consequences for the branching ratios of the ALP to SM particles and for the bounds on ALP couplings derived from flavor-changing processes, which one would miss ignoring these RG effects. Therefore, many of the ALP searches discussed in Section~\ref{sec:quarks} are relevant for a larger class of models than one would naively expect from the coupling structure in the UV. 

In the remainder of this section we briefly discuss a sequence of effective Lagrangians at different scales, describing the most general interactions of an ALP with SM particles, focussing on the operators of lowest dimension ($D=5$). We begin by specifying the effective theory at scales above the weak scale (Section~\ref{sec:eftUVscale}), assuming that the ALP theory respects the SM gauge group and that the ALP is the only new particle below the scale of global symmetry breaking. We then evolve this Lagrangian to the weak scale, integrate out the heavy SM particles, and discuss the evolution of the effective Lagrangian below the weak scale (Section~\ref{sec:eftweakscale}). If the ALP is very light ($m_a<2$\,GeV), its couplings to light hadrons can be described using a chiral Lagrangian. In particular, we comment on the consistent treatment of weak-interaction processes (Section~\ref{sec:chiral}) and of the ALP--nucleon couplings (Section~\ref{sec:chiralbaryons}) in such a framework. We finally discuss the most important decay channels of the ALP (to leptons and photons in Section~\ref{sec:decaysleptonsphotons}, and to hadrons in Section~\ref{sec:decayshadrons}), and the production of an ALP in exotic decays of Higgs and $Z$ bosons (Section~\ref{sec:higgsandZdecays}). 

\subsection{Effective ALP Lagrangian at the UV scale}
\label{sec:eftUVscale}

We consider a new pseudoscalar resonance, $a$, which transforms as a singlet under the SM and arises as a pseudo Nambu--Goldstone boson in the spontaneous breaking of a global symmetry at some new physics scale $\Lambda$. We will assume that $\Lambda$ is much larger than the weak scale. The ALP couplings to SM fields are protected by an approximate shift symmetry ($a\to a+\,$constant) at the classical level, broken only by the presence of a mass term $m_{a,0}^2$. This parameter would be absent for the classical QCD axion. The most general effective Lagrangian including operators of dimension up to~5 reads \cite{Georgi:1986df}
\begin{equation}\label{Leff}
\begin{aligned}
   {\cal L}_{\rm eff}^{D\le 5}
   &= \frac12 \left( \partial_\mu a\right)\!\left( \partial^\mu a\right) - \frac{m_{a,0}^2}{2}\,a^2
    + \frac{\partial^\mu a}{f}\,\sum_F\,\bar\psi_F\spac\bm{c}_F\,\gamma_\mu\spac\psi_F 
    + c_\phi\,\frac{\partial^\mu a}{f}\, 
    \big( \phi^\dagger i \hspace{-0.6mm}\overleftrightarrow{D}\hspace{-1mm}_\mu\spac\phi \big) \\
   &\quad\mbox{}+ c_{GG}\,\frac{\alpha_s}{4\pi}\,\frac{a}{f}\,G_{\mu\nu}^a\,\tilde G^{\mu\nu,a}
    + c_{WW}\spac\frac{\alpha_2}{4\pi}\,\frac{a}{f}\,W_{\mu\nu}^A\,\tilde W^{\mu\nu,A}
    + c_{BB}\,\frac{\alpha_1}{4\pi}\,\frac{a}{f}\,B_{\mu\nu}\,\tilde B^{\mu\nu} \,.
\end{aligned}
\end{equation}
Here $G_{\mu\nu}^a$, $W_{\mu\nu}^A$ and $B_{\mu\nu}$ are the field-strength tensors of $SU(3)_c$, $SU(2)_L$ and $U(1)_Y$, $\tilde B^{\mu\nu}=\frac12\epsilon^{\mu\nu\alpha\beta} B_{\alpha\beta}$ etc.\ (with $\epsilon^{0123}=1$) are the dual field-strength tensors, and $\alpha_s=g_s^2/(4\pi)$, $\alpha_2=g^2/(4\pi)$ and $\alpha_1=g^{\prime\,2}/(4\pi)$ denote the corresponding coupling parameters. The sum in the first line extends over the chiral fermion multiplets $F$ of the SM, and the Higgs doublet is denoted by $\phi$. The quantities $\bm{c}_F$ are $3\times 3$ hermitian matrices in generation space. 

The shift symmetry of the ALP couplings is manifest in the derivative couplings to the fermions and the Higgs boson, whereas for the couplings of $a$ to the $U(1)_Y$ and $SU(2)_L$ gauge fields the effect of the shift, $a\to a+\,$constant, can be removed by field redefinitions. The ALP coupling to QCD gauge fields is not invariant under a continuous shift transformation because of instanton effects, which, however, preserve a discrete version of the shift symmetry. The suppression scale $f$ of the dimension-5 operators is related to the scale of global symmetry breaking by $\Lambda=4\pi f$. In the literature on QCD axions $f$ is often eliminated in favor of the axion decay constant $f_a$ defined such that $f_a\equiv -f/(2c_{GG})$. This parameter thus governs the ALP coupling to gluons.

When QCD instanton effects are taken into account (for instance in the framework of the chiral Lagrangian which will be discussed in Section~\ref{sec:chiral}), the physical ALP mass following from the Lagrangian \eqref{Leff} is \cite{Bardeen:1978nq,Shifman:1979if,DiVecchia:1980yfw}
\begin{equation}\label{eq:ALPmass}
   m_a^2 = m_{a,0}^2 \left[ 1 + \mathcal{O}\bigg( \frac{f_\pi^2}{f^2}\bigg) \right]
    + c_{GG}^2\,\frac{f_\pi^2\,m_\pi^2}{f^2}\,\frac{2 m_u m_d}{(m_u + m_d)^2} \,,
\end{equation}
where $f_\pi\approx 130.5$\,MeV is the pion decay constant, and the corrections to the first term have been calculated in \cite{Bauer:2020jbp}. The contribution to the mass proportional to $c_{GG}$ is generated non-perturbatively by the breaking of the shift symmetry through QCD dynamics. In the case of the QCD axion this is assumed to be the only contribution to the axion mass, whereas we allow for additional sources of shift-symmetry breaking entering in the form of an explicit mass term~$m_{a,0}^2$. Such additional contributions can be due to explicit, dynamically-generated breaking terms occuring for example in non-abelian extensions of the SM with an enlarged spectrum of colored particles. In such models additional instanton contributions can arise, which can be sizable due to an enhancement of the QCD coupling at high energies in the presence of these particles. Early ideas of introducing extra colored matter at an intermediate scale either led to new hierarchy problems or spoil the solution of the strong CP problem due to new CP-violating phases \cite{Holdom:1982ex,Holdom:1985vx,Dine:1986bg,Flynn:1987rs,Choi:1988sy,Choi:1998ep}. Some more recent realizations included mirror copies of the SM, such that the complete particle spectrum inherits an additional $\mathds{Z}_2$ symmetry, which is broken. The symmetry-breaking scale of the mirror sector can be larger than the electroweak scale, thereby enhancing significantly the axion mass \cite{Rubakov:1997vp,Berezhiani:2000gh,Hook:2014cda,Fukuda:2015ana,Dimopoulos:2016lvn}. Another mechanism explored in \cite{Gaillard:2018xgk} considers an enlarged color sector, which solves the strong CP problem via new massless fermions. The spontaneous breaking of the unified color group $SU(6)\times SU(3')$ into QCD and another confining group provides a source of naturally large axion mass due to small-size instantons, while automatically ensuring a CP-conserving vacuum. A different approach was presented in \cite{Agrawal:2017ksf}, where the $SU(3)_c$ group of the SM is extended to be a diagonal subgroup of a parent $SU(3)\times SU(3)\times\dots$ group, which is broken at a high scale. All SM quarks are charged under a single $SU(3)$ factor of the parent group and an axion is introduced for each one, which independently relaxes the corresponding $\theta$ angle to~0. This allows each of the axions to have a mass significantly larger than in the QCD axion case. These studies show that in suitable extensions of the SM it is possible to generate a genuine ALP mass term while preserving the solution of the strong CP problem.

Together with the ALP mass and the four ALP couplings to the gauge and Higgs bosons, there are $1+4+5\times 9=50$ real parameters in the Lagrangian. The five global $U(1)$ symmetries of the SM (individual lepton numbers, baryon number, and hypercharge) can be used to remove five of these parameters \cite{Georgi:1986df}, resulting in 45 real physical parameters. This can be seen by performing ALP-dependent field redefinitions of the SM fields, weighted by the generators of these global symmetries. We define $\bm{Q}_F$ as the charge matrix of the fermion $F$ and $Q_\phi$ as the charge of the Higgs doublet under one of these symmetries, such that e.g.\ $\bm{Q}^{(B)}_d\psi_d=\frac13\mathbbm{1}\spac\psi_d$ gives the baryon number of the down-type quarks. Then a field redefinition
\begin{equation}
\label{eq:fieldredef}
   \psi_F \to \exp\left( ic\,\frac{a}{f}\,\bm{Q}_F \right) \psi_F \,, \qquad
   \phi \to \exp\left( ic\,\frac{a}{f}\,Q_\phi \right) \phi \,,
\end{equation}
where $c$ is any real number (but equal for all fields involved in the transformation), will have the following effects on the ALP couplings in the effective Lagrangian (\ref{Leff}):
\begin{equation}\label{eq:symmtrans}
\begin{aligned}
   \bm{c}_F &\to \bm{c}_F - c\,\bm{Q}_F \,, \\
   c_\phi &\to c_\phi - c\,Q_\phi \,, \\
   c_{GG} &\to c_{GG} + \frac{c}{2}\,\text{Tr} \left( \bm{Q}_u + \bm{Q}_d - 2\spac\bm{Q}_Q \right) , \\
   c_{WW} &\to c_{WW} - \frac{c}{2}\,\text{Tr} \left( 3\spac\bm{Q}_Q + \bm{Q}_L \right) , \\[-1mm]
   c_{BB} &\to c_{BB} + c\,\text{Tr} \left( \frac43\,\bm{Q}_u + \frac13\,\bm{Q}_d - \frac16\,\bm{Q}_Q 
    + \bm{Q}_e - \frac12\,\bm{Q}_L \right) . 
\end{aligned}
\end{equation}
To be more specific, we now consider each global symmetry in turn. Under a transformation~\eqref{eq:fieldredef} proportional to hypercharge, the ALP--Higgs and ALP--fermion couplings transform as
\begin{equation}
\begin{aligned}
   c_\phi &\to c_\phi - \frac{c}{2} \,, & 
    \bm{c}_Q &\to \bm{c}_Q - \frac{c}{6}\spac\mathbbm{1} \,, & 
    \quad \bm{c}_L &\to \bm{c}_L + \frac{c}{2}\spac\mathbbm{1} \,, \\
   \bm{c}_u &\to \bm{c}_u - \frac{2c}{3}\spac\mathbbm{1} \,, & 
    \quad \bm{c}_d &\to \bm{c}_d + \frac{c}{3}\spac\mathbbm{1} \,, & 
    \bm{c}_e &\to \bm{c}_e + c\spac\mathbbm{1} \,,
\end{aligned}
\end{equation}
while the ALP couplings to gauge bosons remain unchanged. Canonically, this transformation is used to remove the ALP--Higgs coupling from the effective Lagrangian \cite{Georgi:1986df}. This is accomplished by choosing $c=2c_\phi$. We adopt this choice for the remainder of this work and define the ALP--fermion couplings in this particular operator basis. Then there remain four other redundant linear combinations of couplings. Under a transformation~\eqref{eq:fieldredef} proportional to baryon number, the ALP--quark and ALP--gauge-boson couplings transform as
\begin{equation}\label{Btrafo}
\begin{aligned}
   \bm{c}_Q &\to \bm{c}_Q - \frac{c}{3}\spac\mathbbm{1} \,, & 
    \bm{c}_u &\to \bm{c}_u - \frac{c}{3}\spac\mathbbm{1} \,, & 
    \bm{c}_d &\to \bm{c}_d - \frac{c}{3}\spac\mathbbm{1} \,, \\
   c_{GG} &\to c_{GG} \,, &
    c_{WW} &\to c_{WW} - \frac{3c}{2} \,, &
    \quad c_{BB} &\to c_{BB} + \frac{3c}{2} \,,
\end{aligned}
\end{equation}
while the ALP couplings to the leptons and the Higgs remain unchanged. Similarly, under a transformation~\eqref{eq:fieldredef} proportional to the lepton number of the $i^{\rm th}$ lepton flavor (in the basis where the SM Yukawa matrix $\bm{Y}_e$ is diagonal), the ALP--lepton and ALP--gauge-boson couplings transform as 
\begin{equation}\label{Litrafo}
\begin{aligned}
   \bm{c}_L &\to \bm{c}_L - c\spac\mathbbm{1}_i \,, & 
    \bm{c}_e &\to \bm{c}_e - c\spac\mathbbm{1}_i \,, \\
   c_{WW} &\to c_{WW} - \frac{c}{2} \,, &
    \quad c_{BB} &\to c_{BB} + \frac{c}{2} \,,
\end{aligned}
\end{equation}
where $\mathbbm{1}_i$ is a diagonal matrix with a 1 in the $ii$ entry and zeroes otherwise. The ALP couplings to the quarks, the Higgs and gluons remain unchanged. Note that the sum $(c_{WW}+c_{BB})$ is invariant in all cases. The transformations of $c_{WW}$ and $c_{BB}$ shown in (\ref{Btrafo}) and (\ref{Litrafo}) reflect the fact that baryon number and lepton number are individually anomalous in the SM. Under the anomaly-free combination $(B-L)$ the ALP couplings to all three gauge bosons are invariant.

The transformations (\ref{Btrafo}) and (\ref{Litrafo}) can be used to eliminate four coupling parameters (or linear combinations thereof), e.g.\ the three diagonal elements of $\bm{c}_L$ or $\bm{c}_e$ and the ALP--boson couplings $c_{BB}$ or $c_{WW}$ (but not both). In this work we will refrain from making a particular choice about which ALP couplings to remove (apart from setting $c_\phi=0$), mainly because there is a large literature on ALP models in which bounds are derived on $c_{WW}$ or $c_{BB}$ individually. However, it is important to keep these parameter redundancies in mind. Predictions for physical quantities can only depend on linear combinations of ALP couplings which are invariant under all symmetry transformations. In \cite{Bauer:2020jbp}, we have shown that these physical ALP couplings can be chosen as 
\begin{equation}\label{cVVtilde}
\begin{aligned}
   \tilde c_{GG} 
   &= c_{GG} + \frac12\,\text{Tr} \left( \bm{c}_u + \bm{c}_d - 2\spac\bm{c}_Q \right) , \\
   \tilde c_{WW} 
   &= c_{WW} - \frac12\,\text{Tr} \left( 3\spac\bm{c}_Q + \bm{c}_L \right) , \\
   \tilde c_{BB} 
   &= c_{BB} + \text{Tr} \left( \frac43\,\bm{c}_u + \frac13\,\bm{c}_d - \frac16\,\bm{c}_Q 
    + \bm{c}_e - \frac12\,\bm{c}_L \right) ,
\end{aligned}
\end{equation}
and
\begin{equation}\label{tildeYf}
\begin{aligned}
   \tilde{\bm{Y}}_u 
   &= i\spac\big( \bm{Y}_u\,\bm{c}_u - \bm{c}_Q\spac\bm{Y}_u - c_\phi\spac\bm{Y}_u \big) \,, \\
   \tilde{\bm{Y}}_d 
   &= i\hspace{0.3mm} \big( \bm{Y}_d\,\bm{c}_d - \bm{c}_Q\spac\bm{Y}_d + c_\phi\spac\bm{Y}_d \big) \,, \\
   \tilde{\bm{Y}}_e 
   &= i\hspace{0.3mm} \big( \bm{Y}_e\,\bm{c}_e - \bm{c}_L\spac\bm{Y}_e + c_\phi\spac\bm{Y}_e \big) \,.
\end{aligned}
\end{equation}
If the effective theory is extended to energies below the weak scale, then the effects of heavy fermions decouple and need to be removed from the above expressions (see \cite{Bauer:2020jbp} for more details). As stated earlier, from now on we work in a basis where $c_\phi\equiv 0$.

\subsection{Effective ALP Lagrangian at the electroweak scale}
\label{sec:eftweakscale}

The RG evolution of the ALP couplings from the UV scale $\Lambda=4\pi f$ to the electroweak scale modifies the ALP--fermion couplings in significant ways, whereas the ALP--boson couplings $c_{GG}$, $c_{WW}$ and $c_{BB}$ are scale invariant at least to two-loop order \cite{Bauer:2020jbp,Chala:2020wvs}. We will show that these RG effects have a profound impact on the flavor phenomenology of ALP models. In addition, it is important to note that loop diagrams containing virtual ALP exchange require dimension-6 operators built out of SM fields as counterterms. The presence of an ALP thus provides source terms for the Wilson coefficients in the effective Lagrangian of the Standard Model Effective Fields Theory (SMEFT) and has an impact on the scale evolution of these coefficients \cite{Galda:2021hbr}.

At the weak scale, we define the ALP Lagrangian in the broken phase of the electroweak symmetry in terms of the SM mass eigenstates:
\begin{equation}\label{LeffmuW}
\begin{aligned}
   {\cal L}_{\rm eff}(\mu_w)
   &= \frac12 \left( \partial_\mu a\right)\!\left( \partial^\mu a\right) - \frac{m_{a,0}^2}{2}\,a^2
    + {\cal L}_{\rm fermion}(\mu) 
    + c_{GG}\,\frac{\alpha_s}{4\pi}\,\frac{a}{f}\,G_{\mu\nu}^a\,\tilde G^{\mu\nu,a}
    + c_{\gamma\gamma}\,\frac{\alpha}{4\pi}\,\frac{a}{f}\,F_{\mu\nu}\,\tilde F^{\mu\nu} \\
   &\quad\mbox{}+ c_{\gamma Z}\,\frac{\alpha}{2\pi s_w\spac c_w}\,
    \frac{a}{f}\,F_{\mu\nu}\,\tilde Z^{\mu\nu}
    + c_{ZZ}\,\frac{\alpha}{4\pi s_w^2\spac c_w^2}\,\frac{a}{f}\,Z_{\mu\nu}\,\tilde Z^{\mu\nu} 
    + c_{WW}\,\frac{\alpha}{2\pi s_w^2}\,\frac{a}{f}\,W_{\mu\nu}^+\,\tilde W^{-\mu\nu} \,,
\end{aligned}
\end{equation}
where $s_w\equiv\sin\theta_W$ and $c_w\equiv\cos\theta_W$ denote the sine and cosine of the weak mixing angle, and \cite{Bauer:2017ris}
\begin{equation}\label{eq:30}
   c_{\gamma\gamma} = c_{WW} + c_{BB} \,, \qquad
   c_{\gamma Z} = c_w^2\,c_{WW} - s_w^2\,c_{BB} \,, \qquad
   c_{ZZ} = c_w^4\,c_{WW} + s_w^4\,c_{BB} \,.
\end{equation}
The ALP couplings to fermions are defined in the fermion mass basis and read
\begin{align}\label{Lferm}
   {\cal L}_{\rm fermion}(\mu)
   &= \frac{\partial^\mu a}{f}\,\Big[
    \bar u_L\,\bm{k}_U(\mu) \,\gamma_\mu\,u_L + \bar u_R\,\bm{k}_u(\mu)\,\gamma_\mu\,u_R 
    + \bar d_L\,\bm{k}_D(\mu)\,\gamma_\mu\,d_L + \bar d_R\,\bm{k}_d(\mu)\,\gamma_\mu\,d_R \notag \\
   &\hspace{1.45cm}\mbox{}+ \bar\nu_L\,\bm{k}_\nu(\mu)\,\gamma_\mu\,\nu_L 
    + \bar e_L\,\bm{k}_E(\mu)\,\gamma_\mu\,e_L + \bar e_R\,\bm{k}_e(\mu)\,\gamma_\mu\,e_R \Big] \,.
\end{align}
They are related to the flavor matrices $\bm{c}_F$ in \eqref{Leff} by the unitary rotations which diagonalize the SM Yukawa matrices. The two matrices $\bm{k}_U$ and $\bm{k}_D$ are connected via the CKM matrix, such that 
\begin{equation}
   \bm{k}_D = \bm{V}^\dagger\bm{k}_U \bm{V} \,,
\end{equation}
and are therefore not independent. Likewise, the ALP couplings to neutrinos are identical to those to the left-handed charged leptons, i.e.\ $\bm{k}_\nu=\bm{k}_E$.

The flavor-conserving ALP couplings to axial-vector currents of the SM fermions play a particularly important role. We define
\begin{equation}\label{cffdef}
   c_{f_i f_i}(\mu)\equiv \left[ k_f(\mu) \right]_{ii} - \left[ k_F(\mu) \right]_{ii} .
\end{equation}
In strong-interaction and electromagnetic processes, the flavor-conserving vector currents are conserved, and hence the corresponding ALP couplings $[k_f(\mu)]_{ii}+[k_F(\mu)]_{ii}$ are unobservable.\footnote{This is no longer true in weak-interaction processes, where {\em differences\/} of the vectorial couplings to different quark flavors can appear in predictions for weak decay amplitudes \cite{Bauer:2021wjo}.} 
Choosing $f=1$\,TeV as a reference value, one finds that RG evolution effects from the new physics scale $\Lambda=4\pi f$ down to the scale $\mu_w=m_t$ modify the ALP coupling to the top quark according to \cite{Bauer:2020jbp}
\begin{equation}\label{eq:cttatmt}
   c_{tt}(m_t) \simeq 0.826\,c_{tt}(\Lambda) 
    - \big[ 6.17\,\tilde c_{GG}(\Lambda) + 0.23\,\tilde c_{WW}(\Lambda)
    + 0.02\,\tilde c_{BB}(\Lambda) \big]\times 10^{-3} \,,
\end{equation}
where the admixtures from the ALP--boson couplings are expressed in terms of the physical coupling parameters defined in (\ref{cVVtilde}) and therefore involve the ALP--fermion couplings as well. The relevant combinations can be rewritten in the form
\begin{equation}\label{cVVtildesimple}
\begin{aligned}
   \tilde c_{GG}(\Lambda) 
   &= c_{GG} + \frac12\spac\sum_q\spac c_{qq}(\Lambda) \,, \\[-2mm]
   \tilde c_{WW}(\Lambda) 
   &= c_{WW} - \frac12\,\text{Tr}\spac\big[ 3\bm{k}_U(\Lambda) + \bm{k}_E(\Lambda) \big] \,, \\[1mm]
   \tilde c_{BB}(\Lambda) 
   &= c_{BB} + \sum_f\spac N_c^f\spac Q_f^2\,c_{ff}(\Lambda) 
    + \frac12\,\text{Tr}\spac\Big[ 3\bm{k}_U(\Lambda) + \bm{k}_E(\Lambda) \Big] \,,
\end{aligned}
\end{equation}
where the sum extends over all quark and fermion flavors. $N_c^f$ denotes the number of color charges of fermion $f$, while $Q_f$ denotes its electric charge in units of $e$. Even if the ALP coupling to the top quark were absent at the UV scale, it is inevitably generated through RG evolution as long as even a single ALP coupling to a SM particle is present in the UV theory. We will find this to be a general feature of all ALP--fermion interactions.

Let us briefly return to the question of parameter redundancies at this point. In the basis where the SM Yukawa matrices are diagonal, the elements of the matrices $\tilde{\bm{Y}}_f$ in (\ref{tildeYf}) take the form
\begin{equation}
   \big( \tilde{\bm{Y}}_f \big)_{ij} 
   = i \left[ y_{f_i}\!\left[ k_f \right]_{ij} - \left[ k_F \right]_{ij}\spac y_{f_j} \right] ,
\end{equation}
where $y_{f_i}$ denote the eigenvalues of the Yukawa matrices (the physical Yukawa couplings of the quarks and leptons). It follows that $(\tilde{\bm{Y}}_f)_{ii}=i\spac y_{f_i}\,c_{f_i f_i}$, which shows that the diagonal ALP--fermion couplings $c_{f_i f_i}$ in (\ref{cffdef}) are physical parameters. For $i\ne j$, one finds that both $\left( k_f \right)_{ij}$ and $\left( k_F \right)_{ij}$ are physical quantities, since for example $i\spac(\tilde{\bm{Y}}_f^\dagger\spac\bm{Y}_f+\bm{Y}_f^\dagger\spac\tilde{\bm{Y}}_f)$ only involves the off-diagonal elements of $\bm{k}_f$. Moreover, from (\ref{cVVtildesimple}) one sees that $\tilde c_{GG}$ and $c_{GG}$ are both unambiguous, because their difference is a linear combination of the physical parameters $c_{qq}$. The same statement applies for the combinations $\tilde c_{\gamma\gamma}=\tilde c_{WW}+\tilde c_{BB}$ and $c_{\gamma\gamma}=c_{WW}+c_{BB}$, but not to the $c_{WW}$ and $c_{BB}$ individually.

\subsection{Effective ALP Lagrangian below the electroweak scale} 
\label{subsec:LagbelowEW}

Let us now assume that the ALP is significantly lighter than the weak scale. For the analysis of ALP effects on flavor observables, it is then necessary to evolve the effective ALP Lagrangian to lower energies. We can integrate out the heavy SM particles -- the top quark, the Higgs boson and the weak gauge bosons $W^\pm$ and $Z^0$ -- at the scale $\mu_w\sim m_t$ and match the effective Lagrangian (\ref{LeffmuW}) onto a low-energy effective Lagrangian in which these degrees of freedom are no longer present as propagating fields. Just below the scale $\mu_w$, this Lagrangian takes the form
\begin{equation}\label{LlowE}
\begin{aligned}
   {\cal L}_{\rm eff}^{D\le 5}(\mu\lesssim\mu_w)
   &= \frac12 \left( \partial_\mu a\right)\!\left( \partial^\mu a\right) - \frac{m_{a,0}^2}{2}\,a^2
    + {\cal L}_{\rm ferm}'(\mu) \\
   &\quad\mbox{}+ c_{GG}\,\frac{\alpha_s}{4\pi}\,\frac{a}{f}\,G_{\mu\nu}^a\,\tilde G^{\mu\nu,a}
    + c_{\gamma\gamma}\,\frac{\alpha}{4\pi}\,\frac{a}{f}\,F_{\mu\nu}\,\tilde F^{\mu\nu} \,,
\end{aligned}
\end{equation}
where ${\cal L}_{\rm ferm}'$ is given by (\ref{Lferm}) but with the top-quark fields $t_L$ and $t_R$ removed. In general, the Wilson coefficients $c_{GG}$, $c_{\gamma\gamma}$, $\bm{k}_F$ and $\bm{k}_f$ in this effective Lagrangian differ from the corresponding coefficients in the effective Lagrangian above the weak scale by calculable matching contributions, which arise when the weak-scale particles are integrated out. However, one finds that there are no matching contribution to the ALP--boson couplings $c_{GG}$ and $c_{\gamma\gamma}$, if the ALP is much lighter than the weak scale. The matching contributions to the ALP--fermion couplings have been calculated at one-loop order in the ALP vertices in \cite{Bauer:2020jbp}. We now summarize the numerical effects of the combined effects of RG evolution and weak-scale matching for the fermion couplings that will be of relevance to our analysis. All of these couplings are free of parameter redundancies.

\subsubsection*{Flavor-diagonal ALP couplings}

With the top quark integrated out, we are left with the couplings of the ALP to the axial-vector currents of the light SM fermions, as defined in (\ref{cffdef}). The relevant flavor-diagonal ALP--fermion couplings can be written as
\begin{equation}\label{eq:2.19}
   {\cal L}_{\rm fermion}^{\rm diag}(\mu)
   = \frac{\partial^\mu a}{2f}\,\sum_{f\ne t} c_{ff}(\mu)\spac\bar f\spac\gamma_\mu\gamma_5\spac f \,,
\end{equation}
where the sum runs over all light fermion mass eigenstates. For the reference scale $f=1$\,TeV, one obtains \cite{Bauer:2020jbp}
\begin{equation}\label{eq:cffrun}
\begin{aligned}
   c_{uu,cc}(m_t) 
   &\simeq c_{uu,cc}(\Lambda) - 0.116\,c_{tt}(\Lambda) 
    - \Big[ 6.35\,\tilde c_{GG}(\Lambda) + 0.19\,\tilde c_{WW}(\Lambda) 
    + 0.02\,\tilde c_{BB}(\Lambda) \Big]\times 10^{-3} \,, \\
   c_{dd,ss}(m_t) 
   &\simeq c_{dd,ss}(\Lambda) + 0.116\,c_{tt}(\Lambda) 
    - \Big[ 7.08\,\tilde c_{GG}(\Lambda) + 0.22\,\tilde c_{WW}(\Lambda) 
    + 0.005\,\tilde c_{BB}(\Lambda) \Big]\times 10^{-3} \,, \\
   c_{bb}(m_t) 
   &\simeq c_{bb}(\Lambda) + 0.097\,c_{tt}(\Lambda) 
    - \Big[ 7.02\,\tilde c_{GG}(\Lambda) + 0.19\,\tilde c_{WW}(\Lambda)
    + 0.005\,\tilde c_{BB}(\Lambda) \Big] \times 10^{-3} \,, \\
   c_{e_i e_i}(m_t)
   &\simeq c_{e_i e_i}(\Lambda) + 0.116\,c_{tt}(\Lambda) 
    - \Big[ 0.37\,\tilde c_{GG}(\Lambda) + 0.22\,\tilde c_{WW}(\Lambda) 
    + 0.05\,\tilde c_{BB}(\Lambda) \Big]\times 10^{-3} \,.
\end{aligned}
\end{equation}
As mentioned earlier, all ALP--fermion couplings are generated radiatively even if only a single ALP coupling to a SM field is non-zero at the UV scale $\Lambda$. To obtain these solutions (from \cite{Bauer:2020jbp}), we have solved the RG equations in leading logarithmic approximation, thereby resumming logarithmically enhanced contributions to all loop orders. We use the two-loop expression for the running QCD coupling $\alpha_s(\mu)$ and one-loop expressions for the running electroweak couplings $\alpha_1(\mu)$ and $\alpha_2(\mu)$ as well as for the running top-quark Yukawa coupling.

\begin{figure}
\begin{center}
\includegraphics[width=\textwidth]{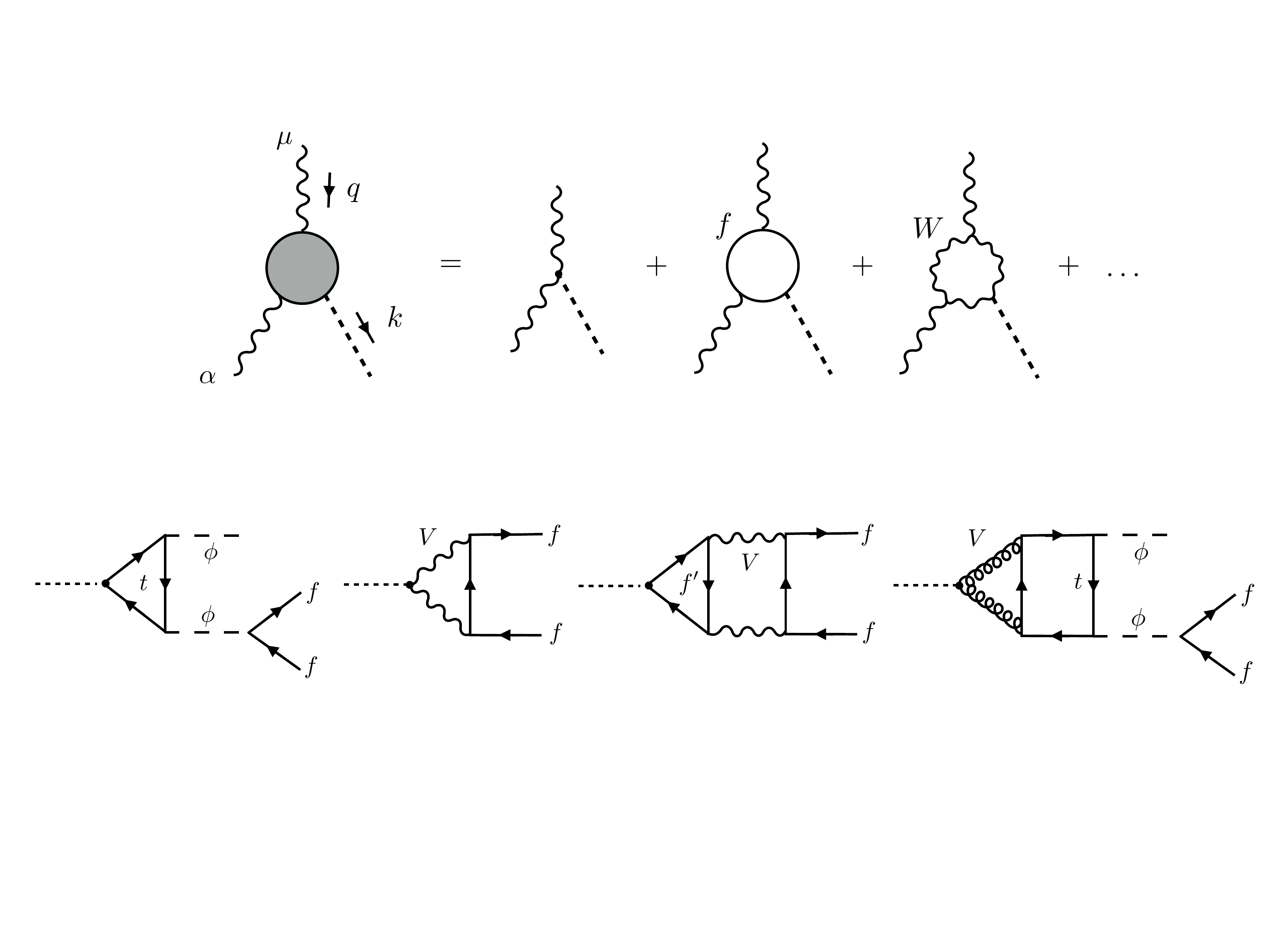}
\vspace{-4mm}
\caption{\label{fig:running} 
Logarithmically enhanced loop diagrams (in lowest order) responsible for the RG evolution effects proportional to $c_{tt}$ (first graph), $\tilde c_{VV}$ (middle two graphs), and the $c_{GG}$ contribution to $c_{e_i e_i}$ (last graph) in the results shown in (\ref{eq:cffrun}).}
\end{center}
\end{figure}

The most important evolution effect is the contribution of the ALP--top-quark coupling $c_{tt}(\Lambda)$ to all fermionic couplings in the low-energy theory. This effect is due to a logarithmically enhanced one-loop contribution of order (here and below we only quote the lowest-order logarithmic terms)
\begin{equation}
   c_{tt}\,\frac{\alpha_t}{\pi}\,\ln\frac{\Lambda^2}{m_t^2} \,,
\end{equation}
where $\alpha_t=y_t^2/(4\pi)$. It arises from the first diagram shown in Figure~\ref{fig:running}.\footnote{The diagram shown in the figure yields a contribution to the ALP--fermion couplings $\tilde{\bm{Y}}_f$ defined in (\ref{tildeYf}).} 
The fact that this contribution generates ALP couplings to all SM fermions has profound consequences for the phenomenology discussed in Sections~\ref{sec:quarks} and \ref{sec:leptons}. 

The contributions from the ALP--boson couplings $\tilde c_{VV}$ (with $V=G,W,B$) have smaller coefficients, but they may still yield the dominant effects in model where some or all of the ALP--fermion couplings vanish at the scale $\Lambda$. These effects are due to logarithmically enhanced one- and two-loop contributions of order (analogous contributions exist for $c_{WW}$ and $c_{BB}$)
\begin{equation}
   \left( \frac{\alpha_s}{\pi}\,c_{GG} \right) \frac{\alpha_s}{\pi}\,\ln\frac{\Lambda^2}{m_t^2} \,, \quad
    c_{ff} \left( \frac{\alpha_s}{\pi} \right)^2 \ln\frac{\Lambda^2}{m_t^2} 
   \quad \to \quad \tilde c_{GG} \left( \frac{\alpha_s}{\pi} \right)^2 \ln\frac{\Lambda^2}{m_t^2} \,,
\end{equation}
which arise from the second and third diagrams in Figure~\ref{fig:running}. Note that the first term in this relation correspond to a one-loop graph, because the second factor of $\alpha_s/\pi$ appears due to our choice of the normalization of the ALP--boson couplings in (\ref{Leff}). The contribution of the ALP--gluon coupling to the ALP--lepton couplings in the last line of (\ref{eq:cffrun}) is further suppressed. It arises from a logarithmically enhanced two-loop contribution of order 
\begin{equation}
   \left( \frac{\alpha_s}{\pi}\,c_{GG} \right) \frac{\alpha_t}{\pi}\,\frac{\alpha_s}{\pi}\,
    \ln^2\frac{\Lambda^2}{m_t^2} \,, 
\end{equation}
corresponding to the last diagram in Figure~\ref{fig:running}. The presence of the bosonic ALP couplings in the RG-improved expressions for the ALP--fermion couplings has important implications for ALP models in which the ALP--fermion couplings are absent (or strongly suppressed) at the UV scale $\Lambda$. 

\subsubsection*{Flavor-violating ALP couplings}

The flavor-changing ALP couplings to fermions play a particularly prominent role in our analysis. It is useful to use the equations of motion for the SM fermions to write the off-diagonal ALP--fermion couplings in the form (with $i\ne j$)
\begin{equation}\label{masssuppression}
\begin{aligned}
   {\cal L}_{\rm fermion}^{\rm FCNC}(\mu\lesssim\mu_w) 
   &= - \frac{ia}{2f}\,\sum_f\,\Big[ 
    (m_{f_i}-m_{f_j}) \left[ k_f(\mu) + k_F(\mu) \right]_{ij} \bar f_i\,f_j \\[-3mm]
   &\hspace{2.4cm} + (m_{f_i}+m_{f_j}) \left[ k_f(\mu) - k_F(\mu) \right]_{ij} 
    \bar f_i\,\gamma_5 f_j \Big] \,,
\end{aligned}
\end{equation}
where we suppress the scale dependence of the running quark masses. This form of the Lagrangian makes it evident that the ALP--fermion couplings are suppressed with the fermion masses,\footnote{This is in accordance with the fact that the physical ALP--fermion couplings defined in (\ref{tildeYf}) contain the SM Yukawa matrices.}  
and that flavor off-diagonal couplings can be of scalar and pseudo-scalar nature. 

The RG evolution of the flavor off-diagonal ALP--fermion couplings from the new physics scale $\Lambda$ to the weak scale, and the matching contributions arising when the heavy SM particles are integrated out have been studied in detail in \cite{Bauer:2020jbp}. One finds that
\begin{equation}\label{eq:47}
\begin{aligned}
   \left[ k_u(\mu_w) \right]_{ij} 
   &= \left[ k_u(\Lambda) \right]_{ij} ; \quad i,j\ne 3 \,, \\
   \left[ k_U(\mu_w) \right]_{ij} 
   &= \left[ k_U(\Lambda) \right]_{ij} ; \quad i,j\ne 3 \,, \\
   \left[ k_d(\mu_w) \right]_{ij} 
   &= \left[ k_d(\Lambda) \right]_{ij} , \\
   \left[ k_e(\mu_w) \right]_{ij}
   &= \left[ k_e(\Lambda) \right]_{ij} , \\
   \left[ k_E(\mu_w) \right]_{ij} 
   &= \left[ k_E(\Lambda) \right]_{ij} .
\end{aligned}
\end{equation}
Note that for $\bm{k}_u$ and $\bm{k}_U$ we only need the entries where $i,j\ne 3$, since the top quark has been integrated out in the effective theory below the weak scale. For the off-diagonal elements of the coefficient $\bm{k}_D$ one obtains the more interesting result
\begin{equation}\label{eq:48}
\begin{aligned}
   \left[ k_D(\mu_w) \right]_{ij}
   &= \left[ k_D(\Lambda) \right]_{ij} 
    - V_{mi}^* V_{nj} \left( \delta_{m3} + \delta_{n3} - 2\spac\delta_{m3}\spac\delta_{n3} \right) 
    \left( 1 - e^{-U(\mu_w,\Lambda)} \right) \left[ k_U(\Lambda) \right]_{mn} \\
   &\quad\mbox{}- \frac16\,V_{3i}^* V_{3j}\,I_t(\mu_w,\Lambda) 
    + \big[ \hat\Delta k_D(\mu_w) \big]_{ij} \,,
\end{aligned}
\end{equation}
where the evolution functions $U(\mu_w,\Lambda)$ and $I_t(\mu_w,\Lambda)$ are defined as
\begin{equation}
   U(\mu_w,\Lambda) 
    = - \int_{\Lambda}^{\mu_w}\!\frac{d\mu}{\mu}\,\frac{y_t^2(\mu)}{32\pi^2} \,, \qquad
   I_t(\mu_w,\Lambda) 
    = \int_\Lambda^{\mu_w}\!\frac{d\mu}{\mu}\,\frac{3 y_t^2(\mu)}{8\pi^2}\,c_{tt}(\mu) \,.
\end{equation}
Explicit analytic expressions for these integrals can be found in eqs.~(3.14) and (3.21) of \cite{Bauer:2020jbp}, while the matching contribution $[\hat\Delta k_D(\mu_w)]_{ij}$ can be found in eq.~(5.7).\footnote{These equation references apply to the published version of the paper.} 
Via these evolution functions, ALP couplings to any SM field at the UV scale will, at some loop order, produce logarithmically-enhanced contributions to flavor-changing down-type quark couplings below the electroweak scale. We will make use of this important point in Section~\ref{sec:quarks} to place new constraints on individual ALP couplings defined at the UV scale, by calculating their flavor effects to leading logarithmic approximation via these equations. 

\begin{figure}
\begin{center}
\includegraphics[width=0.5\textwidth]{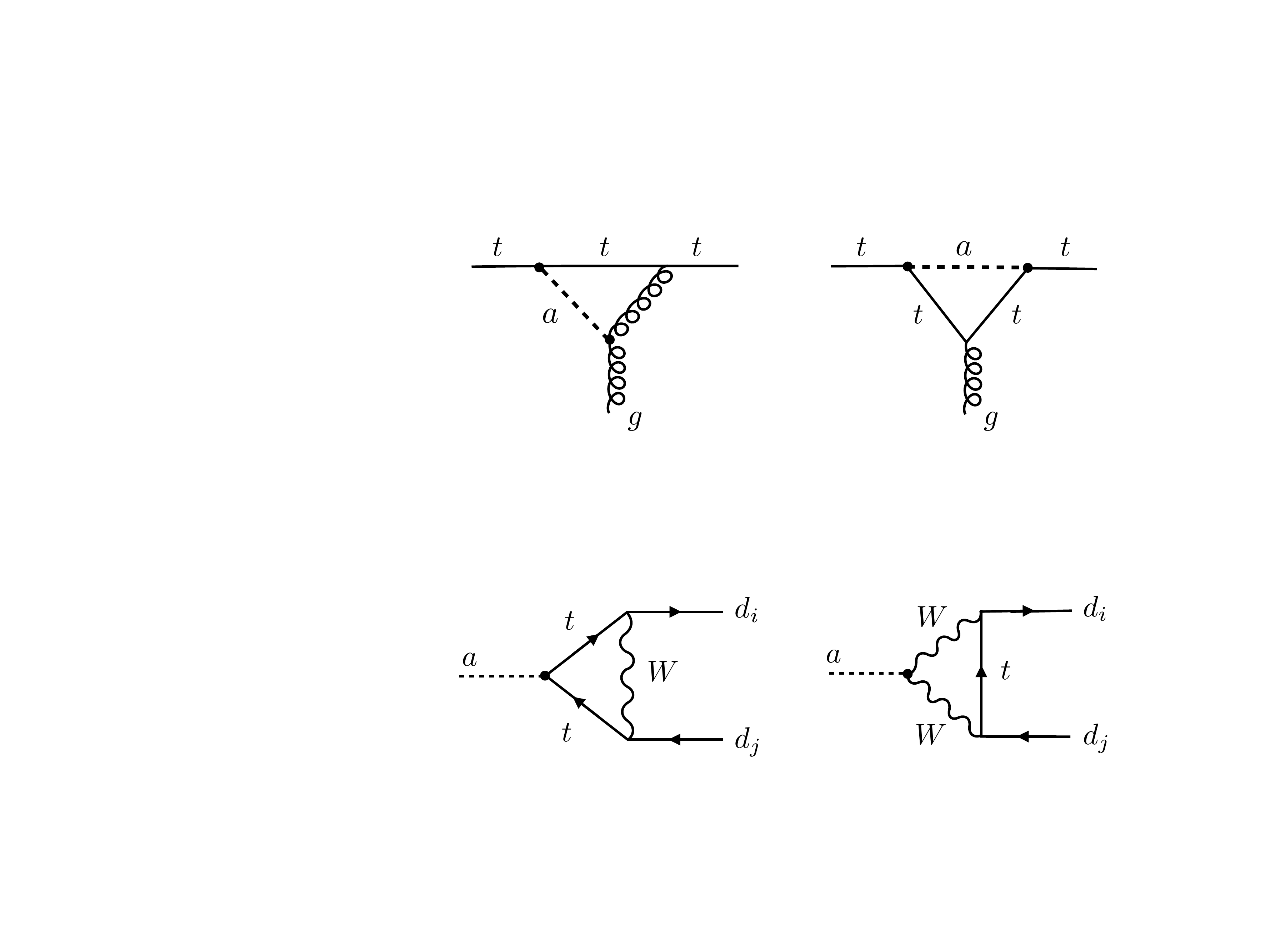}
\vspace{-2mm}
\caption{\label{fig:runningandmatching} 
Representative diagrams contributing to the flavor-changing ALP coupling in \eqref{sonice}.}
\end{center}
\end{figure}

The above results simplify significantly if the ALP Lagrangian at the UV scale $\Lambda$ respects the principle of minimal flavor violation (MFV) \cite{DAmbrosio:2002vsn}. One then finds that \cite{Bauer:2020jbp}
\begin{equation}
    \left[ k_U(\mu_w) \right]_{ij}^{\rm MFV} = \left[ k_u(\mu_w) \right]_{ij}^{\rm MFV}
    = \left[ k_d(\mu_w) \right]_{ij}^{\rm MFV} = \left[ k_E(\mu_w) \right]_{ij}^{\rm MFV} 
    = \left[ k_e(\mu_w) \right]_{ij}^{\rm MFV} = 0 \,,
\end{equation}
whereas
\begin{equation}\label{sonice}
\begin{aligned}
   \left[ k_D(\mu_w) \right]_{ij}^{\rm MFV}
   &= \left[ k_D(\Lambda) \right]_{ij}^{\rm MFV}
    + V_{ti}^* V_{tj}\,\bigg\{ - \frac16\,I_t(\mu_w,\Lambda) \\
   &\quad + \frac{\alpha_t(\mu_w)}{4\pi}\,\bigg[ c_{tt}(\mu_w) 
    \left( \frac12\ln\frac{\mu_w^2}{m_t^2} - \frac14 
    - \frac32\,\frac{1-x_t+\ln x_t}{\left(1-x_t\right)^2} \right) \\
   &\hspace{2.45cm}\mbox{}- \frac{3\alpha}{2\pi\spac s_w^2}\,\tilde c_{WW}(\mu_w)\,
    \frac{1-x_t+x_t\ln x_t}{\left(1-x_t\right)^2} \bigg] \bigg\} \,, 
\end{aligned}
\end{equation}
with $x_t=m_t^2/m_W^2$. Note the important fact that even under the assumption of minimal flavor violation the coefficients $[k_D(\Lambda)]_{ij}$ are not restricted to be flavor-diagonal. Instead,
\begin{equation}\label{UVflavorchange}
   \left[ k_D(\Lambda) \right]_{ij}^{\rm MFV}
   =  V_{ti}^* V_{tj}\,\Big( \left[ k_U(\Lambda) \right]_{33} 
    - \left[ k_U(\Lambda) \right]_{11} \!\Big) 
   \equiv V_{ti}^* V_{tj}\,\Delta k_U(\Lambda) \,,
\end{equation}
which can be non-zero because minimal flavor violation allows the possibility that $[k_U(\Lambda)]_{33}\ne[k_U(\Lambda)]_{11}$, and hence $\Delta k_U(\Lambda)\ne 0$. 

Relation (\ref{sonice}) shows explicitly how flavor-changing effects are generated through RG evolution from the new physics scale $\Lambda$ to the weak scale (first line) and matching contributions at the weak scale (second and third lines). In Figure~\ref{fig:runningandmatching} we show some representative one-loop diagrams accounting for the terms proportional to $c_{tt}$ (left graph) and $\tilde c_{WW}$ (right graph). These loop-induced effects should be considered as the minimal effects of flavor violation present in any ALP model, even if the matrix $\bm{k}_D$ is diagonal at the new physics scale $\Lambda$ (which would be a stronger assumption than minimal flavor violation). The results for the evolution effects and the contribution proportional to $c_{tt}(\mu_w)$ have been derived in \cite{Bauer:2020jbp}.\footnote{The logarithm of $(\mu_w^2/m_t^2)$ in the coefficient of $c_{tt}$, but not the $x_t$-dependent remainder, was found in \cite{Gavela:2019wzg}.} 
The terms proportional to $c_{WW}$ in (\ref{sonice}) agree with a corresponding expression derived in \cite{Izaguirre:2016dfi}. In the sum of the contributions from scale evolution and weak-scale matching, the dependence on the matching scale $\mu_w$ drops out. In fact, the flavor off-diagonal Wilson coefficients do not run below the weak scale (in the approximation where the Yukawa couplings of the light quarks are put to zero). Hence, the expressions shown in (\ref{eq:47}), (\ref{eq:48}) and (\ref{sonice}) hold for all values $\mu<\mu_w$.

The explicit solution for the evolution function $I_t(\mu_w,\Lambda)$ involves again the ALP couplings $c_{tt}$ and $\tilde c_{VV}$. For the reference scale $f=1$\,TeV, one finds numerically (for $i\neq j$) 
\begin{equation}\label{eq:FVveryshort}
\begin{aligned}
   \left[ k_D(m_t) \right]_{ij}^{\rm MFV}
   &\simeq V_{ti}^* V_{tj}\,\Big[ \Delta k_U(\Lambda) + 1.9\times 10^{-2}\,c_{tt}(\Lambda) 
    - 6.1\times 10^{-5}\,\tilde c_{GG}(\Lambda) \\
   &\hspace{1.75cm} - 2.8\times 10^{-5}\,\tilde c_{WW}(\Lambda) 
    - 1.8\times 10^{-7}\,\tilde c_{BB}(\Lambda) \Big] \,. 
\end{aligned}
\end{equation}
Besides the possible matching contribution $\Delta k_U(\Lambda)$ at the UV scale, the contribution with the largest coefficient involves the ALP coupling to top quarks, $c_{tt}(\Lambda)$, which enters via one-loop effects from RG evolution and weak-scale matching and scales like
\begin{equation}
   c_{tt}\,\frac{\alpha_t}{\pi}\,\ln\frac{\Lambda^2}{m_t^2} \,.
\end{equation}
Very interestingly, the term with the second-largest coefficient involves the ALP coupling to gluons, $\tilde c_{GG}(\Lambda)$, which contributes at one-loop order to the evolution of $c_{tt}(\mu)$ and is formally a two-loop effect (enhanced by two powers of large logarithms) of order\footnote{The extra pieces included through the replacement $c_{GG}\to\tilde c_{GG}$ are three-loop contributions.}
\begin{equation}
   \left( \frac{\alpha_s}{\pi}\,c_{GG} \right) \frac{\alpha_t}{\pi}\,\frac{\alpha_s}{\pi}\,
    \ln^2\frac{\Lambda^2}{m_t^2} \,.
\end{equation}
The term proportional to $\tilde c_{BB}(\Lambda)$ has an analogous scaling, but it is numerically suppressed due to the fact that instead of two powers of the strong coupling $\alpha_s$ is comes with two powers of $\alpha_1$. The contribution proportional to $\tilde c_{WW}(\Lambda)$ comes with the third-largest coefficient. It corresponds to a one-loop matching contribution at the weak scale, which scales like
\begin{equation}
   \left( \frac{\alpha_2}{\pi}\,c_{WW} \right) \frac{\alpha_t}{\pi}\, \,,
\end{equation}
without a logarithmic enhancement. In \cite{Izaguirre:2016dfi}, the contribution proportional to $c_{WW}$ in (\ref{eq:FVveryshort}) was considered as the only source of flavor violation in ALP-induced interactions, which obviously makes the strong assumption that the remaining couplings in that equation vanish.

In (\ref{cVVtilde}) we have shown how the parameters $\tilde c_{VV}(\Lambda)$ with $V=G,W,B$ can be expressed in terms of the ALP couplings in the original effective Lagrangian (\ref{Leff}). When combined with (\ref{eq:FVveryshort}), these relations show that, no matter to which SM field the ALP couples at the new physics scale $\Lambda$, even a single non-zero coupling will unavoidably lead to flavor-changing ALP–fermion couplings at scales at or below the electroweak scale, even in the context of an ALP model with MFV.

\subsubsection*{RG evolution below the weak scale}

The flavor off-diagonal Wilson coefficients do not run below the weak scale (in the approximation where the Yukawa couplings of the light quarks are set to zero). The flavor-diagonal couplings $c_{ff}(\mu)$ are still scale dependent at low energies due to loop diagrams involving gluons or photons. The evolution of these coefficients from the scale $\mu_w=m_t$ to the low scale $\mu_0=2$\,GeV yields \cite{Bauer:2020jbp}
\begin{equation}\label{eq:lowrun}
\begin{aligned}
   c_{qq}(\mu_0) 
   &= c_{qq}(m_t) - \Big[ 3.0\spac\tilde c_{GG}(\Lambda) - 1.4\spac c_{tt}(\Lambda)
    - 0.6\,c_{bb}(\Lambda) \Big] \times 10^{-2} \\
   &\quad\mbox{}- Q_q^2\,\Big[ 3.9\spac\tilde c_{\gamma\gamma}(\Lambda) 
    - 4.7\spac c_{tt}(\Lambda) - 0.2\spac c_{bb}(\Lambda) \Big] \times 10^{-5} \,, \\
   c_{\ell\ell}(\mu_0) 
   &= c_{\ell\ell}(m_t) - \Big[ 3.9\spac\tilde c_{\gamma\gamma}(\Lambda)
    - 4.7\spac c_{tt}(\Lambda) - 0.2\spac c_{bb}(\Lambda) \Big] \times 10^{-5} \,.
\end{aligned}    
\end{equation}
For an ALP lighter than the scale $\mu_0$, the interactions with hadrons and photons are affected by non-perturbative hadronic effects. These can be studied in a systematic way using an effective chiral Lagrangian.

\subsection{ALP couplings to mesons in the chiral Lagrangian}
\label{sec:chiral}

At the scale $\mu_0\approx 2$\,GeV it is appropriate to match the Lagrangian \eqref{LlowE} to a chiral effective theory \cite{Georgi:1986df,diCortona:2015ldu,Bauer:2017ris,Bauer:2021wjo}. The ALP--gluon coupling in the Lagrangian can be eliminated by performing a chiral rotation of the quark fields,
\begin{equation}\label{eq:chiralrot}
   q(x)\to \exp\left[ -i\,\bm{\kappa}_q\spac\gamma_5 \,c_{GG}\,\frac{a(x)}{f} \right] q(x) \,, 
\end{equation}
where $q(x)$ is a 3-component object containing the light-quark fields $u(x)$, $d(x)$ and $s(x)$. The transformation parameters $\bm{\kappa}_q$ are hermitian matrices, which we choose to be diagonal in the quark mass basis. The condition $\text{Tr}\,\bm{\kappa}_q=1$ is necessary to remove the ALP--gluon coupling from the Lagrangian. As long as this condition is satisfied, any choice of $\bm{\kappa}_q$ leads to an effective chiral Lagrangian describing the same physics. One obtains
\begin{equation}
\begin{aligned}\label{chiPT}
   {\mathcal L}_{\rm eff}^\chi
   &= \frac{f_\pi^2}{8}\,\mbox{Tr}\big[ \bm{D}^\mu\bm{\Sigma}\,(\bm{D}_\mu\bm{\Sigma})^\dagger \big] 
    + \frac{f_\pi^2}{4}\spac B_0\,\mbox{Tr}\big[ \hat{\bm{m}}_q(a)\spac\bm{\Sigma}^\dagger 
    + \text{h.c.} \big] \\
   &\quad + \frac12\,\partial^\mu a\,\partial_\mu a - \frac{m_{a,0}^2}{2}\,a^2
    + \hat c_{\gamma\gamma}\,\frac{\alpha}{4\pi}\,\frac{a}{f}\,F_{\mu\nu}\,\tilde F^{\mu\nu} \spac ,
\end{aligned}
\end{equation}
where $\bm{\Sigma}(x)=\exp\big[\frac{i\sqrt2}{f_\pi}\,\lambda^a\spac\pi^a(x)\big]$, defined with the pion decay constant $f_\pi\approx 130.5$\,MeV and the Gell-Mann matrices $\lambda_a$, contains the pseudoscalar meson fields, 
\begin{equation}\label{mhatq}
   \hat{\bm{m}}_q(a) 
   = \exp\left( - 2i\bm{\kappa}_q\,c_{GG}\,\frac{a}{f} \right) \bm{m}_q 
\end{equation}
with $\bm{m}_q=\text{diag}(m_u,m_d,m_s)$ is the modified mass matrix, and the derivative ALP couplings to fermions enter in the covariant derivative \cite{Bauer:2021wjo} 
\begin{equation}\label{eq:covD}
   i\bm{D}_\mu\bm{\Sigma} 
   = i\partial_\mu\bm{\Sigma} + e\spac A_\mu\spac[\bm{Q},\bm{\Sigma}]
    + \frac{\partial_\mu a}{f} \left( \hat{\bm{k}}_Q\spac\bm{\Sigma} 
    - \bm{\Sigma}\spac\spac\hat{\bm{k}}_q \right) . 
\end{equation}
ALP couplings with a hat differ from the couplings in the original ALP Lagrangian through terms induced by the chiral rotation. Explicitly, one finds
\begin{equation}\label{eq:2.37}
\begin{aligned}
   \hat{c}_{\gamma\gamma} 
   &= c_{\gamma\gamma} - 2N_c\,c_{GG}\,\text{Tr}\big[\bm{Q}^2\spac\bm{\kappa}_q\big] \,, \\
   \hat{\bm{k}}_Q 
   &= e^{- i\bm{\kappa}_q\spac c_{GG}\spac\frac{a}{f}}\,\big( \bm{k}_Q - \bm{\kappa}_q\,c_{GG} \big)\,
    e^{i\bm{\kappa}_q\spac c_{GG}\spac\frac{a}{f}} \,, \\
   \hat{\bm{k}}_q 
   &= e^{i\bm{\kappa}_q\spac c_{GG}\spac\frac{a}{f}}\,\big( \bm{k}_q + \bm{\kappa}_q\,c_{GG} \big)\,
    e^{- i\bm{\kappa}_q\spac c_{GG}\spac\frac{a}{f}} \,,
\end{aligned}
\end{equation}
where $\bm{Q}=\text{diag}(Q_u,Q_d,Q_s)$ contains the electric charges of the quarks. The matrices $\bm{k}_Q$ and $\bm{k}_q$ have the texture
\begin{equation}
   \bm{k}_Q = \left( \begin{array}{ccc}
    ~\left[k_U\right]_{11} & 0 & 0 \\
    0 & ~\left[k_D\right]_{11} & ~\left[k_D\right]_{12} \\
    0 & ~\left[k_D\right]_{21} & ~\left[k_D\right]_{22} \\
   \end{array} \right) , \qquad
   \bm{k}_q = \left( \begin{array}{ccc}
    ~\left[k_u\right]_{11} & 0 & 0 \\
    0 & ~\left[k_d\right]_{11} & ~\left[k_d\right]_{12} \\
    0 & ~\left[k_d\right]_{21} & ~\left[k_d\right]_{22}
   \end{array} \right) , 
\end{equation}
where the various entries refer to the ALP--fermion couplings in the mass basis defined in (\ref{masssuppression}). We recall that the off-diagonal couplings $[k_D]_{ij}$ and $[k_d]_{ij}$ with $i\ne j$ do not run below the weak scale, and their values at the scale $\mu_w$ have been given in (\ref{eq:47}) and (\ref{eq:48}).

For the case of the QCD axion (with $m_{a,0}^2=0$), the chiral effective ALP Lagrangian was first introduced in \cite{Georgi:1986df} and has been explored in great detail in \cite{diCortona:2015ldu}. By studying the ALP potential following from this Lagrangian, one finds that QCD dynamics generates a mass for the ALP, see (\ref{eq:ALPmass}), thereby breaking the continuous shift symmetry of the classical Lagrangian to the discrete subgroup $a\to a+n\pi f/c_{GG}$. The first term in the first line of (\ref{chiPT}) leads to a kinetic mixing of the ALP with the pseudoscalar mesons $\pi^0$, $\eta_8$, $K^0$ and $\bar K^0$, while the second term gives rise to a mass mixing. In order to eliminate the mass mixing, one can choose the matrix $\bm{\kappa}_q$ in such a way that $\bm{\kappa}_q\spac\bm{m}_q\propto\mathbbm{1}$; however, eliminating both types of mixings requires a different choice \cite{Bauer:2020jbp}. Since all predictions for physical quantities must be independent of the choice of the auxiliary parameters $\kappa_q$, we will refrain from adopting a particular choice in this paper.  

Applying the Noether procedure to the effective Lagrangian (\ref{chiPT}), one finds that the chiral representation of the left-handed quark currents $\bar q^i\gamma_\mu P_L\spac q^j$ is given by \cite{Bauer:2021wjo}
\begin{equation}\label{eq:Lmu}
\begin{aligned}
   L_\mu^{ji} 
   &= - \frac{i f_\pi^2}{4}\,e^{i(\kappa_{q_j}-\kappa_{q_i})\spac c_{GG}\spac\frac{a}{f}}\,
    \big[ \bm{\Sigma}\,(\bm{D}_\mu\bm{\Sigma})^\dagger \big]_{ji} \\
   &\ni - \frac{i f_\pi^2}{4} \left[ 1 + i(\kappa_{q_j}-\kappa_{q_i})\spac c_{GG}\spac\frac{a}{f} \right] 
    \big[ \bm{\Sigma}\,\partial_\mu\bm{\Sigma}^\dagger \big]_{ji} 
    + \frac{f_\pi^2}{4}\,\frac{\partial^\mu a}{f}\,\big[ \hat{\bm{k}}_Q
    - \bm{\Sigma}\,\hat{\bm{k}}_q\spac\bm{\Sigma}^\dagger \big]_{ji} \,.
\end{aligned}
\end{equation}
The derivative ALP couplings in the last term have been omitted in previous treatments of the effective chiral ALP Lagrangian, but they are crucial to ensure the independence of physical amplitudes from the choice of the auxiliary parameters $\bm{\kappa}_q$ \cite{Bauer:2021wjo}. 

The Lagrangian \eqref{chiPT} contains flavor-conserving ALP couplings to mesons, which govern the decays of ALPs into light QCD resonances. It also comprises flavor-changing neutral current couplings, which are due to the off-diagonal elements in the matrices $\hat{\bm{k}}_Q$ and $\hat{\bm{k}}_q$ and will play a role in our discussion of $K\to\pi a$ decays below. For a consistent analysis of weak-interaction decay processes involving ALPs, it is however necessary to also include the SM effective weak interactions at low energies. For the leptonic pion decay $\pi^-\to e^-\bar\nu_e\spac a$ the weak transition is a charged-current process mediated by the effective Lagrangian
\begin{equation}
   {\cal L}_{u\to d} 
   = - \frac{4 G_F}{\sqrt2}\,V_{ud}\,L_\mu^{21}\,\bar e\,\gamma^\mu P_L\spac\nu_e \,.
\end{equation}
The decay amplitude for this process obtained from the chiral Lagrangian (neglecting contributions suppressed by the electron mass) reads~\cite{Bardeen:1986yb,Bauer:2021wjo}
\begin{equation}
\begin{aligned}\label{eq:21}
   i {\cal A}(\pi^-\to e^-\bar\nu_e\spac a)
   &= - \frac{i\spac G_F}{\sqrt2}\,V_{ud}\,\frac{f_\pi}{2f}\,
    \bar u_e\spac\gamma_\mu(1-\gamma_5)\,v_{\bar\nu_e}\,\\
   &\qquad  \times(p_\pi+p_a)^\mu \left[ 2c_{GG}\,\frac{m_d-m_u}{m_d+m_u} + \left[ k_u - k_d \right]_{11}
    + \frac{m_a^2}{m_\pi^2-m_a^2}\,\Delta c_{ud} \right]  \,,
\end{aligned}
\end{equation}
where $k_{u,d}$ denotes the ALP couplings to the right-handed up- and down-quark currents, respectively, and
\begin{equation}\label{eq:cud}
   \Delta c_{ud}\equiv c_{uu} - c_{dd} + 2c_{GG}\,\frac{m_d-m_u}{m_d+m_u} \,.
\end{equation}
All quantities are evaluated at the scale $\mu_0$. 

\begin{figure}
\begin{center}
\includegraphics[width=0.7\textwidth]{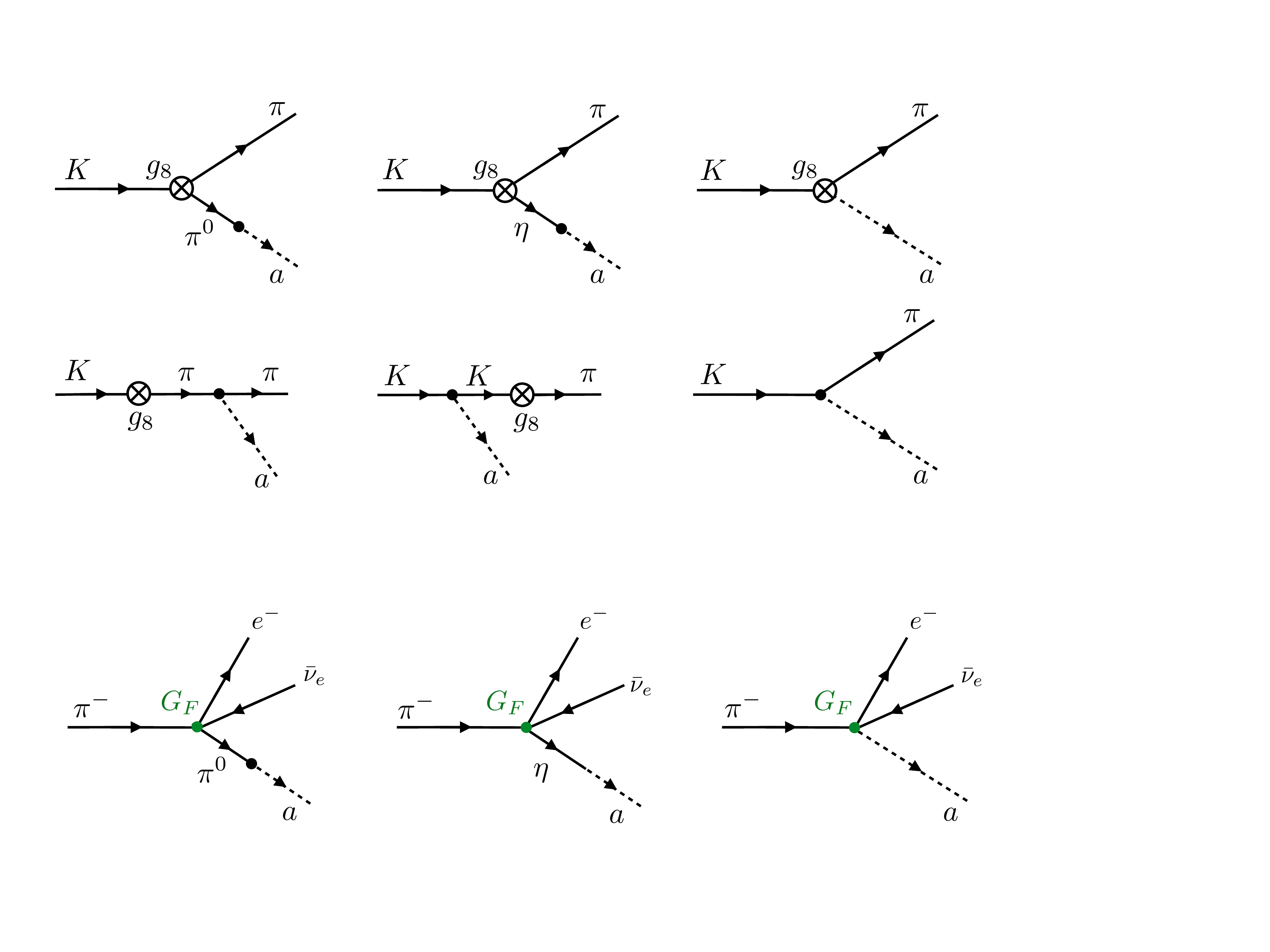}
\vspace{2mm}
\caption{\label{fig:kaondecays} 
Feynman graphs contributing to the $K^-\to\pi^- a$ and $\bar K^0\to\pi^0 a$ decay amplitudes at leading order in the chiral expansion. Weak-interaction vertices mediated by the $SU(3)$ octet operator ${\cal O}_8$ are indicated by a crossed circle, while dots refer to vertices from the Lagrangian (\ref{chiPT}). Analogous graphs exist for the two 27-plet operators. The first two diagrams in the second row vanish for the case of neutral mesons.}
\end{center}
\end{figure}

The leading-order operators mediating flavor-changing non-leptonic meson decays such as $K^-\to\pi^-\pi^0$, $K_S\to\pi^+\pi^-$ and $K_S\to\pi^0\pi^0$ read \cite{Bernard:1985wf,Crewther:1985zt,Kambor:1989tz} 
\begin{equation}\label{eq:weakcpt}
   \mathcal{L}_{s\to d} 
   = - \frac{4G_F}{\sqrt2}\,V_{ud}^* V_{us}
    \left( g_8\,\mathcal{O}_8 + g_{27}^{1/2}\,\mathcal{O}_{27}^{1/2} 
    + g_{27}^{3/2}\,\mathcal{O}_{27}^{3/2} \right) ,
\end{equation}
where the effective chiral operators are classified according to their transformation properties under $SU(3)$ and isospin. The $SU(3)$ octet operator $\mathcal{O}_8$ mediates weak transitions with isospin change $\Delta I=\frac12$, while the 27-plet operators $\mathcal{O}_{27}^{1/2}$ and $\mathcal{O}_{27}^{3/2}$ mediates transitions with $\Delta I=\frac12$ and $\Delta I=\frac32$, respectively. These operators can be expressed in terms of products of the left-handed operators $L_\mu^{ji}$ defined in (\ref{eq:Lmu}). One finds
\begin{equation}
\begin{aligned}
   \mathcal{O}_8 
   &= \sum_i\,L_{3i}\spac L_{i2} \,, \\[-2mm]
   \mathcal{O}_{27}^{1/2}
   &= L_{32}\spac L_{11} + L_{31}\spac L_{12} + 2 L_{32}\spac L_{22} - 3 L_{32}\spac L_{33} \,, \\[1mm]
   \mathcal{O}_{27}^{3/2}
   &= L_{32}\spac L_{11} + L_{31}\spac L_{12} - L_{32}\spac L_{22} \,,
\end{aligned}
\end{equation}
where contraction over the Lorentz indices is implied. The coefficient of the octet operator, $|g_8|\approx 5.0$ \cite{Cirigliano:2011ny}, is larger than the coefficient $|g_{27}^{3/2}|$ by about a factor of 30, and in the $SU(3)$ symmetry limit the coefficient $|g_{27}^{1/2}|$ is smaller than $|g_{27}^{3/2}|$ by a factor of 5 \cite{Neubert:1991zd}. The strong dynamical enhancement of $\Delta I=\frac12$ over $\Delta I=\frac32$ transitions is known as the $\Delta I=\frac12$ selection rule, and in our numerical analysis we will only consider the dominant octet contributions to the decay amplitudes. For completeness, the contributions from the two 27-plet operators are collected in Appendix~\ref{app:A}.

We have calculated the $K^-\to\pi^- a$ and $\bar K^0\to\pi^0 a$ decay amplitudes from the Lagrangians (\ref{chiPT}) and (\ref{eq:weakcpt}), evaluating the Feynman graphs shown in Figure~\ref{fig:kaondecays}. The first two diagrams account for the ALP--meson mixing contributions, while the third graph contains the ALP interactions at the weak vertex derived from (\ref{eq:Lmu}). The following two graphs describe ALP emission of an initial or final state meson. They only exist for the case of the charged mesons $K^-$ and $\pi^-$ and give nonzero contributions if the ALP has non-universal vector-current interactions with down and strange quarks. The last diagram contains possible flavor-changing ALP--fermion couplings, as parameterized by the off-diagonal elements of the matrices $\bm{k}_Q$ and $\bm{k}_q$ in (\ref{masssuppression}). The amplitudes for $K\to\pi a$ decays are therefore sensitive to flavor-changing ALP--quark couplings as well as flavor-conserving ALP couplings to gluons and to up, down and strange quarks. To simplify the analysis we set $m_u=m_d\equiv\bar m$ in order to eliminate the $\pi^0$--\spac$\eta$ mass mixing. The meson masses are then given by $m_\pi^2=2B_0\spac\bar m$, $m_K^2=B_0\spac(m_s+\bar m)$, and $3m_\eta^2=4m_K^2-m_\pi^2$. Corrections to the decay amplitudes proportional to the mass difference $(m_u-m_d)$ are suppressed by a factor $1/m_s$ and hence are very small. We also neglect mixing with the $\eta'$ meson, which is an effect of higher order in the chiral expansion. We then obtain \cite{Bauer:2021wjo}
\begin{equation}\label{eq:AKplus}
\begin{aligned}
   i\spac{\cal A}(K^-\to\pi^- a) 
   &= \frac{N_8}{4f}\,\bigg[
    16\spac c_{GG}\,\frac{(m_K^2-m_\pi^2)(m_K^2-m_a^2)}{4m_K^2-m_\pi^2-3m_a^2} 
    + (2c_{uu}+c_{dd}+c_{ss})\,(m_K^2-m_\pi^2) \\
   &\quad - (2c_{uu}+c_{dd}-3c_{ss})\,m_a^2 
    + 6\spac(c_{uu}+c_{dd}-2c_{ss})\,\frac{m_a^2\,(m_K^2-m_a^2)}{4m_K^2-m_\pi^2-3m_a^2} \\
   &\quad + \big( \left[ k_d + k_D \right]_{11} - \left[ k_d + k_D \right]_{22} \big)\,
    (m_K^2+m_\pi^2-m_a^2) \bigg] \\
   &\quad- \frac{m_K^2-m_\pi^2}{2f} \left[ k_d + k_D \right]_{12} ,
\end{aligned}
\end{equation}
and 
\begin{equation}\label{eq:AK0}
\begin{aligned}
   - i\sqrt2\spac{\cal A}(\bar K^0\to\pi^0 a)
   &= \frac{N_8}{4f}\,\bigg[
    16\spac c_{GG}\,\frac{(m_K^2-m_\pi^2)(m_K^2-m_a^2)}{4m_K^2-m_\pi^2-3m_a^2} 
    + (3c_{dd}+c_{ss})\,(m_K^2-m_\pi^2) \\
   &\quad + (2c_{uu}-c_{dd}-c_{ss})\,m_a^2 
    - 2\spac(c_{uu}+c_{dd}-2c_{ss})\,\frac{m_a^2\,(m_K^2-m_\pi^2)}{4m_K^2-m_\pi^2-3m_a^2} \\
   &\quad - 2\spac(c_{uu}-c_{dd})\,\frac{m_a^2\,(m_K^2-m_a^2)}{m_\pi^2-m_a^2} \\
   &\quad + \big( \left[ k_d + k_D \right]_{11} - \left[ k_d + k_D \right]_{22} \big)\,
    (m_K^2+m_\pi^2-m_a^2) \bigg] \\
   &\quad - \frac{m_K^2-m_\pi^2}{2f} \left[ k_d + k_D \right]_{12} ,
\end{aligned}
\end{equation}
where 
\begin{equation}\label{N8def}
   N_8 = - \frac{G_F}{\sqrt2}\,V_{ud}^* V_{us}\,g_8\,f_\pi^2
   \equiv |N_8|\,e^{i\delta_8} \,,  
\end{equation}
with $|N_8|\approx 1.53\times 10^{-7}$. Here $\delta_8$ denotes the strong-interaction phase of the phenomenological parameter $g_8$, and we adopt the standard phase convention for the CKM matrix, in which the matrix elements $V_{ud}$ and $V_{us}$ are real \cite{Zyla:2020zbs}. Note that the flavor-diagonal ALP--fermion couplings $c_{qq}$ in the above relations are evaluated at the low scale $\mu_0\approx 2$\,GeV. 

\subsection{ALP couplings to nucleons in the chiral Lagrangian}
\label{sec:chiralbaryons}

The ALP couplings to nucleons can be derived by extending the effective chiral Lagrangian discussed in the previous section to include baryon fields \cite{Georgi:1985kw,Georgi:1986df,Gasser:1987rb} (see also \cite{Scherer:2002tk} for a more recent review). For the purposes of this discussion we restrict ourselves to the effective theory containing two light quark flavors $u$ and $d$. We describe the nucleons by a spinor field $\psi=(p~n)^T$ containing the proton and the neutron.\footnote{In the extension to three light flavors, the spin-$\frac12$ octet of the ground state baryons is instead described by a traceless $3\times 3$ matrix.} 

In order to describe the interactions of baryons with pions it is convenient to introduce a field $\bm{\xi}(x)$ defined such that $\bm{\xi}^2(x)=\bm{\Sigma}(x)$, where $\bm{\Sigma}(x)=\exp\big[\frac{i\sqrt2}{f_\pi}\,\bm{\sigma}^a\spac\pi^a(x)\big]$. Under an $SU(2)_L\times SU(2)_R$ transformation, the non-linear transformations of the meson fields follow from $\bm{\Sigma}\to\bm{L}\spac\bm{\Sigma}\spac\bm{R}^\dagger$. The quantity $\bm{\xi}$ transforms according to
\begin{equation}
   \bm{\xi} \to \bm{L}\spac\bm{\xi}\,\bm{U}^\dagger 
    = \bm{U}\bm{\xi}\spac\bm{R}^\dagger \,, \qquad
   \bm{\xi}^\dagger \to \bm{R}\,\bm{\xi}^\dagger\spac\bm{U}^\dagger
    = \bm{U}\bm{\xi}^\dagger\bm{L}^\dagger \,.
\end{equation}
This defined the matrix $\bm{U}$ as a non-linear function of $\bm{L}$, $\bm{R}$ and the pion fields. Without loss of generality, one can choose the nucleon field to transform as $\psi\to\bm{U}\psi$. The covariant derivative of the nucleon field takes the form (neglecting electromagnetic interactions for simplicity)
\begin{equation}
   i\bm{D}_\mu\spac\psi
   = i\big( \partial_\mu + \bm{\Gamma}_\mu \big)\spac\psi 
\end{equation}
with the connection
\begin{equation}
\begin{aligned}
   i\spac\bm{\Gamma}_\mu 
   &= \frac12 \left[ \bm{\xi} \left( i\partial_\mu + \frac{\partial_\mu a}{f}\,\hat{\bm{k}}_q \right) \bm{\xi}^\dagger
    + \bm{\xi}^\dagger \left( i\partial_\mu + \frac{\partial_\mu a}{f}\,\hat{\bm{k}}_Q \right) \bm{\xi} \right] \\
   &\equiv \frac12\,\Big[ \bm{\xi} \left( i\partial_\mu + \bm{r}_\mu \right) \bm{\xi}^\dagger
    + \bm{\xi}^\dagger \left( i\partial_\mu + \bm{l}_\mu \right) \bm{\xi} \Big] + v_\mu^{(s)}\,\mathbbm{1} \,, 
\end{aligned}
\end{equation}
where $\hat{\bm{k}}_q=\text{diag}(\hat k_u,\hat k_d)$ and $\hat{\bm{k}}_Q=\text{diag}(\hat k_U,\hat k_D)$ are diagonal matrices containing the modified ALP--quark couplings defined in (\ref{eq:2.37}), restricted to the case of two flavors. In the second step we have defined the iso-vector chiral couplings
\begin{equation}\label{rmulmudef}
\begin{aligned}
  \bm{r}_\mu 
  &= \frac{\partial_\mu a}{f} \left( \frac{\left[k_u-k_d\right]_{11}}{2} 
   + c_{GG}\,\frac{\kappa_u-\kappa_d}{2} \right) \bm{\sigma}^3 \,, \\ 
  \bm{l}_\mu 
  &= \frac{\partial_\mu a}{f} \left( \frac{\left[k_U-k_D\right]_{11}}{2}
   - c_{GG}\,\frac{\kappa_u-\kappa_d}{2} \right) \bm{\sigma}^3 \,, 
\end{aligned}
\end{equation}
and the iso-scalar vector coupling
\begin{equation}
  v_\mu^{(s)} = \frac{\partial_\mu a}{2f} \left( \frac{\left[k_u+k_d\right]_{11}}{2}
   + \frac{\left[k_U+k_D\right]_{11}}{2} \right) , 
\end{equation}
which is invariant under $SU(2)_L\times SU(2)_R$. 

There exist two additional hermitian building blocks called vielbeins \cite{Ecker:1994gg}, which are defined by
\begin{equation}
   \bm{\xi} \left( i\partial_\mu + \frac{\partial_\mu a}{f}\,\hat{\bm{k}}_q \right) \bm{\xi}^\dagger
    - \bm{\xi}^\dagger \left( i\partial_\mu + \frac{\partial_\mu a}{f}\,\hat{\bm{k}}_Q \right) \bm{\xi} 
   = \bm{u}_\mu + \bm{u}_\mu^{(s)} \,, 
\end{equation}
with
\begin{equation}
\begin{aligned}
   \bm{u}_\mu 
   &= \bm{\xi} \left( i\partial_\mu + \bm{r}_\mu \right) \bm{\xi}^\dagger
    - \bm{\xi}^\dagger \left( i\partial_\mu + \bm{l}_\mu \right) \bm{\xi} \,, \\
   \bm{u}_\mu^{(s)} 
   &= \frac{\partial_\mu a}{f} \left[ \frac{\left[k_u+k_d\right]_{11}}{2} - \frac{\left[k_U+k_D\right]_{11}}{2} 
    + c_{GG} \left( \kappa_u + \kappa_d \right) \right] \mathbbm{1}
   \equiv 2 a_\mu^{(s)}\,\mathbbm{1} \,.
\end{aligned}
\end{equation}
These quantities transform as axial vectors under parity. Note that the iso-scalar axial-vector coupling $a_\mu^{(s)}$ is invariant under $SU(2)_L\times SU(2)_R$. The condition $\kappa_u+\kappa_d=1$ ensures that this quantity is independent of the auxiliary parameters $\kappa_q$. A dependence on these parameters remains in the expressions for the chiral couplings $\bm{r}_\mu$ and $\bm{l}_\mu$ in (\ref{rmulmudef}), but it must cancel in all predictions for physical quantities.

\begin{figure}
\begin{center}
\includegraphics[width=0.7\textwidth]{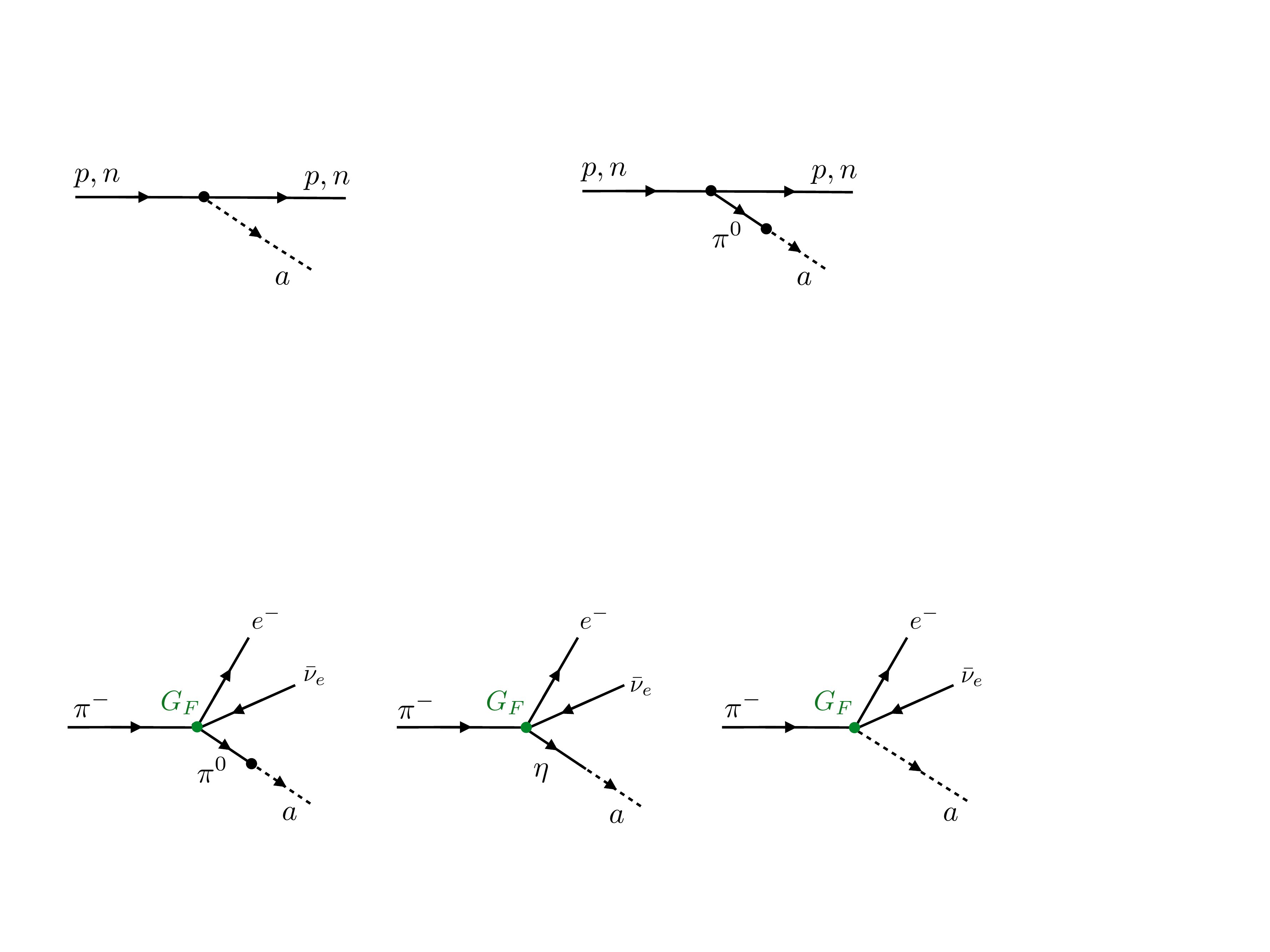}
\caption{\label{pndiagram} 
Diagrams contributing to the effective ALP--nucleon coupling in chiral effective theory.}
\end{center}
\end{figure}

Using these definitions, and working at leading order in the chiral expansion, the most general two-flavor chiral Lagrangian coupling baryons to pions and an ALP can be written in the form 
\begin{equation}\label{eq:ALPPN}
   \mathcal{L}_{\pi N}
   = \bar\psi \left( i\bm{\slashed D} - m_N + \frac{g_A}{2}\,\gamma^\mu\gamma_5\,\bm{u}_\mu
    + \frac{g_0}{2}\,\gamma^\mu\gamma_5\,\bm{u}_\mu^{(s)} \right) \psi \,,
\end{equation}
where $m_N$ is the (leading-order) nucleon mass, and $g_A$ and $g_0$ denote the couplings to external iso-vector and iso-scalar sources. Corrections arising in higher orders of the chiral expansion have been studied in \cite{Vonk:2020zfh} but will be ignored in our discussion. We now derive the effective ALP--nucleon couplings following from the above Lagrangian. As shown in Figure~\ref{pndiagram} there are two diagrams to consider: one in which the ALP is radiated off the nucleon, and one in which the nucleon emits a neutral pion, which then mixes into the ALP. We find that in the sum of the two diagrams the dependence on the auxiliary parameters $\kappa_u$ and $\kappa_d$ cancels, as it should be. We obtain
\begin{equation}\label{ALP-nucleon}
\begin{aligned}
   i\spac\mathcal{A}\big(p(k)\to p(k')+a(q)\big)
   &= - \frac{g_{pa}}{4f}\,\bar{u}_N(k')\spac\rlap/q\spac\gamma_5\,u_N(k) 
    = \frac{m_N\spac g_{pa}}{2f}\,\bar{u}_N(k')\spac\gamma_5\,u_N(k) \,, \\
   i\spac\mathcal{A}\big(n(k)\to n (k')+a(q)\big)
   &= - \frac{g_{na}}{4f}\,\bar{u}_N(k')\spac\rlap/q\spac\gamma_5\,u_N(k) 
    = \frac{m_N\spac g_{na}}{2f}\,\bar{u}_N(k')\spac\gamma_5\,u_N(k) \,,
\end{aligned}
\end{equation}
with 
\begin{equation}\label{gpagna}
\begin{aligned}
   g_{pa} &= g_0 \left( c_{uu} + c_{dd} + 2c_{GG} \right)
    + g_A\,\frac{m_\pi^2}{m_\pi^2-m_a^2}\,\Delta c_{ud} \,, \\
   g_{na} &=g_0 \left( c_{uu} + c_{dd} + 2c_{GG} \right) 
    - g_A\,\frac{m_\pi^2}{m_\pi^2-m_a^2}\,\Delta c_{ud} \,,
\end{aligned}
\end{equation}
where $\Delta c_{ud}$ has been defined in \eqref{eq:cud}. Note that the iso-vector contributions depend in a non-trivial way on the ALP mass, which is an effect not considered in the literature until now. We stress that the expressions on the very right in (\ref{ALP-nucleon}), which are frequently used in the literature on QCD axions, can be misleading, because they seem to suggest that the $N\to N+a$ amplitudes scale with the nucleon mass. This is, however, not the case; rather, the spinor product $\bar{u}_N(k')\spac\gamma_5\,u_N(k)$ scales like $s\cdot(k-k')/m_N$ in the limit where $k'\to k$, where the spin vector $s^\mu$ will be defined below.

The phenomenological coupling $g_A$ can be determined with very good precision from nucleon $\beta$ decay, with the result that $g_A=1.2754(13)$ \cite{Zyla:2020zbs}. In order to determine the parameter $g_0$ we exploit the fact that the ALP--nucleon couplings can also be derived directly from the effective ALP Lagrangians in (\ref{LlowE}) and (\ref{eq:2.19}), without recourse to a chiral effective theory. For the proton one obtains
\begin{equation}
\begin{aligned}
   \mathcal{A}\big(p(k)\to p(k')+a(q)\big)
   &= \sum_q \frac{c_{qq}(\mu_0)}{2f}\,i q_\mu\,
    \langle p(k')|\,\bar q\spac\gamma^\mu\gamma_5\spac q\,|p(k)\rangle_{\mu_0} \\
   &\quad + \frac{c_{GG}}{f}\,\frac{\alpha_s(\mu_0)}{4\pi}\,
    \langle p(k')|\,G_{\mu\nu}^a\,\tilde G^{\mu\nu,a}\,|p(k)\rangle_{\mu_0} \,,
\end{aligned}
\end{equation}
where the sum in the first term runs over the (light) quark flavors, and the hadronic matrix elements are renormalized at the scale $\mu_0$. An analogous expression holds for the neutron. Note that the proton matrix elements of the axial-vector quark currents are scale independent, whereas the matrix element of $G\tilde G$ mixes into the current matrix elements under scale evolution. This mixing is the source of the scale dependence of the parameters $c_{qq}(\mu_0)$. Matching the above expression with (\ref{ALP-nucleon}), we find that 
\begin{equation}
\begin{aligned}
   \left( g_0 + g_A\,\frac{m_\pi^2}{m_\pi^2-q^2} \right) \bar{u}_N(k')\spac\rlap/q\spac\gamma_5\,u_N(k)
   &= 2\,\langle p(k')|\,\bar u\,\rlap/q\spac\gamma_5\spac u\,|p(k)\rangle \,, \\
   \left( g_0 - g_A\,\frac{m_\pi^2}{m_\pi^2-q^2} \right) \bar{u}_N(k')\spac\rlap/q\spac\gamma_5\,u_N(k)
   &= 2\,\langle p(k')|\,\bar d\,\rlap/q\spac\gamma_5\spac d\,|p(k)\rangle \,,
\end{aligned}
\end{equation}
where $q^2=(k-k')^2$. A third relation relates the gluon matrix element of the proton to $g_0$ and $g_A$. Considering the limit $q^\mu\to 0$ (i.e.\ $k'\to k$) in these relations, defining the expectation value of the nucleon spin as
\begin{equation}
   s^\mu \equiv \frac12\,\bar u_N(k)\,\gamma^\mu\spac\gamma_5\,u_N(k) \,; \quad k\cdot s=0 \,,
\end{equation}
and introducing hadronic quantities $\Delta q$ by \cite{Aoki:2019cca}
\begin{equation}
   \langle p(k)|\,\bar q\spac\gamma^\mu\gamma_5\spac q\,|p(k)\rangle
   \equiv 2\spac s^\mu\Delta q \,,
\end{equation}
we obtain
\begin{equation}
   g_0 + g_A = 2\spac\Delta u \,, \qquad
   g_0 - g_A = 2\spac\Delta d \,.
\end{equation}
Solving these equations we obtain $g_0=\Delta u+\Delta d$. The matrix elements $\Delta u$ and $\Delta d$ can be determined using lattice gauge theory (see \cite{Aoki:2019cca} for a comprehensive compilation of relevant results). Since in our analysis the effects of heavy-quark flavors have been integrated out, we use a calculation of the quantities $\Delta q$ in lattice QCD with $N_f=2+1$ dynamical fermions performed by the $\chi$QCD collaboration \cite{Liang:2018pis}, which achieves a pion mass of 171\,MeV close to the physical value. This study reports the values $\Delta u=0.847(18)(32)$ and $\Delta d=-0.407(16)(18)$, from which we obtain $g_0=0.440(44)$. The reported value $g_A=\Delta u-\Delta d=1.254(16)(30)$ is in good agreement with the experimentally determined value quoted above.
\begin{figure}
\begin{center}
\includegraphics[width=.6\textwidth]{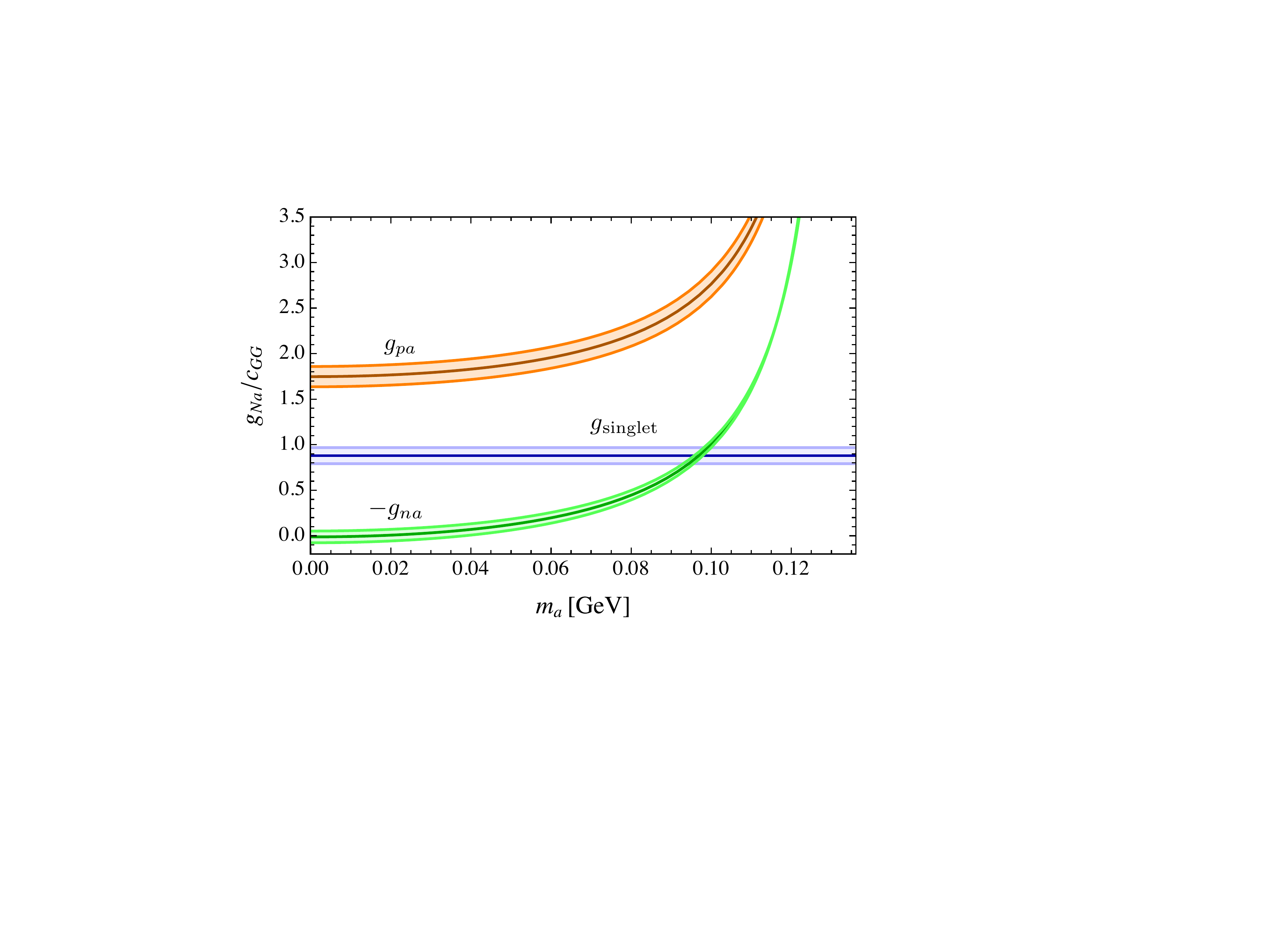}
\caption{\label{gpa} 
Mass dependence of ALP--nucleon couplings for the proton (orange), the neutron (green), and an iso-singlet nucleus with coupling $g\spac_\text{singlet}\equiv(g_{pa}+g_{na})/2$ (blue), in units of the ALP--gluon coupling $c_{GG}$. The ALP couplings to quarks are set to zero ($c_{uu}=c_{dd}=0$).}
\end{center}
\end{figure}

The fact that the effective Lagrangian contains the nucleon mass, which is a large external scale not relevant to chiral dynamics, can be avoided by matching the effective Lagrangian (\ref{eq:ALPPN}) onto a heavy-baryon chiral effective Lagrangian by replacing
\begin{equation}
   \psi(x) \to e^{-i m_N\spac v\cdot x}\,\frac{1+\rlap/v}{2}\,N(x) \,,
\end{equation}
where $v^\mu$ is the 4-velocity of the nucleon. At leading order in the expansion in $1/m_N$, one then obtains from (\ref{eq:ALPPN})
\begin{equation}\label{eq:nonrel}
   \mathcal{L}_{\pi N}
   \to \bar N \left( iv\cdot\bm{D} + g_A\,S\cdot\bm{u} + g_0\,S\cdot\bm{u}^{(s)} \right) N \,,
\end{equation}
where $S^\mu=\frac{i}{2}\spac\sigma^{\mu\nu}\gamma_5\,v_\nu$ denotes the Pauli--Lubanski spin operator.

The effective ALP--nucleon couplings in (\ref{gpagna}) depend on the ALP mass, and the corresponding results for the QCD axion are recovered in the limit $m_a\to 0$. For an ALP with a mass not much smaller than the pion mass, this effect can become relevant, especially in models where the ALP--gluon coupling is much larger than the ALP couplings to the up and down quarks. For the case where $c_{uu}=c_{dd}=0$ at the low scale $\mu_0$, we show in Figure~\ref{gpa} the mass dependence of the effective ALP couplings to the proton, the neutron and an iso-singlet nucleus with equal numbers of protons and neutrons. The mass dependence cancels for iso-singlet nuclei, but can change the ALP interaction strength with non-singlet nuclei significantly. For neutrons the accidental cancellation between the terms proportional to $c_{GG}$ in the second relation in (\ref{gpagna}) is broken by the mass of the ALP.

\subsection{ALP decays into leptons or photons}
\label{sec:decaysleptonsphotons}

For ALP masses below the GeV scale, some of the most important ALP decay modes are those into two charged leptons or two photons. The expressions for the corresponding decay rates have been derived in \cite{Bauer:2017ris}. They are sensitive to the effects of scale evolution and weak-scale matching discussed above. For the leptonic decay modes, one finds
\begin{equation}
   \Gamma(a\to\ell^+\ell^-)
   = \frac{m_a\spac m_\ell^2}{8\pi f^2}\,c_{\ell\ell}^2(m_a)\,\sqrt{1-\frac{4m_\ell^2}{m_a^2}} \,,
\end{equation}
where $\ell=e$, $\mu$ or $\tau$ and we assume that $m_a>2m_\ell$. 

The $a\to\gamma\gamma$ decay rate receives important contributions from loop graphs involving light fermions and gluons and is thus sensitive to strong-interaction effects. If the ALP mass lies far above the QCD scale, then all loop corrections, including those involving colored particles, can be evaluated in perturbation theory. Their contributions can be taken into account by defining an ``effective coupling'' $C_{\gamma\gamma}^{\rm eff}$, such that
\begin{equation}
   \Gamma(a\to\gamma\gamma)
   = \frac{\alpha^2\spac m_a^3}{64\pi^3 f^2} \left| C_{\gamma\gamma}^{\rm eff} \right|^2
\end{equation}
with
\begin{equation}\label{eq:Cgaga}
   C_{\gamma\gamma}^\text{eff}(m_a) 
   = c_{\gamma\gamma} + \sum_{f\ne t}\spac N_c^f Q_f^2\,c_{ff}(m_a)\,B_1(\tau_f) \,; \quad
    \text{for} \quad m_a\gg\Lambda_{\rm QCD} \,.
\end{equation}
Here $\tau_f\equiv 4m_f^2/m_a^2$, and we have defined
\begin{equation}\label{eq:B1}
   B_1(\tau) = 1 - \tau\,f^2(\tau) \,, \qquad
   f(\tau) = \left\{ \begin{array}{ll} 
    \arcsin\frac{1}{\sqrt{\tau}} \,; &~ \tau\ge 1 \,, \\[1mm]
    \frac{\pi}{2} + \frac{i}{2} \ln\frac{1+\sqrt{1-\tau}}{1-\sqrt{1-\tau}} \,; &~ \tau<1 \,.
   \end{array} \right. 
\end{equation} 
This function satisfies $B_1(\tau_f)\approx 1$ for $m_f\ll m_a$ and $B_1(\tau_f)\approx -m_a^2/(12m_f^2)$ for $m_f\gg m_a$, meaning that each electrically charged fermion lighter than the ALP makes a significant contribution to $C_{\gamma\gamma}^\text{eff}$. Note that for the light quarks the running coefficients $c_{ff}(m_q)$ contain important contributions proportional to the ALP--gluon coupling $c_{GG}$ from RG evolution effects. For example, with $m_a=2$\,GeV we find $c_{qq}(m_a)\simeq c_{qq}(\Lambda)\pm 0.1\spac c_{tt}(\Lambda)-0.04\,c_{GG}$, where the plus (minus) sign of the second term refers to down-type (up-type) quarks. 

For ALP masses below the QCD scale this gluon-induced contribution is further enhanced. Hadronic contributions to the effective ALP--photon coupling can be calculated using the effective chiral Lagrangian \eqref{chiPT} and can be expressed as \cite{Georgi:1986df,Bardeen:1986yb, diCortona:2015ldu,Bauer:2017ris}
\begin{equation}\label{eq:CgagaQCD}
\begin{aligned}
   C_{\gamma\gamma}^\text{eff}(m_a)
   &= c_{\gamma\gamma} - (1.92\pm 0.04)\,c_{GG} - \frac{m_a^2}{m_\pi^2-m_a^2} 
    \left[ c_{GG}\,\frac{m_d-m_u}{m_d+m_u} + \frac{c_{uu}-c_{dd}}{2} \right] \\
   &\quad\mbox{}+ \sum_{q=c,b}\spac 3\spac Q_q^2\,c_{qq}(\mu_0)\,B_1(\tau_q)
    + \sum_{\ell=e,\mu,\tau} c_{\ell\ell}\,B_1(\tau_\ell) \,; \quad
    \text{for} \quad m_a<\mu_0 \,,
\end{aligned}
\end{equation}
where we neglect small corrections of order $m_{u,d}/m_s$. In this expression, the running quark masses and the ALP--fermion couplings are evaluated at $\mu_0\approx 2$\,GeV.

\subsubsection*{Off-shell ALP--photon coupling}

\begin{figure}
\begin{center}
\includegraphics[width=0.8\textwidth]{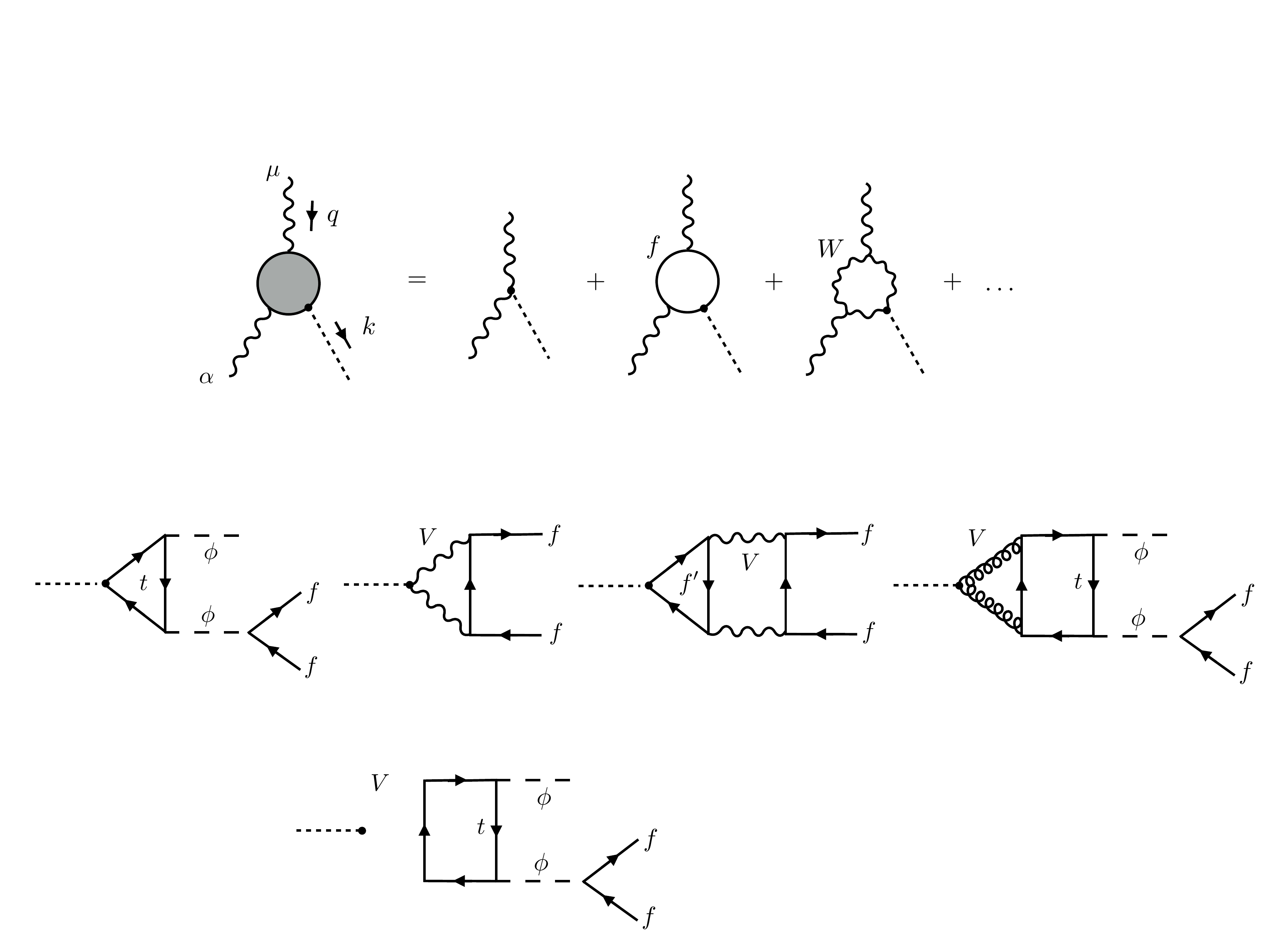}
\vspace{1mm}
\caption{\label{fig:offshell} 
Representative diagrams contributing to the off-shell ALP--photon vertex at one-loop order in ALP interactions.}
\end{center}
\end{figure}

For the discussion of the anomalous magnetic moments of the muon and the electron, it will be useful to define the off-shell ALP--photon vertex function $\Gamma_{\gamma\gamma^* a^*}^{\mu\alpha}(q,k)$ shown in Figure~\ref{fig:offshell}, where $q$ denotes the inflowing momentum of the external on-shell photon ($q^2=0$) with polarization index $\mu$, $k$ is the outflowing momentum of the off-shell ALP ($k^2\ne m_a^2$), and $p=k-q$ is the inflowing momentum of the off-shell photon with polarization index $\alpha$. We have calculated this vertex function at one-loop order in the ALP theory, ignoring the very small contribution of $W$-boson loops (last graph and similar diagrams), which is proportional to two powers of $(\alpha/\pi)$. We find that the vertex function is finite and can be expressed in terms of the parameter integral
\begin{align}
   \Gamma_{\gamma\gamma^* a^*}^{\mu\alpha}(q,k)
   &= \frac{i\alpha}{\pi f}\,\epsilon^{\mu\alpha\beta\gamma}\,q_\beta\,k_\gamma \\
   &\times\left\{ c_{\gamma\gamma} + \sum_f\spac N_c^f\spac Q_f^2\,c_{ff} \left[ 1
    - \int_0^1\!dx \int_0^1\!dy\,
    \frac{m_f^2}{m_f^2-x(1-x) \left( k^2-2y\spac k\cdot q \right) - i\epsilon} 
    \right] \right\} . \notag
\end{align}
The term involving $k\cdot q$ must be kept for flavor off-diagonal dipole transitions such as $\mu\to e\gamma$, but this term can be dropped in the calculation of the anomalous magnetic moments. In this case the integral over $y$ evaluates to~1, and performing the remaining integral over $x$ yields
\begin{equation}
   \Gamma_{\gamma\gamma^* a^*}^{\mu\alpha}(q,k)
   \to \frac{i\alpha}{\pi f}\,\epsilon^{\mu\alpha\beta\gamma}\,q_\beta\,k_\gamma\,
    \bigg[ c_{\gamma\gamma} + \sum_f\spac N_c^f\spac Q_f^2\,c_{ff}\,B_1\bigg(\frac{4m_f^2}{k^2}\bigg) 
    \bigg] \,.
\end{equation}
In the limit where the ALP is taken on-shell ($k^2\to m_a^2$), one recovers the expression for the effective coefficient $C_{\gamma\gamma}^{\rm eff}$ entering the $a\to\gamma\gamma$ decay rate in (\ref{eq:Cgaga}). The off-shell vertex function exhibits the same familiar behavior, that heavy fermions with $m_f^2\gg|k^2|$ decouple from the vertex function, whereas light fermions with $m_f^2\ll|k^2|$ contribute~1 inside the rectangular bracket. 

\subsection{ALP decays into hadrons}
\label{sec:decayshadrons}

If the ALP mass is in the perturbative regime (i.e.\ for $m_a\gg\Lambda_{\rm QCD}$), the inclusive decay rate into light-flavored hadrons can be calculated under the assumption of quark--hadron duality \cite{Poggio:1975af,Shifman:2000jv}. Including the one-loop QCD corrections to the decay rate as calculated in \cite{Spira:1995rr}, one obtains \cite{Bauer:2017ris} 
\begin{equation}\label{Gamma_had}
   \Gamma(a\to\mbox{light-flavored hadrons}) 
   = \frac{\alpha_s^2(m_a)\spac m_a^3}{8\pi^3 f^2}
    \left[ 1 + \frac{83}{4}\,\frac{\alpha_s(m_a)}{\pi} \right] 
    \left| C_{GG}^\text{eff}(m_a)\right|^2 ,
\end{equation}
where 
\begin{equation}
   C_{GG}^\text{eff}(m_a) = c_{GG} + \frac12\spac\sum_{q\ne t}\spac c_{qq}(m_a)\,B_1(\tau_q) \,; \quad
    \text{for} \quad m_a\gg\Lambda_{\rm QCD} \,.
\end{equation}
The decay rate for an ALP into a pair of bottom quarks is given by (working at lowest order in $\alpha_s$, but including the effects of RG evolution)
\begin{equation}
   \Gamma(a\to b\bar b) = \frac{3\spac m_a\spac m_b^2(m_a)}{8\pi f^2}\,
    |c_{bb}(m_a)|^2\,\sqrt{1-\tau_b} \,,
\end{equation}
and an analogous expression holds for $\Gamma(a\to c\bar c)$. 

If the ALP is lighter than 2\,GeV the number of kinematically accessible hadronic decay channels is limited. The two-body decays $a\to\pi\pi$ and $a\to\pi^0\gamma$ are forbidden by parity invariance and angular momentum conservation, and the three-body modes $a\to\pi\pi\gamma$, $a\to\pi^0\gamma\gamma$ and $a\to\pi^0 e^+ e^-$ are strongly suppressed by phase space and powers of the fine-structure constant $\alpha$. The dominant decay modes in this region are $a\to3\pi^0$ and $a\to\pi^+\pi^-\pi^0$ induced by the ALP couplings to pions in the effective chiral Lagrangian \eqref{chiPT}. At leading order in the chiral expansion, one obtains \cite{Bauer:2017ris,Bauer:2020jbp} 
\begin{equation}
   \Gamma(a\to\pi^a\pi^b\pi^0) 
   = \frac{m_a\spac m_\pi^4}{6144\pi^3 f^2f_\pi^2}\,(\Delta c_{ud})^2\, 
    g_{ab}\bigg(\frac{m_\pi^2}{m_a^2}\bigg) \,,
\end{equation}
with $\Delta c_{ud}$ as defined in \eqref{eq:cud}, and (with $0\le r\le 1/9$)
\begin{equation}
\begin{aligned}
   g_{00}(r) &= \frac{2}{(1-r)^2} \int_{4r}^{(1-\sqrt{r})^2}\!\!dz\,\sqrt{1-\frac{4r}{z}}\,
    \lambda^{1/2}(z,r)\,, \\
   g_{+-}(r) &= \frac{12}{(1-r)^2} \int_{4r}^{(1-\sqrt{r})^2}\!\!dz\,\sqrt{1-\frac{4r}{z}}\,(z-r)^2\,
    \lambda^{1/2}(z,r) \,,
\end{aligned}
\end{equation}
where $\lambda(z,r)=(1-z-r)^2-4zr$. Both functions are normalized such that $g_{ab}(0)=1$, and they vanish at the threshold $r=1/9$.

\subsection[$Z$-boson and Higgs decays into ALPs]{$\bm{Z}$-boson and Higgs decays into ALPs}
\label{sec:higgsandZdecays}

The ALP couplings to the top quark and to electroweak gauge bosons can induce exotic decays of $Z$ and Higgs bosons, such as $Z\to\gamma a$, $h\to Za$ and $h\to aa$. In \cite{Bauer:2017ris} the corresponding decay rates were calculated at one-loop order in the effective ALP interactions. Setting the matching scale $\mu_w$ equal to the mass of the decaying particle, one finds \cite{Bauer:2017ris}
\begin{equation}\label{rates}
\begin{aligned}
   \Gamma(Z\to\gamma a) 
   &= \frac{m_Z^3}{96\pi^3 f^2}\,
    \frac{\alpha\spac\alpha(m_Z)}{s_w^2\spac c_w^2}\,\big| C_{\gamma Z}^{\rm eff} \big|^2
    \left( 1 - \frac{m_a^2}{m_Z^2} \right)^3 , \\
   \Gamma(h\to Za) 
   &= \frac{m_h^3}{16\pi f^2}\,\big| C_{Zh}^{\rm eff} \big|^2\,    
    \lambda^{3/2}\bigg(\frac{m_Z^2}{m_h^2},\frac{m_a^2}{m_h^2}\bigg) \,, \\
   \Gamma(h\to aa) 
   &= \frac{m_h^3\,v^2}{32\pi f^4}\,\big| C_{ah}^\text{eff} \big|^2
     \left( 1 - \frac{2m_a^2}{m_h^2} \right)^2 \sqrt{1-\frac{4m_a^2}{m_h^2}} \,.
\end{aligned}
\end{equation}
In the case of $Z\to\gamma a$ decay we have defined
\begin{equation}\label{eq:cgZ}
   C_{\gamma Z}^{\rm eff}
   = c_{\gamma Z} + \sum_f\spac N_c^f\spac Q_f \left( \frac12\,T_3^f - Q_f\spac s_w^2 \right) 
    c_{ff}(m_Z)\,B_3\bigg(\frac{4m_f^2}{m_a^2},\frac{4m_f^2}{m_Z^2}\bigg) \,,
\end{equation}   
where $T_3^f$ denotes the weak isospin of fermion $f$, and 
\begin{equation}
   B_3(\tau_1,\tau_2) = 1 + \frac{\tau_1\spac\tau_2}{\tau_1-\tau_2}
    \left[ f^2(\tau_1) - f^2(\tau_2) \right] ,
\end{equation}
with $f(\tau)$ as given in (\ref{eq:B1}). This function is approximately equal to~1 for all light fermions other than the top quark, for which $B_3\big(\frac{4m_t^2}{m_a^2},\frac{4m_t^2}{m_Z^2}\big)\approx-0.024$ is very small. 

The decay $h\to Za$ is interesting, because the effective ALP Lagrangian (\ref{Leff}) does not contain an interaction that mediates this mode at tree-level. Note, in particular, that the redundant operator involving the Higgs current does not contribute to the decay amplitude, and that a tree-level contribution first arise from the dimension-7 operator \cite{Bauer:2016ydr,Bauer:2016zfj}
\begin{equation}
   {\cal L}_{\rm eff}^{D=7}
   \ni c_\phi^{(7)}\,\frac{\partial^\mu a}{f^3}\,\phi^\dagger\phi\, 
    \big( \phi^\dagger i\hspace{-0.6mm}\overleftrightarrow{D}\hspace{-1mm}_\mu\spac\phi \big) \,.
\end{equation}
The effective coupling $C_{Zh}^{\rm eff}$ is defined as
\begin{equation}\label{Czheff}
   C_{Zh}^{\rm eff} = - \frac{3\spac\alpha_t(m_h)}{4\pi}\,c_{tt}(m_h)\,F
     + \frac{v^2}{2f^2}\,c_\phi^{(7)}(m_h) \,,
\end{equation}
where 
\begin{equation}
   F = \int_0^1\!d[xyz]\,\frac{2m_t^2-x m_h^2-z m_Z^2}{m_t^2-xy m_h^2-yz m_Z^2-xz m_a^2} 
   \approx 0.930 + 2.64\times 10^{-6}\,\frac{m_a^2}{\mbox{GeV}^2} \,,
\end{equation}
with $d[xyz]\equiv dx\,dy\,dz\,\delta(1-x-y-z)$. In (\ref{Czheff}) a loop-suppressed contribution competes with a power-suppressed term, and which of the two dominates depends on the relative size of the Wilson coefficients $c_{tt}$ and $c_\phi^{(7)}$ and on the value of the ratio $v/f$.

The decay amplitude for the process $h\to aa$ starts at ${\cal O}(1/f^2)$. It needs two insertions of ALP vertices from the effective Lagrangian (\ref{Leff}). At the same order, there is a potential contribution from the dimension-6 operator
\begin{equation}
   \mathcal{L}_\text{eff}^{D=6}
   \ni \frac{c_{ah}}{f^2}\,(\partial_\mu a)\spac(\partial^\mu a)\,\phi^\dagger\phi \,.
\end{equation}
The coefficient $C_{ah}^{\rm eff}$ reads
\begin{equation}\label{eq:cah}
\begin{aligned}
   C_{ah}^{\rm eff} 
   &= c_{ah}(m_h) + \frac{3\spac\alpha_t(m_h)}{\pi}\,c_{tt}^2(m_h) 
    \left[ \ln\frac{m_h^2}{m_t^2} - g_1\bigg(\frac{4m_t^2}{m_h^2}\bigg) \right] \\
   &\quad - \frac{3\spac\alpha^3(m_h)}{32\pi^3 s_w^6}\,\tilde c_{WW}^2(m_h)
    \left[ \ln\frac{m_h^2}{m_W^2} + \delta_1 - g_2\bigg(\frac{4m_W^2}{m_h^2}\bigg) \right] \\
   &\quad - \frac{3\spac\alpha^3(m_h)}{64\pi^3 s_w^6\spac c_w^6}\,\tilde c_{ZZ}^2(m_h)
    \left[ \ln\frac{m_h^2}{m_Z^2} + \delta_1 - g_2\bigg(\frac{4m_Z^2}{m_h^2}\bigg) \right] ,
\end{aligned}
\end{equation}
with the loop functions 
\begin{equation}
\begin{aligned}
   g_1(\tau) &= \tau\spac f^2(\tau) + 2\sqrt{\tau-1}\,f(\tau) - 2 \,, \\
   g_2(\tau) &= \frac{2\tau}{3}\,f^2(\tau) + 2\sqrt{\tau-1}\,f(\tau) - \frac83 \,.
\end{aligned}
\end{equation}
The parameter $\delta_1$ in (\ref{eq:cah}) is a scheme-dependent constant related to the treatment of the Levi--Civita symbol in $D=4-2\epsilon$ spacetime dimensions. One finds $\delta_1=-\frac{11}{3}$ is a scheme where $\epsilon^{\mu\nu\alpha\beta}$ is treated as a $D$-dimensional object (our default choice), and $\delta_1=0$ is a scheme where it is treated as a four-dimensional quantity \cite{Bauer:2017ris}. The dimension-6 Wilson coefficient $c_{ah}(m_h)$ in the above expression must be evaluated at the weak scale. The RG evolution equation for this coefficient has not yet been derived in the literature. At lowest logarithmic order, one finds that 
\begin{equation}
   c_{ah}(m_h) 
   = c_{ah}(\Lambda) + \left[ \frac{3\spac\alpha_t(m_h)}{\pi}\,c_{tt}^2(m_h) 
    - \frac{3\spac\alpha^3(m_h)}{32\pi^3 s_w^6}\,\tilde c_{WW}^2(m_h)
    - \frac{3\spac\alpha^3(m_h)}{64\pi^3 s_w^6\spac c_w^6}\,\tilde c_{ZZ}^2(m_h) \right]
    \ln\frac{\Lambda^2}{m_h^2} \,.
\end{equation}

\clearpage

%%%%%%%%%%%%%%%%%%%%%%%%%%%%%%%%%%%%%%%%%%%%%%%%%%%%%%%%%%%%%%%%%%%%%%%%%%%%%%%
\section{Probes of flavor-changing ALP couplings to quarks}
\label{sec:quarks}

The focus of this section is on deriving experimental constraints on the ALP couplings from observables sensitive to flavor-changing interactions in the quark sector, as well as discussing possible ALP explanations for experimental anomalies. 
In the first part of this section we derive general predictions for a number of observables in terms of the elements of the hermitian coupling matrices $\bm{k}_U$, $\bm{k}_D$, $\bm{k}_u$ and $\bm{k}_d$ \eqref{Lferm}, the flavor-diagonal ALP--fermion couplings $c_{ff}$ defined in \eqref{cffdef}, and the ALP--boson couplings $c_{VV}$ defined in \eqref{LeffmuW}. 
We distinguish processes in which ALPs are produced on-shell, processes with virtual ALP exchange and processes that do not involve any flavor change.  
The most sensitive probes of flavor-violating ALP couplings are rare meson decays into mono-energetic final state mesons and ALPs produced on-shell, such as $K\to \pi a$.
We begin by deriving constraints from an extensive list of experimental searches for ALPs in exotic meson decays and show these constraints for various ALP decay modes and lifetimes. This is followed by a discussion of the impact of flavor symmetries of the UV theory on these observables. For the case of minimal flavor violation and universal ALP couplings we give the RG induced flavor off-diagonal ALP couplings explicitly 
 and explain how the constraints depend on the ALP couplings at the scale $\Lambda=4\pi f$.  We proceed with observables sensitive to virtual ALP exchange which lead to weaker constraints because the ALP contribution to the amplitude is suppressed by $v/f$ compared to on-shell ALP decays. 
For observables that are not sensitive to flavor-changing ALP couplings at all, such as vector meson decays $V \to\gamma a$ and the chromomagnetic moment of the top quark we derive the relevant expressions for the ALP contributions and discuss their dependence on the ALP couplings at the scale $\Lambda$. The observables that we use, their measured values, and SM predictions, are collected in tables in Appendix~\ref{app:measurements}.

In the second part of this section we study in detail eight benchmark scenarios based on a theory with flavor-universal ALP couplings in the UV in which 
any flavor violation arises from loop corrections involving SM particles, as described in Section~\ref{subsec:LagbelowEW}. Each benchmark is defined by assuming that either one of the ALP couplings to gauge bosons $c_{GG}, c_{WW}$ and $c_{BB}$ or a single ALP flavor universal coupling to SM fermions $\bm{c_\psi}=c_\psi \mathbbm{1}$ with $\psi= u, d, Q, e, E$ is non-zero at the scale $\Lambda$. 
For the case of a coupling to left-handed down type quarks, our assumption of flavor universality is, in fact, stronger than the hypothesis of minimal flavor violation, which would allow for flavor off-diagonal couplings at the new physics scale $\Lambda$. For these benchmarks we compare constraints from processes in which ALPs are produced on-shell, processes sensitive to virtual ALP exchange and flavor-conserving processes for a range of ALP masses. We further compare these constraints with astrophysical observables and the reach of collider searches for rare $Z$ and Higgs boson decays into ALPs. 
In the last part of this section we use these results to explore the viable parameter space for a possible explanation of experimental anomalies observed in lepton flavor non-universality in rare $B$ meson decays, in nuclear Beryllium and Helium transitions and in the decay rate of the neutral pion $\pi^0\to e^+e^-$.

There has been a lot of recent work studying the constraints on ALPs from quark flavor-changing processes (see e.g.,\ \cite{Batell:2009jf,Dolan:2014ska,Izaguirre:2016dfi,Arias-Aragon:2017eww,Choi:2017gpf,Dobrich:2018jyi,Gavela:2019wzg,Ishida:2020oxl,Gori:2020xvq,MartinCamalich:2020dfe,Chakraborty:2021wda,Carmona:2021seb,Bertholet:2021hjl}), and it is worth outlining what our current work adds to these studies: 

\begin{itemize}
\item 
The ALP can have macroscopic decay lengths, which can critically affect the sensitivity of many of the most important flavor constraints that rely on the ALP being produced on-shell and decaying promptly (or conversely escaping the detector altogether). We provide individual plots for each such constraint (Figures~\ref{fig:kd12plots}, \ref{fig:kd13plots}, and \ref{fig:kd23plots}) showing explicitly the dependence of the constraint on the ALP decay width, within the plane of the ALP mass and the relevant flavor-changing coupling. To calculate these dependences we account for the specifics of the particular experimental setup, and event selection criteria such as kinematic cuts. These ALP width effects are also taken careful account of when plotting the constraints on simplified scenarios in Section~\ref{sec:discussion1}.
\item
We study in general the contributions of a light ALP to $B_{d,s}$\,--\,$\bar B_{d,s}$ mixing, including RG evolution effects and subleading terms in the heavy-quark expansion, which turn out to be non-negligible.
\item 
We derive bounds on the ALP from recent measurements of $K\to \pi$ observables using our recent calculation of the $K\to\pi a$ decay amplitudes in chiral perturbation theory \cite{Bauer:2021wjo}. We take into account one-loop running and matching contributions from the high scale $\Lambda$ down to the chiral scale $\sim 2$\,GeV, and we find that for most scenarios involving flavor-universal ALP couplings at the scale $\Lambda$, the neutral-current flavor-changing ALP couplings induced by RG evolution and weak-scale matching produce a larger effect than the SM weak interactions.
\item 
While we calculate observables in a fully general way in terms of ALP couplings at the scale of the measurements, we also interpret constraints in terms of $SU(2)_L\times U(1)_Y$ invariant couplings defined at the UV scale $\Lambda$, taking account of the dominant RG evolution and matching contributions down to the scale of relevance for the processes considered. This allows us to see at a glance the flavor constraints in various simplified scenarios, and compare them directly with constraints from measurements performed at different energy scales (e.g.,\ LHC and LEP measurements, beam dumps, astrophysical constraints, etc). When plotting bounds on these simplified scenarios, we take into account all decay modes of the ALP, and calculate effects from finite lifetimes. For some ranges of ALP mass, the strongest constraints arise from observables which occur only at loop-level in both the production and decay of the ALP. To give an example, we find that searches for the rare decay $B\to K^{(*)} a (\mu \mu)$ at LHCb provide the strongest constraints on a $1$\,GeV ALP which at the scale $\Lambda$ couples only to right handed up-type quarks.
\item 
Since we take the approach of relating flavor effects back to fundamental ALP couplings at the scale $\Lambda$, observables involving purely flavor-conserving quark couplings can constrain some of the same parameter spaces as flavor-changing observables, and we compare these different types of constraints on the same axes. In this spirit, we also calculate in this section the contributions of the ALP to the chromomagnetic dipole moment of the top quark, and to radiative $J/\psi$ and $\Upsilon$ decays, which we calculate to one-loop order in QCD.
\item 
We consider the possibility of an ALP explanation of some intriguing experimental anomalies, including the observation of lepton non-universality in $b\to s \ell\ell$ decays as measured at LHCb \cite{Aaij:2017vbb,Aaij:2021vac}, discrepancies from SM expectations in excited Beryllium and Helium transitions measured by the ATOMKI collaboration \cite{Krasznahorkay:2018snd,Krasznahorkay:2019lyl}, and an excess in the branching ratio of $\pi^0\to e^+ e^-$ measured at the KTeV experiment \cite{Abouzaid:2006kk}. When confronted with constraints from other measurements, we find that an ALP could explain the deviation in the low-$q^2$ bin of the $R_K$ observable, but not the high-$q^2$ bin of $R_K$ and not $R_{K^*}$. We find that an ALP could in principle provide a joint explanation of the Beryllium and Helium transitions measured by ATOMKI, however, the relevant parameter space is already ruled out by $K\to \pi a$ searches. We show a small viable region of parameter space that could explain the Helium transition. Furthermore, we show that an ALP with couplings to electrons as well as quarks or gluons could explain the KTeV anomaly. 
\end{itemize}

\subsection{ALP production in exotic two-body decays of mesons}
\label{sec:ExoticK}

The most promising decay processes for the discovery of ALPs are those in which the ALP is produced as an on-shell resonance. Indeed, some of the strongest constraints on the couplings of a light ALP can be derived from exotic two-body decays of pseudoscalar mesons, such as $K\to\pi a$, $D\to\pi a$, $B\to\pi a$ etc. We will discuss this in detail for the kaon decays $K^-\to\pi^- a$ and $K_L\to\pi^0 a$, illustrating our general approach for analyzing ALP effects on flavor observables. The extension to other decay modes is then straightforward.

The key signature of $K\to\pi a$ decays is a mono-energetic final state pion with energy
\begin{equation}
   E_{\pi} = \frac{m_{K}^2+m_{\pi}^2-m_a^2}{2m_{K}}
\end{equation} 
in the kaon rest frame. The decay rates for the charged and neutral kaon decays are given by 
\begin{equation}\label{Kpiarates}
   \Gamma(K\to\pi a)
   = \frac{1}{16\pi\spac m_K} \left| \mathcal{A}(K\to\pi a) \right|^2 
    \lambda^{1/2}\bigg(\frac{m_\pi^2}{m_K^2},\frac{m_a^2}{m_K^2}\bigg) \,, 
\end{equation}
where 
\begin{equation}\label{eq:PhaseSpaceLambda}
   \lambda(r_i,r_j) = 1 + r_i^2 + r_j^2 - 2r_i - 2r_j - 2r_i\spac r_j \,.
\end{equation}
As discussed in Section~\ref{sec:chiral}, there are contributions to the decay amplitudes involving both flavor-violating and flavor-conserving ALP couplings. The decay amplitude for the charged mode $K^-\to\pi^- a$ and the neutral mode $\bar K^0\to\pi^0 a$ have been given in \eqref{eq:AKplus} and \eqref{eq:AK0}, respectively, in terms of the ALP mass and the ALP couplings to gluons and quarks. For $m_a=0$, one finds numerically 
\begin{align}\label{Kpianumerics}
   i\mathcal{A}(K^-\to\pi^- a)
   &\simeq - 1.12\times 10^{-4}\,\text{GeV} \left[ \frac{1\,\text{TeV}}{f} \right] \big[k_d+k_D\big]_{12} \notag\\
   &\hspace{-3.1cm} + 10^{-11}\,\text{GeV} \left[ \frac{1\,\text{TeV}}{f} \right] e^{i\delta_8}\spac 
    \Big[ 3.50\,c_{GG} + 0.86\,(2c_{uu}+c_{dd}+c_{ss}) \notag\\
   &\hspace{1.63cm} + 1.01\,\big( \left[ k_d + k_D \right]_{11} - \left[ k_d + k_D \right]_{22} \big) \Big] \,, \\
   -i\sqrt2\,\mathcal{A}(\bar K^0\to\pi^0 a)
   &\simeq - 1.15\times 10^{-4}\,\text{GeV} \left[ \frac{1\,\text{TeV}}{f} \right] \big[k_d+k_D\big]_{12} \notag\\
   &\hspace{-3.1cm}  + 10^{-11}\,\text{GeV} \left[ \frac{1\,\text{TeV}}{f} \right] e^{i\delta_8}\spac 
    \Big[ 3.58\,c_{GG} + 0.88\,(3 c_{dd}+c_{ss}) 
    + 1.02\,\big( \left[ k_d + k_D \right]_{11} - \left[ k_d + k_D \right]_{22} \big) \Big] \spac, \notag
\end{align}
where the ALP couplings are defined at the low scale $\mu_0=2$\,GeV. Note, however, that the flavor-changing ALP--fermion couplings do not run below the weak scale. For different values of the ALP mass the coefficients change, but the general pattern remains the same. The amplitudes for the CP-conjugate decay modes $K^+\to\pi^+ a$ and $K^0\to\pi^0 a$ can be obtained from these expressions by reversing the overall sign and replacing $\left[k_d+k_D\right]_{12}\to\left[k_d+k_D\right]_{21}=\left[k_d+k_D\right]_{12}^*$. (One should also take the complex conjugate of the product $V_{ud}^* V_{us}$ of CKM matrix elements in the definition of the quantity $N_8$ in \eqref{N8def}, which has no effect since these parameters are real in the standard convention for the CKM matrix.) The amplitude for the decay $K_L\to\pi^0 a$, on which constraints can be derived using existing searches for $K_L\to\pi^0\nu\bar\nu$ and $K_L\to\pi^0 X$, is then obtained using the relation~\cite{Buras:1998raa}
\begin{equation}
   K_L = \frac{(1+\epsilon)\spac K^0 + (1-\epsilon)\spac\bar K^0}{\sqrt{2\spac(1+|\epsilon|^2)}}\,,
\end{equation}
where $\epsilon=2.228(11)\times 10^{-3}\,e^{i\phi_\epsilon}$ with $\phi_\epsilon\approx 43.5^\circ$ is the parameter measuring CP violation in $K^0$\,--\,$\bar K^0$ mixing \cite{Zyla:2020zbs}. We find numerically
\begin{equation}\label{eq:KLpi0a}
\begin{aligned}
   i\mathcal{A}(K_L\to\pi^0 a)
   &\simeq 10^{-4}\,\text{GeV} \left[ \frac{1\,\text{TeV}}{f} \right] 
    \Big[ 1.15\spac i\,\text{Im}\spac\big[k_d+k_D\big]_{12} 
    - 2.56\times 10^{-3}\,e^{i\phi_\epsilon}\,\text{Re}\spac\big[k_d+k_D\big]_{12} \Big] \\
   &\hspace{-2.7cm} + 10^{-14}\,\text{GeV} \left[ \frac{1\,\text{TeV}}{f} \right] e^{i(\delta_8+\phi_\epsilon)}\spac 
    \Big[ 7.97\,c_{GG} + 1.95\,(3c_{dd}+c_{ss}) 
    + 2.27\spac\big(\! \left[ k_d + k_D \right]_{11} - \left[ k_d + k_D \right]_{22} \!\big) \Big] \spac.
\end{aligned}
\end{equation}
A comparison with the result for the charged mode in \eqref{Kpianumerics} shows that the decay $K_L\to\pi^0 a$ is useful primarily for probing the imaginary part of the flavor-changing ALP--fermion coupling $\left[k_d+k_D\right]_{12}$. The sensitivity to all other ALP couplings is reduced, compared to the decay $K^-\to\pi^- a$, by a factor $|\epsilon|\approx 2\times 10^{-3}$.

When one squares the decay amplitudes to obtain the decay rates in \eqref{Kpiarates}, the interference terms involving one flavor-changing and one flavor-diagonal ALP coupling are sensitive to the strong-interaction phase $\delta_8$, which cannot be calculated reliably. In practice this is not a limitation, because the two types of terms come with coefficients that differ by many orders of magnitude. It would require a strong fine tuning to zoom in on a region of parameter space where the interference terms would matter phenomenologically.

The above results show that searches for the exotic $K\to\pi a$ decay modes can constrain the flavor off-diagonal ALP couplings $\left[k_d\right]_{12}$ and $\left[k_D\right]_{12}$ with seven orders of magnitude higher sensitivity compared with the flavor-conserving ALP couplings to quarks and gluons. The reason is that FCNC processes in the SM are loop and GIM suppressed, whereas they can arise at tree-level if the ALP has flavor-changing couplings to quarks. In our analysis in this section we therefore exclusively focus on the bounds derived on the flavor-changing ALP couplings, finding that these are in general very strong if the corresponding decays are kinematically allowed. 

Similarly to ALP production in kaon decays, an ALP can be produced by decays of $B$ and $D$ mesons together with pions or kaons. In terms of the flavor-changing ALP couplings, we find the decay rates
\begin{align}\label{eq:mesondecays}
   \Gamma(B^-\to\pi^- a)
   &= \frac{m_B^3}{64\pi f^2} \left| \left[k_D+k_d\right]_{13} \right|^2 
    \left| F_0^{B\to \pi}(m_a^2) \right|^2 \left( 1 - \frac{m_\pi^2}{m_B^2} \right)^2 
    \lambda^{1/2}\bigg(\frac{m_\pi^2}{m_B^2},\frac{m_a^2}{m_B^2}\bigg) \,, \notag\\
   \Gamma(\bar B^0\to\pi^0 a)
   &= \frac12\,\Gamma(B^-\to\pi^- a) \,, \notag\\
   \Gamma(B^-\to K^- a)
   &= \frac{m_B^3}{64\pi f^2} \left| \left[k_D+k_d\right]_{23} \right|^2 
    \left| F_0^{B\to K}(m_a^2) \right|^2 \left( 1 - \frac{m_K^2}{m_B^2} \right)^2 
    \lambda^{1/2}\bigg(\frac{m_K^2}{m_B^2},\frac{m_a^2}{m_B^2}\bigg) \,, \notag\\
   \Gamma(B^-\to K^{*-} a)
   &= \frac{m_B^3}{64\pi f^2} \left| \left[k_D-k_d\right]_{23} \right|^2 
    \left| A_0^{B\to K^*}(m_a^2) \right|^2 
    \lambda^{3/2}\bigg(\frac{m_K^{*2}}{m_B^2},\frac{m_a^2}{m_B^2}\bigg) \,, \\[1mm]
   \Gamma(\bar B^0\to\bar K^{(*)0}\spac a)
   &= \Gamma(B^-\to K^{(*)-} a) \,, \notag\\
   \Gamma(D^+\to\pi^+ a)
   &= \frac{m_D^3}{64\pi f^2} \left| \left[k_U+k_u\right]_{12} \right|^2 
    \left| F_0^{D\to \pi}(m_a^2) \right|^2 \left( 1 - \frac{m_\pi^2}{m_D^2} \right)^2 
    \lambda^{1/2}\bigg(\frac{m_\pi^2}{m_D^2},\frac{m_a^2}{m_D^2}\bigg) \,, \notag\\
   \Gamma(D^0\to\pi^0 a)
   &= \frac12\,\Gamma(D^+\to\pi^+ a) \,, \notag\\
   \Gamma(D_s^+\to K^+ a)
   &= \frac{m_{D_s}^3}{64\pi f^2} \left| \left[k_U+k_u\right]_{12} \right|^2 
    \left| F_0^{D_s\to K}(m_a^2) \right|^2 \left( 1 - \frac{m_K^2}{m_{D_s}^2} \right)^2 
    \lambda^{1/2}\bigg(\frac{m_K^2}{m_{D_s}^2},\frac{m_a^2}{m_{D_s}^2}\bigg) \,. \notag
\end{align}
For $B\to K^* a$ decays the $K^*$ meson is longitudinally polarized, since the ALP is a pseudoscalar particle. The quantities $F_0(q^2)$ and $A_0(q^2)$ are scalar form factors defined in \cite{Wirbel:1985ji}. We take $F^{B\to K}_0(q^2)$ and $F^{B\to \pi}_0(q^2)$ from the lattice averages of \cite{Aoki:2019cca}, $A^{B\to K^*}_0(q^2)$ from the light-cone QCD sum-rule calculation of \cite{Straub:2015ica}, $F_0^{D\to\pi}(q^2)$ from the lattice calculation of \cite{Lubicz:2017syv}, and $F_0^{D_s\to K}(q^2)$ from the covariant light-front calculation of \cite{Wang:2008ci}. 

In the above expressions for the decay rates we focus only on the contributions to the decay amplitudes mediated by the flavor-changing ALP-quark couplings. In all cases these couplings are renormalized at the scale of the measurement, but because the flavor-changing ALP couplings do not run below the weak scale, it is equivalent to use couplings renormalized at the weak scale. Contributions involving the SM weak interactions in combination with flavor-conserving ALP couplings are only included to the extent that these contribute to the flavor-changing ALP couplings at low energies, see (\ref{eq:FVveryshort}). This is justified by the observation we made for $K\to\pi a$ decays, that contributions to the amplitude involving the SM weak transition $s\to u\spac\bar u\spac d$ are strongly suppressed. We expect a similar statement to hold for the decays of heavy $B$ and $D$ mesons. For example, we expect that subprocesses of the type $B^-\to\pi^-\pi^0\to\pi^- a$ via ALP--pion mixing give rise to subdominant contributions to the $B^-\to\pi^- a$ rate. It would be interesting to work out such effects in detail in future work, for instance using the framework of QCD factorization for non-leptonic $B$ decays \cite{Beneke:1999br,Beneke:2001ev}.

\begin{table}[!htbp]
\scalebox{0.6}{
\begin{tabular}{l c c c c c c}
\toprule
Observable & Mass range [MeV] & ALP decay mode & Constrained & Limit (95\% CL) on 
 & Limit (95\% CL) on & Figure\\
 &  &  & coupling $c_{ij}$ & $c_{ij}\cdot \left(\frac{\text{TeV}}{f}\right)\cdot \sqrt{\mathcal{B}}$
 & $c_{ij}/|V_{ti}^* V_{tj}|\cdot \left(\frac{\text{TeV}}{f}\right)\cdot \sqrt{\mathcal{B}}$ &\\
\midrule
$\text{Br}(K^-\to\pi^-a (\text{inv}))$ & $0<m_a<261^{\,(\ast)}$ & long-lived & $|k_D+k_d|_{12}$
 & $1.2\times 10^{-9}$ & $3.9\times 10^{-6}$ & \ref{fig:kd12plots} a)\\
 $\text{Br}(K_L\to\pi^0a (\text{inv}))$ & $0<m_a<261$ & long-lived & $|\text{Im}[[k_D+k_d]_{12}|$
 & $8.1\times 10^{-9}$ & $7.0\times 10^{-5}$ & \ref{fig:kd12plots} b)\\
 $\text{Br}(K^-\to \pi^-\gamma\gamma)$ & $m_a<108$ & $\gamma\gamma$ & $|k_D+k_d|_{12}$
 & $2.1\times 10^{-8}$ & $6.9\times 10^{-5}$ & \ref{fig:kd12plots} c)\\
$\text{Br} (K^-\to\pi^-\gamma\gamma)$ & $220<m_a<354$ & $\gamma\gamma$ & $|k_D+k_d|_{12}$
 & $2.0\times 10^{-7}$ & $6.5\times 10^{-4}$ & \ref{fig:kd12plots} d)\\
$\text{Br}(K_L\to\pi^0\gamma\gamma)$ & $m_a<110$ & $\gamma\gamma$ & $|\text{Im}[[k_D+k_d]_{12}]|$ 
 & $1.3\times 10^{-8}$ & $1.1\times 10^{-4}$ & \ref{fig:kd12plots} e)\\
$\text{Br}(K_L\to\pi^0\gamma\gamma)$ & $m_a<363^{(\maltese\maltese)}$ & $\gamma\gamma$ & $|\text{Im}[[k_D+k_d]_{12}]|$ 
 & $1.3\times 10^{-7}$ & $1.1\times 10^{-3}$ & \ref{fig:kd12plots} f)\\
 $\text{Br}(K^+\to\pi^+ a(e^+e^-))$ & $1<m_a<100$ & $e^+e^-$ & $|k_D+k_d|_{12}$ 
 & $3.4\times 10^{-7}$ & $1.1\times 10^{-3}$ &\ref{fig:kd12plots} g) \\
$\text{Br}(K_L\to\pi^0 e^+e^-)$ & $140<m_a<362$ & $e^+e^-$ & $|\text{Im}[[k_D+k_d]_{12}]|$ 
 & $3.1\times 10^{-9}$ & $2.6\times 10^{-5}$ &\ref{fig:kd12plots} h) \\
 $\text{Br}(K_L\to\pi^0\mu^+\mu^-)$ & $210<m_a<350$ & $\mu^+\mu^-$ & $|\text{Im}[[k_D+k_d]_{12}]|$
 & $4.0\times 10^{-9}$ & $3.4\times 10^{-5}$ &\ref{fig:kd12plots} i)\\
 \midrule
 $\text{Br}(B^+ \to \pi^+ e^+ e^-)$ & $140<m_a<5140$ & $e^+e^-$ & $|k_D+k_d|_{13}$ &$7.0 \times 10^{-7}$ & $8.7\times 10^{-5}$& \ref{fig:kd13plots} a)\\
 $\text{Br}(B^+ \to \pi^+ \mu^+ \mu^-)$ & $211<m_a <5140^{\,(\ddag\ddag)}$ & $\mu^+\mu^-$ & $|k_D+k_d|_{13}$ &$1.2 \times 10^{-7}$ & $1.4\times 10^{-5}$& \ref{fig:kd13plots} b)\\
 \midrule
$\text{Br}(B^-\to K^-\nu\bar\nu)$ & $0<m_a<4785$ & long-lived & $|k_D+k_d|_{23}$ 
 & $6.2\times 10^{-6}$ & $1.6\times 10^{-4}$ & \ref{fig:kd23plots} a)\\
$\text{Br}(B\to K^*\nu\bar\nu)$ & $0<m_a<4387$ & long-lived & $|k_D-k_d|_{23}$
 & $4.1\times 10^{-6}$ & $1.1\times 10^{-4}$ & \ref{fig:kd23plots} b)\\
$d\text{Br}/dq^2(B^0\to K^{*0} e^+e^-)_{[0.0,0.05]}$ & $1<m_a<224$ & $e^+e^-$ & $|k_D-k_d|_{23}$
 & $6.4\times 10^{-7}$ & $1.6\times 10^{-5}$& \ref{fig:kd23plots} c) \\
$d\text{Br}/dq^2(B^0\to K^{*0} e^+e^-)_{[0.05,0.15]}$ & $224<m_a<387$ & $e^+e^-$ & $|k_D-k_d|_{23}$
 & $9.3\times 10^{-7}$ & $2.4\times 10^{-5}$ & \ref{fig:kd23plots} d)\\
 $\text{Br}\big(B^-\to K^-\,a(\mu^+\mu^-)\big)$ & $250<m_a<4700^{\,(\dagger)}$ & $\mu^+\mu^-$
 & $|k_D+k_d|_{23}$ & $4.4\times 10^{-8}$ & $1.1\times 10^{-6}$ & \ref{fig:kd23plots} e)\\
$\text{Br}\big(B^0\to K^{*0}\,a(\mu^+\mu^-)\big)$ & $214<m_a<4350^{\,(\dagger)}$ & $\mu^+\mu^-$
 & $|k_D-k_d|_{23}$ & $5.1\times 10^{-8}$ & $1.3\times 10^{-6}$ & \ref{fig:kd23plots} f)\\
 $\text{Br}(B^-\to K^-\tau^+\tau^-)$ & $3552<m_a<4785$ & $\tau^+\tau^-$ & $|k_D+k_d|_{23}$
 & $8.2\times 10^{-5}$ & $2.1\times 10^{-3}$ & \ref{fig:kd23plots} g)\\
 \midrule
$\text{Br}(D^0\to \pi^0 e^+ e^-)$ & $1<m_a< 1730^{(\ddag)}$ & $e^+e^-$ &$|k_U+k_u|_{12}$ & $2.8 \times 10^{-5}$ & $-$ & \ref{fig:ku12plots} a)\\
$\text{Br}(D^+\to \pi^+ e^+ e^-)$ & $200<m_a< 1730^{(\dagger\dagger)}$ & $e^+e^-$ &$|k_U+k_u|_{12}$ & $8.4 \times 10^{-6}$ & $-$& \ref{fig:ku12plots} b)\\
$\text{Br}(D_s^+ \to K^+ e^+ e^-)$ & $200<m_a< 1475^{(\maltese)}$ & $e^+e^-$ &$|k_U+k_u|_{12}$ &$2.4 \times 10^{-5}$ & $-$ & \ref{fig:ku12plots} c)\\
$\text{Br}(D^+\to \pi^+ \mu^+ \mu^-)$ & $250 < m_a < 1730^{(**)}$  & $\mu^+\mu^-$&$|k_U+k_u|_{12}$ & $2.1 \times 10^{-6}$ & $-$ & \ref{fig:ku12plots} d)\\
$\text{Br}(D_s^+ \to K^+ \mu^+ \mu^-)$ & $200 < m_a < 1475^{(***)}$ & $\mu^+\mu^-$ &$|k_U+k_u|_{12}$ & $5.7 \times 10^{-5}$ &$-$ & \ref{fig:ku12plots} e)\\
\bottomrule
\end{tabular}
}
\caption{Summary of indicative constraints on quark flavor-violating ALP couplings renormalized at the scale $\mu_w=m_t$, derived from measurements of branching fractions (first column) for various decays of kaons and $B$ mesons in a mass range where an on-shell ALP can be produced. The relevant measurements and SM predictions (where appropriate) are given in~\Cref{tab:longlivedobs,tab:gamgamobs,tab:electronicobs,tab:muonicobs,tab:tauonicobs,tab:hadronicobs} in Appendix~\ref{app:measurements}. In each line, the limit cited is the strongest limit 
found within the mass range probed by the measurement. In the sixth and seventh columns the symbol ${\cal B}$ denotes the ALP branching ratio to the relevant final state, while in the seventh column the constraints are divided by $|V_{ti}^* V_{tj}|$ as an estimate of the strength of the bounds on the MFV case (since this is only relevant for left-handed down type couplings, the up-type decays are not included in this column). The final column refers to figures showing the dependence of the bound on the ALP mass and lifetime. Asterisks next to the mass range mean that cuts are applied within the mass range to exclude resonance regions, and therefore the corresponding measurement is insensitive to an ALP with mass in the excluded ranges. The excluded regions are as follows. $(\ast)$: $100<m_{\nu \bar{\nu}}<161$ MeV; $(\ast\ast)$: $525<m_{\mu\mu}< 1250$ MeV; ($\ast$$\ast$$\ast$): $990<m_{\mu\mu}< 1050$ MeV; $(\maltese)$: $950<m_{ee}< 1050$ MeV; $(\maltese\maltese)$: $100<m_{\gamma\gamma}<160$ MeV; $(\ddag)$: $935<m_{ee}< 1053$ MeV; $(\ddag\ddag)$: $8.0<m_{\mu\mu}^2<11.0 \,\text{GeV}^2$ and $12.5<m_{\mu\mu}^2<15.0 \,\text{GeV}^2$; $(\dagger)$: various cuts are applied to exclude the regions around the $J/\psi$, $\psi(2S)$ and $\psi(3370)$ resonances; $(\dagger\dagger)$: $525<m_{\mu\mu}< 1250$ MeV.
}
\label{tab:hadronbounds}
\end{table}

\begin{figure}
\centering
\includegraphics[width=1.\textwidth]{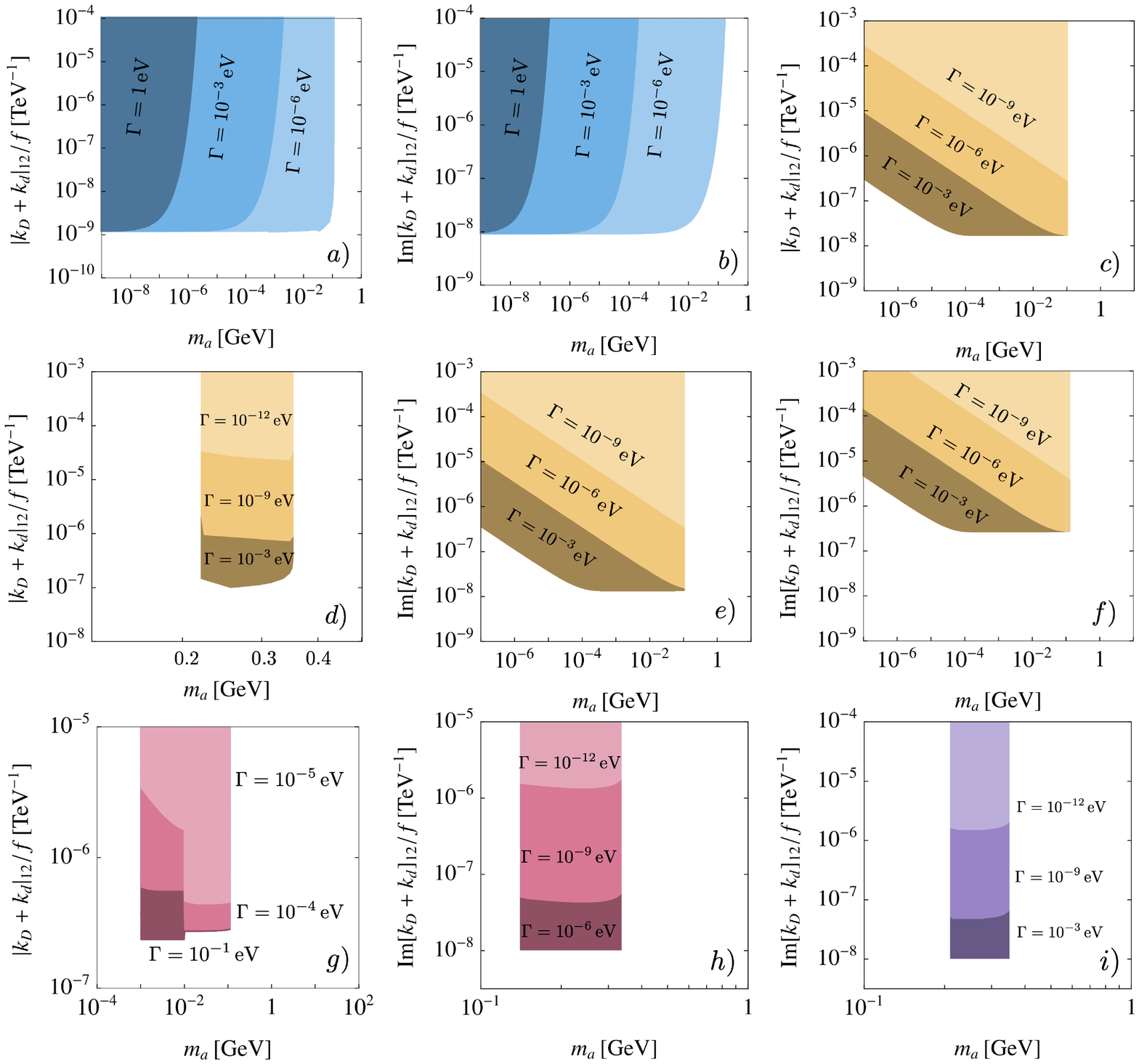}
\caption{\label{fig:kd12plots} Constraints on the flavor-violating ALP couplings $[k_d]_{12}$ and $[k_D]_{12}$ from kaon decays, collected in Table~\ref{tab:hadronbounds}, for different values of the total ALP width. The observables are  $\text{Br}(K^-\to\pi^-\nu\bar\nu)$ (top left and center), $\text{Br}(K^-\to\pi^-\gamma\gamma)$  (top right and middle left), $\text{Br}(K_L^0\to\pi^0\gamma\gamma)$ (middle center and right), $\text{Br}(K_L\to \pi^{0}e^+e^-)$ (bottom left and center) and  $\text{Br}(K_L\to \pi^{0}\mu^+\mu^-)$ (bottom right). }
\end{figure}

\begin{figure}
\centering
\includegraphics[width=0.7\textwidth]{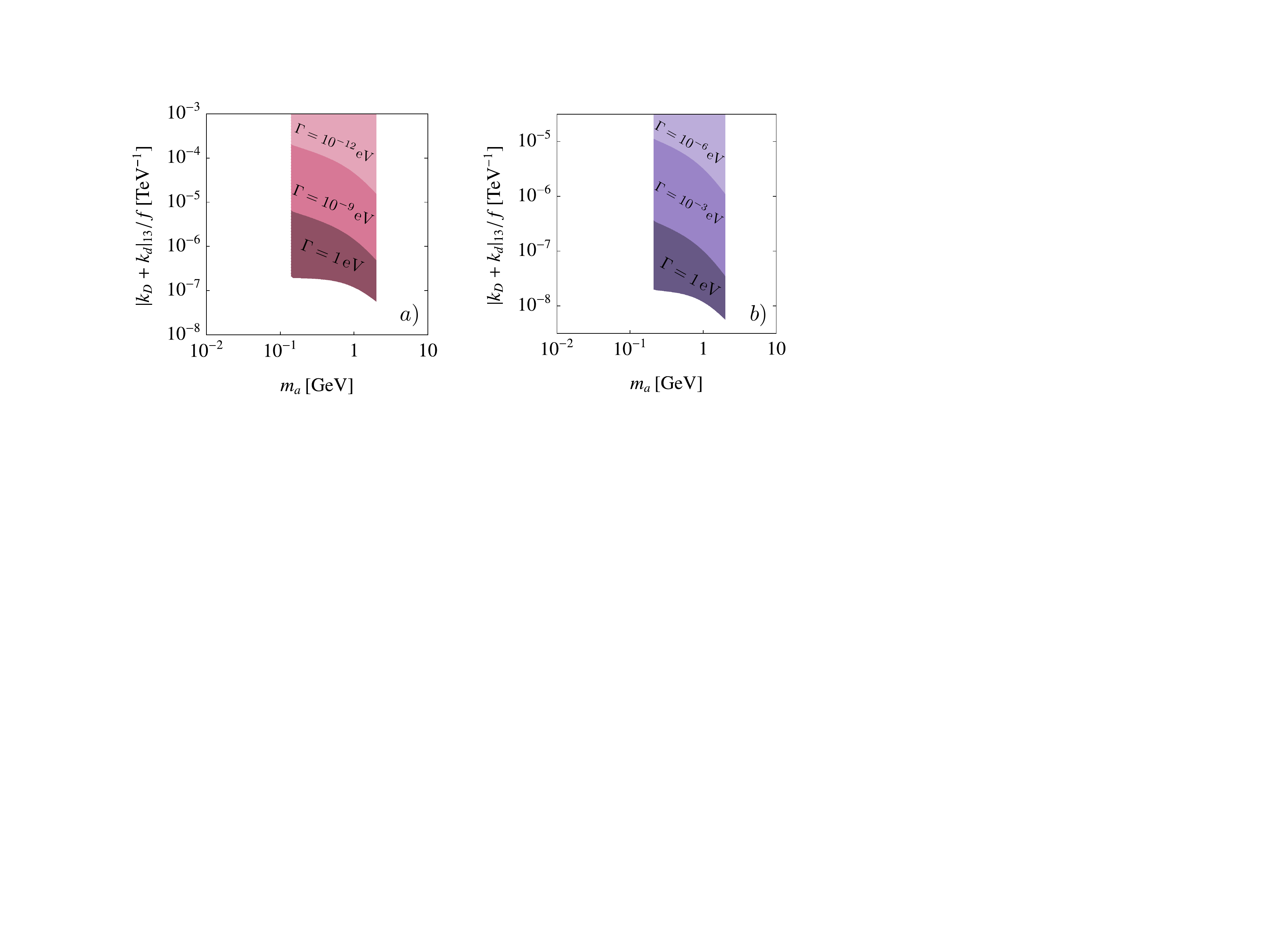}
\caption{\label{fig:kd13plots} Constraints on the flavor-violating ALP couplings $[k_d]_{13}$ and $[k_D]_{13}$ from $B$ meson decays, collected in Table~\ref{tab:hadronbounds}, for different values of the total ALP width. The observables are $\text{Br}(B^+\to\pi^+e^+e^-)$ (left) and $\text{Br}(B^+\to\pi^+\mu^+\mu^-)$ (right).}
\end{figure}

\begin{figure}
\centering
\includegraphics[width=1.\textwidth]{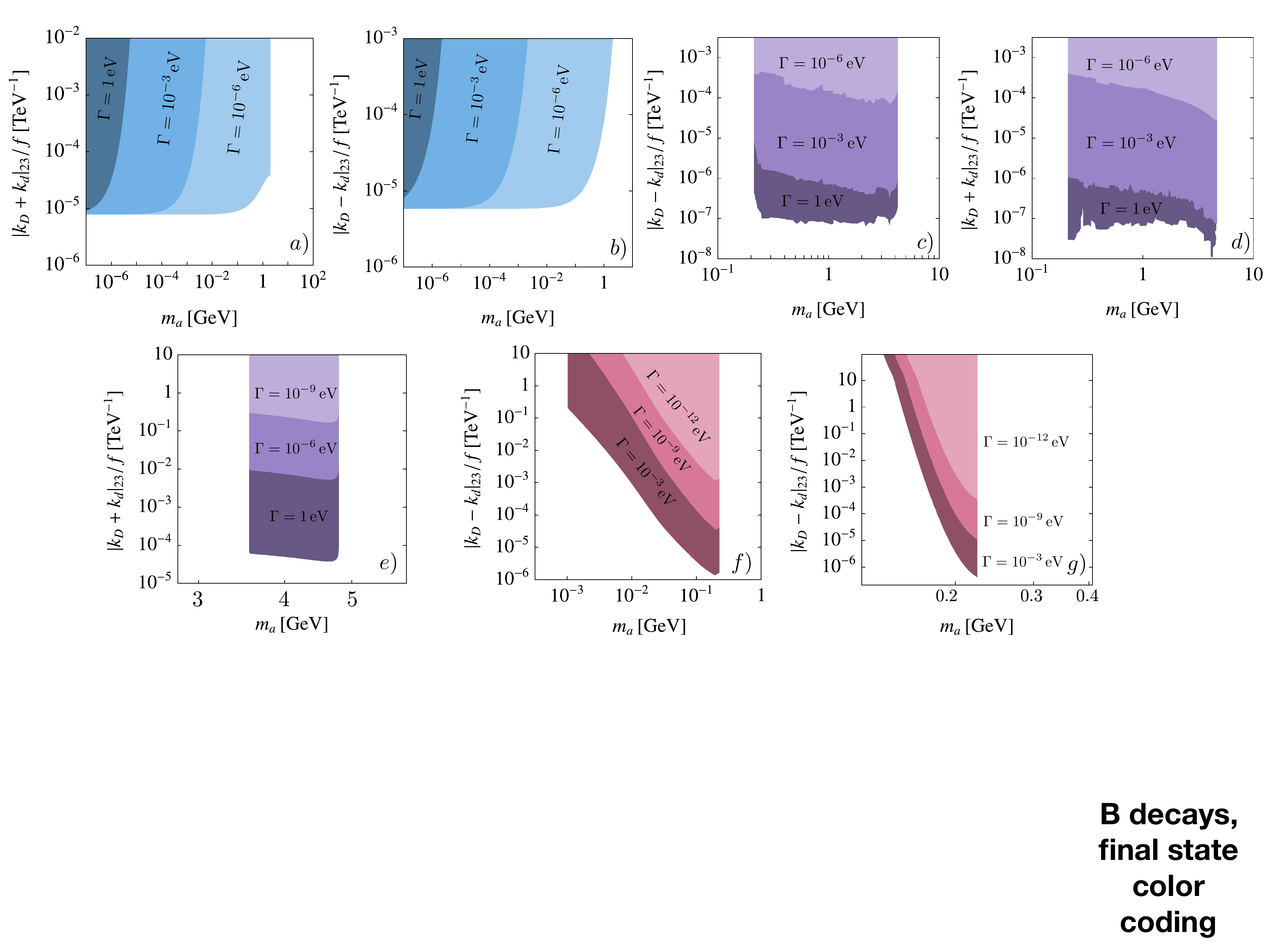}
\caption{\label{fig:kd23plots} Constraints on the flavor-violating ALP couplings  $[k_d]_{23}$ and $[k_D]_{23}$ from $B$ meson decays, collected in Table~\ref{tab:hadronbounds}, for different values of the total ALP width. The observables are  $\text{Br}(B^-\to K^-\nu\bar\nu)$ (top left), $\text{Br}(B\to K^*\nu\bar\nu)$ (top center left), $\text{Br}(B^-\to K^- a (\mu^+\mu^-))$ (top center right), $\text{Br}(B^0\to K^{*,0} a (\mu^+\mu^-))$ (top right), $\text{Br}(B^-\to K^{-} \tau^+\tau^-)$ (bottom left), $d\text{Br}/dq^2(B^0\to K^{*0} e^+e^-)_{[0,0.05]}$ (bottom center) and $d\text{Br}/dq^2(B^0\to K^{*0} e^+e^-)_{[0.05,0.15]}$ (bottom right).}
\end{figure}

\begin{figure}[t]
\centering
\includegraphics[width=.85\textwidth]{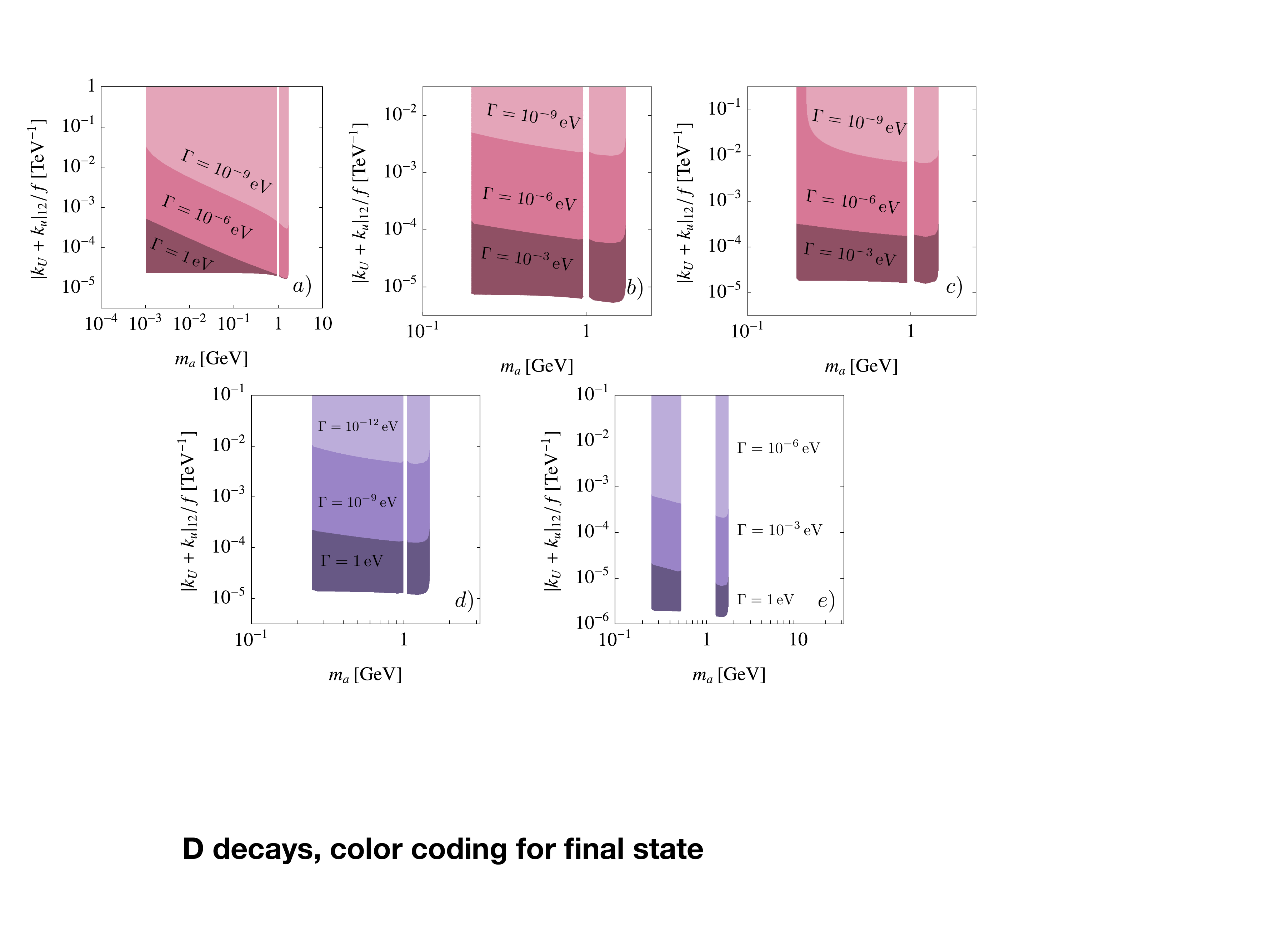}
\caption{\label{fig:ku12plots} Constraints on the flavor-violating ALP couplings $[k_u]_{12}$ and $[k_U]_{12}$ from D meson decays, collected in Table~\ref{tab:hadronbounds}, for different values of the total ALP width. The observables are  $\text{Br}(D^0\to\pi^0 e^+e^-)$ (top left), 
$\text{Br}(D^+\to\pi^+ e^+e^-)$ (top center), 
$\text{Br}(D^+_s\to K^+ e^+e^-)$ (top right),  $\text{Br}(D^+\to\pi^+\mu^+\mu^-)$ (bottom left) and  $\text{Br}(D_s^+\to K^+\mu^+\mu^-)$ (bottom right).  }
\end{figure}

While ALP production in two-body meson decays provides a particularly sensitive probe of flavor-changing ALP couplings, the phenomenology of these processes depends very sensitively on the ALP lifetime (i.e., on whether the ALP decays promptly in the detector, has a macroscopic decay length or is long lived) and the branching fractions for the various ALP decay modes, such as $a\to\gamma\gamma$ and $a\to\ell^+\ell^-$. Constraints on a long-lived ALP can be derived from searches for rare decays such as $K\to\pi\nu\bar\nu$ and $B\to K^{(*)}\nu\bar\nu$, whereas bounds on a short-lived ALP can be obtained by recasting searches for meson decays into a final state meson accompanied by a pair of photons or leptons, or by dedicated searches for new light resonances in the final state. An extensive list of experimental searches and the respective limits on the ALP couplings $[k_d]_{ij}$ and $[k_D]_{ij}$ with $ij=12,13,23$ and $[k_u]_{12}$ and $[k_U]_{12}$ from exotic decays of kaons, $B$ mesons and $D_{(s)}$ mesons are compiled in Table~\ref{tab:hadronbounds}. For ALPs with ${\cal O}(1)$ flavor off-diagonal couplings, these searches can probe new physics scales of $f\lesssim 10^9\,\text{TeV}\times\sqrt{\mathcal{B}}$. The constraints scale with the square root of the branching ratio of the ALP into the signal final state, $\mathcal{B}=\text{Br}(a\to\gamma\gamma, e^+e^-, \mu^+\mu^-, \tau^+\tau^-)$, which depend on the ALP mass and couplings. This underlines the importance of searches for resonances in different decay channels, even though they probe the same flavor off-diagonal ALP couplings. Further, the lifetime of the ALP $\tau_a$ changes the strength of the constraints shown in Table~\ref{tab:hadronbounds}, because the fraction of ALPs decaying in the range of sensitivity of the experiment depends on the ALP decay length 
\begin{equation}
   \ell_a = c\spac\tau_a = \frac{\hbar\spac c}{\Gamma} 
   \simeq 0.197\spac\mu\text{m}\,\frac{1\,\text{eV}}{\Gamma} \,, 
\end{equation}
where $\Gamma$ is the total decay width of the ALP. The effect of the ALP lifetime depends on the experimental setup, because of the boost of the initial state meson used in the experiment. Details about the lifetime effects and experimental parameters can be found in Appendix~\ref{appendix:decaylengths}. For the limits in Table~\ref{tab:hadronbounds}, we show the effect of a finite ALP lifetime in Figure~\ref{fig:kd12plots} (kaon decays), Figure~\ref{fig:kd13plots} and Figure~\ref{fig:kd23plots} ($B$-meson decays), and Figure~\ref{fig:ku12plots} ($D$-meson decays). The color coding distinguishes constraints on invisible ALPs (blue), ALPs decaying into photons (yellow), electrons (red), and muons or taus (purple). In all cases, lighter colors correspond to smaller decay widths and darker colors to larger decay widths. We show constraints for three different values of the ALP width $\Gamma$ for each experimental constraint in Table~\ref{tab:hadronbounds} and we assume a branching ratio of $100\%$ for ALPs decaying into the respective final state for the purpose of these plots. Searches for missing energy from long-lived ALPs are most sensitive if the ALP has a small decay width, since the fraction of ALPs which escape the detector is suppressed by $\text{exp}(-m_a\spac\Gamma)$. ALPs with a decay width of $\Gamma>10^{-6}$\,eV and a corresponding lifetime of $\ell_a\gtrsim 0.2$\,m are effectively stable on detector scales for all experiments and ALP masses considered. If the ALP width is larger, only ALPs with smaller masses are constrained by searches for missing-energy signatures. In the plots in Figure~\ref{fig:kd12plots}a), \ref{fig:kd12plots}b) and Figure~\ref{fig:kd23plots}a), \ref{fig:kd23plots}b) the blue areas therefore always extend towards ALPs with smaller masses. If the ALP width is larger the fraction of ALPs escaping the detector decreases and searches for missing energy constrain only smaller ALP masses.
In the case of searches for ALPs decaying into photons or leptons, the dark shaded regions corresponding to shorter lifetimes lead to the most stringent constraints. A smaller ALP mass requires a larger ALP width for the ALPs to decay inside the detector. This explains the slope towards lighter ALP masses and smaller flavor-violating couplings in the plots corresponding to decays with visible final states. For resonant ALPs decaying into leptons, the range of constrained ALP masses is limited either by experimental cuts or the kinematic window $2m_\ell\le m_a\le m_{M_1}-m_{M_2}$ for a meson decay $M_1\to M_2\spac a$. Complementary experimental searches for ALPs with macroscopic decay lengths and ALPs that decay promptly are important to fully constrain the parameter space.  

\subsection{Flavor symmetries in the UV theory}\label{sec:UVsymmetry}

The excessively strong bounds shown in Table~\ref{tab:hadronbounds} can be relaxed very efficiently by imposing a flavor symmetry. If the UV theory respects minimal flavor violation (MFV), the flavor-changing ALP couplings to left-handed down-type quarks satisfy $\left[k_D\right]_{ij}^{\rm MFV}\propto V_{ti}^* V_{tj}$ (with $i\ne j$) \cite{Bauer:2020jbp}, whereas all other flavor-changing ALP couplings vanish. An explicit expression of the couplings $\left[k_D\right]_{ij}$ in terms of ALP parameters defined at the new physics scale $\Lambda=4\pi f$ has been given in (\ref{eq:FVveryshort}). The relevant CKM suppression factors for the various transitions are $|V_{td}^* V_{ts}|\approx 3.1\times 10^{-4}$ for $s\to d$ transitions, $|V_{td}^* V_{tb}|\approx 8.0\times 10^{-3}$ for $b\to d$ transitions, and $|V_{ts}^* V_{tb}|\approx 3.9\times 10^{-2}$ for $b\to s$ transitions. The constraints from on-shell ALP production in ALP models respecting MFV are shown in the last column of Table~\ref{tab:hadronbounds}. Flavor-changing transitions in the up-quark sector are forbidden altogether in MFV scenarios (in the approximation where the Yukawa couplings of the down-type quarks are set to zero). The suppression factor is most efficient for the decays $K\to\pi a$, where it reduces the relevant flavor-changing ALP coupling $\left[k_D\right]_{12}$ by almost four orders of magnitude. A further suppression by about two orders of magnitude can be achieved if one assumes that the flavor-changing couplings are induced at one-loop order. Assuming MFV, the flavor-diagonal ALP--quark couplings satisfy the relations \cite{Bauer:2020jbp}
\begin{equation}
   c_{ss} = c_{dd} \,, \qquad
   \left[ k_d \right]_{11} - \left[ k_d \right]_{22}
   = \left[ k_D \right]_{11} - \left[ k_D \right]_{22} = 0 \,,
\end{equation}
which further simplify the expressions for the $K\to\pi a$ decay amplitudes given in (\ref{Kpianumerics}). In the $K_L\to\pi^0 a$ decay rate one primarily probes the imaginary part of the flavor-changing ALP coupling, see (\ref{eq:KLpi0a}). With the standard phase convention for the CKM matrix, this coupling satisfies 
\begin{equation}
   \big[k_D\big]_{12}^{\rm MFV}\propto V_{td}^* V_{ts}
   = - |V_{td}^* V_{ts}|\,e^{i\beta} \left[ 1 + {\cal O}(\lambda^5) \right] ,
\end{equation}
where $\beta$ is one of the angles of the unitarity triangle (with $\sin 2\beta\approx 0.70$), and $\lambda\approx 0.2$ denotes the Wolfenstein parameter. We observe that this coupling has a sizable CP-violating phase, and hence its imaginary part does not receive a further suppression beyond that of the CKM matrix elements.

For the discussion of our benchmark scenarios in Section~\ref{sec:discussion1}, we go one step beyond the MFV hypothesis and assume a {\em flavor-universal ALP\/} at the new physics scale $\Lambda$, for which all ALP--fermion coupling matrices $\bm{c}_F\equiv c_F\spac\mathbbm{1}$ (with $F=u,d,Q,e,L$) are proportional to the unit matrix. It is then useful to express the ALP couplings defined at the low scale in terms of the couplings at the scale $\Lambda$. Throughout our analysis we use $f=1$\,TeV as a reference scale (corresponding to a new physics scale $\Lambda\approx 12.6$\,TeV), unless indicated otherwise. We can then express all coupling parameters in the decay amplitudes in terms of the three ALP--boson couplings and the five ALP--fermion couplings at the scale $\Lambda$. For the flavor-changing couplings, we obtain from (\ref{eq:FVveryshort})
\begin{align}\label{eq:Knum}
   \left[ k_D(m_t) \right]_{ij}^{\rm univ}
   &\simeq 10^{-5}\,V_{ti}^* V_{tj}\,\Big[ - 6.1\,c_{GG} - 2.8\,c_{WW} - 0.02\,c_{BB} \\
   &\quad + 1.9\times 10^3\,c_u(\Lambda) - 9.2\,c_d(\Lambda) -1.9\times 10^3\,c_Q(\Lambda)  
    - 0.05\,c_e(\Lambda) + 4.2\,c_L(\Lambda) \Big] \,. \notag
\end{align}
This result shows the minimal amount of low-scale flavor violation present in any ALP model, and it clearly demonstrates a key observation of our analysis: even a single non-zero ALP coupling at the new physics scale $\Lambda$ will unavoidably lead to flavor-changing ALP--quark couplings below the weak scale, irrespective of whether or not the UV theory is flavor blind. For the flavor-conserving ALP couplings, we find from \eqref{eq:cffrun} and \eqref{eq:lowrun} 
\begin{equation}\label{eq:cuucdd}
\begin{aligned}
   \left[ c_{uu}(\mu_0) \right]^{\rm univ}
   &\simeq 0.84\,c_u(\Lambda) - 0.049\,c_d(\Lambda) - 0.79\,c_Q(\Lambda) - 0.037\,c_{GG} \\
   &\quad - 10^{-4}\,\Big[\, 1.0\,c_e(\Lambda) - 3.7\,c_L(\Lambda) + 2.1\,c_{WW} + 0.34\,c_{BB} \Big] \,, \\
   \left[ c_{dd}(\mu_0) \right]^{\rm univ}
   &\simeq 0.074\,c_u(\Lambda) + 0.95\,c_d(\Lambda) - 1.02\,c_Q(\Lambda) - 0.037\,c_{GG} \\
   &\quad - 10^{-4}\,\Big[\, 0.30\,c_e(\Lambda) 
    - 3.6\,c_L(\Lambda) + 2.3\,c_{WW} + 0.10\,c_{BB} \Big] \,.
\end{aligned}
\end{equation}

Let us return one last time to the $K\to\pi a$ decay amplitudes, now in the context of a flavor-universal ALP model. Expressing all ALP couplings in terms of the eight parameters in the UV Lagrangian, we find for the $K^-\to\pi^- a$ decay amplitude evaluated for $m_a=0$ 
\begin{align}\label{eq:KAamp2}
   & i\mathcal{A}(K^-\to\pi^- a)
   = 10^{-11}\,\text{GeV} \left[ \frac{1\,\text{TeV}}{f} \right] \notag\\
   &\quad\times \bigg\{ e^{i\beta}\,\Big[ - 0.21\,c_{GG} - 0.10\,c_{WW} - 6.4\times 10^{-4}\,c_{BB} 
    + 67\spac c_u(\Lambda) \notag\\
   &\hspace{2.0cm} - 0.32\,c_d(\Lambda) - 66\,c_Q(\Lambda) - 1.9\times 10^{-3}\,c_e(\Lambda) 
    + 0.15\,c_L(\Lambda) \Big] \\
   &\hspace{0.65cm} + e^{i\delta_8}\,\Big[\, 3.4\,c_{GG} - 7.5\times 10^{-4}\,c_{WW} - 7.5\times 10^{-5}\,c_{BB} 
    + 1.6\,c_u(\Lambda) \notag\\
   &\hspace{2.0cm} + 1.5\,c_d(\Lambda) - 3.1\,c_Q(\Lambda) - 2.2\times 10^{-4}\,c_e(\Lambda) 
    + 1.2\times 10^{-3}\,c_L(\Lambda) \Big] \bigg\} \,, \notag
\end{align}
which makes it explicit that the coefficients in the contribution associated with flavor-changing ALP couplings (terms proportional to $e^{i\beta}$) are now more or less commensurate with the coefficients in the contribution to the amplitude mediated by the weak transition $s\to u\spac\bar u\spac d$ of the SM (terms proportional to $e^{i\delta_8}$). For an ALP coupling only to gluons or right-handed down-type quarks at the new physics scale the main contributions arise via the SM weak interactions, while in all other scenarios the dominant contributions arise via the RG-induced flavor-violating ALP coupling $\left[k_D\right]_{12}$ in the low-energy theory.

For the $K_L\to\pi^0 a$ decay amplitude, we obtain in the flavor-universal ALP scenario
\begin{align}\label{eq:KLAamp2}
   & i\mathcal{A}(K_L\to\pi^0 a)
   = 10^{-11}\,\text{GeV} \left[ \frac{1\,\text{TeV}}{f} \right] \notag\\
   &\quad\times \bigg\{ i\spac e^{i\xi_\epsilon}\,\Big[\, 
     0.083\,c_{GG} + 0.037\,c_{WW} + 2.5\times 10^{-4}\,c_{BB} 
    - 26\,c_u(\Lambda) \notag\\
   &\hspace{2.4cm} + 0.12\,c_d(\Lambda) + 26\,c_Q(\Lambda) + 7.4\times 10^{-4}\,c_e(\Lambda) 
    - 0.056\,c_L(\Lambda) \Big] \\
   &\hspace{0.95cm} + e^{i(\delta_8+\phi_\epsilon)}\,\Big[ 
    7.7\times 10^{-3}\,c_{GG} - 1.8\times 10^{-6}\,c_{WW} - 7.8\times 10^{-8}\,c_{BB} 
    + 5.8\times 10^{-4}\,c_u(\Lambda) \notag\\
   &\hspace{2.4cm} + 7.4\times 10^{-3}\,c_d(\Lambda) - 8.0\times 10^{-3}\,c_Q(\Lambda) - 2.4\times 10^{-7}\,c_e(\Lambda) 
    + 2.8\times 10^{-6}\,c_L(\Lambda) \Big] \bigg\} \,, \notag
\end{align}
where $\xi_\epsilon\approx-0.226^\circ$. In this case the contribution shown in the last two lines, which arises from the diagrams in Figure~\ref{fig:kaondecays} in which the flavor-changing transition is mediated by the effective weak Lagrangian of the SM, gives rise to subdominant contributions for all eight ALP couplings. The rather different dependence of the two amplitudes on the ALP couplings would be of great help in the case of a discovery. For instance, an ALP coupling only to gluons at the scale $\Lambda$ would give a 40 times larger contribution to the $K^-\to\pi^- a$ amplitude than to the $K_L\to\pi^0 a$ amplitude, whereas for an ALP coupling only to right-handed down-type quarks the ratio of the two amplitudes would be 13. If any of the other ALP couplings is dominant at the scale $\Lambda$, then the $K^-\to\pi^- a$ amplitude is about 2.6 times larger than the $K_L\to\pi^0 a$ amplitude.

\subsection[Three-body pion decay $\pi^-\to a\spac e^-\bar\nu_e$]{\boldmath Three-body pion decays $\pi^-\to a\spac e^-\bar\nu_e$}
\label{subsec:Pienu}

ALPs can also be discovered in exotic three-body decays of mesons. Leptonic decays of charged mesons mediated by the weak force are particularly interesting, because they are insensitive to the flavor-violating ALP couplings and thus can be used to probe the ALP couplings to gluons and light quarks. In particular, the charged pion decay $\pi^-\to a\spac e^-\bar\nu_e$ can be used to search for ALPs with masses $m_a<m_\pi-m_e$. The amplitude for this decay is given in \eqref{eq:21}. Neglecting contributions suppressed by $m_e^2/(m_\pi^2-m_a^2)$, one finds the decay rate 
\begin{equation}
   \Gamma(\pi^-\to a\spac e^-\bar\nu_e)
   = \frac{G_F^2\left|V_{ud}\right|^2}{24576\pi^3}\,\frac{f_\pi^2}{f^2}\,m_\pi^5\,g(x_a)
    \left[ 2c_{GG}\,\frac{m_d-m_u}{m_d+m_u} + \left[ k_u \right]_{11} - \left[ k_d \right]_{11} 
    + \frac{m_a^2}{m_\pi^2-m_a^2}\,\Delta c_{ud} \right]^2\! ,
\end{equation}
where $x_a=m_a^2/m_\pi^2$, and the phase-space function is given by 
\begin{equation}
   g(x) = 1 - 8x - 12x^2\ln x + 8x^3 - x^4 \,.
\end{equation}
This result agrees with corresponding expressions derived in \cite{Altmannshofer:2019yji} and \cite{Bardeen:1986yb} (for $m_a=0$). The PIENU collaboration has recently put a limit on the branching ratio $\text{Br}(\pi^-\to a\spac e^-\bar\nu_e)<10^{-6}\,\text{Br}(\pi^-\to\mu^-\bar\nu_\mu)$ \cite{Aguilar-Arevalo:2021utl}. Three-body pion decays are insensitive to flavor-changing ALP couplings, and so the corresponding constraints are shown in Section~\ref{sec:discussion1}, where we consider the parameter space for ALPs with flavor-conserving couplings at the new physics scale.

\subsection[Modification of $B_{s,d}\to\mu^+\mu^-$]{\boldmath Modification of $B_{s,d}\to\mu^+\mu^-$}
\label{subsec:Bstoll}

Because of their chiral suppression in the SM, the decays $B_s\to\mu^+\mu^-$ and $B_d\to\mu^+\mu^-$ are sensitive probes of flavor-changing ALP couplings. In the SM the first of these processes is induced by the operator $O_{10}=\bar s_L\gamma_\mu b_L\,\bar\ell\gamma^\mu\gamma_5\ell$ in the effective weak Hamiltonian (see e.g.,~\cite{Hiller:2014yaa})
\begin{equation} \label{eq:Hbtosll}
   {\cal H}_{\rm eff}^{b\to s\ell\ell} 
   = - \frac{4G_F}{\sqrt2}\,V_{ts}^* V_{tb}\,\frac{\alpha}{4\pi}\,
    \Big[ C_{10}(\mu)\,O_{10}(\mu) + \dots \Big] \,.
\end{equation}
A corresponding Hamiltonian with $s\to d$ holds for the second process. A flavor-changing ALP contributes to the decay amplitude at tree-level. For the case of a heavy ALP ($m_a\gg m_b$), the corresponding operators in the effective Lagrangian are (neglecting the strange-quark mass for simplicity) 
\begin{equation}\label{LefftreeheavyALP}
   {\cal L}_{\rm eff}^{\rm heavy\,ALP}
   = \sum_\ell\,c_{\ell\ell}\,\frac{m_b\spac m_\ell}{f^2\spac m_a^2}\, 
    \Big[ [k_d]_{23}\,\bar s_R\spac b_L + [k_D]_{23}\,\bar s_L\spac b_R \Big]\,\bar\ell\spac\gamma_5\ell \,,
\end{equation}   
where $\ell=\mu$ in the present case. 
For a light ALP ($m_a\sim m_b$ or lighter), the ALP propagator cannot be integrated out and the ALP contribution must instead be computed as part of the decay amplitude. In the present case, however, this distinction is irrelevant, because in both cases the relevant hadronic information is contained in the $B_s$-meson decay constant. After taking the $B_s\to\mu^+\mu^-$ matrix element, the SM contribution and the ALP contribution to the decay amplitude have the same structure. Taking their interference properly into account, we find that the ALP contribution modifies the branching ratios according to
\begin{equation}\label{eq:Bsmumu}
   \frac{\mbox{Br}(B_s\to\mu^+\mu^-)}{\mbox{Br}(B_s\to\mu^+\mu^-)_{\rm SM}}
   = \left| 1 - \frac{c_{\mu\mu}(\mu_b)}{C_{10}^{\rm SM}(\mu_b)}\,\frac{\pi}{\alpha(\mu_b)}\,
    \frac{v^2}{f^2}\,\frac{1}{1-m_a^2/m_{B_s}^2}\,\frac{[k_D-k_d]_{23}}{V_{ts}^*\,V_{tb}} \right|^2 .
\end{equation}
An analogous expression holds for the case of $B_d\to\mu^+\mu^-$ decay. Here $\mu_b\sim m_b$ is an appropriate choice of the renormalization scale. In the SM, one finds $C_{10}^{\rm SM}(m_b)\simeq -4.2$ \cite{Hiller:2014yaa}. According to (\ref{eq:cffrun}) and (\ref{eq:lowrun}), the value of $c_{\mu\mu}$ at the scale $\mu_b$ is approximately equal to $c_{\mu\mu}(\Lambda)+0.12c_{tt}(\Lambda)$. The above formula exhibits the decoupling ($\sim 1/m_a^2$) for a heavy ALP as mentioned above. It becomes singular if the ALP is degenerate in mass with the $B_s$ (or $B_d$) meson. This case can be safely excluded, because it would lead to a significant mixing of the ALP with the pseudoscalar $(\bar s b)$ or $(\bar d b)$ flavor eigenstates, in which case all precision flavor observables of $B$ mesons tested at the $B$ factories would be strongly affected.

\begin{figure}[t]
\centering
\includegraphics[width=0.8\textwidth]{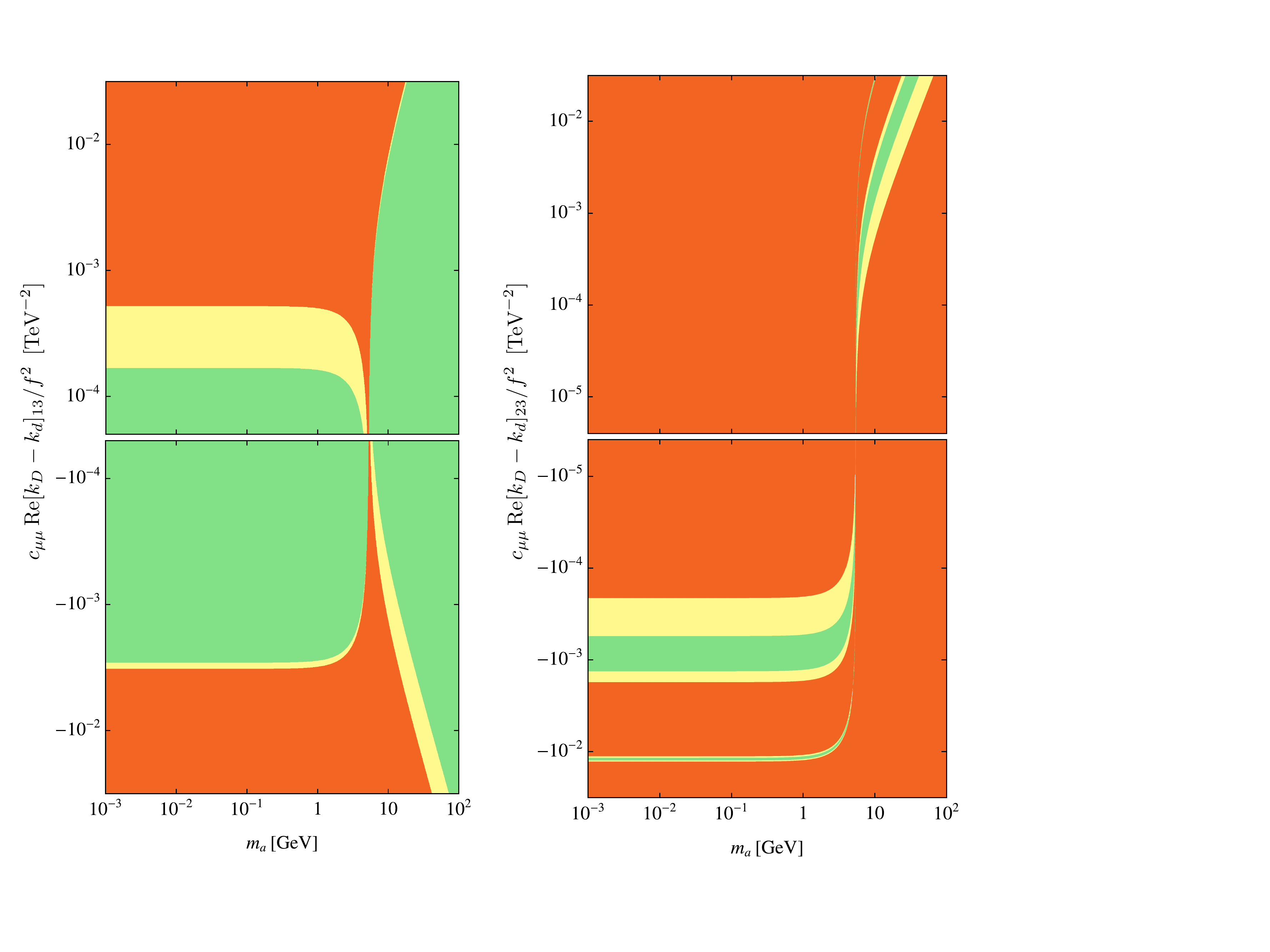}
\caption{\label{fig:bsmumuplot} Left: Constraints on the ALP parameter space from the measurement of $\text{Br}(B_d\to \mu^+\mu^-)$ in the $c_{\mu\mu}(\mu_b)\,[k_D-k_d]_{13}$ vs $m_a$ plane. The green and yellow areas respectively lie within $1\sigma$ and $2\sigma$ of the experimental value and the orange region is excluded at $2\sigma$. 
Right: Parameter space of $c_{\mu\mu}(\mu_b)\,[k_D-k_d]_{23}$ vs $m_a$ where the tension in $B_s\to \mu^+\mu^-$ can be explained by an ALP with flavor-violating couplings.}
\end{figure}

A combination of results from ATLAS, CMS and LHCb finds the values \cite{ATLAS-CONF-2020-049}
\begin{align}
\text{Br}_{\text{exp}}(B_d\to \mu^+\mu^-) &= (0.6^{+0.7}_{-0.7}) \times 10^{-10} \,, \\
\text{Br}_{\text{exp}}(B_s\to \mu^+\mu^-) &= (2.69^{+0.37}_{-0.35}) \times 10^{-9} \,, 
\end{align}
These measurements differ from the SM prediction~\cite{Beneke:2019slt} 
\begin{align}
\text{Br}_{\text{SM}}(B_d\to \mu^+\mu^-) &= (1.03\pm 0.05) \times 10^{-10} \,, \\
\text{Br}_{\text{SM}}(B_s\to \mu^+\mu^-) &= (3.66\pm 0.14) \times 10^{-9} \,, 
\end{align}
by $0.64\sigma$ and $2.4\sigma$, respectively.
The measurements provide model-independent constraints on the coupling product $c_{\mu\mu}(\mu_b)\,[k_D-k_d]_{13}/f^2$ and $c_{\mu\mu}(\mu_b)\,[k_D-k_d]_{23}/f^2$ as a function of $m_a$ as shown in the left ($B_d$) and right ($B_s$) panel of Figure~\ref{fig:bsmumuplot}, respectively.\footnote{Note that an ALP with $m_a < 300\,$MeV and sizeable couplings to quarks could also be discovered in $B_q \to \mu\mu a$ decays where the ALP can be produced resonantly \cite{Albrecht:2019zul}.} Green (yellow) indicates the region where the ALP contribution is within $1\sigma$ ($2\sigma$) from the theory prediction and the experimentally measured value. The orange regions are excluded at $2\sigma$.

The left panel of Figure~\ref{fig:bsmumuplot} depicts the constraints from $B_d\to \mu^+\mu^-$
for $c_{\mu\mu}\text{Re}[k_D-k_d]_{13} > 0$ (top) and $c_{\mu\mu}\text{Re}[k_D-k_d]_{13} < 0$ (bottom), with all couplings defined at the scale of the measurement. Since the experimentally measured value of the branching ratio agrees well with the SM expectation, the ALP contribution needs to be small to lie within one standard deviation: $0 < \mbox{Br}(B_d\to\mu^+\mu^-)/\mbox{Br}(B_d\to\mu^+\mu^-)_{\rm SM} < 1.3$. For $m_a > m_{B_d}$, the $1\,\sigma$ region is mass dependent. Larger ALP masses allow for larger couplings to lie within the $1\,\sigma$ region.
The top right panel depicts the constraints from $B_s\to \mu^+\mu^-$ for $[k_D-k_d]_{23} > 0$. We find that the presence of an ALP can only alleviate the tension in $B_s\to \mu^+\mu^-$ for $m_a > m_{B_s}$. The ALP contribution to the branching ratio must be such that $0.64 < \mbox{Br}(B_s\to\mu^+\mu^-)/\mbox{Br}(B_s\to\mu^+\mu^-)_{\rm SM} < 0.84$ at $1\sigma$. The quadratic form of (\ref{eq:Bsmumu}) leads to the appearance of two $1\sigma$ branches. As the ALP mass approaches $m_{B_s}$, ever smaller couplings are required to compensate for the growing denominator. For $m_a < m_{B_s}$, the sign of the ALP contribution in (\ref{eq:Bsmumu}) flips and the branching ratio becomes too large to be within the $2\sigma$ region, even for vanishing coupling values. 

The situation is reversed for $c_{\mu\mu}(\mu_b)\,\text{Re}[k_D-k_d]_{23} < 0$ shown in the bottom left panel of Figure~\ref{fig:bsmumuplot}, since here the SM prediction is in tension with the data, and a sizeable ALP contribution is needed to bring the prediction in line with measurement. The branching ratio is thus too large if the ALP mass is big, but can be within the $1\,\sigma$ region for $m_a < m_{B_s}$. Again we find two branches of the $1\sigma$ region due to the quadratic nature of (\ref{eq:Bsmumu}). 
As $m_a$ gets closer to $m_{B_s}$, smaller couplings compensate for the large denominator.

\subsection[Modification of $B_{d,s}\,\mbox{--}\,\bar B_{d,s}$ mixing]{\boldmath Modification of $B_{d,s}\,\mbox{--}\,\bar B_{d,s}$ mixing}
\label{sec:Bsmixing}

\begin{figure}
\centering
\includegraphics[width=.6\textwidth]{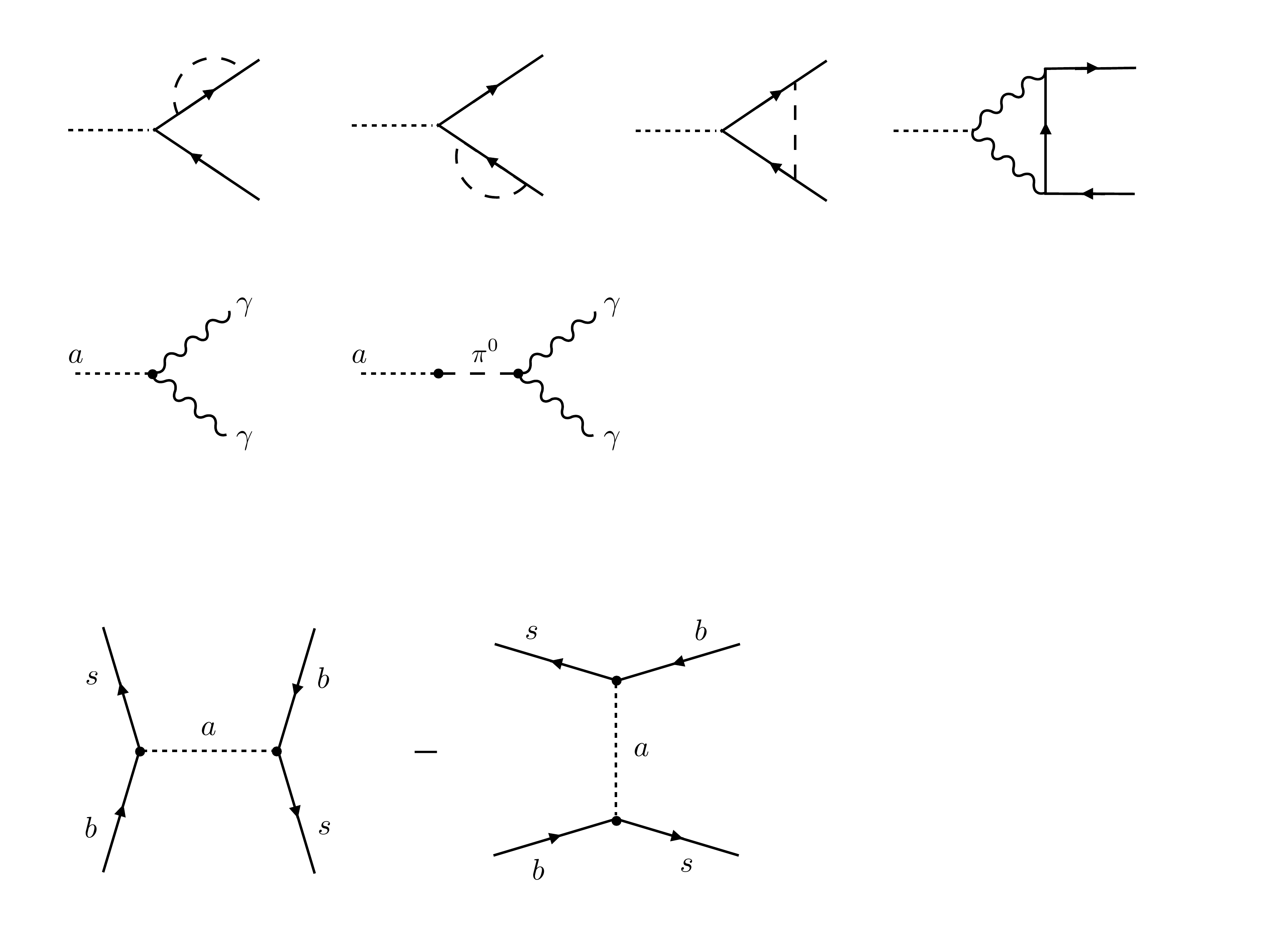}
\caption{\label{fig:bsmixing} ALP tree-level diagrams contributing to $B_s-\bar B_s$ mixing. The relative signs between the s-channel and t-channel exchange is relevant for the case of a light ALP $m_a\ll m_b$.}
\end{figure}

Mixing of neutral $B_{s,d}$ mesons with their anti-particles can be induced by the exchange of a flavor-changing ALP via both $s$- and $t$-channel diagrams, as shown in Figure~\ref{fig:bsmixing} \cite{Feng:1997tn,MartinCamalich:2020dfe}. There is a relative minus sign between the contributions from the two graphs, because they are related by an odd number of fermion exchanges. We first evaluate these diagrams for the case of a light ALP with mass $m_a\sim m_b$ or $m_a\ll m_b$. The case of a heavy ALP ($m_a\gg m_b$) will be considered later. Throughout, we will neglect the masses of the light $d$ and $s$ quarks, which is a very good approximation.

\subsubsection[Light ALP ($m_a\sim m_b$ or $m_a\ll m_b$)]{\boldmath Light ALP ($m_a\sim m_b$ or $m_a\ll m_b$)}
\label{subsec:3.4.1}

The propagator of the $s$-channel diagram carries the full momentum of the $B_q$ meson. For the $t$-channel graph, we assign incoming momenta $p_b^\mu=m_b v^\mu+k_1^\mu$, $p_{\bar q}^\mu=\bar\Lambda v^\mu-k_1^\mu$ to the $b$ quark and light anti-quark, respectively, where $\bar\Lambda=m_{B_q}-m_b$ and $m_b$ denotes the pole mass of the $b$ quark. Similarly, we label the outgoing momenta as $p_{\bar b}^\mu=m_b v^\mu+k_1^\mu$, $p_q^\mu=\bar\Lambda v^\mu-k_2^\mu$. Here $v^\mu$ denotes the 4-velocity of the $B$ mesons, which equals $v^\mu=(1,\bm{0})$ in the meson rest frame. The $t$-channel propagator can then be expanded as
\begin{equation}\label{propa}
   \frac{1}{(p_b-p_q)^2-m_a^2}
   = \frac{1}{(m_b-\bar\Lambda)^2-m_a^2}
    \left[ 1 - \frac{2(m_b-\bar\Lambda)}{(m_b-\bar\Lambda)^2-m_a^2}\,v\!\cdot\!(k_1+k_2)
    + \frac{{\cal O}(\Lambda_{\rm QCD}^2)}{\left[ (m_b-\bar\Lambda)^2-m_a^2 \right]^2} \right] ,
\end{equation}
where we neglect terms of quadratic order in the soft momenta of the light quarks. The expansion is valid only if $\big|(m_b-\bar\Lambda)^2-m_a^2\big|\gg\Lambda_{\rm QCD}^2$, which we assume to be the case for the purposes of this discussion. In this case the length scale resolved by the propagating ALP is much smaller than the size of the $B$ meson, which is set by the inverse of $\Lambda_{\rm QCD}$. We can thus describe the decay amplitude in terms of hadronic matrix elements of local 4-quark operators defined in heavy-quark effective theory (HQET), where the $b$-quark field is replaced by an effective field $b_v$ satisfying $\rlap/v\,b_v=b_v$ and $iv\cdot D\,b_v=0$ \cite{Neubert:1993mb}. The first-order correction term on the right-hand side of (\ref{propa}) corresponds to higher-dimensional HQET operators in which one of the two heavy-quark fields is replaced by $iv\cdot D\,b_v$. The matrix elements of these operators vanish by virtue of the equations of motion of HQET. At the scale $\mu_b\sim\big|(m_b-\bar\Lambda)^2-m_a^2\big|^{1/2}$, we thus define the effective Hamiltonian
\begin{equation}
   \mathcal{H}_{\rm eff}^{\Delta B=2}
   = \sum_{i=1}^5\,C_i(\mu_b)\,\mathcal{O}_i 
    + \sum_{i=1}^3\,\tilde{C}_i(\mu_b)\,\mathcal{\tilde{O}}_i \,,
\end{equation}
where the basis of local operators is (with $q=d,s$) \cite{Ciuchini:1998ix}
\begin{equation}\label{eq:Bmixingops}
\begin{aligned}
   \mathcal{O}_1 
   &= \bar q_L^i\gamma_\mu b_{v,L}^i\,\bar q_L^j\gamma^\mu b_{v,L}^j \,, &\quad
   \mathcal{O}_2 &= \bar q_R^i b_{v,L}^i\,\bar q_R^j b_{v,L}^j \,, &\quad
   \mathcal{O}_3 &= \bar q_R^i b_{v,L}^j\,\bar q_R^j b_{v,L}^i \,, \\
   \mathcal{O}_4 &= \bar q_R^i b_{v,L}^i\,\bar q_L^j b_{v,R}^j \,, &\quad
   \mathcal{O}_5 &= \bar q_R^i b_{v,L}^j\,\bar q_L^j b_{v,R}^i \,, &
\end{aligned}
\end{equation}
as well as operators $\tilde{\mathcal{O}}_{1,2,3}^q$ obtained by exchanging $L\leftrightarrow R$ in $\mathcal{O}_{1,2,3}^q$. Here $i$ and $j$ are color indices, and $b_v$ denotes the effective heavy-quark field in HQET. For the Wilson coefficients relevant for $B_s-\bar B_s$ mixing we obtain
\begin{equation}\label{CiBBmix}
\begin{aligned}
   C_1(\mu_b) &= C_1^{\rm SM}(\mu_b) \,, \\
   C_2(\mu_b) &= - \frac{m_b^2(\mu_b)}{2f^2}\,\frac{N_c^2 A_+ - A_-}{N_c^2-1}
    \left( [k_d]_{23} \right)^2 , \\
   C_3(\mu_b) &= - \frac{m_b^2(\mu_b)}{2f^2}\,\frac{N_c\hspace{0.3mm}(A_- - A_+)}{N_c^2-1}
    \left( [k_d]_{23} \right)^2 , \\
   \tilde C_2(\mu_b) &= - \frac{m_b^2(\mu_b)}{2f^2}\,\frac{N_c^2 A_+ - A_-}{N_c^2-1}
    \left( [k_D]_{23} \right)^2 , \\
   \tilde C_3(\mu_b) &= - \frac{m_b^2(\mu_b)}{2f^2}\,\frac{N_c\hspace{0.3mm}(A_- - A_+)}{N_c^2-1}
    \left( [k_D]_{23} \right)^2 , \\
   C_4(\mu_b) &= - \frac{m_b^2(\mu_b)}{f^2}\,\frac{N_c^2 A_+ - A_-}{N_c^2-1}\,
    [k_d]_{23}\,[k_D]_{23} \,, \\
   C_5(\mu_b) &= - \frac{m_b^2(\mu_b)}{f^2}\,\frac{N_c\hspace{0.3mm}(A_- - A_+)}{N_c^2-1}\,
    [k_d]_{23}\,[k_D]_{23} \,,
\end{aligned}
\end{equation}
where $A_\pm=1/[(m_b\pm\bar\Lambda)^2-m_a^2]$ and $N_c=3$ is the number of colors. Analogous expressions involving the 13 entries of the flavor-changing ALP couplings hold for the case of $B_d-\bar B_d$ mixing. 
Note that in the strict heavy-quark limit $m_b\to\infty$ one has $A_+=A_-=1/m_b^2$, in which case the Wilson coefficients $C_3$, $\tilde C_3$ and $C_5$ vanish. In practice, however, the corrections governed by the parameter $\bar\Lambda$ are rather significant. With $m_b=4.8$\,GeV, we find $(A_- - A_+)/A_+\simeq 0.61$ for $m_a=0$ and 0.78 for $m_a=2$\,GeV.

The mass difference observable from the $B_q-\bar B_q$ oscillation frequency is given by \cite{Buras:1998raa}
\begin{equation}
   \Delta M_q = \frac{1}{m_{B_q}}
    \left| \langle B_q|\,\mathcal{H}^{\Delta B=2}\,|\bar B_q \rangle \right| .
\end{equation}
The relevant hadronic matrix elements of the 4-quark operators defined in HQET are related to the corresponding matrix elements in QCD by perturbative matching coefficients, which equal~1 at tree-level. Since we work to zeroth order in the QCD coupling in this section, we will consistently neglect these matching effects. The hadronic matrix elements of the relevant operators in \eqref{eq:Bmixingops} can then be written in terms of hadronic parameters $B^{(i)}_{B_q}$ defined as
\begin{equation}
   \frac{1}{m_{B_q}}\,\langle B_q|\,\mathcal{O}_i\,|\bar B_q\rangle 
   \equiv f_{B_q}^2 m_{B_q}\,\eta_i^q(\mu_b)\,B^{(i)}_{B_q}(\mu_b) \,,
\end{equation}
where $f_{B_q}$ is the decay constant of the $B_q$ meson. The normalization factors $\eta_i^q(\mu_b)$ are conventionally obtained using the naive vacuum insertion approximation for the matrix elements. One obtains (neglecting $m_q$, meaning that the superscript on $\eta_i^q(\mu_b)$ can also be dropped since $\eta_i^s(\mu_b)=\eta_i^d(\mu_b)\equiv \eta_i(\mu_b)$) \cite{Bagger:1997gg} 
\begin{equation}
\begin{aligned}
   \eta_2(\mu_b) = \tilde\eta_2(\mu_b)
   &= - \frac12 \left( 1 - \frac{1}{2N_c} \right)
    \left( \frac{m_{B_q}}{m_b(\mu_b)} \right)^2 , \\
   \eta_3(\mu_b) = \tilde\eta_3(\mu_b)
   &= - \frac12 \left( \frac{1}{N_c} - \frac{1}{2} \right)
    \left( \frac{m_{B_q}}{m_b(\mu_b)} \right)^2 , \\
   \eta_4(\mu_b) 
   &= \frac12 \left[ \left( \frac{m_{B_q}}{m_b(\mu_b)} \right)^2 + \frac{1}{2N_c} \right] , \\
   \eta_5(\mu_b) 
   &= \frac12 \left[ \frac{1}{N_c} \left( \frac{m_{B_q}}{m_b(\mu_b)} \right)^2
    + \frac{1}{2} \right] .
\end{aligned}
\end{equation}
Note that the factors $m_b^2(\mu_b)$ contained in the Wilson coefficients in (\ref{CiBBmix}) cancel against corresponding factors of $1/m_b^2(\mu_b)$ contained in the definitions of the $\eta_i$ parameters except for $\eta_{4,5}$, where some extra terms remain. We take values of the hadronic parameters $B^{(i)}_{B_q}(\mu_b)$ at the scale $\mu_b=m_b(m_b)$ from the lattice calculations of \cite{Dowdall:2019bea}. Parity invariance of QCD implies that the parameters $\tilde B^{(i)}_{B_q}$ with $i=1,2,3$ are equal to $B^{(i)}_{B_q}$. Eventually, the mass difference observable is obtained as 
\begin{align}
   \Delta M_q^{\text{SM+ALP}} 
   &= f_{B_q}^2 m_{B_q} \bigg|
    \eta_1(m_b)\,B^{(1)}_{B_q}(m_b)C_{1}^\text{SM}(m_b)+ \sum_{i=2}^5\,C_i^q(m_b)\,\eta_i(m_b)\,B^{(i)}_{B_q}(m_b) \notag
    \\&+ \sum_{i=2,3}\,\tilde C_i^q(m_b)\,\eta_i(m_b)\,B^{(i)}_{B_q}(m_b) \bigg| ,\label{eq:generalBsmixingexpression}
\end{align}
where the first term contains the SM contribution, and all other terms are due to the ALP. Using the hadronic parameters given in \cite{Dowdall:2019bea} and decay constants computed in \cite{Hughes:2017spc}, we obtain the numerical expressions for a light ALP
\begin{align}
\label{eq:numericalBsmixingexpression}
\Delta M_d^{\text{SM+ALP}}  &= \bigg|\Delta M_d^{\text{SM}} -0.07\,\text{GeV}^2\,\left( C_2^d(m_b) +\tilde{C}_2^d(m_b)\right) +0.01\,\text{GeV}^2\,\left( C_3^d(m_b) +\tilde{C}_3^d(m_b)\right)  \notag\\ &\qquad +0.14\,\text{GeV}^2\,C_4^d(m_b)+0.08\,\text{GeV}^2\,C_5^d(m_b)\bigg|\, \text{GeV}\,,\\
\Delta M_s^{\text{SM+ALP}}  &= \bigg| \Delta M_s^{\text{SM}} -0.12\,\text{GeV}^2\,\left( C_2^s(m_b) +\tilde{C}_2^s(m_b)\right)
+0.02\,\text{GeV}^2\,\left( C_3^s(m_b) +\tilde{C}_3^s(m_b)\right) \notag\\ &\qquad +0.20\,\text{GeV}^3\,C_4^s(m_b)+0.12\,\text{GeV}^2\,C_5^s(m_b)\bigg|\, \text{GeV}\,.
\end{align}

\subsubsection[Heavy ALP ($m_a\gg m_b$)]{\boldmath Heavy ALP ($m_a\gg m_b$)}
\label{subsec:3.4.1}

Let us now discuss the case where the ALP is much heavier than the $b$-quark mass, $m_a\gg m_b$. In this case, the propagators in both the $s$- and $t$-channel diagrams can be approximated as $-1/m_a^2$, and hence one generates the local 4-quark operators in (\ref{eq:Bmixingops}) -- with QCD $b$-quark fields rather than HQET fields -- at the scale $\mu_a\sim m_a$. In analogy with (\ref{CiBBmix}), we find that the relevant Wilson coefficients are given by
\begin{equation} \label{CiBBmixLargeMass}
\begin{aligned}
   C_2(\mu_a) &= \frac{m_b^2(\mu_a)}{2m_a^2\hspace{0.3mm} f^2} \left([k_d]_{23} \right)^2 , \\
   \tilde C_2(\mu_a) &= \frac{m_b^2(\mu_a)}{2m_a^2\hspace{0.3mm} f^2} \left( [k_D]_{23} \right)^2 , \\
   C_4(\mu_a) &= \frac{m_b^2(\mu_a)}{m_a^2\hspace{0.3mm} f^2}\,[k_d]_{23}\,[k_D]_{23} \,,
\end{aligned}
\end{equation}
whereas $C_3$, $\tilde C_3$ and $C_5$ vanish in this limit.
For the case of a heavy ALP, 
these coefficients should be run down to the scale $\mu_b\sim m_b$ using the evolution equations 
\cite{Bagger:1997gg}\footnote{$C_3$ and $\tilde{C}_3$ are also generated by mixing from $C_2$ and $\tilde{C}_2$, but their values remain numerically very small, so we neglect them.}
\begin{align}
C_2(\mu_b) &= \left(0.983\, \eta^{-2.42} +0.017 \,\eta^{2.75} \right) C_2(\mu_a) \,,\notag \\
C_4(\mu_b) &= \eta^{-4} C_4(\mu_a) \,,
\label{eq:Bsmixingrunning}
\end{align}
where $\eta=[\alpha_s(\mu_a)/\alpha_s(\mu_b)]^{6/23}$. In this way, large logarithms of the scale ratio $m_a/m_b$ are resummed at leading logarithmic order. Since QCD preserves parity, the equation for $\tilde{C}_2$  is obtained simply by replacing $C_{2}\to \tilde{C}_{2}$ on both sides of the equality.

\begin{figure}[t]
\centering
\includegraphics[width=1.\textwidth]{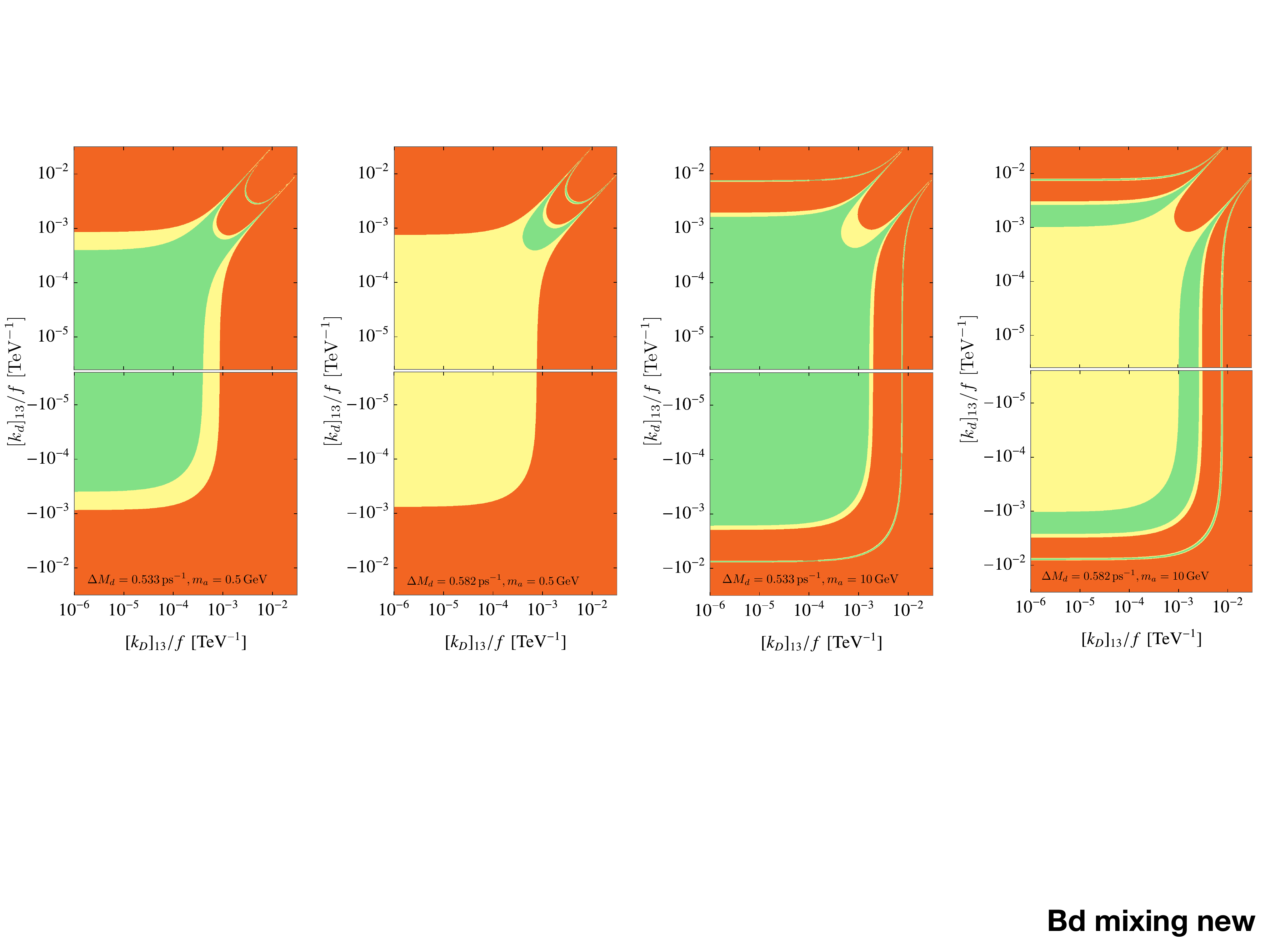}
\caption{\label{fig:bdmixingplot} ALP couplings for which the mass difference $\Delta M_d$ is reproduced within $1\sigma$ (green) or $2\sigma$ (yellow) for $m_a=0.5$\,GeV (two left panels) and $m_a=10$\,GeV (two right panels) using recent weighted averages (first and third panels) and FLAG 2019 values (second and fourth panels) for the SM prediction. The orange region is excluded at $2\sigma$.}
\end{figure}

\begin{figure}[t]
\centering
\includegraphics[width=\textwidth]{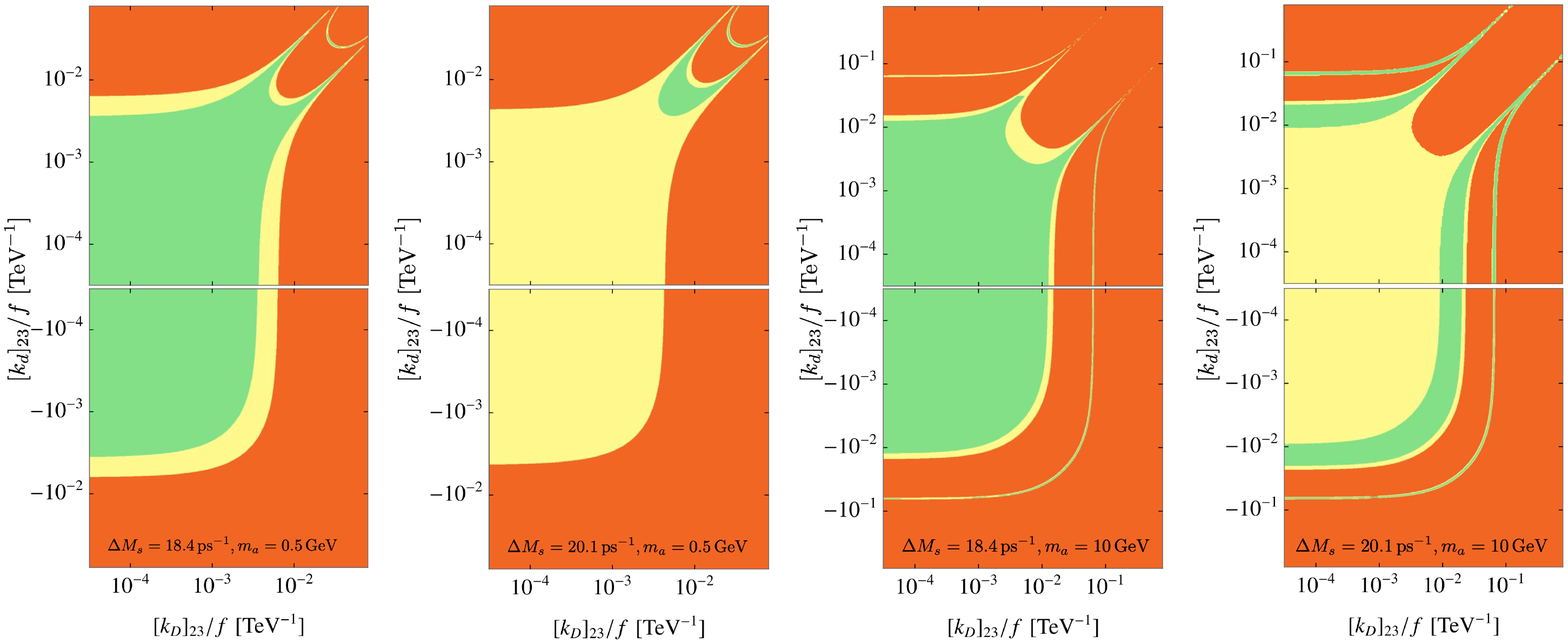}
\caption{\label{fig:bsmixingplot} ALP couplings for which the mass difference $\Delta M_s$ is reproduced within $1\sigma$ (green) or $2\sigma$ (yellow) for $m_a=0.5$\,GeV (two left panels) and $m_a=10$\,GeV (two right panels) using recent weighted averages (first and third panels) and FLAG 2019 values (second and fourth panels) for the SM prediction. The orange region is excluded at $2\sigma$.}
\end{figure}

For $m_a = 10\,$GeV we obtain the numerical expressions 
\begin{align}
\Delta M_d &= \left|\Delta M_d^{\text{SM}} -0.08\,\text{GeV}^2\,\left( C_2^d(m_a) +\tilde{C}_2^d(m_a)\right) +0.17\,\text{GeV}^2\,C_4^d(m_a)\right|\, \text{GeV}, 
\label{eq:numericalBdmixingexpression}\\
\Delta M_s &= \left|\Delta M_s^{\text{SM}} -0.13\,\text{GeV}^2\,\left( C_2^s(m_a) +\tilde{C}_2^s(m_a)\right) +0.24\,\text{GeV}^2\,C_4^s(m_a)\right|\, \text{GeV}, 
\label{eq:numericalBsmixingexpression}
\end{align}
where the Wilson coefficients here have units of GeV$^{-2}$ and $\Delta M_{s,d}^{\text{SM}}$ are the SM predictions as given below. 
\subsubsection{Bounds on ALP couplings}

Recent weighted averages give \cite{DiLuzio:2019jyq}
\begin{align}
   \Delta M_d^{\text{SM}}
   &= \left( 0.533^{+0.022}_{-0.036} \right) \text{ps}^{-1}
    = \left( 1.05^{+0.04}_{-0.07} \right) \Delta M_d^{\text{exp}} \,,\label{eq:MdSMavg}\\
    \Delta M_s^{\text{SM}}
   &= \left( 18.4^{+0.7}_{-1.2} \right) \text{ps}^{-1}
    = \left( 1.04^{+0.04}_{-0.07} \right) \Delta M_s^{\text{exp}} \,, \label{eq:MsSMavg}
\end{align}
and we also consider the values reported by FLAG in 2019~\cite{Aoki:2019cca,Bazavov:2016nty},
\begin{align}
       \Delta M_d^{\text{SM}} 
   &= \left( 0.582^{+0.049}_{-0.056} \right) \text{ps}^{-1}
    = \left( 1.15^{+0.1}_{-0.11} \right) \Delta M_d^{\text{exp}}  \label{eq:MdSM19} \,, \\
    \Delta M_s^{\text{SM}} 
   &= \left( 20.1^{+1.2}_{-1.6} \right) \text{ps}^{-1}
    = \left( 1.13^{+0.07}_{-0.09} \right) \Delta M_s^{\text{exp}}  \label{eq:MsSM19}\,, 
\end{align}    
which imply a $1.26\sigma$ and $1.17\sigma$ tension with the following measured values
\begin{equation}
\begin{aligned}
 \Delta M_d^{\text{exp}} &= \left( 0.5064\pm 0.0019 \right)\text{ps}^{-1}\, \text{\cite{Amhis:2016xyh}}\,, \\
 \Delta M_s^{\text{exp}} &= \left( 17.7656\pm 0.0057 \right) \text{ps}^{-1} \, \text{\cite{Aaij:2021jky}}\,. 
\end{aligned}
\end{equation}

Figures~\ref{fig:bdmixingplot} and ~\ref{fig:bsmixingplot} show constraints from $B_d-\bar B_d$ and $B_s-\bar B_s$ mixing, respectively, on flavor-violating ALP couplings for $m_a=0.5$\,GeV in the first and second panels from the left and for $m_a=10$\,GeV in the right two panels. The first and third panels show the constraints using the average values in~\eqref{eq:MdSMavg} and~\eqref{eq:MsSMavg} for the SM prediction while the values reported by FLAG in 2019 in~\eqref{eq:MdSM19} and~\eqref{eq:MsSM19} are depicted in the second and fourth panels. 
Parameter space shown in red is excluded at 2$\sigma$ while the regions in green and yellow are within 1$\sigma$ and 2$\sigma$ of the theoretical and experimental central values. For small ALP masses, the tension in $\Delta M_d$ and $\Delta M_s$ can be lifted for $|[k_d]_{13}/f| = |[k_D]_{13}/f| < 6.3 \times 10^{-4}$\,TeV$^{-1}$ and $|[k_d]_{23}/f| = |[k_D]_{23}/f| < 3.2 \times10^{-3}$\,TeV$^{-1}$. 
The shape of the 2$\sigma$ region is symmetric in $k_d$ and $k_D$ as can be seen from the form of the Wilson coefficients in \eqref{CiBBmix}. Switching the sign of $k_d$, however, changes the relative importance of the last two terms in \eqref{eq:numericalBdmixingexpression} and \eqref{eq:numericalBsmixingexpression}. Since the numerical factor in $C_4(\mu_b)$ is one order of magnitude larger than the corresponding factor in $C_5(\mu_b)$, large negative values of $k_d$ lead to an ALP contribution that is too big to satisfy the constraints. For positive couplings, a second solution appears due to the absolute value in \eqref{eq:numericalBsmixingexpression}. This is not present for negative values of $k_d$ as the ALP contribution becomes too large.
The allowed regions look different when the SM prediction obtained by FLAG in 2019 is used. In this case, the experimentally measured value is more than $1\sigma$ away from the theoretical prediction, which implies that arbitrarily small coupling values do not satisfy the $1\sigma$ but only the $2\sigma$ constraint. 

A heavy ALP can also lift the tension between experimental and theoretical values. For a mass of $m_a = 10\,$GeV the ALP contribution lies within $1\sigma$ when $|[k_d]_{13}/f| = |[k_D]_{13}/f| < 1.8 \times 10^{-3}$\,TeV$^{-1}$ and $|[k_d]_{23}/f| = |[k_D]_{23}/f| < 6.3 \times10^{-3}$\,TeV$^{-1}$ as shown in the third panels of Figures~\ref{fig:bdmixingplot} and ~\ref{fig:bsmixingplot} respectively. The $1\sigma$ and $2\sigma$ regions are again symmetric in the couplings $k_d$ and $k_D$ but differ when the sign of $k_d$ is reversed. Large negative couplings turn the ALP contribution in \eqref{eq:numericalBdmixingexpression} and \eqref{eq:numericalBsmixingexpression} negative which makes $\Delta M_d$ and $\Delta M_s$ too small to lie within the $2\sigma$ region. A second solution, where the ALP contribution is larger than the SM part within the absolute value, appears for even larger couplings. Here, either $k_d$ or $k_D$ is small enough to render $C_4^{s,d}$ in \eqref{eq:numericalBdmixingexpression} or \eqref{eq:numericalBsmixingexpression} negligible which is why the solution exists for positive and negative values of $k_d$. The second solution looks different at large and small ALP mass due to the opposite sign in the Wilson coefficients in \eqref{CiBBmixLargeMass} and \eqref{CiBBmix}. The FLAG 2019 values for the SM prediction are used in the rightmost plots in Figures~\ref{fig:bdmixingplot} and ~\ref{fig:bsmixingplot}. As for small masses, arbitrarily small couplings only lie within $2\sigma$ of the central value.

Outside of the region for which $m_a \approx m_b$, the results depend only mildly on the ALP mass. From Table~\ref{tab:hadronbounds} it is clear that large couplings are ruled out if the ALP can decay into charged leptons or if it is stable and does not decay. Small couplings, however, are currently unconstrained. ALPs decaying predominantly into photons are also still allowed. This motivates a search for $B\to K^{(*)} a\to K^{(*)} \gamma\gamma $, which could discover an ALP contributing to $B_d-\bar B_d$ and $B_s-\bar B_s$ mixing.

\subsection[Radiative $J/\psi$ and $\Upsilon$ decays]{\boldmath Radiative $J/\psi$ and $\Upsilon$ decays}
\label{sec:JPsiandUpsilon}

\begin{figure}
\centering
\includegraphics[width=.6\textwidth]{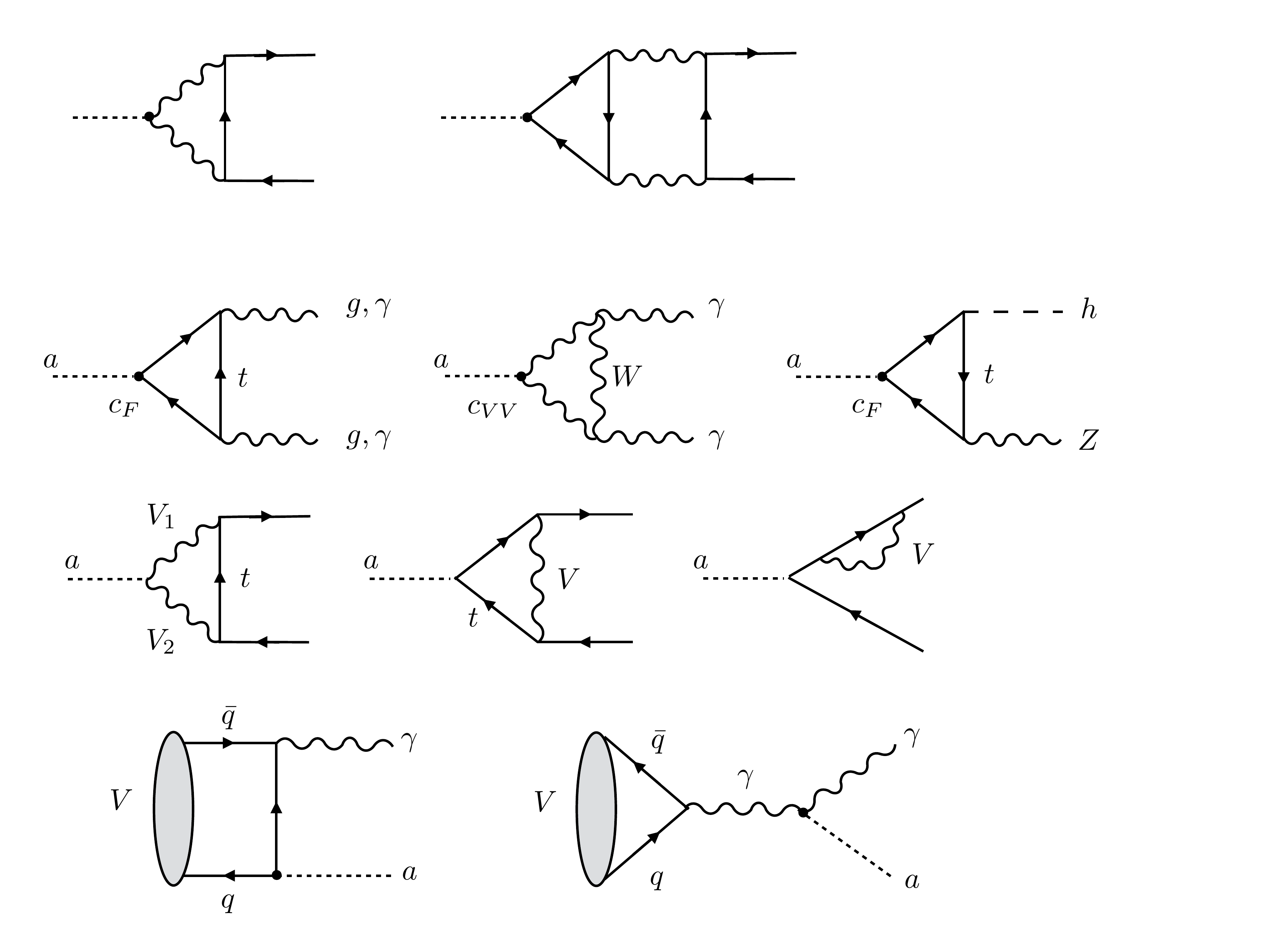}
\caption{\label{fig:Upsagamma} 
Diagrams contributing to the decays $V\to\gamma a$ of heavy vector mesons.}
\end{figure}

Searches for decays of charmonium or bottomonium states to an ALP and a photon can be employed to place interesting constraints on the ALP couplings to bottom quarks \cite{Wilczek:1977zn,Nason:1986tr,Mangano:2007gi,McKeen:2008gd} and photons \cite{Masso:1995tw,CidVidal:2018blh}. 
Although these observables do not involve a flavor change, they nevertheless constrain the quark couplings of the ALP within the same mass range relevant for many of the flavor-changing observables we consider.
The first combined analysis of the contributions from both the ALP-photon coupling and the ALP-quark coupling was performed in \cite{Merlo:2019anv}, neglecting however important QCD corrections. The relevant diagrams are depicted in Figure~\ref{fig:Upsagamma}. Including one-loop radiative corrections, we find for the decay rate
\begin{equation}\label{Upsilonrate}
   \Gamma(V\to\gamma a)
   = \frac{m_V f_V^2}{6 f^2}\,Q_q^2\,\alpha \left( 1 - \frac{m_a^2}{m_V^2} \right)
    \left| c_{qq}(\mu_q) \left[ 1 - \frac{2\alpha_s(\mu_q)}{3\pi}\,a_P(x) \right]
    - \frac{\alpha}{2\pi}\,c_{\gamma\gamma} \left( 1 - \frac{m_a^2}{m_V^2} \right) \right|^2 ,
\end{equation}
where $q=c,b$ as appropriate, and $\mu_q\sim m_q$ is an appropriate matching scale. \\The scale dependence of the coefficients $c_{bb}$ and $c_{cc}$ are such that, if $\Lambda=4\pi f$ TeV, 
\begin{align}
   c_{bb}(m_b) &\simeq c_{bb}(\Lambda) +0.09 c_{tt}(\Lambda) -0.02\,c_{GG}, \\
   c_{cc}(m_c) &\simeq c_{cc}(\Lambda) -0.13 c_{tt}(\Lambda) -0.04\,c_{GG} \,.
\end{align}
In the strict non-relativistic limit, where each of the two heavy quarks in the quarkonium state carries one half of its momentum, the QCD radiative corrections give rise to \cite{Nason:1986tr}
\begin{align}
   a_P(x)
   &= \frac{3-7x}{1-2x} + \frac{1-7x+8x^2}{\left(1-2x\right)^2}\,\ln 2x 
    + 4 \sqrt{\frac{1-x}{x}}\,\arctan\sqrt{\frac{1-x}{x}} \notag\\
   &\quad\mbox{}+ \frac{2(1-2x)}{x}\,\arctan^2\sqrt{\frac{1-x}{x}} 
    - \frac{1-4x}{2x}\,\text{Li}_2(1-2x) - \frac{5-8x}{2x}\,\frac{\pi^2}{6} \,,
\end{align}
where $x=E_\gamma/E_\gamma^{\text{max}}=1-m_a^2/m_V^2$. This is an increasing function of its argument, which varies between $a_P(0)=2$ and $a_P(1)=\frac{\pi^2}{8}+2\ln 2+4\approx 6.62$, thus giving rise to a rather large correction. Note that the contribution proportional to the coefficient $c_{\gamma\gamma}$ in (\ref{Upsilonrate}) does not receive any QCD radiative corrections. In the calculation of the decay amplitude we have used the identity 
\begin{equation}
   \left\langle 0|\,\bar b\,\Gamma\,b\,|V(p,\varepsilon)\right\rangle 
   = \frac{i f_V m_V}{2}\,\text{tr}\left[\slashed{\varepsilon}\,\Gamma\,
    \frac{(1+\slashed{v})}{2} \right]
\end{equation}
based on Heavy Quark Effective Theory \cite{Neubert:1993mb}, where $v^\mu=p^\mu/m_V$ denotes the 4-velocity of the quarkonium state and $\varepsilon^\mu$ is its polarisation vector. This identity also serves to define the decay constant $f_V$. The ${\cal O}(\alpha_s^0)$ part of our result agrees with \cite{Merlo:2019anv}. 

Many experimental results are quoted as a ratio with the SM decay width to electrons, which is given by
\begin{align}
   \Gamma(V\to e^+ e^-) 
   = \frac{\alpha^2 \pi Q_q^2}{3}\frac{f_V^2}{m_V} \left[ 1 - \frac{\alpha_s(\mu_q)}{3\pi} \right] .
\end{align}
Searches have been done in the dimuon final state for radiative $J/\psi$ decays \cite{Ablikim:2015voa}, and in the invisible \cite{delAmoSanchez:2010ac}, dimuon \cite{Lees:2012iw}, ditau \cite{Lees:2012te} and hadronic \cite{Lees:2011wb} final states for radiative $\Upsilon$ decays. Radiative vector meson decays are not sensitive to flavor-changing ALP couplings and so we show the corresponding constraints in Section~\ref{sec:discussion1}, where we consider the parameter space for ALPs with flavor-conserving couplings at the new physics scale.
%

%%%%%%
%%%%%%%%%%%%%%%%%%
\begin{figure}
\centering
\includegraphics[width=.5\textwidth]{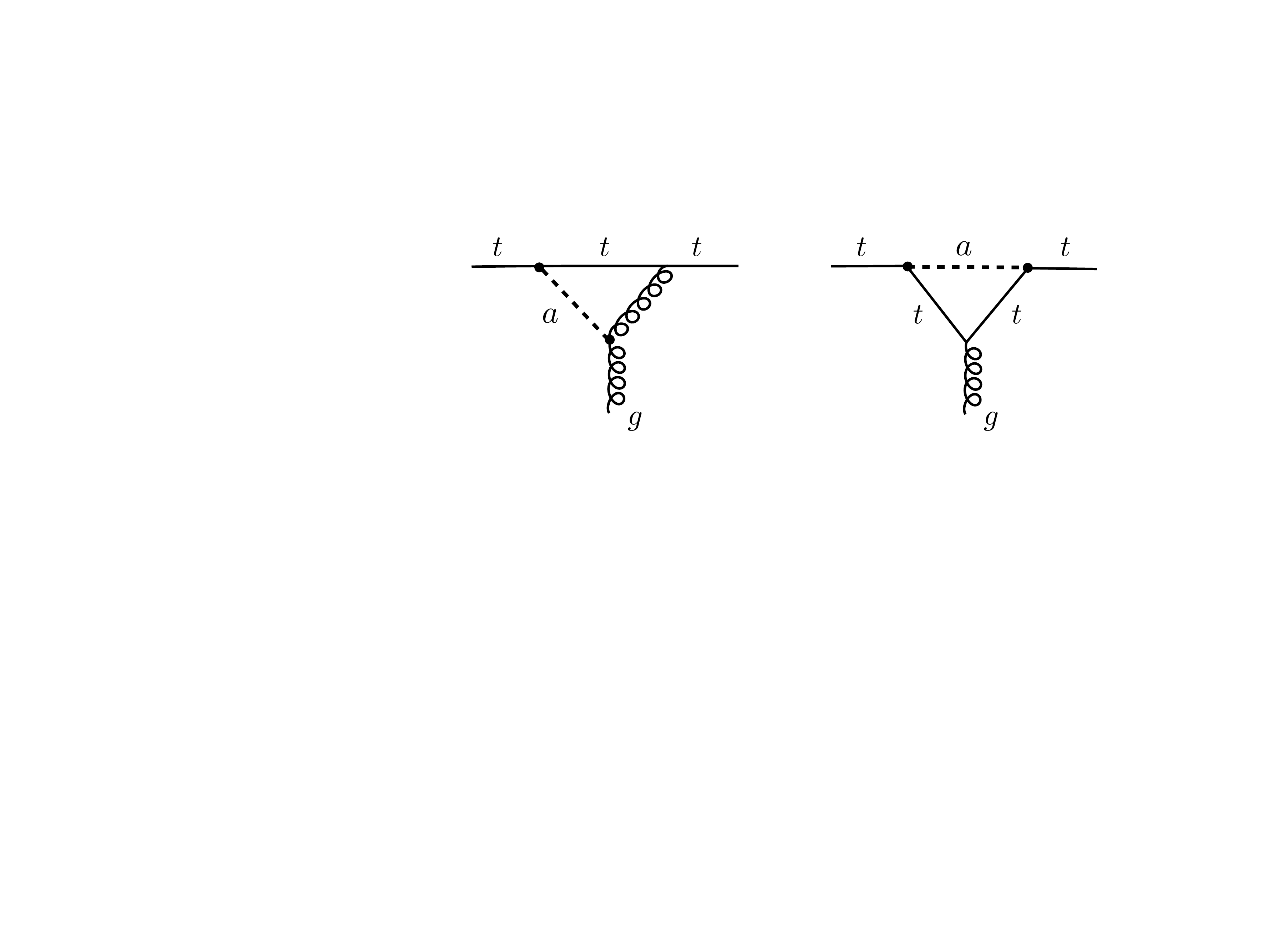}
\caption{\label{fig:chromomagneticmoment} Diagrams contributing to the chromomagnetic dipole moment of the top.}
\end{figure}
%%%%%%%%%%%%%%%%%%
%%%%%%

\subsection{The chromomagnetic dipole moment of the top quark}
\label{subsec:chromo}

The CMS collaboration has recently published bounds on the chromomagnetic dipole moment of the top quark, $\hat{\mu}_t$. At 95\% CL the limit is~\cite{Sirunyan:2019lnl}
\begin{equation}
-0.014 \le \text{Re}(\hat{\mu}_t) < 0.004.
\end{equation}
The chromomagnetic dipole moment is defined as the coefficient of the following effective operator \cite{Franzosi:2015osa}
\begin{equation}
\mathcal{L} \supset -\hat{\mu}_t\frac{g_s}{2m_t}\, \bar{t}\sigma^{\mu\nu} T^a t\, G^a_{\mu\nu}.
\end{equation}
At one-loop order, the ALP contributes to this operator via the two diagrams shown in Figure~\ref{fig:chromomagneticmoment} and is given by

\begin{equation}
\hat{\mu}_t = \frac{m_t^2}{f^2} \frac{1}{32 \pi^2}\bigg\{c_{tt}^2 \, h_1\left(x_t \right) +\frac{2\alpha_s}{\pi}c_{tt}c_{GG}\left[\log\frac{\Lambda^2}{m_t^2}- h_2\left(x_t\right) \right] -\frac{25\alpha_s^3}{16\pi^3}c_{GG}^2 \log^2\frac{\Lambda^2}{m_t^2}\bigg\}.
\end{equation}
where $x_t=m_a^2/m_t^2$ and where the last term is found via the RGEs for dimension six operators in the presence of an ALP, see~\cite{Galda:2021hbr}. The loop functions are given explicitly in Eqns.~\eqref{eq:h1h2} and \eqref{eq:h2} and satisfy $h_{1,2}(x) \to 1$ in the limit that $x \ll 1$, which applies in the mass range we focus on in this paper $m_a\lesssim$ 10 GeV.\footnote{The ALP contribution to $\hat{\mu_t}$ was studied in the opposite limit, $m_a^2/m_t^2 \gg 1$, in Ref.~\cite{Ebadi:2019gij}.}  Since this observable is (approximately) independent of flavor-changing ALP couplings, we show the constraints from the chromomagnetic moment of the top quark in the next section, where we consider the parameter space of ALPs with flavor-conserving couplings.

%%%
%%%%%%%%%%%%%
\begin{figure}[t]
\begin{center}
\includegraphics[width=1\textwidth]{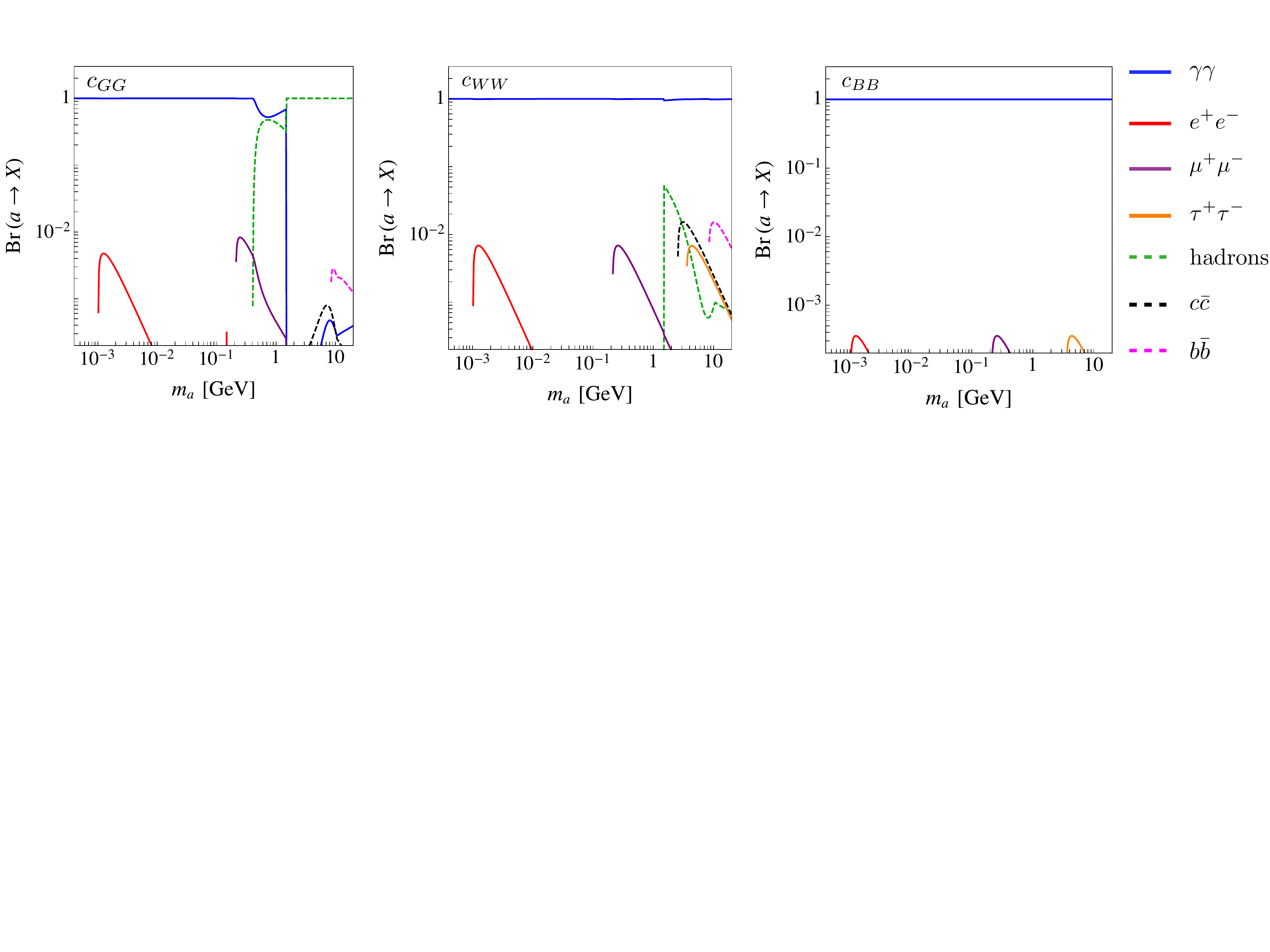}
\caption{Branching ratio for an ALP with couplings to either $c_{GG}$ (left), $c_{WW}$ (central) and $c_{BB}$ (right) at $\Lambda = 4\pi f$ and $f=1$\,TeV.  The branching ratios of the ALP into photons and charged leptons are indicated by solid lines and the branching ratios of the ALP into hadronic states are given by dashed lines. \label{fig:alp_Brs_bosons}}
\end{center}
\end{figure}
%%%%%%%%%%%%
%%%

%%%
%%%%%%%%%%%%%
\begin{figure}[t]
\begin{center}
\includegraphics[width=.8\textwidth]{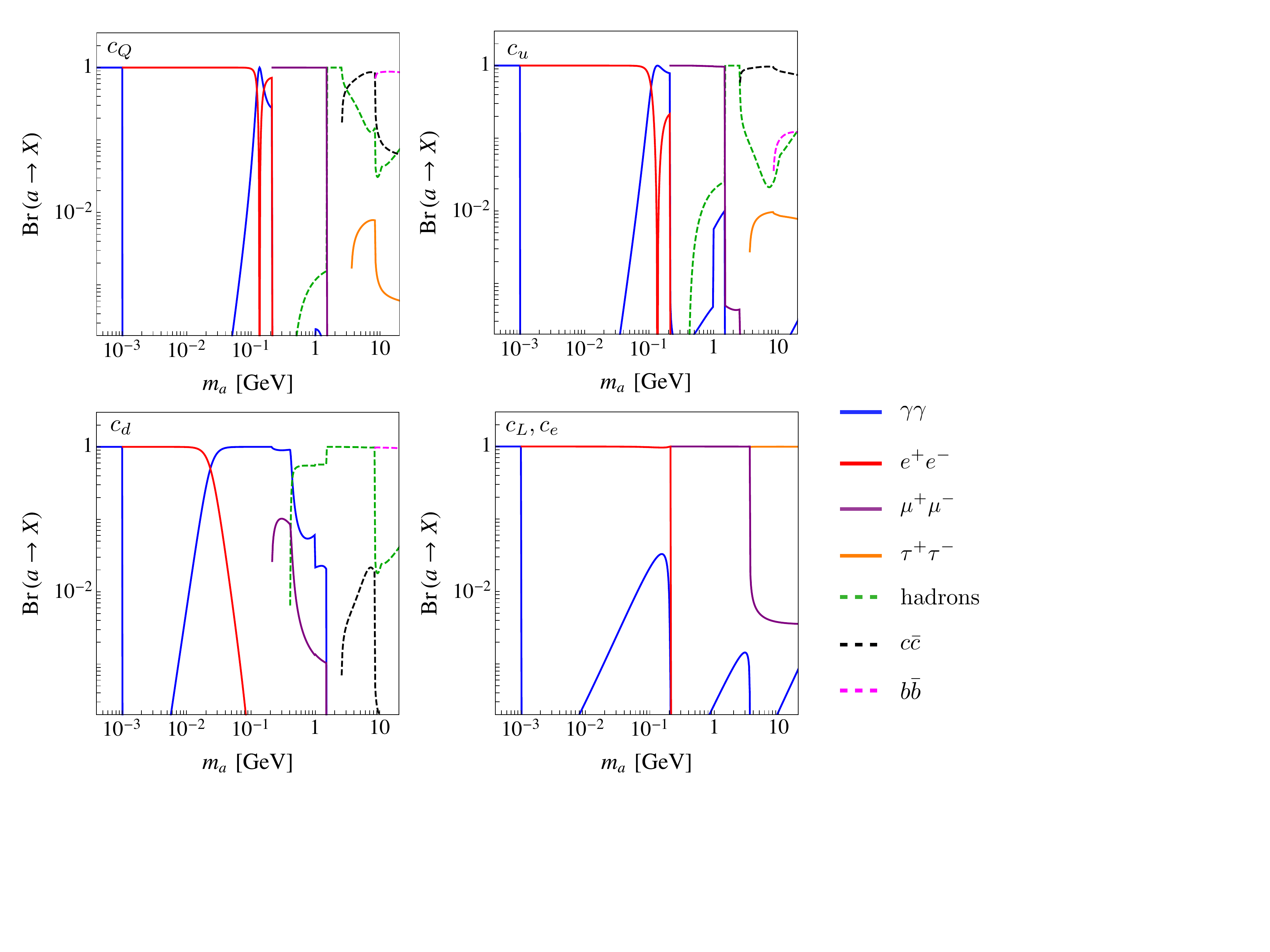}
\caption{Branching ratio for an ALP with couplings to either $c_{Q}$ (top left), $c_{u}$ (top right), $c_{d}$ (bottom left), $c_L$ and $c_e$ (bottom right) at $\Lambda = 4\pi f$ and $f=1$\,TeV.  The branching ratios of the ALP into photons and charged leptons are indicated by solid lines and the branching ratios of the ALP into hadronic states are given by dashed lines. \label{fig:alp_Brs_fermions}}
\end{center}
\end{figure}
%%%%%%%%%%%%
%%%

%%%%%%%%%%%%%
\subsection{Constraints on flavor universal UV benchmarks}
\label{sec:discussion1}
%%%%%%%%%%%%%

In this subsection we consider scenarios in which the ALP has a dominant flavorless or flavor-universal coupling at the UV scale $\Lambda=4\pi f$ with $f=1$ TeV. 
Flavor-violating couplings are induced at a lower scale through RG running and matching using the equations in Section~\ref{subsec:LagbelowEW}. Effects induced at two- or higher-loop order are calculated to leading logarithm. A UV structure with a single dominant ALP coupling can be motivated by the underlying global symmetry giving rise to the ALP pseudo Nambu-Goldstone boson. For example, a QCD axion of KSVZ type, where the new heavy quarks are $SU(2)_L\times U(1)_Y$ singlets, can be described by an ALP with a coupling to gluons, $c_{GG}$, at $\Lambda$ and vanishing ALP couplings to all other SM fields at that scale. Alternatively, one could consider a global symmetry which gives rise to an ALP with only lepton couplings in the UV which is completely unrelated to the strong CP problem. Flavor-violating couplings are then induced through RG mixing and matching \eqref{sonice}, arising from the diagrams in Figure~\ref{fig:runningandmatching}, and by the chiral Lagrangian given in \eqref{eq:weakcpt}. 
Horizontal global symmetries, under which the different quark flavors transform differently, typically lead to ALPs with flavor-changing tree-level couplings \cite{Wilczek:1982rv}. These couplings do not respect the MFV structure of the SM and are strongly constrained, as shown in Table~\ref{tab:hadronbounds}.

\begin{figure}[t]
\begin{center}
\includegraphics[width=1.\textwidth]{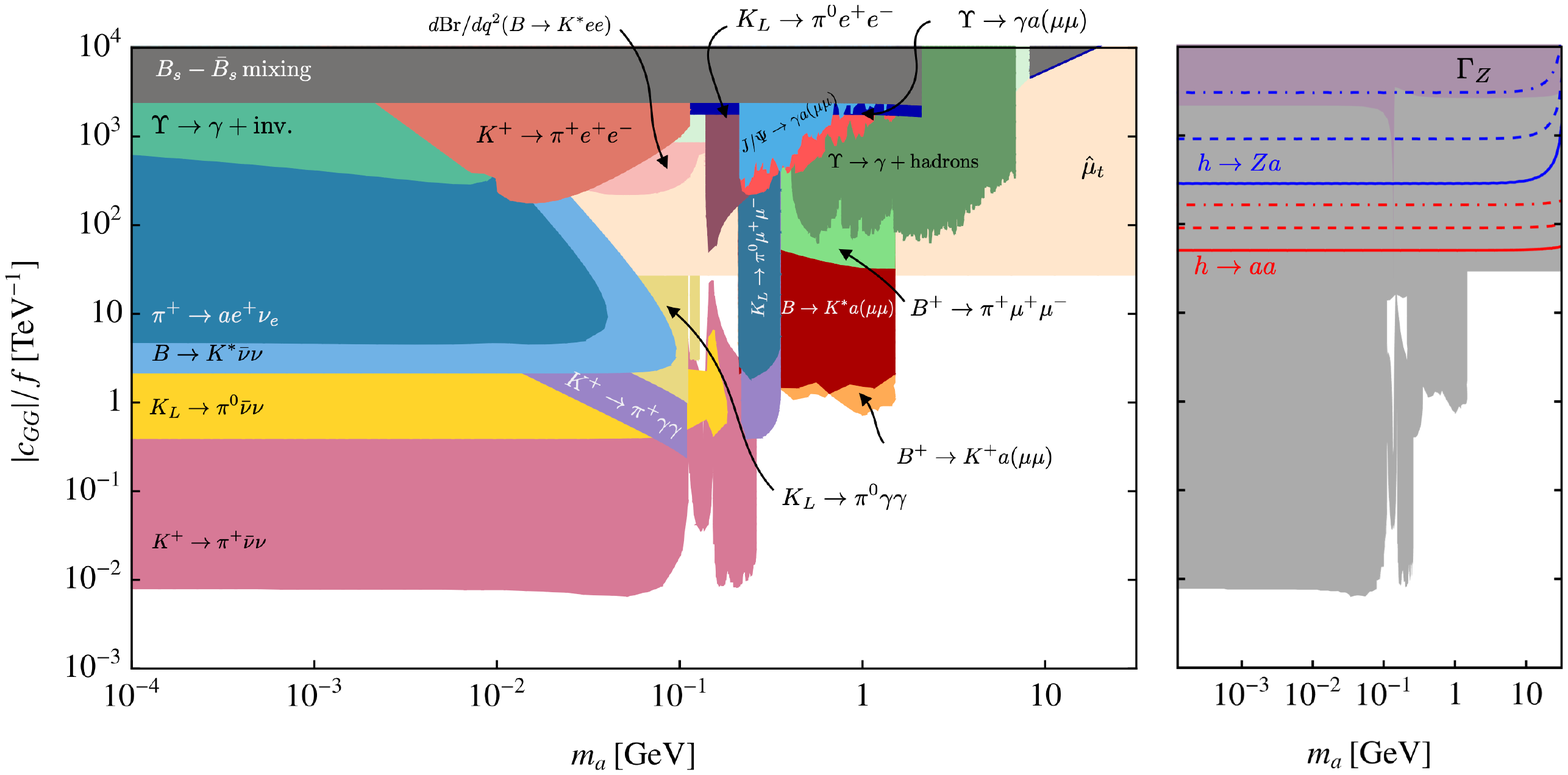}
\caption{Left: Flavor bounds on ALP couplings to gluons with all other Wilson coefficients set to zero at $\Lambda=4\pi f$ and $f=1$\,TeV. Right: Comparison of the same flavor constraints (light gray) with the constraints on $Z\to a\gamma$ decays from the LEP measurement of the $Z$ boson width (violet), contours of constant $\text{Br}(h \to aa)=10^{-1}, 10^{-2} $ and $10^{-3}$, depicted as red dotted, dashed and solid lines, and contours of constant $\text{Br}(h \to Za)=10^{-1}, 10^{-2} $ and $10^{-3}$, shown as blue dotted, dashed and solid lines, respectively. \label{fig:flavorconstraintscGG}}
\end{center}
\end{figure}

\begin{figure}[t]
\begin{center}
\includegraphics[width=1.\textwidth]{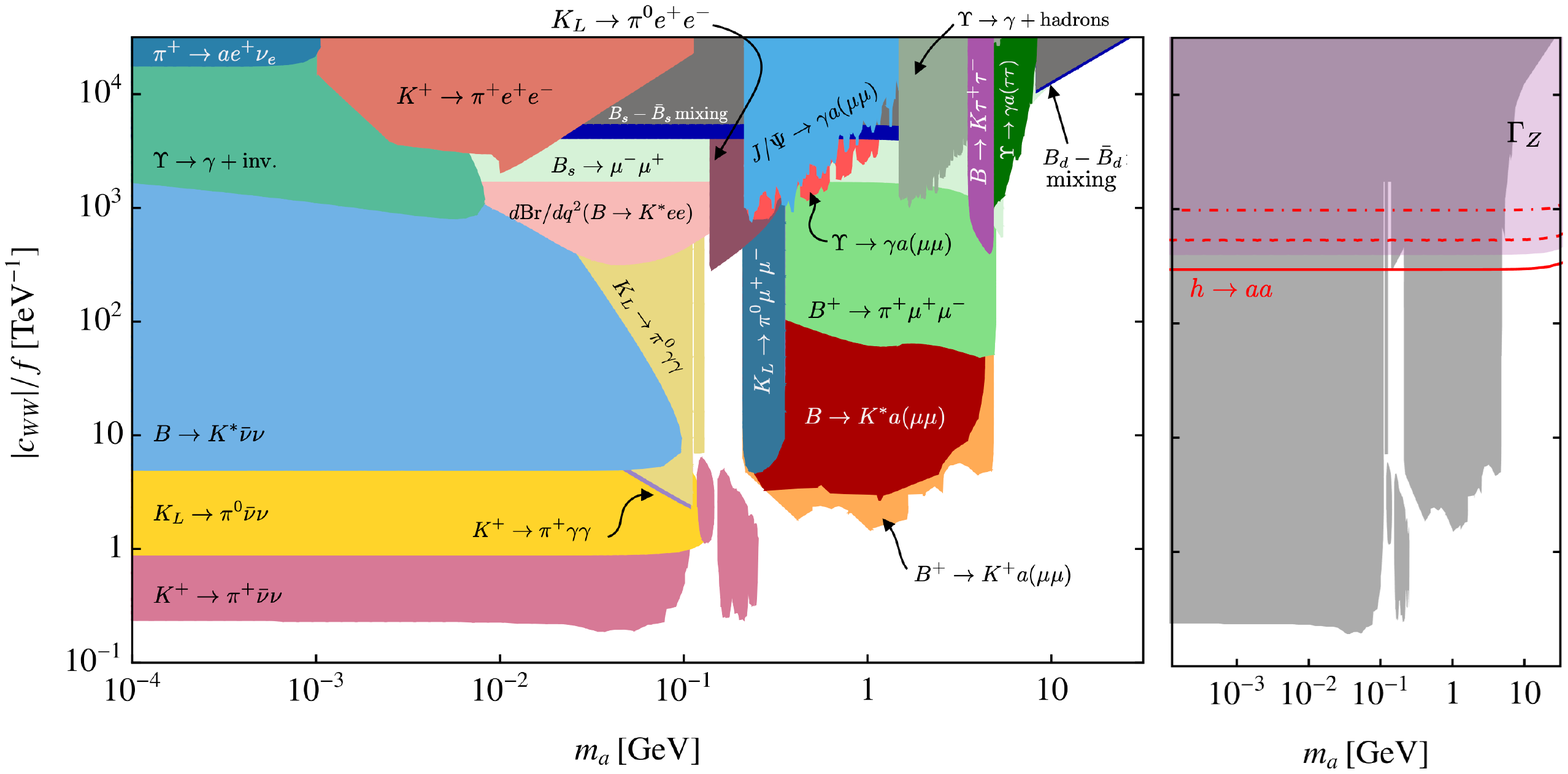}
\caption{Left: Flavor bounds on ALP couplings to $SU(2)_L$ gauge bosons with all other Wilson coefficients set to zero at $\Lambda=4\pi f$ and $f=1$\,TeV. Right: Comparison of the same flavor constraints (light gray) with the constraints on $Z\to a\gamma$ decays from the LEP measurement of the $Z$ boson width (violet) and contours of constant $\text{Br}(h \to aa)=10^{-1}, 10^{-2} $ and $10^{-3}$ depicted as red dotted, dashed and solid lines. \label{fig:flavorconstraintscWW}}
\end{center}
\end{figure}

We first discuss ALPs with couplings to vector bosons in the UV.  The corresponding ALP branching ratios  into SM particles are shown in Figure~\ref{fig:alp_Brs_bosons}. Constraints from flavor observables on an ALP with a single non-vanishing coupling to gluons, $c_{GG}$, at the UV scale $\Lambda$ are shown in Figure~\ref{fig:flavorconstraintscGG}, constraints on an ALP with couplings only to $SU(2)_L$ gauge bosons, $c_{WW}$, at  $\Lambda$ are shown in Figure~\ref{fig:flavorconstraintscWW}, and constraints on an ALP with couplings only to the hypercharge gauge boson, $c_{BB}$, at $\Lambda$ are shown in Figure~\ref{fig:flavorconstraintscBB}. In all cases, the effect of the ALP lifetime has been taken into account by carefully considering the detector layouts and initial state boosts for the different experiments. Details on the ALP lifetime effects are given in Appendix~\ref{app:E}.

The branching ratios of an ALP with couplings to SM fermions are shown in Figure~\ref{fig:alp_Brs_fermions}. 
Constraints for an ALP with a flavor-universal coupling to singlet up-type quarks ($\bm{c_u}=c_u\mathbbm{1}$), singlet down-type quarks ($\bm{c_d}=c_d\mathbbm{1}$) or $SU(2)_L$ quark doublets ($\bm{c_Q}=c_Q\mathbbm{1}$) at the scale $\Lambda$ are shown in Figures \ref{fig:flavorconstraints2}, \ref{fig:flavorconstraints3} and \ref{fig:flavorconstraints4}, respectively. Limits on ALPs with either flavor-universal $SU(2)_L$ doublet ($\bm{c_L}=c_L\mathbbm{1}$) or singlet ($\bm{c_e}=c_e\mathbbm{1}$) lepton couplings are displayed in Figures \ref{fig:flavorconstraints5} and \ref{fig:flavorconstraints6}.

Note that we consider the relatively low UV scale $\Lambda=4\pi f$ with $f=1$\,TeV in all scenarios discussed here. However, the numerical impact of the RG running from a much larger scale $\Lambda$ is small. For example, choosing a UV scale of $f=10^{12}$\,TeV changes the coefficients in \eqref{eq:Knum} and \eqref{eq:cuucdd} by less than one order of magnitude~\cite{Bauer:2020jbp}. This implies that the constraints on the ALP couplings, in the combination $c/f$, derived below depend only weakly on the UV scale and that the exclusion plots would change only minimally for different values of $\Lambda$. 

In the rest of the subsection, we describe the features of the different constraint plots in turn.

%%%
%%%%%%%%%%%%%
%%%%%%%%%%%%
%%%

\subsubsection{ALP coupling to gluons}
First we consider an ALP which only has a coupling to gluons in the UV, i.e., only $c_{GG}$ is non-zero at the scale $\Lambda$. We focus on ALPs with masses $m_a< \mathcal{O}(10)$ GeV, which obtain sizeable couplings to hadrons and photons as well as flavor off-diagonal couplings to down-type quarks from RG evolution. As a result, an ALP which only couples to gluons at the UV scale decays mostly into hadrons for ALP masses above the QCD scale, and dominantly into photons for $m_a< \Lambda_\text{QCD}$, as shown in Figure~\ref{fig:alp_Brs_bosons}. The branching ratios of such an ALP into leptons are $\text{Br}(a\to \ell^+\ell^-)< 1\%$. Since the ALP-gluon coupling for $m_a< \Lambda_\text{QCD}$ induces an order one ALP coupling to photons $c_{\gamma\gamma}(m_a=0)= -(1.92\pm 0.04)\,c_{GG}$, one can estimate the ALP lifetime as $\tau_a\propto 1/(c_{GG}^{2}m_a^{3})$, so that the lifetime exceeds the typical size of the experiment ($\ell_\text{det}\lesssim 10$ m) for $m_a\approx 0.05\,c_{GG}^{3/2}$ GeV, and hence lighter ALPs are more likely to decay outside the detector.  

As a consequence, the strongest bounds for $m_a< m_\pi$ and small $c_{GG}$ arise from the NA62 constraint on $\text{Br}(K^+\to \pi^+ X)$, where $X$ either decays invisibly or escapes the detector, which constrains ALPs long-lived enough to escape the NA62 detector before decaying \cite{CortinaGil:2021nts}. The parameter space excluded by this constraint is shown in pink in Figure~\ref{fig:flavorconstraintscGG} and corresponds to values of $c_{GG}/ f \gtrsim 0.072 /\text{TeV} $.
 Constraints from the neutral mode $K_L\to \pi^0 a$ are considerably weaker, because of the suppression of the CP conserving part in \eqref{eq:KLpi0a} by $\epsilon=2.228\times 10^{-3}$.  
The parameter space ruled out by the KOTO~\cite{Ahn:2018mvc} search for $K_L\to \pi^0 X$ is shown in yellow. Other searches for invisible final states lead to weaker constraints, and we show the excluded parameter space by the measurements of $B\to K^* \nu\bar \nu$ from Belle \cite{Grygier:2017tzo} in light blue, and $\pi^+ \to a e^+ \nu_e$ from the PIENU collaboration in dark blue \cite{Aguilar-Arevalo:2021utl}. 
For larger ALP masses, decays into photons become relevant and constraints from searches for $K^+\to \pi^+ \gamma\gamma$ and $K^0\to \pi^0 \gamma\gamma$ performed at E949, NA48, NA62 and KTeV exclude the parameter space for larger values of $c_{GG}/f$ \cite{Artamonov:2005ru,Ceccucci:2014oza, Lai:2002kf, Abouzaid:2008xm}. The corresponding parameter space is shown in purple and yellow in Figure~\ref{fig:flavorconstraintscGG}.  These searches provide important constraints even for $m_a>2m_e$ when decays to electrons are allowed, because of the dominant ALP branching ratio $\text{Br}(a\to \gamma\gamma)> 99\%$ at $m_a< 3m_\pi$.\footnote{Refs.~\cite{Chakraborty:2021wda,Bertholet:2021hjl} calculate projected limits from Belle II searches for axions decaying into hadronic or photonic final states.}

Leptonic ALP decay channels lead to comparatively weak constraints. The excluded parameter space from the LHCb measurement of $B\to K^* e^+e^-$ decays \cite{Aaij:2015dea} is shown in peach, LHCb searches~\cite{Aaij:2016qsm,LHCb:2015nkv} for the charged and neutral $B$ meson decays $B^+\to K^+ a(\mu^+\mu^-)$ and  $B^0\to K^* a(\mu^+\mu^-)$ provide the dominant constraints for $m_a> 2m_\mu$ and rule out couplings of the order of $c_{GG}/f \gtrsim 1/ \text{TeV} $ 
for ALP masses $m_a< m_B$. The parameter regions excluded by these searches are shown in light orange and red. 
The weaker constraint from the measurement of the $B^+\to \pi^+ \mu^+\mu^-$ decay rate by LHCb is shaded in green in Figure~\ref{fig:flavorconstraintscGG}~\cite{Aaij:2015nea}. For $m_a> m_B$, the dominant constraints come from a search for flavor diagonal ALP production through $\Upsilon\to \gamma + a$ with subsequent $a\to \text{hadrons}$ decays by BaBar~\cite{Lees:2011wb} shown in dark green in Figure~\ref{fig:flavorconstraintscGG}. 

Non-resonant ALP contributions to $B_s-\bar B_s$ meson mixing and $B_s\to \mu^+\mu^-$ decays lead to very weak constraints because they require two flavor-violating ALP couplings or are suppressed by the small ALP-lepton coupling induced by $c_{GG}$. For ALPs with masses $m_a> m_\Upsilon$ the ALP contribution to the chromomagnetic dipole moment of the top quark leads to a universal constraint of $c_{GG}/ f \gtrsim 30 /\text{TeV}$.  
%checked

On the right panel in Figure~\ref{fig:flavorconstraintscGG} we compare all the aforementioned constraints from flavor physics (in gray) with the constraints on $Z\to a\gamma$ decays from the LEP measurement of the $Z$ boson width shown in purple. The red dotted, dashed and solid contours show constant values of $\text{Br}(h \to aa)=10^{-1}, 10^{-2} $ and $10^{-3}$, respectively and the blue dotted, dashed and solid contours show constant values for Higgs decays into $Z$ bosons and ALPs with $\text{Br}(h \to Za)=10^{-1}, 10^{-2} $ and $10^{-3}$, respectively. 

For light ALPs with couplings to gluons, there are very strong constraints from beam dump searches and astrophysical observables below $m_a\leq 100$ MeV. These constraints are shown for an effective coupling of ALPs to photons in Figure~\ref{fig:flavorconstraintsandphotons}. The large values of effective photon couplings induced by the scenario considered here are ruled out in this parameter space by these constraints, and flavor bounds are only competitive for higher ALP masses. A direct comparison as in the case of an ALP coupled to $SU(2)_L$ or $U(1)_Y$ gauge bosons described below is difficult because of the sizable ALP coupling to nuclei induced by $c_{GG}$ which is not taken into account in the derivation of the constraints shown in Figure~\ref{fig:flavorconstraintsandphotons}. An analysis of astrophysical observables considering the presence of ALP couplings to photons and nucleons simultaneously would be very welcome in order to enable a direct comparison of these constraints in the future.

%%%
%%%%%%%%%%%%%
%%%%%%%%%%%%
%%%

\subsubsection[ALP coupling to $SU(2)_L$ gauge bosons]{\boldmath ALP coupling to $SU(2)_L$ gauge bosons}
An ALP with couplings only to $SU(2)_L$ gauge bosons in the UV, so that only $c_{WW}$ is non-zero at the scale $\Lambda$, obtains flavor-diagonal couplings to quarks and charged leptons as well as flavor off-diagonal couplings to down-type quarks through loop diagrams containing a $W$ boson. The coupling $c_{WW}$ induces a tree-level coupling of the ALP to photons which implies a dominant ALP decay width into photons $\text{Br} (a\to \gamma\gamma)\approx 1$ for all of the parameter space we consider. The loop-induced decays into fermions do not exceed $1\%$ for most of the parameter space, as shown in Figure~\ref{fig:alp_Brs_bosons}. The ALP lifetime is therefore well approximated by $\tau_a\propto 1/(c_{WW}^{2}m_a^{3})$, so that the ALP has a decay length of more than $10\,$m for $m_a\approx 0.1\,c_{WW}^{3/2}$\,GeV and hence lighter ALPs are likely to decay outside the detector.  
As for the ALP with a gluon coupling in the UV, the search for $\text{Br}(K^+\to \pi^+ X)$~\cite{CortinaGil:2021nts}, with $X$ decaying invisibly or escaping the NA62 detector, provides the strongest constraint for $m_a< m_\pi$ and small $c_{WW}$.  The parameter space excluded by this constraint is shown in pink in Figure~\ref{fig:flavorconstraintscWW} and excludes values of $c_{WW}/ f \gtrsim 0.25 /\text{TeV} $. 
The bound on $|c_{WW}|/f$ from the equivalent neutral mode search $K_L\to \pi^0 X$ \cite{Ahn:2018mvc} is only about an order of magnitude smaller than the bound from the charged mode, in contrast to the much larger hierarchy between the corresponding bounds on $|c_{GG}|/f$ discussed in the previous subsection. The reason for this is that the $K^+\to \pi^+ a$ and the $K_L\to \pi^0 a$ amplitudes have similar numerical dependences on $c_{WW}$, as discussed in Section~\ref{sec:UVsymmetry}. The three-body decay $\pi^+\to e^+ a \nu_e$ provides only a weak constraint on $|c_{WW}|/f$ compared to the constraint in the scenario with ALP couplings to gluons, because the coupling $c_{WW}$ only enters the amplitude~\eqref{eq:21} through RG effects.  
The constraint from  $B\to K^* \nu\bar \nu$ from Belle \cite{Grygier:2017tzo} is shown in light blue in Figure~\ref{fig:flavorconstraintscWW}. 

For a sizeable $c_{WW}$ coupling, the branching ratio of ALP decays into photon pairs can be large enough to allow prompt ALP decays. Constraints from searches for $K^+\to \pi^+ \gamma\gamma$ and $K^0\to \pi^0 \gamma\gamma$ exclude the parameter space shown in purple and yellow in Figure~\ref{fig:flavorconstraintscWW} \cite{Artamonov:2005ru,Ceccucci:2014oza, Lai:2002kf, Abouzaid:2008xm}. 

Currently, there is no published search for the decay $B \to K^{(*)}  \gamma\gamma$, which would be sensitive to an ALP decaying into photons and could provide an important constraint that would probe the unconstrained parameter space for the mass range $m_K< m_a < m_B$. Our estimate based on the search for $B^+ \to K^+\pi^0\to K^+ \gamma\gamma$ at Belle \cite{Duh:2012ie} and Babar \cite{Lees:2012mma} results in a constraint $c_{WW}/f \lesssim 6/\text{TeV}$. 
We expect that a dedicated search for resonances in this channel could yield much better sensitivity than this estimate, in particular for ALP masses larger than the pion mass.

The ALP-lepton coupling is induced at one-loop, so constraints on $SU(2)_L$-coupled ALPs decaying into leptons are comparatively stronger than for an ALP with only a gluon coupling in the UV.
Above the muon threshold, LHCb searches for the charged and neutral $B$ meson decays $B^+\to K^+ a(\mu^+\mu^-)$ and  $B^0\to K^* a(\mu^+\mu^-)$ therefore provide the dominant constraints~\cite{Aaij:2016qsm,LHCb:2015nkv} and rule out couplings of the order of $c_{WW}/f \gtrsim 2/ \text{TeV} $ 
for ALP masses $m_a< m_B$. The parameter regions excluded by these searches are shown in light orange and red.  

ALPs with stronger couplings are also constrained by the measurement of $B_s \to \mu^+\mu^-$. Radiative $\Upsilon \to \gamma\mu^+\mu^-, \Upsilon \to \gamma\tau^+\tau^-$ and $J/\Psi \to \gamma\mu^+\mu^-$ decays yield constraints for $m_a> 2m_\mu$ and $m_a> 2m_\tau$ respectively, which are of similar strength to the constraint from $B_s \to \mu^+\mu^-$~\cite{Ablikim:2015voa, Lees:2012te}. Even weaker limits arise from the virtual exchange of ALPs in $B$-meson mixing, which is suppressed by two flavor-changing vertices. 

In the right panel of Figure~\ref{fig:flavorconstraintscWW} we compare the aforementioned constraints from flavor observables (in gray) with the constraints on $Z\to a\gamma$ decays from the LEP measurement of the $Z$ boson width, excluding $c_{WW}/f \gtrsim 400 /\text{TeV}$ throughout the ALP mass range. 

The red dotted, dashed and solid contours show constant values of $\text{Br}(h \to aa)=10^{-1}, 10^{-2} $ and $10^{-3}$, respectively, which are mostly ruled out by the width measurement of the $Z$ boson as well.  Higgs decays into $Z$ bosons and ALPs, $h \to a Z $, are not induced by the Wilson coefficient $c_{WW}$. 

%%%
%%%%%%%%%%%%%
\begin{figure}[t]
\begin{center}
\includegraphics[width=1.\textwidth]{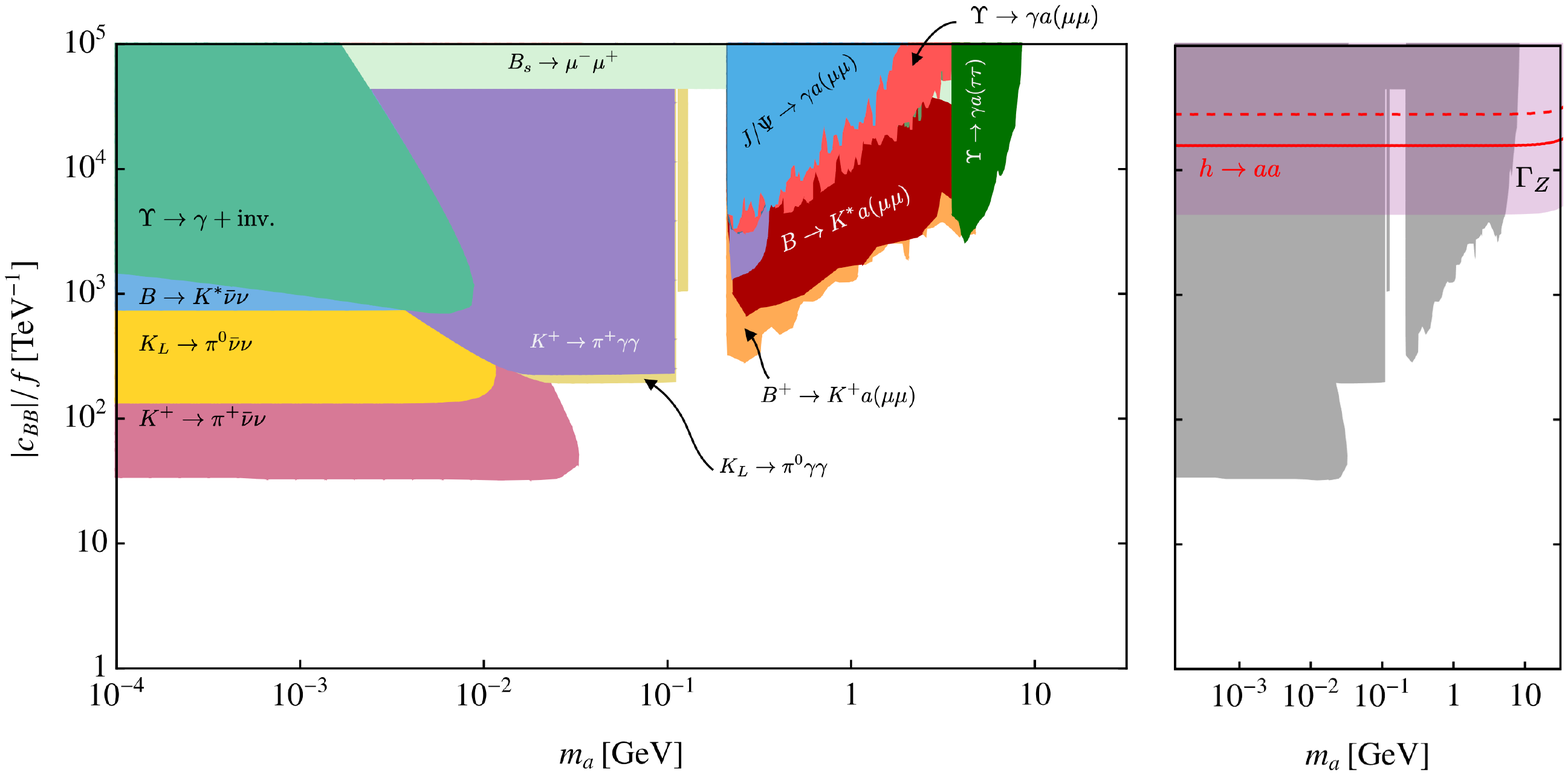}
\caption{Left: Flavor bounds on ALP couplings to the hypercharge gauge boson, with all other Wilson coefficients set to zero at $\Lambda=4\pi f$ and $f=1$\;TeV. Right: Comparison of the same flavor constraints (light gray) with the constraints on $Z\to a\gamma$ decays from the LEP measurement of the $Z$ boson width (violet) and  contours of constant $\text{Br}(h \to aa)=10^{-1} $ and $10^{-2} $ as depicted in red dotted, dashed and solid lines. \label{fig:flavorconstraintscBB}}
\end{center}
\end{figure}
%%%%%%%%%%%%
%%%

\subsubsection{ALP coupling to hypercharge gauge bosons}

Flavor constraints on ALPs with a coupling to hypercharge gauge bosons, $c_{BB}$, at the UV scale $\Lambda$ are shown in Figure~\ref{fig:flavorconstraintscBB}. The constraints are considerably weaker compared to both Figure~\ref{fig:flavorconstraintscGG} and Figure~\ref{fig:flavorconstraintscWW} because an ALP with couplings to the hypercharge gauge boson in the UV does not have any flavor-changing couplings to quarks at one-loop.
Any flavor-changing couplings involving the ALP are generated by two-loop diagrams in which the ALP coupling to top quarks is induced by $c_{BB}$ (see~\eqref{eq:cttatmt}). Similarly to an ALP with couplings to $SU(2)_L$ gauge bosons in the UV, the only tree-level coupling relevant for light ALPs is the coupling to photons. The dominant ALP branching ratio is therefore $\text{Br}(a\to\gamma\gamma)\approx 1$ throughout the parameter space as shown in Figure~\ref{fig:alp_Brs_bosons}.    
The dominant constraint for ALPs with small masses is induced by the search $\text{Br}(K^+\to \pi^+ \bar X)$~\cite{CortinaGil:2021nts} with $X$ decaying invisibly or escaping the NA62 detector, shown in pink in Figure~\ref{fig:flavorconstraintscBB}, which rules out couplings larger than $c_{BB}/ f \gtrsim 30 /\text{TeV}$.
Constraints from searches for $K_L\to \pi^0 X$ by KOTO~\cite{Ahn:2018mvc} shown in yellow and $B\to K^* \nu\bar \nu$ from Belle \cite{Grygier:2017tzo} are significantly weaker.

The constraint from searches for $\Upsilon \to \gamma+\text{invisible}$ shown in turquoise is similar in strength to the corresponding constraint shown in Figure~\ref{fig:flavorconstraintscWW}, because of the contribution of the ALP coupling to photons induced by the righthand diagram of Figure \ref{fig:Upsagamma}. The constraints from kaon decays $K_L\to \pi^0 \gamma \gamma$ and $K^+\to \pi^0 \gamma \gamma$ are the strongest flavor constraints for promptly decaying ALPs with masses up to the muon threshold. For $m_a> 2m_\mu$, the strongest constraints arise from $B^+\to K^+ a(\mu^+\mu^-)$ and  $B^0\to K^* a(\mu^+\mu^-)$, but only rule out very large ALP hypercharge couplings of order $c_{BB}/ f \gtrsim 320 /\text{TeV}$.
ALPs with masses $m_a> m_\Upsilon$ are unconstrained by mesonic observables in this scenario. The ALP coupling to the hypercharge gauge boson induces a rather strong constraint from the $Z$-boson width shown in purple in the right panel of Figure~\ref{fig:flavorconstraintscBB}. Branching ratios of $\text{Br}(h\to aa)=10^{-1} $ and $10^{-2} $ shown by solid and dashed red contours are excluded by this constraint.

\begin{figure}[t]
\begin{center}
\includegraphics[width=1\textwidth]{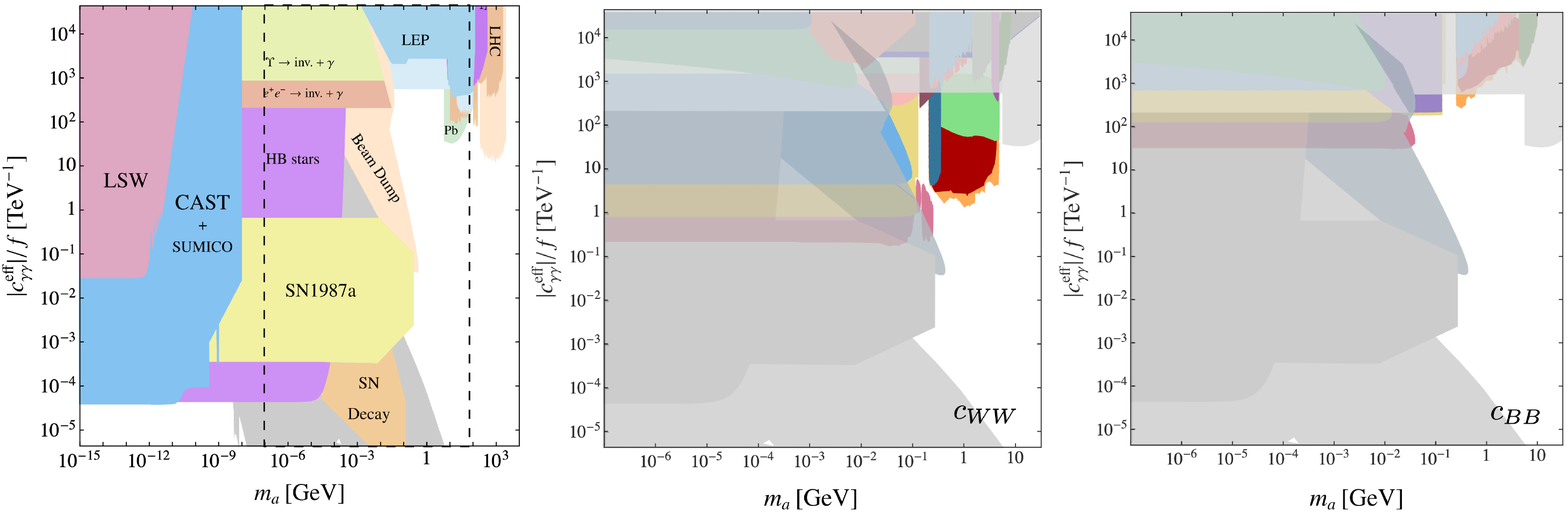}
\caption{Left: Bounds on ALP couplings to photons \cite{Bauer:2017ris}. The dashed contours indicate the part of the plot shown in various shades of gray in the center and right panels. Center and right: In color, we show flavor bounds on ALPs coupling only to $SU(2)_L$ gauge bosons (same as in Figure~\ref{fig:flavorconstraintscWW} above) and the $U(1)_Y$ gauge boson (as in Figure~\ref{fig:flavorconstraintscBB}), respectively. They are compared to the gray astrophysical, beam dump and collider constraints on ALP couplings to photons with $c_{\gamma\gamma}^\text{eff}=c_{WW}$ (center) and $c_{\gamma\gamma}^\text{eff}=c_{BB}$ (right). \label{fig:flavorconstraintsandphotons}}
\end{center}
\end{figure}

It is instructive to compare the constraints from flavor observables with the constraints from 
helioscopes CAST \cite{Arik:2008mq} and SUMICO \cite{Inoue:2008zp,Graham:2015ouw}, 
cosmological and astrophysical observables \cite{Raffelt:1985nk,Raffelt:1987yu,Raffelt:2006cw,Cadamuro:2011fd,Millea:2015qra,Ertas:2020xcc, Depta:2020wmr}, 
the Supernova SN1987a observation \cite{Payez:2014xsa,Jaeckel:2017tud},
collider experiments \cite{Balest:1994ch,delAmoSanchez:2010ac,Mimasu:2014nea,Jaeckel:2015jla,Casolino:2015cza,Knapen:2016moh,Haisch:2016hzu,Brivio:2017ije,Bauer:2017ris,Haisch:2018kqx,BuarqueFranzosi:2021kky} 
and beam dump searches \cite{Riordan:1987aw, Bjorken:1988as,Alekhin:2015byh,Dobrich:2015jyk} for ALPs that couple to photons. 
In Figure~\ref{fig:flavorconstraintsandphotons}, we show the flavor observables superimposed with the results from these searches for the case of an ALP photon coupling given by $c_{\gamma\gamma}^\text{eff}=c_{WW}$ (centre) and $c_{\gamma\gamma}^\text{eff}=c_{BB}$ (right). For light ALPs and very small couplings, bounds from astrophysical observables are much stronger than flavor constraints, and for ALPs with masses $m_a\gtrsim 10$ GeV collider observables are more sensitive. For the case of an ALP with a $c_{WW}$ coupling, flavor observables, in particular B meson decays, constrain precisely the ALP masses and couplings in the ``gap'' for which astrophysical observables and colliders lose sensitivity, because the ALP is too short-lived to be detected in beam-dumps and too light and weakly coupled to be produced and efficiently reconstructed at colliders. This comparison motivates a dedicated search for $B \to K  a$ with subsequent $a\to \gamma\gamma$ decays, which could provide the most sensitive probe of ALPs in the parameter space unconstrained by either astrophysical, beam dump or collider constraints.   \\

%%%
%%%%%%%%%%%%%
%%%%%%%%%%%%
%%%

%%%
%%%%%%%%%%%%%
\begin{figure}[t]
\begin{center}
\includegraphics[width=1.\textwidth]{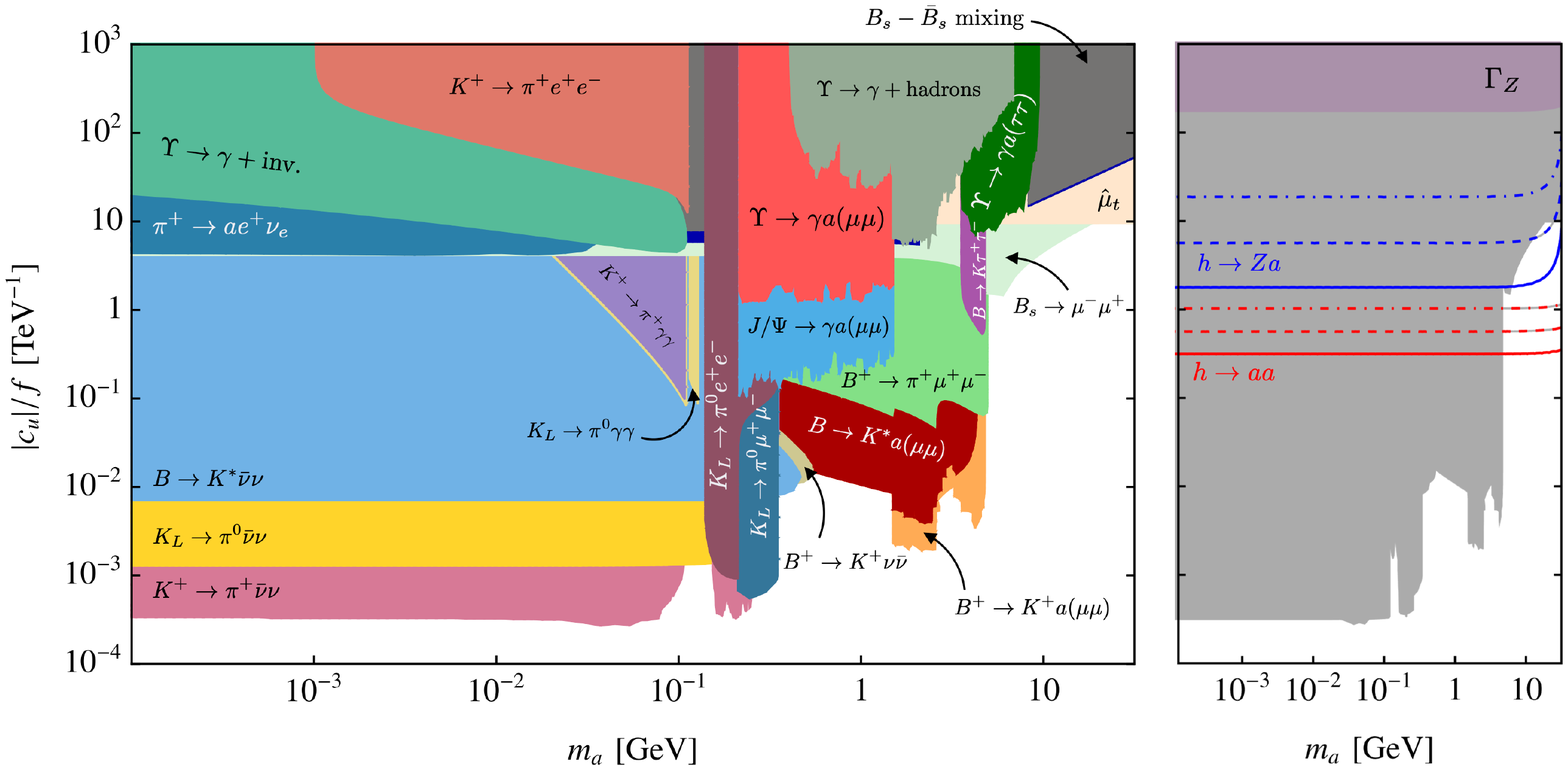}
\caption{Left: Flavor bounds on universal ALP couplings to right-handed up-type quarks with $\bm{c_u}=c_u\mathbbm{1}$, with all other Wilson coefficients set to zero at $\Lambda=4\pi f$ and $f=1$\,TeV. Right: Comparison of flavor constraints (light gray) with the constraint on $Z\to a\gamma$ decays from the LEP measurement of the $Z$ boson width, contours of constant $\text{Br}(h \to aa)=10^{-1}, 10^{-2} $ and $10^{-3}$  depicted as red dotted, dashed and solid lines and contours of constant $\text{Br}(h \to Za)=10^{-1}, 10^{-2} $ and $10^{-3}$ shown as blue dotted, dashed and solid lines, respectively.\label{fig:flavorconstraints2}}
\end{center}
\end{figure}
%%%%%%%%%%%%
%%%

\subsubsection{ALP coupling to right-handed up-type quarks}
In Figure~\ref{fig:flavorconstraints2}, we collect the constraints on ALPs with universal couplings to right-handed up-type quarks $\bm{c_u}(\Lambda)=\bm{k_u}(\Lambda)=c_u \mathbbm{1}$. The branching ratios of such an ALP are given in the upper right panel of Figure~\ref{fig:alp_Brs_fermions}. Leptonic decay channels dominate for ALPs with masses $2m_e<m_a< $ few GeV, because of the sizeable contribution of $c_{tt}$ to all fermionic couplings in \eqref{eq:cffrun}. Above a few GeV, hadronic ALP decay channels and, in particular, the tree-level induced $\text{Br}(a\to c\bar c)$ dominate over all other branching ratios.   
In contrast to the scenarios with ALP couplings to gauge bosons, the ALP-photon coupling is one-loop suppressed. For light ALPs $m_a< 2m_e$, constraints which rely on the ALP escaping the detector are therefore important.

For masses $m_a\lesssim m_\pi$, ALP couplings of $|c_{u}|/f \gtrsim 4\times 10^{-4} / \, \text{TeV}$ 
are excluded by the measurement of $\text{Br}(K^+\to \pi^+ X)$ \cite{CortinaGil:2021nts}, with $X$ decaying invisibly or escaping the NA62 detector shown in pink. Constraints from searches for $K_L\to \pi^0 X$ by KOTO~\cite{Ahn:2018mvc}, shown in yellow, and $B\to K^* \nu\bar \nu$ from Belle \cite{Grygier:2017tzo}, shown in light blue, are weaker. The three-body decay $\pi^+ \to a e^+ \nu_e$ from the PIENU collaboration shown in dark blue \cite{Aguilar-Arevalo:2021utl} only constrains large values of $|c_{u}|/f \gtrsim 5 / \, \text{TeV}$.
Similarly to an ALP with couplings to $SU(2)_L$ gauge bosons, larger couplings $c_{u}$ are excluded by constraints from searches for $K^+\to \pi^+ \gamma\gamma$ and $K^0\to \pi^0 \gamma\gamma$ decays shown in purple and yellow.  While the constraints on $c_{u}$ from $K^+\to \pi^+ \bar \nu \nu$ decays are stronger compared to the constraints on $c_{WW}$ in Figure~\ref{fig:flavorconstraintscWW}, constraints from photon decays are relatively weaker. This is due to the fact that the ALP coupling to photons is generated at the one-loop level (second line of \eqref{eq:CgagaQCD}) or suppressed by $m_a^2/m_pi^2$ (first line of \eqref{eq:CgagaQCD}).
The tree-level couplings to up-type quarks induce sizeable lepton couplings at the low scale \eqref{eq:cffrun} and therefore relatively large leptonic ALP branching ratios explain the strength of the  
constraints from the vector meson decays $\Upsilon \to \gamma a\to \gamma \mu^+\mu^-$, $\Upsilon \to \gamma a\to \gamma \tau^+\tau^-$ and $J/\Psi \to \gamma a\to \gamma \mu^+\mu^-$, which are considerably stronger than constraints from $B$-meson mixing and $B_s\to \mu^+\mu^-$ decays. The dominance of hadronic decay channels above $m_a\gtrsim 1$\,GeV explains why constraints from $K_L\to \pi^0 e^+ e^-$ and  $K_L\to \pi^0 \mu^+ \mu^-$ are stronger than the constraints from $B\to K^* \mu^+\mu^-$ and $B^+\to K^+ \mu^+\mu^-$ decays relevant for larger values of $m_a$, where the branching ratio to muons is correspondingly suppressed. 
Couplings $|c_{u}|/f \gtrsim 6.5 /\,\text{TeV}$ are ruled out throughout the parameter space by the measurement of the chromomagnetic dipole moment of the top quark $\hat\mu_t$, shown in magenta in Figure~\ref{fig:flavorconstraints2}. Constraints of similar strength arise from the contribution of virtual ALP exchange in $B$-meson mixing, even though the measurements are significantly more precise than in the case of the chromomagnetic dipole moment of the top quark, because it requires two loop-induced flavor-changing ALP vertices. In plotting the limits from $B$-meson mixing, we have excluded the parameter space $m_b/2 < m_a < 2m_b$ as discussed in Section~\ref{sec:Bsmixing}.

The horizontal purple region in the right panel of Figure~\ref{fig:flavorconstraints2} indicates the parameter space excluded by    
the contribution of $\Gamma(Z\to a\gamma)$ to the total $Z$ width, $|c_{u}|/f \gtrsim 146 /\,\text{TeV}$,  which represents a significantly weaker constraint relative to the constraints from measurements of flavor transitions compared to Figure~\ref{fig:flavorconstraintscWW}, because the ALP coupling to photons and $Z$- bosons are only induced at one-loop here.
Contours of constant $\text{Br}(h \to aa)=10^{-1}, 10^{-2} $ and $10^{-3}$ are depicted as red dotted, dashed and solid lines, respectively. The ALP coupling to top quarks also induces the exotic Higgs decay $h \to Z a$, and the corresponding contours of constant $\text{Br}(h \to Za)=10^{-1}, 10^{-2} $ and $10^{-3}$ are shown as blue dotted, dashed and solid lines. In contrast to ALPs coupled to $SU(2)_L$ gauge bosons, neither flavor constraints nor the measurement of the $Z$ width exclude large branching ratios for exotic Higgs decays for $m_a \gtrsim 5$\,GeV, but $\text{Br}(h \to Za)\gtrsim 1\%$ is in conflict with the measurement of the chromomagnetic dipole moment of the top quark in this scenario.  \\

%%%
%%%%%%%%%%%%%
\begin{figure}[t]
\begin{center}
\includegraphics[width=1\textwidth]{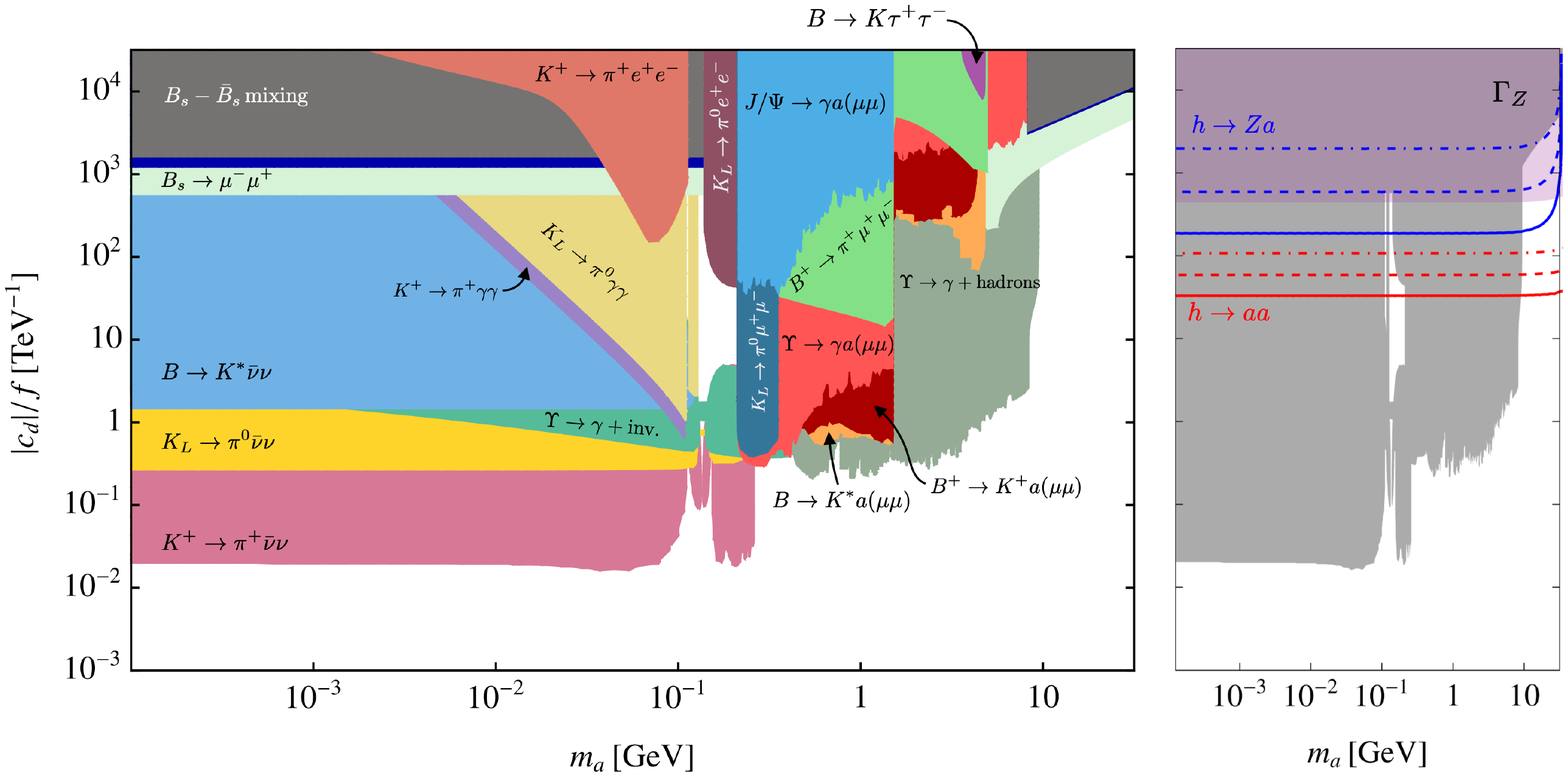}
\caption{Left; Flavor bounds on universal ALP couplings to down-type quarks with $\bm{c_d}=c_d\mathbbm{1}$, with all other Wilson coefficients set to zero at $\Lambda=4\pi f$ and $f=1$\,TeV. Right: Constraints from flavor observables (light gray) are compared to the constraint on $Z\to a\gamma$ decays from the LEP measurement of the $Z$ boson width. Contours of constant $\text{Br}(h \to aa)=10^{-1}, 10^{-2} $ and $10^{-3}$ are depicted as red dotted, dashed and solid lines, respectively. Contours of constant $\text{Br}(h \to Za)=10^{-1}, 10^{-2} $ and $10^{-3}$ are shown as blue dotted, dashed and solid lines, respectively. \label{fig:flavorconstraints3}}
\end{center}
\end{figure}
%%%%%%%%%%%%
%%%

\subsubsection{ALP coupling to right-handed down-type quarks}
For universal ALP couplings to right-handed down-type quarks $\bm{c_{d}}(\Lambda)=\bm{k_d}(\Lambda)=c_d \mathbbm{1}$, the ALP branching ratios are shown in the lower left panel of Figure~\ref{fig:alp_Brs_fermions} and the constraints from flavor observables are shown in Figure~\ref{fig:flavorconstraints3}. Since only flavor-universal couplings to down-quarks are present at $\Lambda$, flavor-violating couplings of the ALP to down-type quarks \eqref{eq:FVveryshort} are only generated by RG evolution via $\tilde c_{GG}$ and $\tilde c_{BB}$ \eqref{cVVtildesimple} or from the chiral Lagrangian in \eqref{eq:AKplus} and \eqref{eq:AK0}.
Flavor-violating ALP couplings are therefore 2-loop effects. Relatively strong limits result from searches for invisible ALPs because of the lack of a tree-level ALP coupling to photons. The bounds from the search for $\text{Br}(K^+\to \pi^+ X)$\cite{CortinaGil:2021nts}, with invisible $X$  by NA62 is shown in pink, the $K_L\to \pi^0 \nu \bar \nu$ limit by KOTO shown is yellow \cite{Ahn:2018mvc}, and the bound from $B\to K^* \nu\bar \nu$ decays observed by Belle \cite{Grygier:2017tzo} shown in light blue.

Constraints from $B$-meson mixing that require two flavor-changing ALP couplings are almost irrelevant in this scenario and in fact all flavor constraints that rely on a down-type flavor-changing transition are substantially weaker compared to the scenarios that allow for ALP couplings to up-type quarks at the UV scale. ALP decays into photons are mediated at one-loop, whereas ALP-lepton couplings are two-loop effects. As a result, observables with photon final states such as $K^+\to \pi^+ \gamma\gamma$ and $K_L\to \pi^0 \gamma\gamma$ are stronger relative to other constraints compared to the scenario in which ALPs couple through $c_u$ in the UV. 
Radiative $\Upsilon$ decays lead to important constraints because of the tree-level coupling of the ALP to $b$-quarks. Searches for resonances in $\Upsilon \to \gamma +\text{invisible}$ \cite{Lees:2012iw} and $\Upsilon \to \gamma +\text{hadrons}$~\cite{Lees:2011wb} by BaBar provide the strongest limit for ALPs with masses $m_a\gtrsim m_\pi$. The corresponding decays of $J/\Psi\to \gamma a$ are strongly suppressed because of the small ALP coupling to charm quarks induced only by RGE effects.
Couplings below $|c_{d}|/f \lesssim 10^{-2} / \text{TeV}$ are almost unconstrained by flavor observables. This does not mean that this parameter space is unconstrained in this scenario. Astrophysical and cosmological constraints, such as energy loss of red giants \cite{Raffelt:1985nk, Raffelt:1987yu, Raffelt:2006cw} and supernova observations \cite{Payez:2014xsa,Jaeckel:2017tud,Chang:2018rso,Ertas:2020xcc,Camalich:2020wac} are sensitive to long-lived particles with couplings to photons or nuclei and lead to strong constraints for $m_a<m_\pi$. We leave it to future work to quantify these constraints in this particular scenario.

The contribution of $\Gamma(Z\to a\gamma)$ to the total $Z$ width results in the constraint $|c_{d}|/f \gtrsim 442 / \text{TeV}$. The excluded parameter space is shown in the right panel of Figure~\ref{fig:flavorconstraints3}. Higgs decays are strongly suppressed for ALP couplings to down-type quarks, because the amplitudes are proportional to the Yukawa coupling of the $b$-quark. 
The corresponding sensitivity on $h\to aa$ and $h\to aZ$ are therefore orders of magnitude weaker compared to Figure~\ref{fig:flavorconstraints2}.

%%%
%%%%%%%%%%%%%
\begin{figure}[t]
\begin{center}
\includegraphics[width=1.\textwidth]{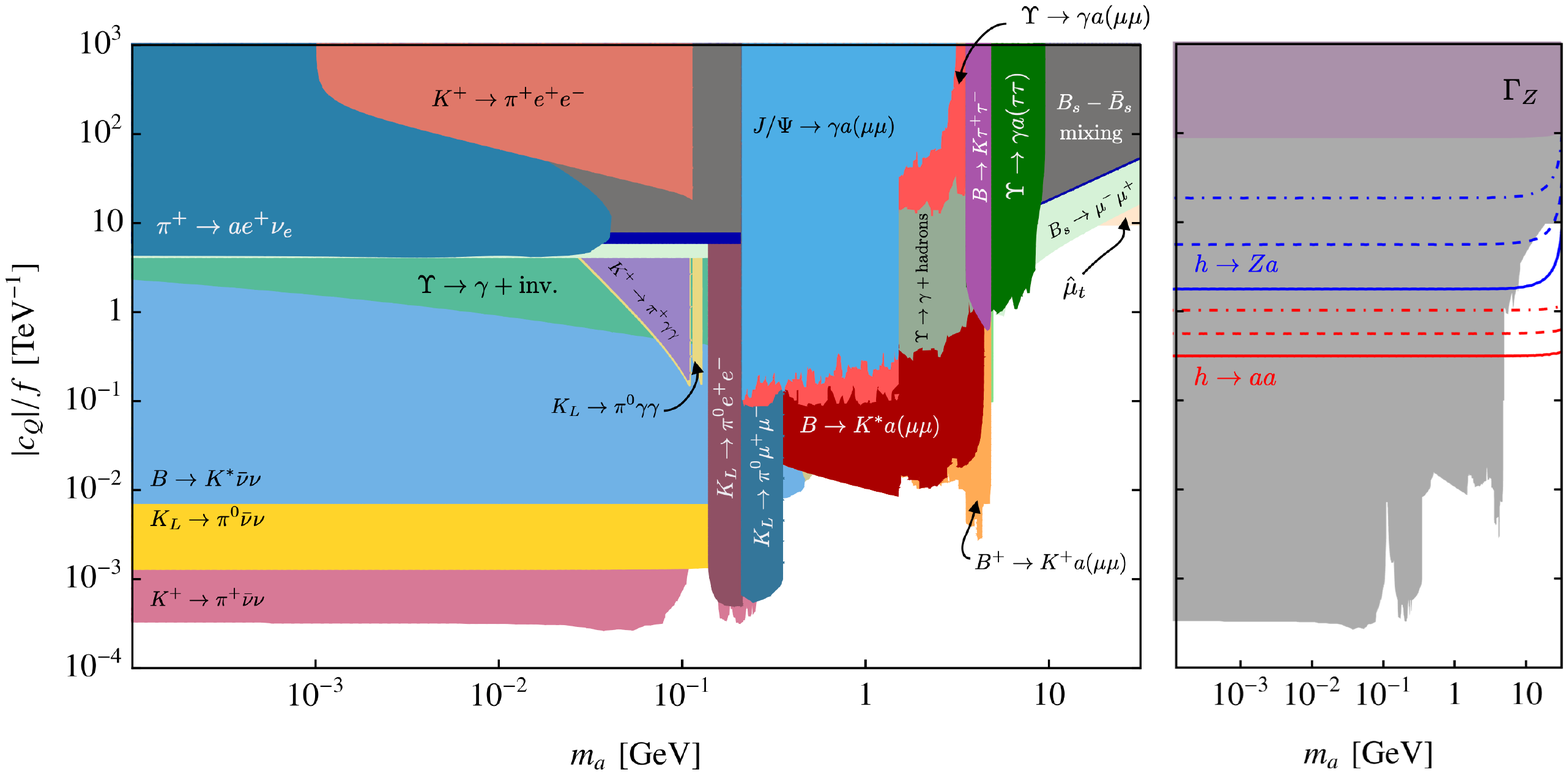}
\caption{Left: Flavor bounds on universal ALP couplings to quark doublets with $\bm{c_Q}=c_Q\mathbbm{1}$, and all other Wilson coefficients set to zero at the scale $\Lambda=4\pi f$ and $f=1$\,TeV. Right:  Constraints from flavor observables (light gray) are compared to the constraint from $Z\to a\gamma$ decays from the LEP measurement of the $Z$ boson width. Contours of constant $\text{Br}(h \to aa)=10^{-1}, 10^{-2} $ and $10^{-3}$ are depicted as red dotted, dashed and solid lines, respectively. Contours of constant $\text{Br}(h \to Za)=10^{-1}, 10^{-2} $ and $10^{-3}$ are shown as blue dotted, dashed and solid lines, respectively.\label{fig:flavorconstraints4}}
\end{center}
\end{figure}
%%%%%%%%%%%%
%%%

\subsubsection{ALP coupling to left-handed quark doublets}
Universal ALP couplings to quark doublets, $\bm{c_{Q}}(\Lambda)=\bm{k_U}(\Lambda)=\bm{k_D}(\Lambda)=c_Q \mathbbm{1}$, lead to the ALP branching ratios shown in the upper left panel of Figure~\ref{fig:alp_Brs_fermions} and constraints from flavor observables shown in Figure~\ref{fig:flavorconstraints4}. In this scenario, the ALP branching ratios are very similar to the case in which only couplings to right-handed up-quarks are present in the UV, apart from the $a\to b\bar b$ decay rate which dominates for $m_a> 2m_b$ here. As a result, similar constraints to those seen in both Figure~\ref{fig:flavorconstraints2} and Figure~\ref{fig:flavorconstraints3} appear in Figure~\ref{fig:flavorconstraints4}, because ALP couplings to both down-type and up-type quarks are present. There are however some important differences with respect to ALPs coupling only to right-handed up- or down-quarks.

%%
%%%%%%%%%%%%
\begin{figure}[t]
\begin{center}
\includegraphics[width=1.\textwidth]{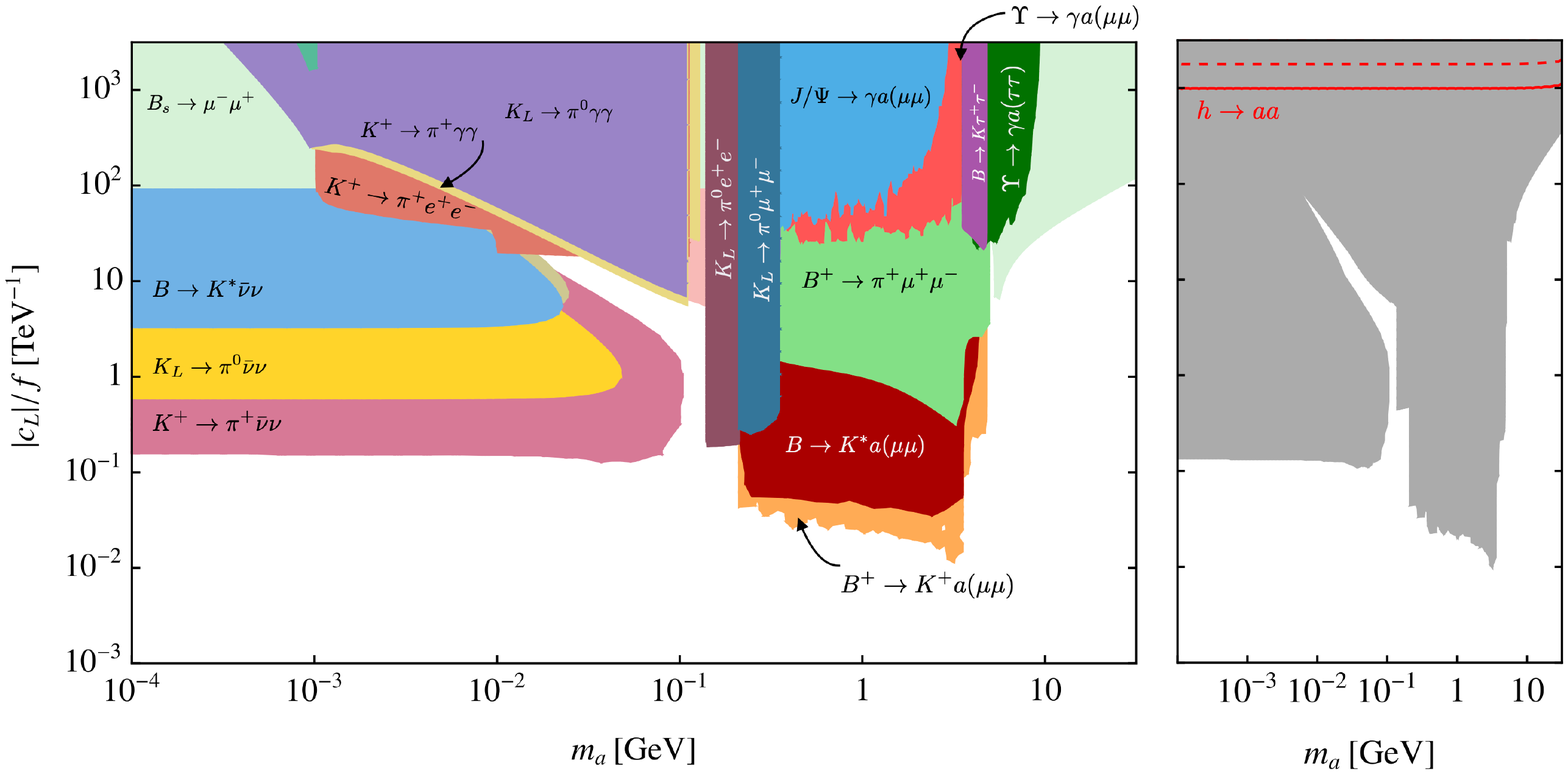}
\caption{Left: Flavor bounds on universal ALP couplings to lepton doublets with $\bm{c_L}=c_L\mathbbm{1}$, and all other Wilson coefficients zero at the scale $\Lambda=4\pi f$ and $f=1$ TeV. Right:  Contours of constant $\text{Br}(h \to aa)=10^{-1}, 10^{-2} $ and $10^{-3}$ are depicted as red dotted, dashed and solid lines, respectively. Contours of constant $\text{Br}(h \to Za)=10^{-1}$ and $ 10^{-2}$ are shown as blue dashed and solid lines, respectively. \label{fig:flavorconstraints5}}
\end{center}
\end{figure}
%%%%%%%%%%%
%%

%%%
%%%%%%%%%%%%%
\begin{figure}[t]
\begin{center}
\includegraphics[width=1.\textwidth]{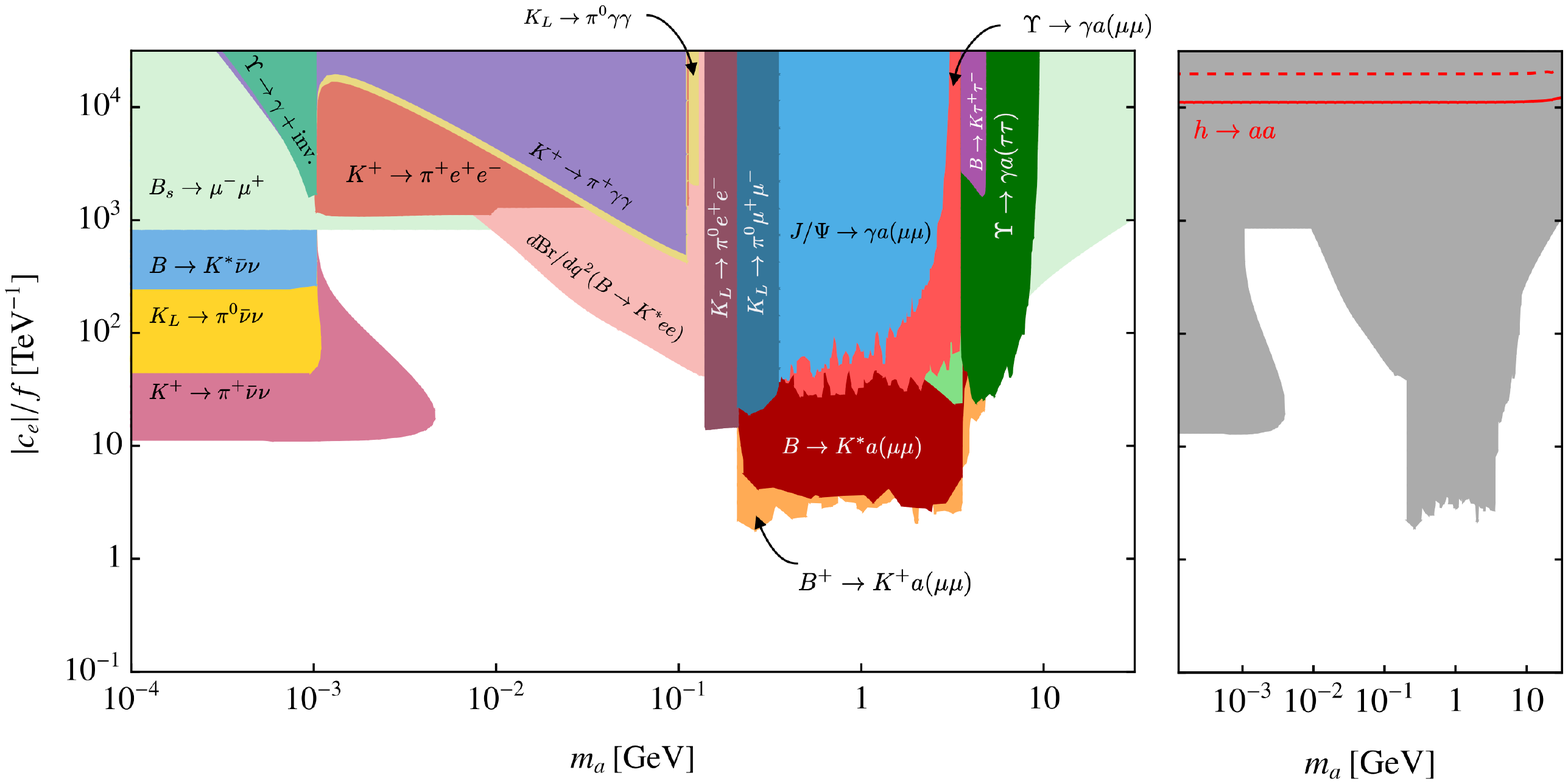}
\caption{Left: Flavor bounds on universal ALP couplings to lepton doublets with $\bm{c_e}=c_e\mathbbm{1}$, and all other Wilson coefficients zero at the scale $\Lambda=4\pi f$ and $f=1$ TeV. Right:  Contours of constant $\text{Br}(h \to aa)=10^{-1}$ and $10^{-2} $ are depicted as red dashed and solid lines, respectively.\label{fig:flavorconstraints6}}
\end{center}
\end{figure}
%%%%%%%%%%%%
%%%
In this scenario, any isospin violating effect is a consequence of running and matching from $\Lambda$ to the scale of the measurement. The ALP coupling to photons \eqref{eq:CgagaQCD}, when induced by isospin conserving ALP couplings such as $c_Q$, is proportional to the isospin breaking term $(m_d-m_u)/(m_u+m_d)\approx 0.35$ and therefore suppressed compared to a scenario where the ALP has isospin breaking ALP couplings, $c_u$ or $c_d$. The decay width $\Gamma(a\to \gamma\gamma)$ is thus suppressed, the ALP branching ratios into leptons are larger and the corresponding constraints from, e.g., $K_L\to \pi^0e^+e^-$ are slightly stronger compared to scenarios in which the ALP has isospin breaking couplings in the UV. Since $c_Q$ couples the ALP to both left-handed up-type and down-type quarks in the UV, constraints from $J/\Psi$ and $\Upsilon$ decays are comparable to the $c_u$ and $c_d$ scenarios.

The constraint from $Z\to a \gamma$ is slightly stronger than for ALP couplings to right-handed up and down-type quarks, because all flavors contribute to the loop-induced coupling, whereas the sensitivity to Higgs decays into ALPs is the same as in Figure~\ref{fig:flavorconstraints2}.

\subsubsection{ALP coupling to leptons} \label{sec:ALPtolep}
Constraints on ALPs with universal couplings to lepton doublets and singlets are shown in Figure~\ref{fig:flavorconstraints5} and  Figure~\ref{fig:flavorconstraints6}, respectively. For these scenarios, we set either $\bm{c_L}=c_L\mathbbm{1}$ or $\bm{c_e}=c_e\mathbbm{1}$, with couplings to all other SM fields set to zero at the scale $\Lambda=4\pi f$ with $f=1$\,TeV. In these scenarios ALPs dominantly decay into leptons if kinematically allowed, or into photons if $m_a< 2m_e$, as shown in the bottom right panel of Figure~\ref{fig:alp_Brs_fermions}. Hadronic ALP decay modes are irrelevant, because ALP couplings to quarks are suppressed by at least two loops.
As a consequence, only observables with ALP decays into lepton and photon final states are sensitive for the parameter space shown in Figure~\ref{fig:flavorconstraints5} and Figure~\ref{fig:flavorconstraints6}.  ALPs with couplings to lepton doublets induce quark flavor-changing amplitudes at the two-loop level. Due to the normalisation of the ALP gauge boson couplings, this leads to constraints on $|c_L|/f$ similar in strength to the constraints on $|c_{WW}|/f$ in Figure~\ref{fig:flavorconstraintscWW}. For ALP couplings to right-handed leptons, these 2-loop contributions are absent and quark flavor-changing transitions are generated by the $\tilde c_{BB}$ contribution entering the amplitude through RG running~\eqref{eq:FVveryshort}. This leads to universally weaker constraints from all observables sensitive to flavor-changing ALP couplings in Figure~\ref{fig:flavorconstraints6} compared to Figure~\ref{fig:flavorconstraints5}. 
In  contrast, meson decays in which the ALP is produced through flavor-conserving couplings to quarks, such as $J/\Psi$ and $\Upsilon$ decays, are equally sensitive to both scenarios. In both cases, all constraints allow values of $|c_L|/f, |c_e|/f<0.01/$TeV for all ALP masses. Exotic Higgs decays are more sensitive to ALP couplings to lepton doublets, which result in larger values of $c_{tt}$ at the electroweak scale. The measurement of the $Z$ decay width does not provide a strong bound, because of the suppressed lepton coupling to $Z$ bosons.

%%%
%%%%%%%%%%%%%
\begin{figure}[t]
\begin{center}
\includegraphics[width=1\textwidth]{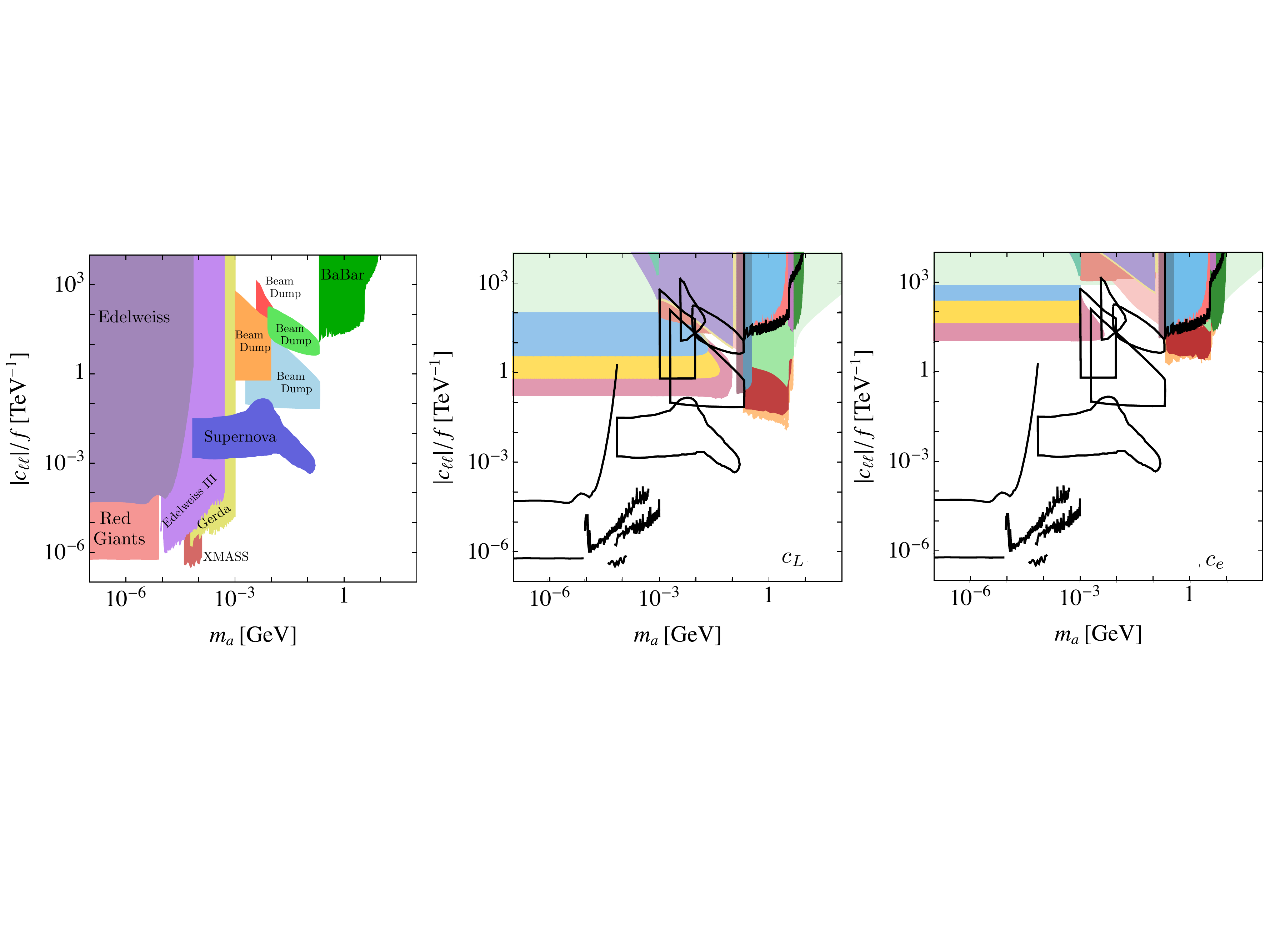}
\caption{Left: Astrophysical, beam dump, and collider constraints on ALP couplings to leptons $c_{\ell\ell}=c_{e}-c_L$ (see text for further details).  Center and right: In color, we show flavor constraints on ALPs coupling to either $SU(2)_L$ lepton doublets (central, as in Figure~\ref{fig:flavorconstraints5} above) and lepton singlets (right, as in Figure~\ref{fig:flavorconstraints6}), respectively. For easy comparison, the black contours depict the bounds from the left panel. \label{fig:flavorconstraintsandleptons}}
\end{center}
\end{figure}
%%%%%%%%%%%%
%%%

Finally, we compare the constraints from flavor observables with the constraints from cosmological observables, collider and beam dump searches for ALPs that couple to leptons in Figure~\ref{fig:flavorconstraintsandleptons}. The constraints in the left panel are: 
searches by the Edelweiss and Edelweiss III collaborations (dark and light purple respectively)~\cite{Armengaud:2013rta, EDELWEISS:2018tde} for ALPs produced in the Sun; 
observations of red giants (red)~\cite{Raffelt:2006cw}; 
searches by the neutrinoless double-beta decay experiment GERDA \cite{GERDA:2020emj};
searches by dark matter direct detection experiment XMASS (red-brown) \cite{XMASS:2018pvs};
beam dump searches at KEK, SLAC and Fermilab in orange~\cite{Konaka:1986cb}, lighter blue, light green~\cite{Essig:2010gu} and red~\cite{Riordan:1987aw,Bross:1989mp}; 
SN1987A supernova bounds (dark blue) \cite{Lucente:2021hbp} 
and a dark photon search at BaBar (green)~\cite{BaBar:2014zli}. Note that the light green beam dump constraint assumes the presence of ALP-muon and ALP-electron couplings while the BaBar bound applies only to ALP-muon couplings. All other constraints have been derived for the ALP-electron coupling. The ALP-tau coupling still remains unconstrained. In this section we assume $c_{ee} = c_{\mu \mu} = c_{\tau\tau}$ and show the combined experimental constraints in the left panel of Figure~\ref{fig:flavorconstraintsandleptons}.
For comparison these constraints are then overlaid with the flavor bounds on ALPs coupling only to $SU(2)_L$ lepton doublets (as in Figure~\ref{fig:flavorconstraints5} above) and lepton singlets (as in Figure~\ref{fig:flavorconstraints6}), respectively. It can be seen that flavor constraints can provide competitive and complementary constraints on ALP couplings to leptons in the MeV-GeV mass range, particularly when the ALP couples to left-handed leptons at the UV scale $\Lambda$. Astrophysical constraints dominate at smaller values of $m_a$.

\subsection{ALPs and low-energy anomalies}
\label{sec:Anomalies}
%%%%%%%%%%%%%%%%%%%%%%%%%%%%%

Various measurements of quark flavor-changing transitions show deviations from the SM predictions. Here we discuss whether an ALP can explain hints of lepton flavor universality violation in $b\to s$ transitions, the excess observed in the excited Beryllium and Helium decays, $^8\text{Be}^*\to\,^8\text{Be}+ e^+ e^-$ and $^4\text{He}^* \to {^4\text{He}} + e^+\!e^-$, or the longstanding KTeV anomaly in $\pi^0 \to e^+ e^-$.

%%%%%%%%%%%%%%%%%%%%%%%%%%%%%
\subsubsection[Anomalies in rare $B$ decays]{\boldmath Anomalies in rare $B$ decays}
\label{sec:Banomalies}
%%%%%%%%%%%%%%%%%%%%%%%%%%%%%

%%
%%%%%%%%%%%%
\begin{figure}[t]
\begin{center}
\includegraphics[width=0.45\textwidth]{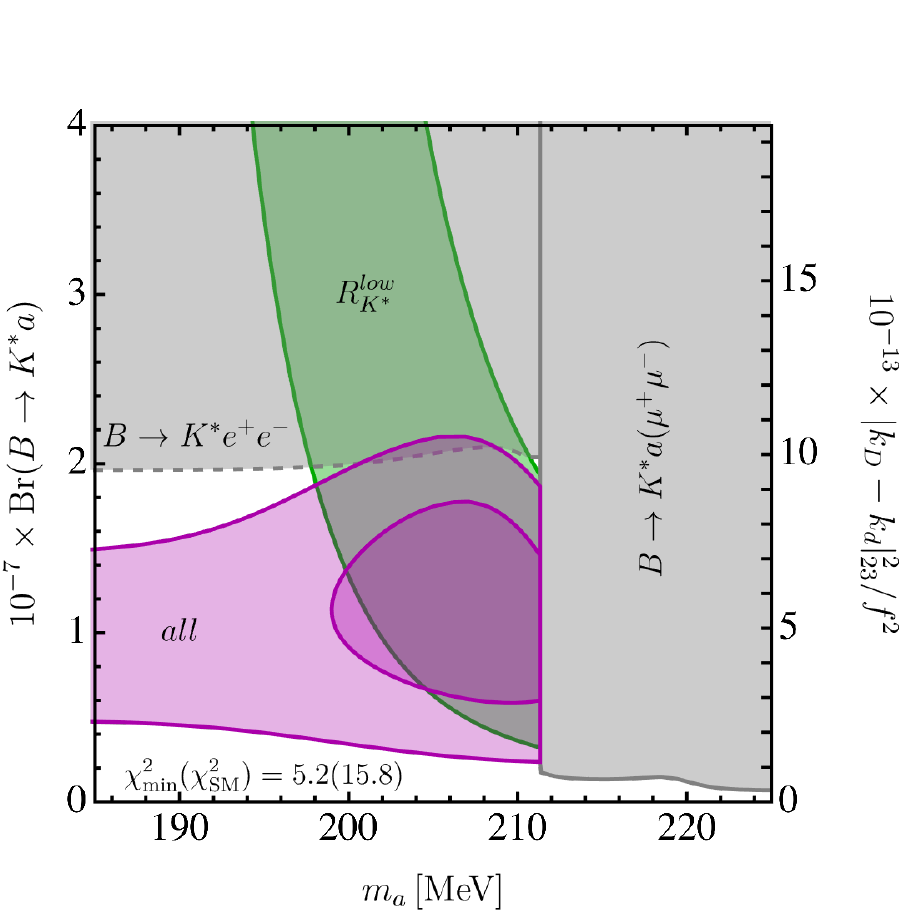} \label{Fig:RKstar}
\caption{The parameter space where a light ALP resonance with flavor universal couplings $c_{ee}/f=c_{\mu\mu}/f=c_{\tau\tau}/f$ can explain the low-$q^2$ bin of the $R_{K^*}$ measurement at $68.27\%\,$CL (green).  The bounds from $B \to K^* e^+ e^-$ \cite{Aaij:2015dea} (dashed) and from searches for peaks in the di-muon mass spectrum \cite{Aaij:2015tna} (solid) are shown at $95\%\,$CL in grey. The preferred regions of $\Delta\chi^2 = \chi^2 - \chi^2_\text{min}$ corresponding to $68.27\%\,$CL and $95.45\%\,$CL are shown in light and dark purple, respectively. \label{fig:RKstar}}
\end{center}
\end{figure}
%%%%%%%%%%%%
%%%

The ratios of two neutral-current $B$ meson decays have been measured by the LHCb collaboration to be (where $q^2$ is the invariant mass squared of the final state lepton pair)
\begin{align}
R_K &= \frac{\text{Br}(B^+ \to K^+ \mu^+ \mu^-)}{\text{Br}(B^+ \to K^+ e^+e^-)}=0.846^{\,+0.042\,\, + 0.013}_{\,-0.039\,\,-0.012}\,\qquad \text{for}\,\, 1.1 \,\text{GeV}^2 < q^2 <6 \,\text{GeV}^2 \,,\quad\text{\cite{Aaij:2021vac}}\\
R_{K^*} &= \frac{\text{Br}(B^0 \to K^{*0} \mu^+ \mu^-)}{\text{Br}(B^0 \to K^{*0} e^+e^-)}=\begin{cases}0.66^{\,+0.11}_{\,-0.07}\pm0.03\,\qquad \text{for}\,\, 0.045 \,\text{GeV}^2 < q^2 < 1.1 \,\text{GeV}^2 \\0.69^{\,+0.11}_{\,-0.07}\pm0.05\,\qquad \text{for}\,\, 1.1 \,\text{GeV}^2 < q^2 <6 \,\text{GeV}^2 \end{cases}\!\!\!\!\!\!\!\!\!\!\,,\quad \text{\cite{Aaij:2017vbb}}\label{eq:RKstar}
\end{align}
which deviate from the SM expectation by $3.1\sigma$ ($R_K$), $2.3\sigma$ ($R_K^*$ low $q^2$-bin) and $2.5\sigma$ ($R_K^*$ high $q^2$-bin), respectively.
Overall, these measurements seem to indicate a deviation from the SM prediction of lepton flavor universality.

\paragraph{Heavy ALPs}
In principle, ALPs could address this discrepancy as they can mediate the decays $B \to K^{(*)} \ell^+\ell^-$ with different interaction strengths for $\ell=e$ and $\ell=\mu$. ALP couplings to different lepton flavors are naturally non-universal due to the fact that the ALP-fermion coupling in \eqref{masssuppression} is proportional to the corresponding fermion masses after using the equations of motion. 
A heavy ALP ($m_a^2\gg q^2$) can, in principle, provide an explanation of $R_K$ for \cite{Bobeth:2007dw, Hiller:2014yaa}  
\begin{align}\label{eq:RKcond}
11 \lesssim \, \text{Re}[C^\mu_{P+}] - \big|C^\mu_{P+}\big|^2 
+ \big|C^e_{P+}\big|^2
\lesssim 20\,,
\end{align}
where $C^\ell_{P+} = C^\ell_{P} + C^{\ell \prime}_{P}$ and $C^\ell_{P} $ is the Wilson coefficient corresponding to the operator $O_{P}=\bar s_R b_R\,\bar\ell\gamma_5\ell$ in the normalisation of \ref{eq:Hbtosll}. $C^{\ell \prime}_{P}$ denotes the Wilson coefficient of the chirality flipped operator. Matching onto our notation we find
\begin{align}
C^\ell_{P+} = c_{\ell \ell} \frac{\pi}{\alpha(\mu_b)} \frac{v^2}{f^2} \frac{2 m_\ell m_b}{m_a^2} \frac{[k_D+k_d]_{32}}{V_{ts}^*\,V_{tb}} \,.
\end{align}
The ALP-muon coupling alone can not explain the discrepancy in $R_K$, as it contributes with the wrong sign~\cite{Hiller:2014yaa}. For an ALP coupling only to electrons the tension can be explained if
\begin{align}\label{eq:RKcondee}
2.7 \times 10^{-3} \, \text{TeV}^{-2} \lesssim
\bigg|  \frac{c_{ee}[k_D+k_d]_{32}}{f^2} \frac{m_e m_b}{m_a^2} \,\bigg|
\lesssim 3.7 \times 10^{-3} \, \text{TeV}^{-2}\,.
\end{align}
The limit on $\text{Br}(B_s\to e^+e^-)< 9.4 \times 10^{-9}$ \cite{Zyla:2020zbs}, however, results in the constraint
\begin{align}\label{eq:Bsee}
\bigg|  \frac{c_{ee}[k_D-k_d]_{32}}{f^2} \frac{m_e m_b}{m_a^2} \,\bigg|
\lesssim 1.9 \times 10^{-4}  \, \text{TeV}^{-2} \,.
\end{align}
An explanation of $R_K$ is thus only possible if $[k_D]_{32}$ and $[k_d]_{32}$ are similar in size and mostly cancel in $\text{Br}(B_s\to e^+e^-)$. To achieve this we must assume tree-level flavor-violating couplings; in the scenarios discussed in Section~\ref{sec:discussion1}, where flavor off-diagonal couplings are induced at one-loop, only $[k_D]_{32}$ is non-zero. The ratio $R_{K^*}$, on the other hand, is a function of  $[k_D-k_d]_{32}$ and cannot be explained at all by a heavy ALP, given the constraint~\eqref{eq:Bsee}. 

The coupling combination $[k_D+k_d]_{32}$ entering \eqref{eq:RKcondee} can be constrained by $B_s-\bar B_s$ mixing. Assuming that $[k_D-k_d]_{32} =0$ and demanding that the ALP contribution to  $B_s-\bar B_s$ mixing in \eqref{eq:numericalBsmixingexpression} does not exceed the $2\sigma$ limit gives 
\begin{align}
\frac{\big|[k_D+k_d]_{32}\big|}{f} \leq 0.017\, \text{TeV}^{-1} \,.
\end{align}
for $m_a = 10\,$GeV. For a value of $[k_D+k_d]_{32}$ which saturates this bound, it follows from \eqref{eq:RKcondee} that an explanation of $R_K$ requires an ALP coupling to electrons in the range 
\begin{align}
6.5 \times 10^{3}\,\text{TeV}^{-1}  \lesssim \frac{|c_{ee}|}{f} \lesssim 8.8 \times 10^{3}\, \text{TeV}^{-1} \,.
\end{align}
The required coupling values get even larger with increasing ALP mass.
Such large values are however in conflict with the measurement of the anomalous magnetic moment of the electron, see the discussion in Section~\ref{sec:gminus2}. 
As a result,  an explanation of $R_K$ by an ALP coupling only to electrons is ruled out by a combination of $B_s-\bar B_s$ mixing constraints and the ALP contribution to the anomalous magnetic moment of the electron. 

\paragraph{Light ALPs}

A light resonance with a mass up to $10\,$MeV below the di-muon threshold dominantly decaying to electrons can provide an explanation of the low $q^2$-bin of $R_{K^*}$, as pointed out in Ref.~\cite{Altmannshofer:2017bsz}.\footnote{For further explanations of the low $q^2$-bin of $R_{K^*}$ using a light resonance see, e.g., \cite{Sala:2017ihs,Ghosh:2017ber,Datta:2017ezo}.} 
It can not, however, account for the deviations observed in $R_K$, or $R_{K^*}$ in the $1.1\, \text{GeV}^2 < q^2 < 6.0 \,\text{GeV}^2$ bin. 
The deviation seen in the $0.045\, \text{GeV}^2 < q^2 < 1.1 \,\text{GeV}^2$ bin of $R_{K^*}$ can be reduced to less than $1\sigma$ by an ALP with $6 \times 10^{-8} < \text{Br}(B \to K^* a) < 1.7 \times 10^{-7}$, $m_a\in [200, 210]$\,MeV and a branching ratio $\text{Br}(a \to e^+ e^-)=1$. This $B$ branching ratio corresponds to a flavor-violating ALP coupling of $5.4 \times 10^{-7}<|[k_D-k_d]_{32}|<9.1 \times 10^{-7}$. Figure~\ref{fig:RKstar} shows the best fit region in purple and the constraints from $B \to K^* e^+ e^-$ \cite{Aaij:2015dea} and $B \to K^* a (\mu^+ \mu^-)$ \cite{Aaij:2015tna} in grey. Details of the fitting procedure are given in \cite{Altmannshofer:2017bsz}. Note that universal lepton couplings $c_{ee}=c_{\mu\mu}=c_{\tau\tau}$ are allowed for an ALP of this mass since decays into muon and tau pairs are kinematically forbidden. As seen in Figure \ref{fig:alp_Brs_fermions}, non-zero ALP-lepton couplings in the UV, $c_L$ or $c_e$, lead to $\text{Br}(a \to e^+ e^-)\approx1$ for ALPs with $m_a \sim 200\,$MeV, irrespective of the exact values of the couplings.

%%%%%%%%%%%%%%%%%%%%%%%%%%%%%
\subsubsection[The ATOMKI $^8\text{Be}$ and $^4\text{He}$ anomalies]{\boldmath The ATOMKI $^8\text{Be}$ and $^4\text{He}$ anomalies}
\label{sec:Berillium}
%%%%%%%%%%%%%%%%%%%%%%%%%%%%%

\begin{figure}
\begin{center}
\includegraphics[width=1.\textwidth]{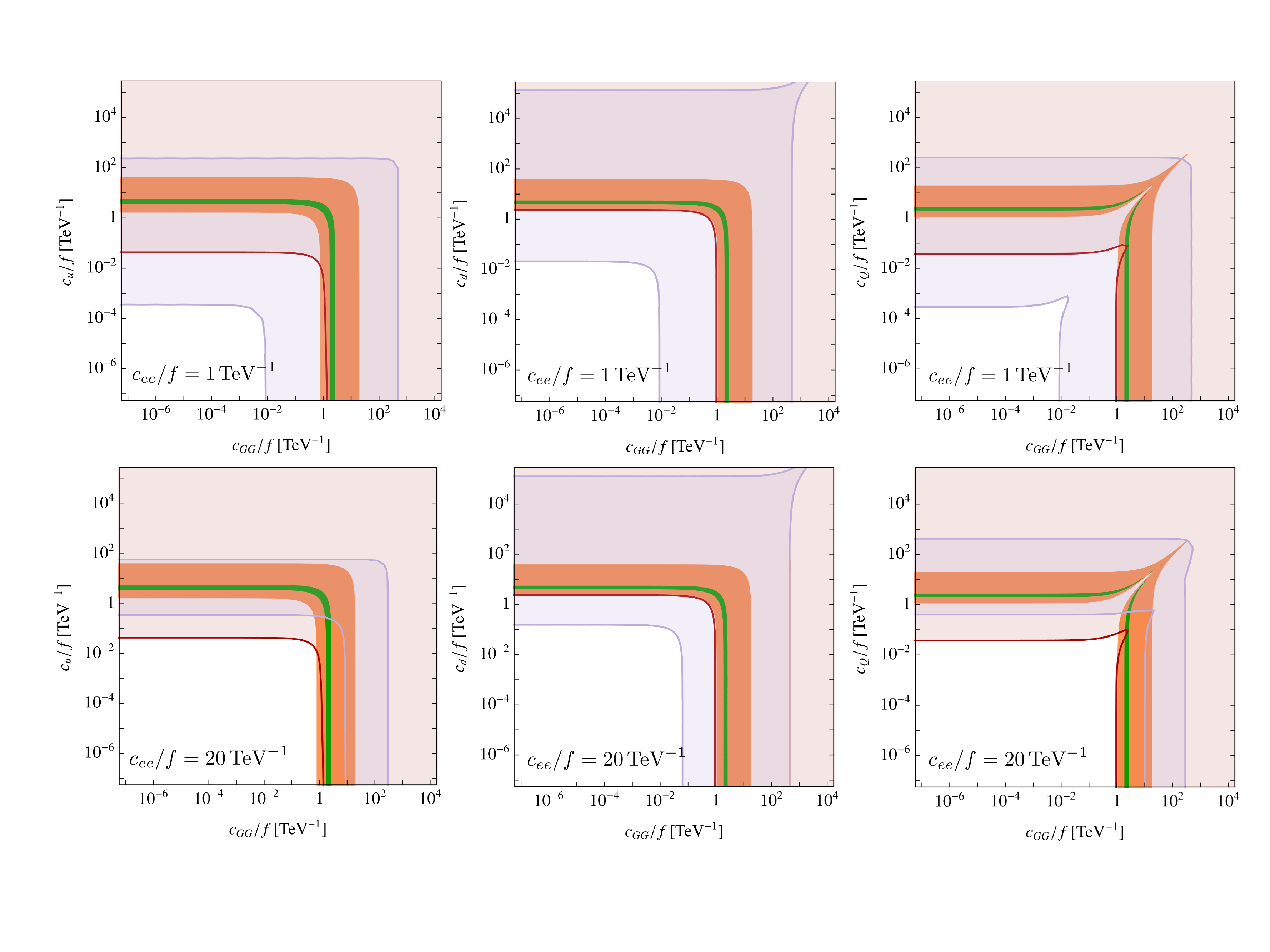}
\end{center}
\vspace{-0.5cm}
\caption{\label{fig:BeandHe} 
Parameter space where an ALP can explain the anomalies in the Beryllium (green) and Helium (orange) nuclear transitions measured by the ATOMKI collaboration at $3\sigma$ for different ALP-couplings to quarks (columns) and different values of the ALP-electron coupling (rows). The light purple regions are ruled out by $K_L \to \pi^0 \nu \bar \nu$ decays \cite{CortinaGil:2021nts} while the light orange regions correspond to limits from $K^- \to \pi^- a (e^+ e^-)$ measurements \cite{Baker:1987gp}. }
\end{figure}

The ATOMKI collaboration measured the transitions of excited Beryllium and Helium nuclei to their respective ground states and finds a discrepancy from the SM expectation of $7.37\sigma$ and $4.9\sigma$ for two independent Beryllium runs~\cite{Krasznahorkay:2018snd} and $7.2\sigma$ for Helium~\cite{Krasznahorkay:2019lyl}. The Helium measurement was not obtained from a single resonant transition but from a collective population of $^4\text{He}(20.21)$ and $^4\text{He}(21.01)$ excited states.
The best fit parameters for these measurements are~\cite{Firak:2020eil}
\begin{align} 
&\Gamma(^8\text{Be}^*(18.15) \to {^8\text{Be}}\, e^+\!e^-)=(1.2\pm 0.2)\times 10^{-5}\,\text{eV}\,, \quad m_{ee}=(17.01\pm 0.16)\,\text{MeV}\,,\\
&\Gamma(^4\text{He}^*(20.49) \to {^4\text{He}}\, e^+\!e^-)=(4.0\pm 1.2)\times 10^{-5}\,\text{eV}\,, \quad m_{ee}=(16.98\pm 0.16_\text{stat}\pm 0.20_\text{sys})\,\text{MeV}\,. \label{eq:WidthHe}
\end{align}
The origin of these large discrepancies has been widely discussed in the literature. For references exploring the possibility of a new resonance being responsible for the anomalous measurements of the Beryllium transition see the recent review \cite{Fornal:2017msy} and references therein. A light pseudoscalar explanation was proposed in \cite{Ellwanger:2016wfe}, while a SM explanation of the Beryllium anomaly was suggested in \cite{Aleksejevs:2021zjw}. Combined new physics explanations of the Beryllium and Helium transitions were discussed in \cite{Feng:2020mbt,Alves:2020xhf}.

In the following we consider an ALP as a possible explanation of the Beryllium and Helium transitions. A potential spin-0 resonance with mass $m_a=m_{ee}\approx 17\,$MeV would need to have a large branching ratio into electron positron pairs,\footnote{See \cite{Banerjee:2019hmi} for constraints on the electron coupling for a dark photon explanation of the $17\,$MeV resonance.} $\text{Br}(a\to e^+ e^-)\approx 1$, to avoid the stringent constraints from $a\to \gamma \gamma$ decays in this mass range. Here, we focus on $^8\text{Be}(18.15)$ and $^4\text{He}(21.01)$ transitions since parity and angular momentum conservation require pseudoscalar couplings for their decays but scalar couplings for the $^4\text{He}(20.21)$ transition~\cite{Feng:2020mbt}. We therefore assume that the measured decay width given in \eqref{eq:WidthHe} is solely due to a $^4\text{He}(21.01)$ transition. Since $^8\text{Be}(18.15)$ is a $J^P=1^+$ state its transition into a pseudoscalar proceeds through a p-wave, whereas $^4\text{He}(21.01)$, as a $0^-$ state, can decay into a pseudoscalar through an s-wave. 

The ALP couplings to nucleons are given in \eqref{gpagna}. Note that the combinations $g_{pa} + g_{na}$ and $g_{pa} - g_{na}$ are the coefficients of the isosinglet and -triplet currents, respectively. 
Since the excited states, $^8\text{Be}(18.15)$ and $^4\text{He}(21.01)$, and the corresponding ground states have the same isospin, we expect an ALP mediating these nuclear transitions to couple to the isosinglet current as opposed to the isotriplet current, which would mediate transitions between states differing by one unit of isospin. Here we neglect the mass splitting between the neutron and the proton, $m_n - m_p = 1.3\,$MeV, and collectively denote the nucleon mass by $m_N \approx 1$\,GeV.
Matching onto the Lagrangian of the form \cite{Alves:2020xhf}
\begin{align}
\mathcal{L} = a \, \bar \psi \, i \gamma_5 (g_{aNN}^{(0)} + g_{aNN}^{(1)} \tau^3) \, \psi\,,
\end{align}
at the amplitude level we find
\begin{align}
g_{aNN}^{(0)} = -\frac{m_N}{4f} (g_{pa} + g_{na})\,, \qquad g_{aNN}^{(1)} = -\frac{m_N}{4f} (g_{pa} - g_{na})\,.
\end{align}
For reference, we give $g_{aNN}^{(0)}$ in terms of the Wilson coefficients at the UV scale
\begin{align} \label{eq:gaNNNum}
 g_{aNN}^{(0)} = & 10^{-4}\, \left[ \frac{1\,\text{TeV}}{f} \right] 
   \times \Big[  - 4.2 \,c_{GG} + 9.7\times 10^{-4}\,c_{WW} + 9.7\times 10^{-5}\,c_{BB}  - 2.0\,c_u(\Lambda) \notag\\ 
    &- 2.0\,c_d(\Lambda) +4.0\,c_Q(\Lambda) + 2.9\times 10^{-4}\,c_e(\Lambda) - 1.6\times10^{-3}\,c_L(\Lambda) \Big]^2\,.
\end{align}

The ratio of the ALP emission rate of $^8\text{Be}(18.15)$ to the corresponding photon emission rate was derived in~\cite{Donnelly:1978ty, Barroso:1981bp, Alves:2020xhf} and is given by
\begin{align} \label{eq:BeWidth}
\frac{\Gamma(^8\text{Be}^*\to ^8\!\!\text{Be}+a )}{\Gamma(^8\text{Be}^*\to ^8\!\!\text{Be}+\gamma )}\approx\frac{1}{2\pi\alpha}
\left| \frac{g_{aNN}^{(0)}}{\mu^{0}-1/2}\right|^2\left(1-\frac{m_a^2}{\Delta E^2}\right)^{3/2}\,,
\end{align}
where the isoscalar magnetic moment is given by $\mu^0\approx 0.88$~\cite{Donnelly:1978ty}. The ratio is independent of the momentum, because both the ALP and the photon emission rates scale with the third power of the ALP and photon momentum, respectively. Note that the ALP emission rate is indeed proportional to $g_{pa} + g_{na}$.\footnote{There is an admixture of the isospin-1 component in ${^8}\text{Be}^*$, for which the relative transition rate depends on unknown nuclear structure-dependent parameters. The effect of this contribution was estimated in~\cite{Alves:2020xhf}.} Numerically, we find
\begin{align} \label{eq:BeWidthNum}
 \frac{\Gamma(^8\text{Be}^*\to^{8}\!\!\text{Be}+a )}{\Gamma(^8\text{Be}^*\to^8\!\!\text{Be}+\gamma )} = 6.5  \, |g_{aNN}^{(0)}|^2.
\end{align}
Since the photon emission rate has been determined to be $\Gamma(^8\text{Be}^*(18.15)\to^8\text{Be}+\gamma ) \approx 1.9\pm 0.4$\,eV~\cite{Tilley:2004zz,Alves:2020xhf}, the experimental constraint on their ratio is given by 
\begin{align}
\frac{\Gamma(^8\text{Be}^*\to^8\!\!\text{Be}+a )}{\Gamma(^8\text{Be}^*\to^8\!\!\text{Be}+\gamma )}\approx (6\pm 1)\times 10^{-6}\,.
\end{align}
The $^4\text{He}(21.01)$ transition to the ground state cannot occur through single photon emission but instead proceeds through two-photon transitions or electron conversion effects. The ALP emission rate scales as $|\vec p_a|^5\approx (\Delta E^2-m_a^2)^{5/2}$ and is given by~\cite{Donnelly:1978ty, Alves:2020xhf} 
\begin{align}\label{eq:HeWidth}
\Gamma(^4\text{He}^*\to^4\!\!\text{He}+a)\approx \frac{ 2 (\Delta E^2-m_a^2)^{5/2}}{m_N^2 Q^2}|a_\text{M0}^0 \, g_{aNN}^{(0)}|^2\,,
\end{align}
where the nuclear momentum scale $Q\approx 1/R_N\approx 250\,$MeV is set by the inverse nuclear radius. The coefficient $a_\text{M0}^0$ is unknown and we vary it between $1/3 < |a_\text{M0}^0| < 3$. Numerically, we find
\begin{align} \label{eq:HeWidthNum}
\Gamma(^4\text{He}^*\to^4\!\!\text{He}+a) = 9.2 \, |a_\text{M0}^0 \, g_{aNN}^{(0)}|^2 \, \text{eV}.
\end{align}

We can now use \eqref{eq:BeWidth} and \eqref{eq:HeWidth} to determine the parameter space allowed by the experimental measurements. The $3\sigma$ regions for an ALP explanation of the Beryllium and Helium anomalies are shown in Figure~\ref{fig:BeandHe} in green and orange, respectively. We assume just two couplings to be present at a time: $c_{GG}$ and either $c_u$ (left), $c_d$ (centre) or $c_Q$ (right), where all couplings are defined at the UV scale $\Lambda$. The top row shows the parameter regions for $c_{ee}/f = 1\,$TeV$^{-1}$ while the bottom row depicts $c_{ee}/f = 20\,$TeV$^{-1}$.  
The different shape of the right hand plots ($c_Q$) compared to the left and middle column ($c_u$, $c_d$) is explained by the different sign of $c_Q(\Lambda)$ compared to $c_u(\Lambda)$ and $c_d(\Lambda)$ in \eqref{eq:gaNNNum} .
As can be seen from the figure and as previously pointed out in~\cite{Alves:2020xhf} it is possible to explain the Helium and Beryllium anomalies simultaneously. 
The required couplings for an ALP explanation are, however, already mostly excluded by $K \to \pi a$ measurements. The light purple regions in Figure~\ref{fig:BeandHe} are ruled out by $K_L \to \pi^0 X$ decays~\cite{CortinaGil:2021nts}, with $X$ decaying invisibly or escaping the NA62 detector, while the light red regions correspond to limits from a $K^- \to \pi^- a (e^+ e^-)$ search \cite{Baker:1987gp}. Constraints from $K^- \to \pi^- \gamma \gamma$ \cite{Artamonov:2005ru} are relevant but subdominant in this region of parameter space and have therefore been omitted in Figure \ref{fig:BeandHe} for clarity. For large ALP-electron couplings, there is still a viable region of parameter space that could account for the ATOMKI Helium anomaly for $c_{GG}(\Lambda) \sim 1\,$TeV$^{-1}$ and $c_{u}(\Lambda), c_{d}(\Lambda),c_{Q}(\Lambda) \ll 1\,$TeV$^{-1}$. The Beryllium anomaly cannot be explained for these parameters. Note, however, that sizeable ALP couplings to electrons are strongly constrained. As shown in the left panel of Figure \ref{fig:flavorconstraintsandleptons}, an ALP with a mass of $17$\,MeV is already excluded by beam dump experiments. The presence of additional ALP couplings, e.g., to gluons as required here, may alter these constraints and a dedicated beam dump analysis would be necessary to understand whether an ALP explanation of the Helium anomaly is still viable.

%%%%%%%%%%%%%%%%%%%%%%%%%%%%%
\subsubsection{\boldmath The KTeV anomaly} 
\label{sec:KTeV}
%%%%%%%%%%%%%%%%%%%%%%%%%%%%%

% 
\begin{figure}
\begin{center}
\includegraphics[width=0.8\textwidth]{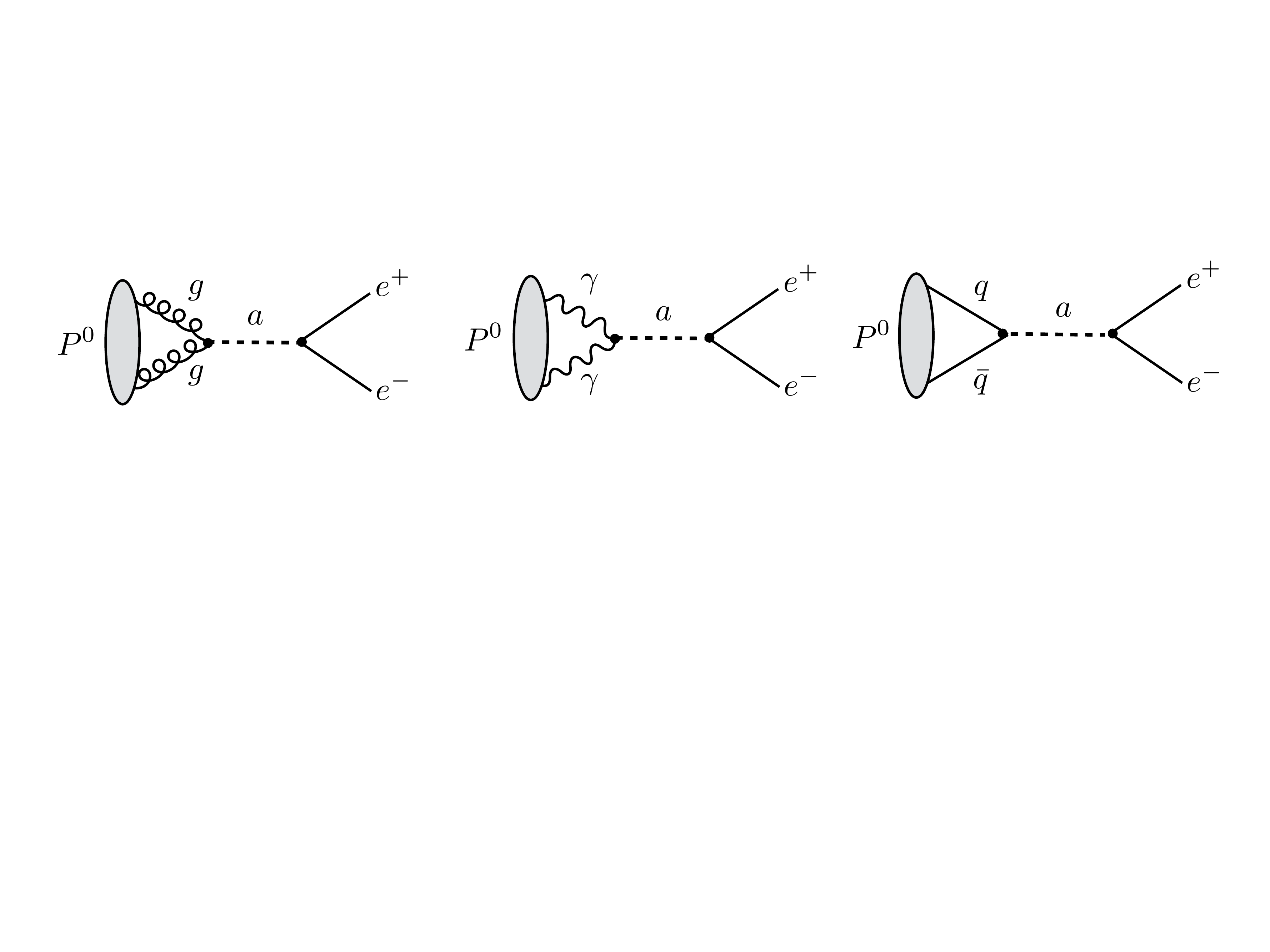}
\end{center}
\caption{\label{fig:P0ee} ALP-induced contributions to the $P^0\to\ell^+\ell^-$ decay amplitude.} 
\end{figure}

\begin{figure}
\begin{center}
\includegraphics[width=.8\textwidth]{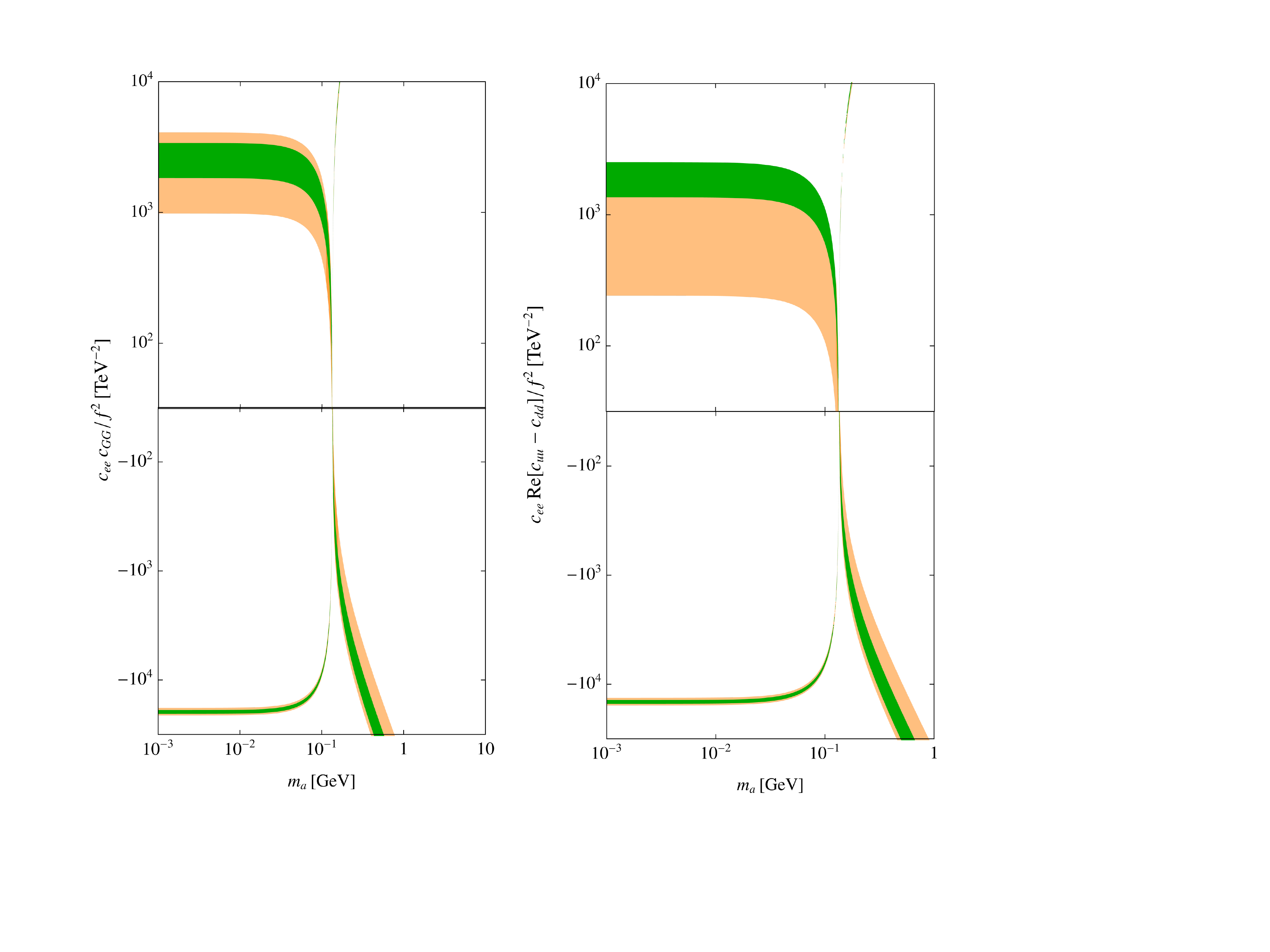}
\end{center}
\vspace{-0.5cm}
\caption{\label{fig:ktev} 
Parameter space for which an ALP can explain the KTeV anomaly by ALP couplings to electrons and first generation quarks (left) or gluons (right) at $2\sigma$ (orange) and $1\sigma$ (green), respectively.}
\end{figure}

The KTeV collaboration has measured the branching ratio \cite{Abouzaid:2006kk}
\begin{equation}
   \mbox{Br}(\pi^0\to e^+ e^-)_{\rm no\!-\!rad} = (7.48\pm 0.29\pm 0.25)\times 10^{-8} \,,
\end{equation}
which is $1.8\sigma$ above the predicted SM value of 
$\mbox{Br}(\pi^0\to e^+ e^-)=(6.25\pm 0.03)\times 10^{-8}$ obtained in \cite{Hoferichter:2021lct}, taking into account the 2-loop QED corrections~\cite{Vasko:2011pi,Husek:2014tna}. The leading contribution in the SM arises from triangle diagrams, in which the pion couples to two virtual photons via the
axial anomaly, which then couple to a lepton pair \cite{Drell1959}. The corresponding decay rates are strongly suppressed, see \cite{Dorokhov:2007bd,Hoferichter:2021lct} for recent analyses. The leading SM contribution to the amplitude can be written as~\cite{Bergstrom:1982wk}
\begin{align}
{\cal A}(\pi^0\to e^+ e^-) =-\frac{\alpha^2}{\sqrt{2}\pi^2}\frac{m_e}{f_{\pi^0}}A^\text{SM}\,\bar u_e(k_1)\gamma_5 v_e(k_2) \,,
\end{align}
where $f_{\pi^0} \approx 130\,$MeV and $\alpha$ is evaluated at $m_\pi$. The imaginary part of the SM contribution is model-independent~\cite{Drell1959}
\begin{align}
\text{Im} A^\text{SM}=-\frac{\pi}{2\beta}\ln \frac{1+\beta}{1-\beta},\qquad \beta=\bigg(1-4\frac{m_e^2}{m_\pi^2}\bigg)^{1/2}\,,
\end{align}
which establishes a unitarity bound on the branching ratio $\text{Br}(\pi^0\to e^+ e^-)$ using $|{\cal A}|^2 \geq (\text{Im}{\cal A})^2$. For the real part of the reduced amplitude we use $\text{Re}\,A^\text{SM}$=10.11(10) from \cite{Hoferichter:2021lct}. 
Including a new physics contribution, the prediction for the $\pi^0\to e^+ e^-$ branching ratio can be expressed as
\begin{align}
\text{Br}(\pi^0\to e^+ e^-)=2\left(\frac{\alpha}{\pi}\frac{m_e}{m_\pi}\right)^2\beta\, |A^\text{SM}+A^\text{ALP}|^2\, \text{Br}(\pi^0\to \gamma\gamma) 
\end{align}
and $\text{Br}(\pi^0\to \gamma\gamma) =0.988$~\cite{Tanabashi:2018oca}. Computing the diagrams in Figure \ref{fig:P0ee}, we find an ALP contribution to the reduced amplitude
\begin{align}
A^\text{ALP}&=c_{ee}\frac{2\pi^2}{\alpha^2}\frac{f_{\pi^0}^2}{f^2}\,\frac{m_{\pi^0}^2}{m_{\pi^0}^2-m_a^2}\bigg[ \frac{\sqrt2}{f_{\pi^0}\,m_{\pi^0}^2} \left( c_{GG}\,a_{\pi^0}^G
    + c_{\gamma\gamma}\,a_{\pi^0}^F \right) - \frac{c_{uu}-c_{dd}}{2}\bigg]\notag \\
   &=c_{ee}\frac{2\pi^2}{\alpha^2}\frac{f_{\pi^0}^2}{f^2}\,\frac{m_{\pi^0}^2}{m_{\pi^0}^2-m_a^2}\bigg[ - \frac{m_d-m_u}{m_d+m_u}\,c_{GG} - \frac{c_{uu}-c_{dd}}{2}\bigg]\,,
\end{align}
where the three terms in the first line correspond to the contributions from the three different diagrams shown in Figure~\ref{fig:P0ee} and we have used 
\begin{align} \label{eq:piGG}
   a_{\pi^{0}}^G& = \big\langle 0 \big|\,\frac{\alpha_s}{4\pi}\,G_{\mu\nu}^A\,\tilde G^{\mu\nu,A}\,\big|
    \pi^0(q) \big\rangle =-\frac{f_{\pi^0}m_{\pi^0}^2 }{\sqrt{2}}\frac{m_d-m_u}{m_d+m_u}\,,
\end{align}
and neglected the contribution from $a_{\pi^0}^F$, which is suppressed by $\alpha/\alpha_s$. In principle, the coupling $c_{ee}$ and $c_{qq}$ for light fermions are not constrained by perturbativity, and values significantly larger than~1 are thus not excluded. Note that the effect would also be enhanced if by chance the masses of the ALP and the pion are close to each other. In Figure~\ref{fig:ktev} we show the parameter space for which the KTeV anomaly can be explained assuming ALP couplings only to electrons and gluons (left panel) or electrons and first generation quarks (right panel). The plot in the right panel  agrees with the recent analysis in \cite{Hoferichter:2021lct}, for $m_a<100\,$MeV. Note that any ALP explanation of the KTeV anomaly requires sizeable ALP couplings to electrons or nucleons which are already severely constrained. An exception is the mass range $m_a\approx m_\pi$ for which the required couplings are smaller.

\clearpage
%%%%%%%%%%%%%%%%%%%%%%%%%%%%%%%%%%%%%%%%%%%%%%%%%%%%%%%%%%%%%%%%%%%%%%%%%%%%%%%

\section{Probes of flavor-changing ALP couplings to leptons}
\label{sec:leptons}

In this section we discuss the phenomenology of ALPs with lepton flavor-violating (LFV) interactions and the experimental constraints on them. In contrast to the quark sector, in the lepton sector there are negligible SM contributions to flavor-changing observables since these effects are proportional to the neutrino masses. For the same reason flavor-violating effects from MFV-type ALP couplings in the lepton sector are either absent or proportional to neutrino masses, meaning that constraints on these couplings will only apply to models in which the ALP inherits explicit lepton flavor-violating interactions from the new physics which generates it.

Here, we extend our work in \cite{Bauer:2019gfk} and present general expressions for lepton flavor-conserving and lepton flavor-changing form factors, and decay rates of lepton flavor-violating decays. 
We also study non-decay observables such as the electron (muon) electric dipole moment e($\mu$)EDM and muonium-antimuonium oscillations. Furthermore a detailed study of how ALPs could underlie the long-lasting discrepancy in the anomalous magnetic moment of the muon (and that of the electron) is performed, with particular attention to lepton flavor-violating scenarios.
We summarize the relevant measurements and constraints, and present exclusion plots for ALPs in the mass range $0.1$\,MeV to $10$\,GeV, including projected reaches of future experiments where available. 

Lepton flavor-changing couplings of ALPs are an active field of research (e.g.,~\cite{Hou:1995dg, Marciano:2016yhf, Bjorkeroth:2018dzu,Bauer:2019gfk,Cornella:2019uxs,Endo:2020mev,Calibbi:2020jvd,Ishida:2020oxl,Bonnefoy:2020llz,Escribano:2020wua,DiLuzio:2020oa,Ma:2021jkp}), so we outline what our current work adds.
\begin{itemize}
	\item The ALP can have macroscopic decay lengths which greatly affect the sensitivity of certain experiments to its parameter space. For example in searches for $\mu\to e\gamma\gamma$ it is mandatory for the ALP to decay before reaching the detector, whereas for $\mu\to ea \,\text{(invisible)}$ the ALP has to escape detection. We derive these effects specifically for each experiment and apply relevant event selection criteria.
	\item When studying limits on $\mu\to e\gamma$, we show that the process $\mu\to e a$ with subsequent $a\to \gamma\gamma$ decay can increase the sensitivity in the parameter region where the two photons are so collimated that they hit the detector at points closer than its spatial resolution, thus mimicking $\mu\to e\gamma$. This is an example of the vastly increased sensitivity that is achieved in regions where resonant ALP decays are kinematically allowed.
	\item We discuss the interplay between flavor-conserving and flavor-violating ALP couplings to charged leptons in detail, and show the complementarity between flavor bounds and constraints from astrophysical, beam dump and collider experiments. Restricting our attention to values of flavor-conserving lepton couplings which are not in conflict with other measurements can have a large impact on the relative strengths of different LFV constraints. For example, it has previously been observed~\cite{Bauer:2019gfk,Cornella:2019uxs} that in the parameter space regions for which an on-shell ALP can be produced in $\mu \to e a$ decays, strong constraints can be set by limits on $\mu\to 3e$, much stronger than those from $\mu \to e \gamma$ limits. We find that this remains true even when the ALP coupling to electrons is small enough to evade other constraints in this mass region, from supernova and beam dump bounds. However, other LFV observables such as $\mu \to e \gamma \gamma$ and $\mu \to e a~\text{(invisible)}$ provide equally stringent limits with current data.
	\item We deliver a comprehensive study of ALP contributions (including the most important 2-loop diagrams with ALP-fermion couplings) to the long-lasting discrepancies of the anomalous magnetic moments of the muon and electron.

	\item To the best of our knowledge, constraints on ALP-lepton flavor-violating couplings derived from limits on the electron (muon) electric dipole moment (e($\mu$)EDM) are presented in this work for the first time.
\end{itemize}
\vspace{.3cm}
%%%%%%%%%%%%%%%%%%%%%%%%%%%%%
\subsection{Form Factors}
\label{sec:FormFactors}
%%%%%%%%%%%%%%%%%%%%%%%%%%%%%

If the ALP has LFV couplings at tree-level, it follows from eq.~\eqref{masssuppression} that these couplings are suppressed by the charged lepton masses. Given the large hierarchy in charged lepton masses, loop-induced contributions to leptonic observables can hence be important if the lepton in the loop is heavier than the external leptons. In lepton flavor-changing decay observables such as $\mu \to e \gamma$, $\mu \to 3e$ or similar tau decays, ALP contributions to electromagnetic form factors may therefore dominate over e.g.,~tree-level ALP-exchange contributions to four-fermion operators.\footnote{Look ahead to Figure~\ref{fig:mueee} for an illustration of these two contributions.} Likewise, if an ALP has lepton flavor-violating couplings, it can induce additional mass-enhanced loop contributions to flavor-conserving observables such as anomalous magnetic moments.

\begin{figure}
\centering
\includegraphics[width=.33\textwidth]{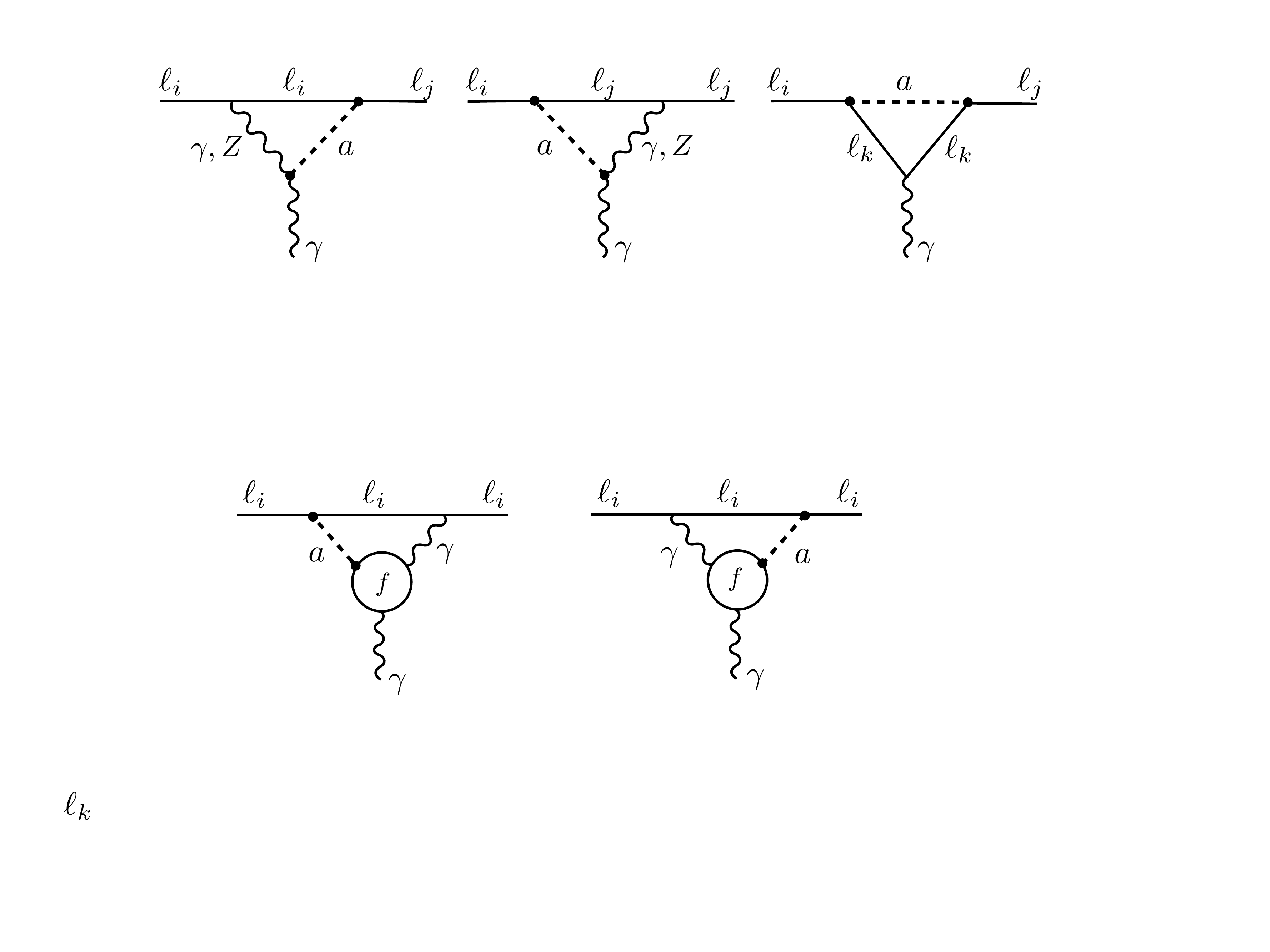}
\caption{\label{fig:formfactor2loop}Two-loop diagram contributing to lepton anomalous magnetic moments, involving only ALP-fermion couplings. The fermion label $f$ runs over all fermion species.}
\end{figure}

Below, we calculate the ALP contributions to the electromagnetic form factors induced by the diagrams shown in Figure~\ref{fig:formfactors}. The expressions below cover the general case in which the external leptons may be different from each other as well as from the lepton in the loop. We further give analytical expressions for the corresponding loop functions, in various limits motivated by the phenomenological applications discussed in the remainder of the paper. For the case of identical leptons in the initial and final state, we additionally provide a calculation of the two-loop form factor contribution shown in Figure~\ref{fig:formfactor2loop}.
\vspace{.3cm}

\subsubsection[Different initial and final state leptons, $\ell_i \neq \ell_j$]{\boldmath{Different initial and final state leptons,  $\ell_i \neq \ell_j$}}\label{sec:formfactors}
%%%%%%
%%%%%%%%%%%%%%%%%%
\begin{figure}
\centering
\includegraphics[width=.7\textwidth]{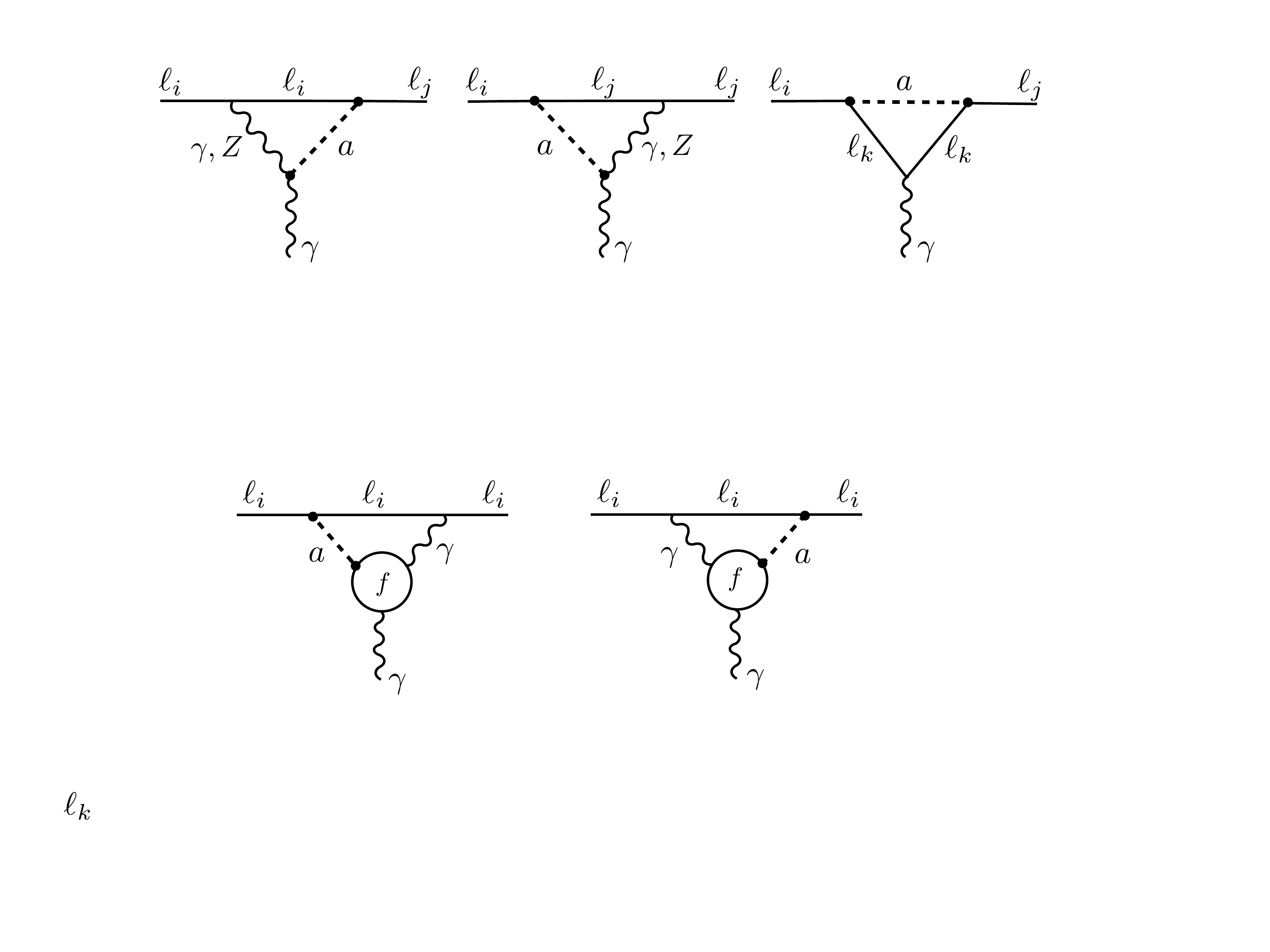}
\caption{\label{fig:formfactors}Diagrams contributing to LFV electromagnetic form factors.}
\end{figure}
%%%%%%%%%%%%%%%%%%
%%%%%%

We first assume that the initial $\ell_i$ and final state $\ell_j$ leptons are not the same. The corresponding form factors are relevant for processes such as $\mu \to e \gamma$, and $\mu \to 3e$.
The ALP-generated contribution to the interaction between initial lepton $\ell_i$, final lepton $\ell_j$ and a photon is defined such that the matrix element $\mathcal{M}^\mu \epsilon_\mu (-q)$ for the interaction between leptons and a photon is found by
\begin{equation} \label{eq:MFF}
\mathcal{M}^\mu= \bar{u}_j (p_2) \Gamma^\mu u_i(p_1),
\end{equation}
which can be parameterised in terms of form factors $F_2^{(5),i\to j}(q^2)$ and $F_3^{(5),i\to j}(q^2)$ as follows, where $p= p_1+p_2$, and $q=p_1-p_2$ is the \emph{outgoing} photon momentum,
\newpage
\begin{align}\label{eq:ffactorsgeneral}
\hspace{-.3cm}\bar \ell_j (p_2)\Gamma^\mu(p_1,p_2)&\,\ell_i(p_1) =\notag\\
& \bar \ell_j (p_2)\bigg[ F_2^{i \to j}(q^2)\big(p^\mu-(m_i+m_j)\gamma^\mu\big)+F_3^{i \to j}(q^2)\Big(q^\mu -\frac{q^2}{m_i-m_j}\gamma^\mu\Big)\\
&\phantom{=}+ F_2^{5,i \to j}(q^2)\big(p^\mu+(m_i-m_j)\gamma^\mu\big)\gamma_5+F_3^{5,i \to j}(q^2)\Big(q^\mu +\frac{q^2}{m_i+m_j}\gamma^\mu\Big)\gamma_5\,
\bigg]\ell_i(p_1)\,.\notag
\end{align}
%
%%%%%%
%%%%%%%%%%%%%%%%%%
%%%%%%%%%%%%%%%%%%
%%%%%%

\noindent
\textbf{One flavor-changing coupling}\\
If only one flavor-changing coupling is present, the form factor is calculated from all the diagrams in Figure~\ref{fig:formfactors}, where in the last diagram the sum is taken over situations in which $l_k=l_i$ and $l_k=l_j$.
Assuming $m_i > m_j$ and keeping only the zeroth order contribution in an expansion in $m_j/m_i$, the ALP contribution to the electromagnetic form factors is given by
\begin{align}
F_2^{i \to j}(q^2)&=- \frac{m_i e Q_i}{16 \pi^2 f^2}\Big([k_E]_{ij}-[k_e]_{ij}\Big)\bigg(\frac{\alpha}{4\pi} c_{\gamma \gamma}g_2(q^2,m_i,m_a)+\frac{1}{4}c_{ii}g_1(q^2,m_i,m_a)\nonumber\\&\phantom{=}+\frac{\alpha}{4\pi}\frac{(4s_w^2-1)}{2(s_wc_w)^2}c_{\gamma Z}\Big(\log \frac{\Lambda^2}{m_Z^2}+\frac{3}{2}+\delta_2 \Big)\bigg) \,,\label{eq:F2one}\\
F_2^{5,i \to j}(q^2)&=- \frac{m_i e Q_i}{16 \pi^2 f^2}\Big([k_E]_{ij}+[k_e]_{ij}\Big)\bigg(\frac{\alpha}{4\pi} c_{\gamma \gamma}g_2(q^2,m_i,m_a)+\frac{1}{4}c_{ii}g_1(q^2,m_i,m_a)\nonumber\\&\phantom{=}+\frac{\alpha}{4\pi}\frac{(4s_w^2-1)}{2(s_wc_w)^2}c_{\gamma Z}\Big(\log \frac{\Lambda^2}{m_Z^2}+\frac{3}{2}+\delta_2 \Big)\bigg)\,, \label{eq:F25one}\\
F_3^{i \to j}(q^2)&= -\frac{m_i e Q_i}{16 \pi^2 f^2}\Big([k_E]_{ij}-[k_e]_{ij}\Big)\left(\frac{\alpha}{4\pi} c_{\gamma \gamma}l_2(q^2,m_i,m_a)+\frac{1}{4}c_{ii}l_1(q^2,m_i,m_a)\right)\,, \\
F_3^{5,i \to j}(q^2)&= -\frac{m_i e Q_i}{16 \pi^2 f^2}\Big([k_E]_{ij}+[k_e]_{ij}\Big)\left(\frac{\alpha}{4\pi} c_{\gamma \gamma}l_2(q^2,m_i,m_a)+\frac{1}{4}c_{ii}l_1(q^2,m_i,m_a)\right) \,,
\end{align}
where the loop functions are given in terms of Feynman integrals in Appendix~\ref{app:formfactors}. The results for the $F_2^{i \to j}$ and $F_2^{5,i \to j}$ loop functions when $q^2=0$ (i.e.,~for an on-shell photon) are
\begin{align}
g_1(0, m_i, m_a)&=2x_i^{3/2} \sqrt{4-x_i}\arccos\frac{\sqrt{x_i}}{2}+1-2x_i+\frac{x_i^2(3-x_i)}{1-x_i}\log x_i,\label{eq:gzeroq2}\\
g_2(0, m_i, m_a)&=2 \log \frac{\Lambda^2}{m_i^2}+2\delta_2+4-\frac{x_i^2 \log x_i}{x_i-1}+(x_i-1)\log (x_i-1),\label{eq:ggamgamzeroq2}
\end{align}
where we have set the scale $\mu=\Lambda= 4\pi f$ and $x_i = m_a^2/m_i^2$. The scheme dependent constant $\delta_2$ arises from the treatment of the Levi Civita symbol in $d$ dimensions, and for us $\delta_2=-3$. The functions $l_1$, $g_1$ and $l_2$ all tend to zero as $m_a^2/m_i^2 \to \infty$, while $l_2(0,m_i,0)=1$ and $g_1(0,m_i,0)=1$. \\
\\
\textbf{Two flavor-changing couplings}\\
Although generically it is expected that flavor-changing couplings should be suppressed relative to flavor-conserving ones, it is possible that diagrams containing two flavor-changing couplings may be enhanced by a heavier mass (relative to diagrams with only one flavor-changing coupling) and should therefore be considered. This occurs if the mass of the fermion in the loop is much larger than that of either of the external fermions, for example $\mu \to e \gamma$ via an internal $\tau$. In this case, the form factor is calculated from the last diagram in Figure~\ref{fig:formfactors} with $\ell_k\neq \ell_i \neq \ell_j$.

Assuming $m_k> m_i > m_j$ (where $k$ is the flavor index of the lepton in the loop) and keeping only the zeroth order contribution in an expansion in $m_i/m_k$, the ALP contribution to the electromagnetic form factors is given by
\begin{align}
F_2^{i \to j}(q^2)&=\frac{m_k e Q_k}{32 \pi^2 f^2}\Big([k_e]_{ik}[k_E]_{k j}+[k_E]_{ik}[k_e]_{k j}\Big) \,g_3(q^2,m_k, m_a) \,,\\
F_2^{5,i \to j}(q^2)&=\frac{m_k e Q_k}{32 \pi^2 f^2}\Big([k_e]_{ik}[k_E]_{k j}-[k_E]_{ik}[k_e]_{k j}\Big) \,g_3(q^2,m_k, m_a) \,,
\end{align}
with
\begin{align}\label{eq:g3loop}
g_3(q^2,m_k, m_a) = \frac{1-3 x_k}{2(x_k-1)^2}+\frac{x_k^2}{(x_k-1)^3}\log x_k \,,
\end{align}
where $x_k = m_a^2/m_k^2$. No terms involving $q^2$ appear in this function, because $q^2\leq m_i^2$, so these terms are suppressed by a factor proportional to $q^2/m_k^2\leq m_i^2/m_k^2$, and have been dropped, along with terms dependent on $m_i^2$.
The $F_3^{(5),i \to j}(q^2)$ form factors are suppressed by a factor $\sim m_i^2/m_k^2$ relative to the $F_2^{(5),i \to j}(q^2)$ form factors, so we do not quote them here.

\subsubsection[Same initial and final state leptons,  $\ell_i = \ell_j$]{\boldmath{Same initial and final state leptons,  $\ell_i = \ell_j$}}
\label{sec:FFsg-2}

The relevant form factors for dipole moments are found if the initial and final state leptons are identical. The gauge invariant form factor parameterisation is now 
\begin{align}\label{eq:ffactorsieqj}
\hspace{-.3cm}\bar \ell_i (p_2)\Gamma^\mu(p_1,p_2)\,\ell_i(p_1) &= \bar \ell_j (p_2)\bigg[ F_2^{i\to i}(q^2)\big(p^\mu-2m_i\gamma^\mu\big)+2m_i F_3^{i\to i}(q^2)\gamma^\mu \notag \\
&\phantom{=}+ F_2^{5,i\to i}(q^2)p^\mu \gamma_5+F_3^{5,i\to i}(q^2)\Big(q^\mu +\frac{q^2}{2m_i}\gamma^\mu\Big)\gamma_5\,
\bigg]\ell_i(p_1)\,,
\end{align}
where $p= p_1+p_2$, and $q=p_1-p_2$ is the \emph{outgoing} photon momentum.
This is defined such that the matrix element $\mathcal{M}^\mu \epsilon_\mu (-q)$ for the interaction between leptons and a photon is found by \eqref{eq:MFF} with $i=j$ and at tree-level in the SM, $\Gamma^\mu_{\text{SM},0}=Q_ie\gamma^\mu$, where $Q_i$ is the charge of $\ell_i$.

Then the anomalous magnetic moment of the lepton $\ell_i$ is defined by
\begin{equation}
a_i = \frac{(g-2)_i}{2} \,,
\end{equation}
and expressed in form factors
\begin{equation}
a_i = \frac{2m_i}{e} F_2^{i \to i}(0) \,.
\end{equation}
If the ALP has purely flavor-conserving interactions then all the diagrams in Figure~\ref{fig:formfactors} contribute with $\ell_i=\ell_j=\ell_k$ and we also include the contribution from the 2-loop diagrams shown in Figure~\ref{fig:formfactor2loop} where we sum over all internal fermions $f$,
\begin{align}\label{eq:F2iii}
F_2^{i \to i}(0)=\frac{e\,Q_i}{32\pi^2}\frac{m_i}{f^2}\bigg\{c_{ii}^2h_1(x_i)& -\frac{2\alpha}{\pi}c_{ii}\bigg[\tilde{c}_{\gamma\gamma}\left(\log\frac{\mu^2}{m_i^2}-h_2(x_i)\right)+\sum_f N_c^f Q_f^2 c_{ff}\int\limits_0^1 dz\,F(y_z,x_f)\bigg]\notag\\
&  -\frac{\alpha}{2\pi}\frac{1-4s_w^2}{s_wc_w}c_{ii}c_{\gamma Z}\left(\log\frac{\mu^2}{m_Z^2} +\delta_2+\frac32\right) \bigg\}
\end{align}
where 
\begin{align}
h_1(x)&=  1+2x+(1-x)x\log x-2x(3-x)\sqrt{\frac{x}{4-x}}\arccos\frac{\sqrt{x}}{2}\,,\label{eq:h1h2}\\
h_2(x)&=1-\frac{x}{3}+x^2\log x+\frac{x+2}{3}\sqrt{(4-x)x}\arccos\frac{\sqrt{x}}{2}-\delta_2-3\,,\label{eq:h2}\\
F(y_z,x_f)&=\frac{1}{1-y_z}\left[h_2\left(\frac{x_f}{y_z}\right)-h_2\left(x_f\right)\right]\,.
\end{align}
Here we have defined
\begin{equation}
 y_z=z(1-z)\frac{m_a^2}{m_f^2}\,,\,\,\,\, \tilde{c}_{\gamma\gamma}=c_{\gamma\gamma}+\sum_f N_c^fQ_f^2c_{ff}\,.
\end{equation}
In the two limits (a) $m_a^2\gg m_\mu^2$ and (b) $m_f^2\gg m_{a}^2,m_{\mu}^2$, the integral of $F(y_z,x_f)$ can be given explicitly
\begin{align}
	&\int\limits_0^1  F(y_z,x_f)\, dz
	\notag\\
	&
	=\left\{\begin{aligned}&\frac{-4}{\sqrt{x_f(x_f-4)}}\left[\frac{\pi^2}{12}+\ln^2\left(\frac{1}{2}(\sqrt{x_f}-\sqrt{x_f-4})\right)+\text{Li}_2\left(-\frac{1}{4}\left(\sqrt{x_f}-\sqrt{x_f-4}\right)^2\right)\right]\,, &\text{(a)} \\&-\ln\frac{m_f^2}{m_\mu^2}+h_2\left(\frac{m_a^2}{m_\mu^2}\right)-\frac{7}{2}+\mathcal{O}\left(\frac{m_{a}^2}{m_f^2},\,\frac{m_{\mu}^2}{m_f^2}\right)\,, &\text{(b)} \end{aligned}\right.
\end{align}
Our results in these limits are in good agreement with Ref.~\cite{Buen-Abad:2021fwq}.
Flavor-violating vertices can contribute via the rightmost diagram in Figure~\ref{fig:formfactors}, when there is a different flavor fermion in the loop, $\ell_k \neq \ell_i$. The relevant form factors are
\begin{align}\label{eq:fullF2same}
F_2^{i \to i}(0)&=-\frac{e Q_k m_k}{32\pi^2 f^2} \bigg\lbrace \frac{m_i^3}{m_k^3} \left(\left|[k_E]_{i k} \right|^2 +\left|[k_e]_{i k} \right|^2 \right)\int_0^1 dx\,  \frac{x(1-x)^2}{\Delta_{i \to i}}\\
&+2  \, \text{Re}\left[[k_E]_{i k}^*[k_e]_{i k}  \right]\int_0^1 dx\,  \frac{(1-x)^2}{\Delta_{i \to i}}\nonumber \\
&+ 2\frac{m_i^2}{ m_k^2} \, \text{Re}\left[[k_E]_{i k}^*[k_e]_{i k}  \right]\int_0^1 dx\,  \frac{(1-x)^2(1-2x)}{\Delta_{i \to i}}\nonumber \\
&+\frac{m_i}{m_k}\left(\left|[k_E]_{i k} \right|^2 +\left|[k_e]_{i k} \right|^2 \right)\int_0^1 dx\,  \frac{(1-x)^2(x-2)}{\Delta_{i \to i}} \bigg\rbrace\nonumber \,,
\end{align}
\begin{align}\label{eq:fullF25same}
F_2^{5,i\to i}(0)&=-\frac{eQ_k}{16\pi^2f^2}\left(1-\frac{m_i^2}{m_k^2}\right)\bigg\lbrace m_i \left(\left|[k_E]_{i k} \right|^2 +\left|[k_e]_{i k} \right|^2 \right)\\
&+2 i m_k\text{Im}\left[[k_E]_{i k}^*[k_e]_{i k}\right]\bigg\rbrace\int_0^1 dx\,  \frac{(1-x)^2}{\Delta_{i \to i}}\nonumber\,,
\end{align}
where 
\begin{equation}
\Delta_{i \to i}= x \frac{m_a^2}{m_k^2}+x(x-1) \frac{m_i^2}{m_k^2} +(1-x)\,.
\end{equation}
There are two important limits, discussed below. \\
\\
{\boldmath $m_k \gg m_i$}
This is the limit where the internal fermion is much heavier than the external fermion, for example in the case of a contribution to the anomalous magnetic moment of the electron via a diagram with an internal muon.
\begin{align}
F_2^{i \to i}(0) &= -\frac{m_k e Q_k}{32\pi^2 f^2}\, \text{Re}\left([k_E]_{k i}^*[k_e]_{k i}  \right) h(x_k) + \mathcal{O}\left(\frac{m_i}{m_k} \right) \,,
%\bigg\lbrace \frac{3x_{k}-1}{2(x_k-1)^2}-\frac{x_k^2}{(x_k-1)^3}\log x_k \bigg \rbrace
\end{align}
where $ x_k = m_a^2/m_k^2$ and
\begin{align}\label{eq:hloop}
h(x)=\frac{2x^2}{(x-1)^3}\log x-\frac{3x-1}{(x-1)^2}\,.
\end{align}
{\boldmath $m_k \ll m_i$}
This is the limit where the internal fermion is much lighter than the external fermion, for example in the case of the anomalous magnetic moment of the muon via an internal electron.
\begin{align}
F_2^{i \to i}(0)&= \frac{m_i e Q_k}{64\pi^2 f^2} \left(\left|[k_E]_{k i} \right|^2 +\left|[k_e]_{k i} \right|^2 \right) j(x_i) + \mathcal{O}\left(\frac{m_k}{m_i} \right)\,,
\end{align}
where $x_i = m_a^2/m_i^2$ and 
\begin{equation}\label{eq:jloop}
j(x)=1+2x -2x^2 \log \frac{x}{x-1} \,.
\end{equation}

\begin{table}
\begin{center}
\scalebox{0.8}{
\begin{tabular}{l c c c c c}
\toprule
Observable & Mass Range [MeV] & ALP decay mode & Constrained & Limit (95\% CL) on & Figure \\
 &  &  & coupling $c$ & $|c| \cdot  \left(\frac{\text{TeV}}{f}\right) \cdot\sqrt{\mathcal{B}}$\\
\midrule
$\text{Br}(\mu \to ea(\text{invisible}))$ & $0<m_a<13$ & \text{Long-lived}  & $ c_{\mu e}$
 & $1.2 \times 10^{-6}$ & \ref{fig:cmueplots}a)\\
$\text{Br}(\mu \to ea(\text{invisible}))$ & $13<m_a<80$ & \text{Long-lived} & $ c_{\mu e}$
& $5.0 \times 10^{-7}$ &  \ref{fig:cmueplots}a)\\
$\text{Br}(\mu \to ea(\text{invisible}) \gamma )$ & $0<m_a<105$ & \text{Long-lived} & $ c_{\mu e}$
& $1.0 \times 10^{-5}$ & \ref{fig:cmueplots}b)\\
$\text{Br}(\mu \to e \gamma \gamma)$ & $0<m_a<105$ & $\gamma\gamma$ & 
$ c_{\mu e}$
& $1.3 \times 10^{-9}$ & \ref{fig:cmueplots}c)\\
$\text{Br}(\mu \to e \gamma_{\text{eff}})$ & $0<m_a<105$ & $\gamma\gamma$ & 
$ c_{\mu e}$
&$1.0 \times 10^{-10}$ & \ref{fig:cmueplots}d)\\
$\text{Br}(\mu \to 3 e)$ & $0<m_a<105$ & $e^+ e^-$ & 
$ c_{\mu e}$
& $1.6 \times 10^{-10}$& \ref{fig:cmueplots}e)\\
\midrule
$\text{Br}(\tau \to e a(\mathrm{invisible}))$ & $0<m_a<1600$ & Long-lived &  
$ c_{\tau e}$
& $3.6 \times 10^{-4}$& \ref{fig:ctaueplots}a)\\
$\text{Br}(\tau \to e \gamma_{\text{eff}})$ & $0<m_a<1776$ & $\gamma\gamma$ & 
$ c_{\tau e}$
&$1.3 \times 10^{-6}$ & \ref{fig:ctaueplots}b)\\
$\text{Br}(\tau \to 3 e)$  & $200<m_a< 1776 $ & $e^+ e^-$ &  
$ c_{\tau e}$
& $1.1 \times 10^{-6}$ & \ref{fig:ctaueplots}c)\\
$\text{Br}(\tau^- \to e^- \mu^+ \mu^-)$  & $211<m_a< 1776 $ & $\mu^+ \mu^-$ &  
$ c_{\tau e}$
& $1.1 \times 10^{-6}$ & \ref{fig:ctaueplots}d)\\
\midrule
$\text{Br}(\tau \to \mu a(\mathrm{invisible}))$ & $0<m_a<1600$  & Long-lived & 
$ c_{\tau \mu}$
& $4.9\times 10^{-4}$& \ref{fig:ctaumuplots}a) \\
$\text{Br}(\tau \to \mu \gamma_{\text{eff}})$ & $0<m_a<1671$ & $\gamma\gamma$ & 
$ c_{\tau \mu}$
& $1.5 \times 10^{-6}$& \ref{fig:ctaumuplots}b)\\
$\text{Br}(\tau^- \to \mu^- e^+ e^-)$  & $200<m_a< 1671$& $e^+ e^-$ & 
$ c_{\tau \mu}$
& $9.3 \times 10^{-7}$ & \ref{fig:ctaumuplots}c)\\
$\text{Br}(\tau \to 3 \mu)$  & $211 <m_a<1671$ & $\mu^+ \mu^-$ & 
$ c_{\tau \mu}$
& $1.0 \times 10^{-6}$& \ref{fig:ctaumuplots}d)\\
\bottomrule
\end{tabular}
}\end{center}
\caption{\label{tab:leptonbounds} Summary of constraints on the lepton flavor-violating ALP couplings derived from measurements of branching fractions (first column) for various muon and tau decays, in which the lepton can decay to an on-shell ALP. The measurements and SM predictions (where appropriate) are given in \Cref{tab:LFVobs} in Appendix \ref{app:measurements}. The limit cited is the strongest limit found within the mass range probed by the measurement. In the fifth column the symbol $\mathcal{B}$ denotes the ALP branching ratio into the relevant final state. The final column refers to figures showing the dependence of the bound on the ALP mass and lifetime.}
\end{table}

%%%%%%
%%%%%%%%%%%%%%%%%%
\begin{figure} 
\centering
\includegraphics[width=1.\textwidth]{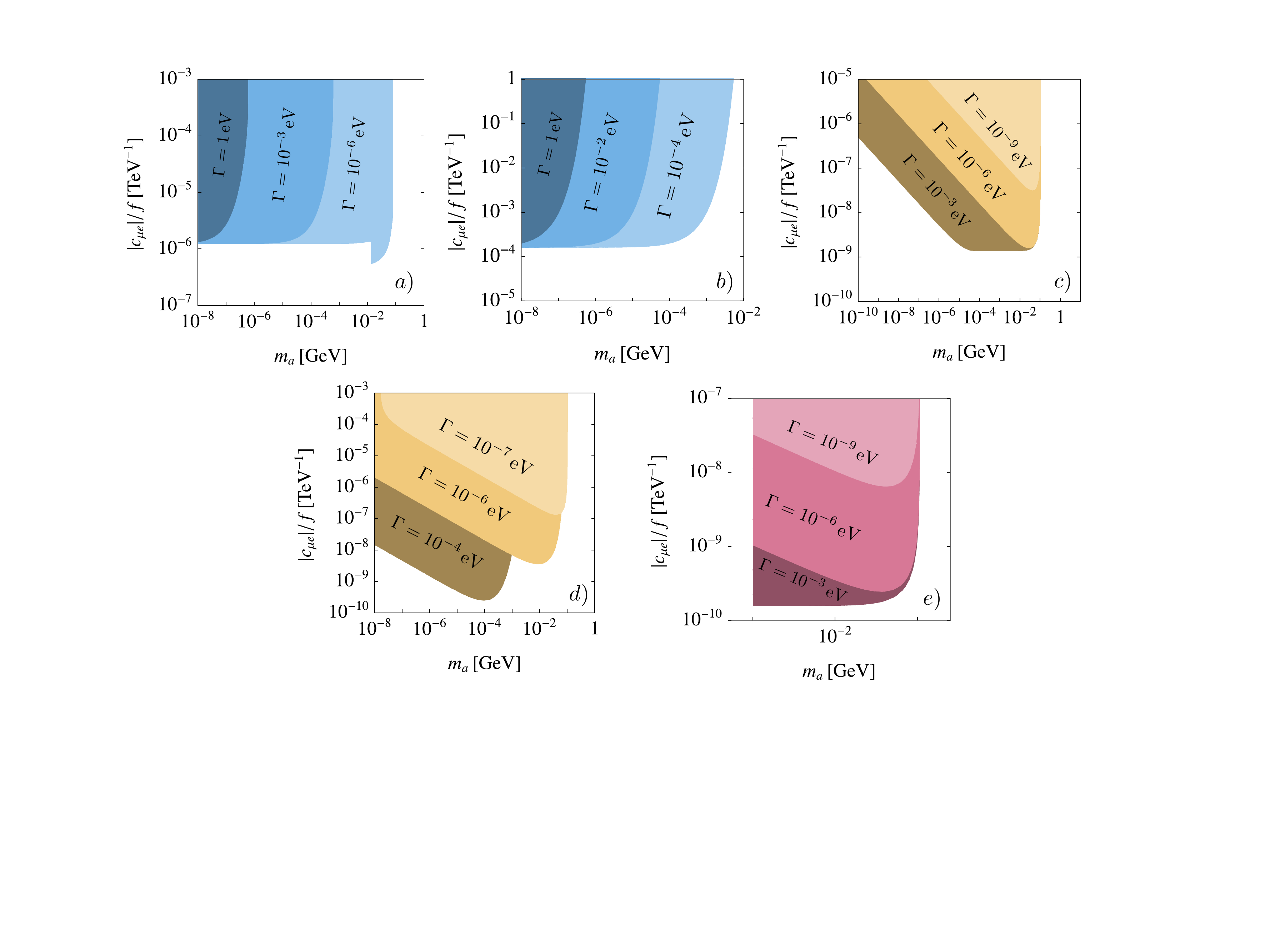}
\caption{\label{fig:cmueplots} { Constraints on the flavor-violating ALP couplings $c_{\mu e}$ from muon decays, collected in Table~\ref{tab:leptonbounds}, for different values of the total ALP width. The observables are $\text{Br}(\mu \to ea(\text{invisible}))$ (upper left), $\text{Br}(\mu \to ea(\text{invisible}) \gamma )$ (upper center), $\text{Br}(\mu \to e \gamma \gamma )$  (upper right),  $\text{Br}(\mu \to e  \gamma_\text{eff} )$  (lower left), and $\text{Br}(\mu \to 3 e)$ (lower right)  }
}\end{figure}
%%%%%%%%%%%%%%%%%%
%%%%%%

%%%%%%
%%%%%%%%%%%%%%%%%%
\begin{figure} 
\centering
\includegraphics[width=1.\textwidth]{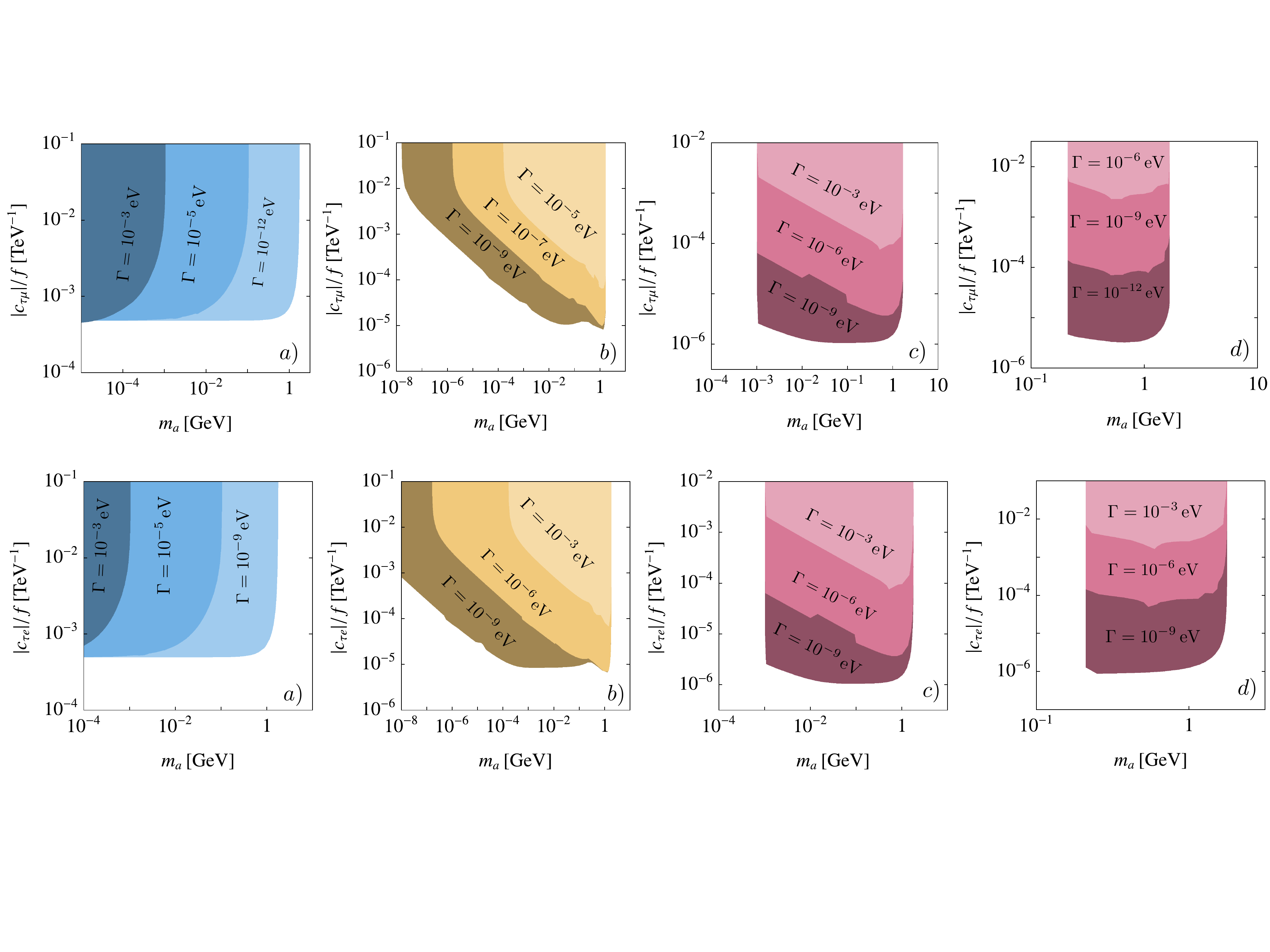}
\caption{\label{fig:ctaueplots} { Constraints on the flavor-violating ALP couplings $c_{\tau e}$ from tau decays, collected in Table~\ref{tab:leptonbounds}, for different values of the total ALP width. The observables are $\text{Br}(\tau \to e a(\mathrm{invisible})) $ (left), $\text{Br}(\tau \to e \gamma\gamma)$ (center left), $\text{Br}(\tau \to 3 e)$ (center right), and $\text{Br}(\tau \to e \mu^+ \mu^-)$  (right).  }
}\end{figure}
%%%%%%%%%%%%%%%%%%
%%%%%%

%%%%%%%%%%%%%%%%%%%%%%%%%%%%%
\subsection[$\mu \to e a$]{\boldmath{$\mu \to e a$}}
\label{sec:lla}
%%%%%%%%%%%%%%%%%%%%%%%%%%%%%
If the ALP is light enough, it can be produced on-shell in LFV decays of muons and taus. The decay rate for the decay of a muon into an electron and an ALP is given by
\begin{align}
&\Gamma(\mu \to e a)=\frac{1}{32\pi m_{\mu}f^2}\,\lambda^{1/2}\left(x_\mu, x_{e/\mu} \right)\times \\
& \Big\{ \!\!\left(|[k_e]_{12}|^2+ |[k_E]_{12}|^2 \right)\left[m_{\mu}^4 \lambda^{1/2}\left(x_\mu, x_{e/\mu} \right)-m_a^4\left(1-x_\mu -x_{e/\mu}\right)  \right]+4 \text{Re}\left[[k_E]_{12}[k_e]^*_{12}\right] m_{\mu}m_{e}m_{a}^2  \Big\} \nonumber
\end{align}
with $x_{e/\mu}=m_e^2/m_\mu^2$ and $\lambda(r_i,r_j)$ defined in \eqref{eq:PhaseSpaceLambda}. Analogous expressions hold for the tau decays $\tau \to \mu a$ and $\tau \to e a$.
In the limit $m_{e}/m_{\mu} \to 0$,
\begin{align}
\Gamma(\mu \to e a)&\approx \frac{m_{\mu}^3}{32\pi f^2}\left(1-\frac{m_a^2}{m_{\mu}^2}\right)^2\left(|[k_e]_{12}|^2 + |[k_E]_{12}|^2 \right) \,.
\end{align}

\begin{figure} 
\centering
\includegraphics[width=1.\textwidth]{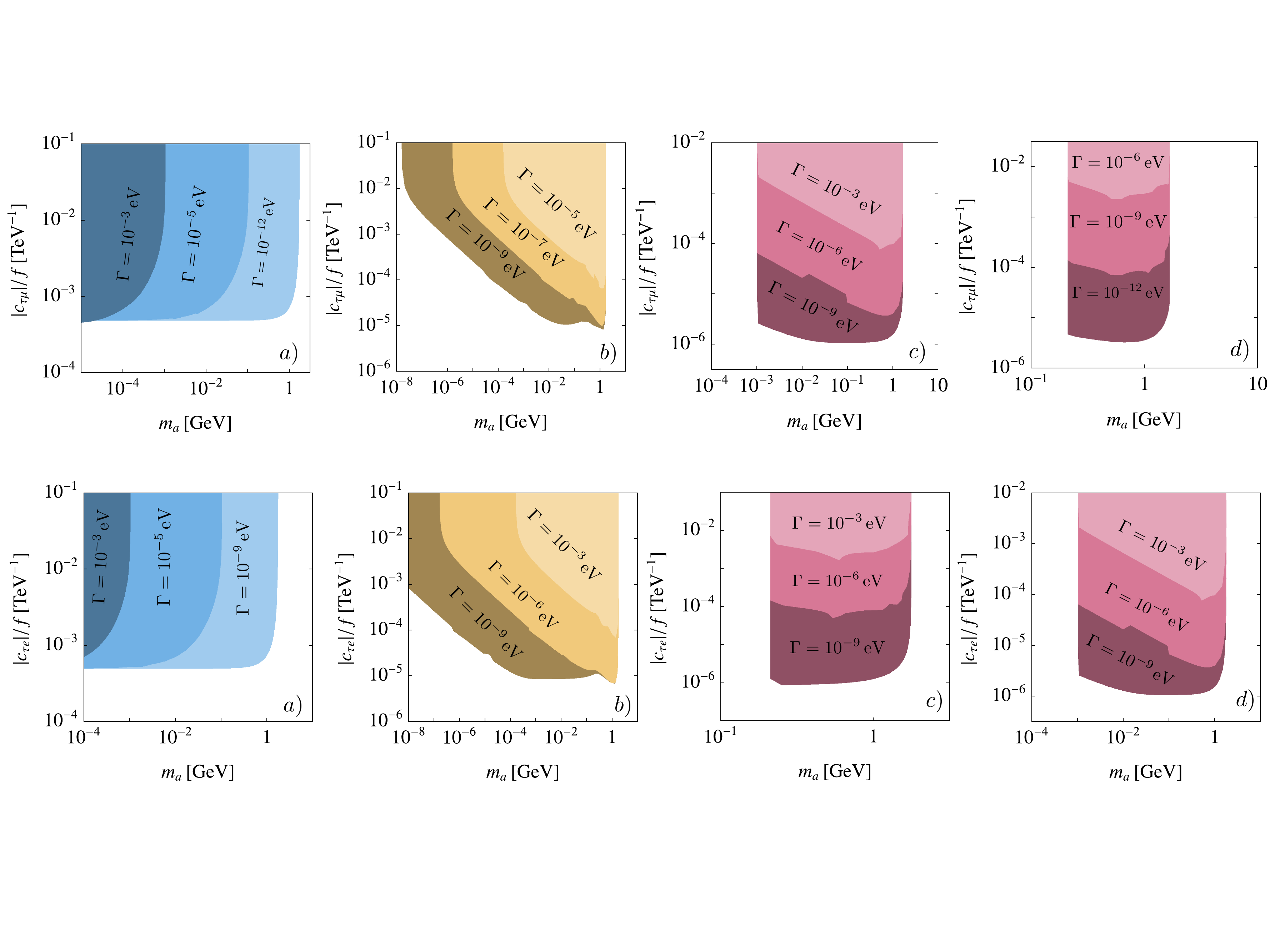}
\caption{\label{fig:ctaumuplots} {Constraints on the flavor-violating ALP couplings $c_{\tau\mu }$ from tau decays, collected in Table~\ref{tab:leptonbounds}, for different values of the total ALP width. The observables are $\text{Br}(\tau \to \mu a(\mathrm{invisible}))$ (left), $\text{Br}(\tau \to \mu \gamma\gamma)$ (center left), $\text{Br}(\tau \to \mu e^+ e^-)$  (center right),  and $\text{Br}(\tau \to 3 \mu)$ (right).  }
}\end{figure}

Due to its resonant nature, the rate of this decay can be enhanced relative to other processes in which the ALP is off-shell, so searches for these processes can be some of the most stringent tests of LFV ALPs. Depending on the ALP lifetime and the branching fractions of the various ALP decay modes, this process can mediate the decays $\mu \to 3e$, $\mu \to e \gamma\gamma$ or $\mu \to e +\text{invisible}$ (and analogous processes with an initial $\tau$ lepton). For ALPs decaying into collimated photons below the experimental angular resolution, the signature $\mu \to e \gamma\gamma$ can be reconstructed as $\mu \to e \gamma$. We discuss this in detail in Section~\ref{sec:discussion2}. A comprehensive list of experimental searches and the respective limits on the flavor-violating ALP couplings $c_{\mu e}, c_{\tau e}$ and $c_{\tau \mu}$ from exotic lepton decays is given in Table \ref{tab:leptonbounds}. For ALPs with $\mathcal{O}(1)$ flavor off-diagonal couplings, these searches can probe new physics scales of up to $f \sim 10^{10}$\,TeV $\times \sqrt{\mathcal{B}}$. Note that the constraints scale with the ALP branching fraction. As in the case of meson decays discussed in Section \ref{sec:ExoticK}, these constraints also crucially depend on the ALP lifetime since the fraction of ALPs decaying within the detector volume of the relevant experiment depends on the decay length of the ALP.
This effect is shown for $\mu \to e$ transitions in Figure \ref{fig:cmueplots}, $\tau \to e$ transitions in Figure \ref{fig:ctaueplots} and $\tau \to \mu$ transitions in Figure \ref{fig:ctaumuplots}. The colors encode different ALP decay modes, where an invisible signature is depicted in blue, while the decays into photons or leptons (electrons or muons) are shown in yellow and red, respectively. Lighter colors correspond to smaller decay widths and darker colors to larger decay widths. For each experimental limit in Table~\ref{tab:leptonbounds}, we show the corresponding exclusion region in the ALP mass vs lepton flavor-violating ALP coupling for three different values of the ALP width $\Gamma$. For the purpose of these plots, we assume a $100\%$ branching ratio for ALPs decaying into the respective final state. Missing energy searches for long-lived ALPs are most sensitive for small decay widths since the fraction of ALPs which escape the detector is suppressed by $\text{exp}(-m_a\spac\Gamma)$. For larger ALP widths the fraction of ALPs escaping the detector decreases and searches for missing energy signatures lose sensitivity. The blue panels in Figure~\ref{fig:cmueplots}a), \ref{fig:cmueplots}b), \ref{fig:ctaueplots}a) and \ref{fig:ctaumuplots}a) therefore extend towards smaller ALP masses. The situation is reversed for ALPs decaying into photons or leptons. The dark shaded regions with shorter lifetimes correspond to the most stringent constraints. The ALP mass range of the constraints is dictated either by the experimental cuts or  by the kinematic window $2m_\ell\le m_a\le m_{\ell_1}-m_{\ell_2}$ for the decay $\ell_1\to \ell_2\spac a$.

%%%%%%
%%%%%%%%%%%%%%%%%%
%%%%%%%%%%%%%%%%%%
%%%%%%

%%%%%%
%%%%%%%%%%%%%%%%%%
\begin{figure}
\centering
\includegraphics[width=.6\textwidth]{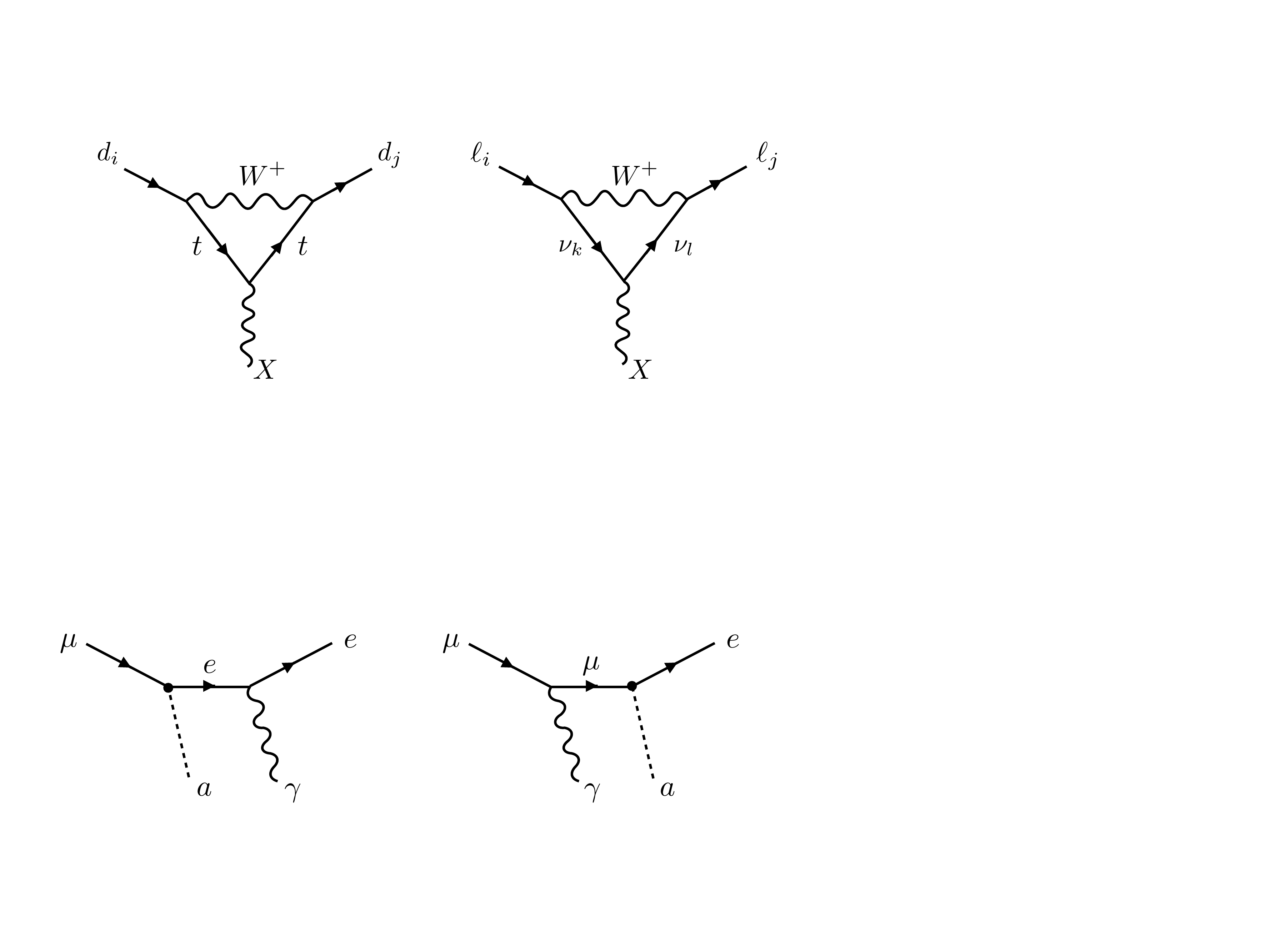}
\caption{\label{fig:mueag} Diagrams contributing to $\mu\to e a \gamma$ decays.}
\end{figure}
%%%%%%%%%%%%%%%%%%
%%%%%%

%%%%%%%%%%%%%%%%%%%%%%%%%%%%%
\subsection[$\mu \to e a \gamma$]{\boldmath{$\mu \to e a \gamma$}}
\label{sec:mueag}
%%%%%%%%%%%%%%%%%%%%%%%%%%%%%
Further constraints arise from the very similar decay $\mu\to e a \gamma$ which can be regarded as  $\mu\to e a$ decay with additional initial or final state radiation as portrayed in the Feynman diagrams Figure~\ref{fig:mueag}. The differential decay rate is given by
\begin{equation}
d\Gamma(\mu\to e a \gamma)=\frac{\alpha_{\text{QED}}}{4\pi^2}\frac{1}{32m_\mu}\frac{|[k_E]_{12}|^2+|[k_e]_{12}|^2}{f^2}\,\mathcal{F}\,\, ds_{12}ds_{23}
\end{equation}
with (in the limit $m_e^2/m_\mu^2\to 0$)
\begin{align}
\mathcal{F}=&\frac{1}{s_{12}(m_a^2-s_{12}-s_{23})^2}\big[m_a^6-s_{23}^2(s_{12}+s_{23})-m_a^4(2m_\mu^2+s_{12}+s_{23})
\nonumber\\ 
&+2m_\mu^2(s_{12}+s_{23})(2s_{12}+s_{23})
-2m_\mu^4(4s_{12}+s_{23})+m_a^2(2m_\mu^4+4m_\mu^2s_{12}+s_{23}^2)\big],
\end{align}
where $s_{ij}=(p_i+p_j)^2$ and the electron carries momentum $p_1$, the photon carries momentum $p_2$ and the ALP carries momentum $p_3$. Up to a prefactor, $\mathcal{F}$ is the squared matrix element summed over electron spins and photon polarisations and averaged over muon spins. Our findings are in good agreement with ~\cite{Calibbi:2020jvd}.

%%%%%%
%%%%%%%%%%%%%%%%%%
\begin{figure}[b]
\centering
\includegraphics[width=.6\textwidth]{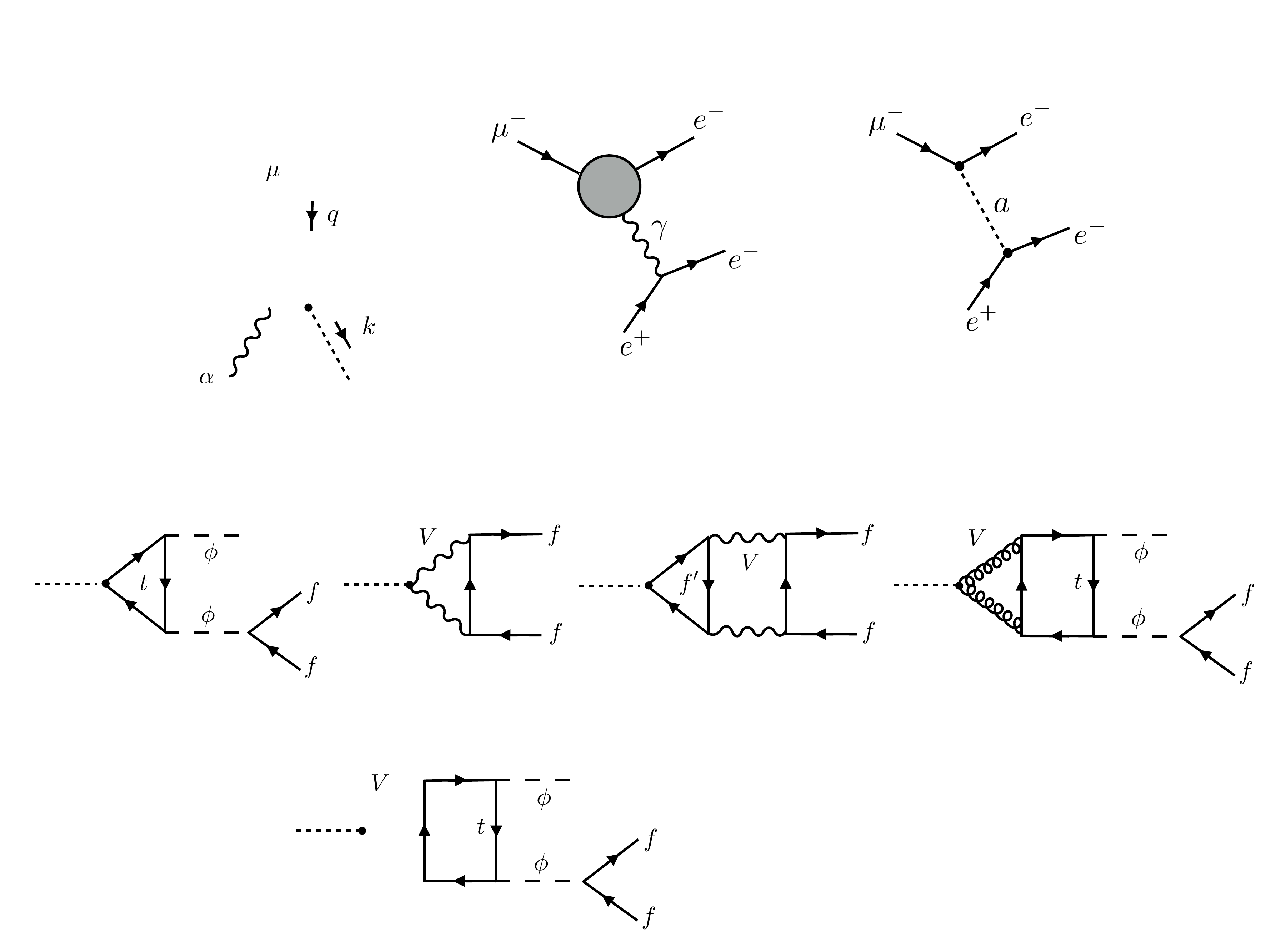}
\caption{\label{fig:mueee} Diagrams contributing to $\mu\to 3e$ decays.}
\end{figure}
%%%%%%%%%%%%%%%%%%
%%%%%%
%%%%%%%%%%%%%%%%%%%%%%%%%%%%%
\subsection[$\mu \to 3e $]{\boldmath{$\mu \to 3e $}}
\label{sec:mu3e}
%%%%%%%%%%%%%%%%%%%%%%%%%%%%%
If the ALP is too heavy to be produced on-shell in lepton decays, it can nevertheless mediate the lepton flavor-violating decays $\mu \to 3 e$ (and similarly $\tau \to 3 e$, $\tau \to 3 \mu$, $\tau \to e\mu\mu$, $\tau \to \mu ee$). In this case both the tree-level exchange of the ALP and the contribution from photon penguins with subsequent $\gamma\to e^+ e^-$ decays contribute. The corresponding diagrams are shown in Figure~\ref{fig:mueee}. The differential decay width for the three-body decay $\mu^+ \to e^+ e^- e^+$ is given by 
\begin{align}
d\Gamma=\frac{1}{(2\pi)^3} \frac{1}{32 m_\mu^2} |\overline {\mathcal M} |^2 ds_{12} ds_{23}\,,
\end{align}
where $s_{ij}=(p_i+p_j)^2$ and the two indistinguishable positrons carry momenta $p_1$ and $p_2$, while the momentum of the $e^-$ is given by $p_3$. The squared matrix element summed over electron and positron spins and averaged over muon spin states is given by
\begin{align}
|\overline{\mathcal{M}}|^2&=\big(|[k_e]_{12}|^2 + |[k_E]_{12}|^2\big)\,|c_{ee}|^2\,\frac{m_e^2m_\mu^2\, }{f^4}\notag\\
&\times\bigg\{\frac{2s_{23}\,(s_{12}+s_{13})}{|s_{23}-m_a^2+im_a\Gamma_a|^2}
-\frac{s_{13}s_{23}}{\text{Re}[(s_{23}-m_a^2+im_a\Gamma_a)(s_{13}-m_a^2-im_a\Gamma_a)]}\bigg\}\,\,
\notag\\
&+4e^2\bigg[2\,(s_{12}+s_{13})\,\,\text{Re}\left[F_2^*(s_{23})F_3(s_{23})+F_2^{5*}(s_{23})F_3^{5}(s_{23})\right]\notag\\
&+\frac{1}{s_{23}}(m_\mu^2\,(s_{12}+s_{13})-2s_{12}s_{13})(|F_2(s_{23})|^2+|F_2^5(s_{23})|^2)\notag\\
&+\frac{1}{m_\mu^2}(s_{23}\,(s_{12}+s_{13})+2s_{12}s_{13})(|F_3(s_{23})|^2+|F_3^5(s_{23})|^2)\notag\\
&+s_{12}\Big(F_2^*(s_{23}) F_2(s_{13}) + F_2^{5\,*}(s_{23}) F_2^5(s_{13}) + F_2^*(s_{23}) F_3(s_{13})\notag\\
&+F_2^{5\,*}(s_{23}) F_3(s_{13}) + F_2^5(s_{13}) F_3^{5\,*}(s_{23}) + F_2(s_{13}) F_3^{*}(s_{23})\Big)\notag\\
&+\frac{s_{12}\,(s_{13}+s_{23})}{m_\mu^2}\,\,\left(F_3(s_{23})F_3^*(s_{13})+F_3^5(s_{23})F_3^{5\,*}(s_{13})\right)\bigg]\notag\\
&+\frac{2 e s_{23}m_e}{f^2}c_{ee}\,\text{Re}\bigg[\frac{[k_e]_{21}+[k_E]_{21} }{s_{23}-m_a^2-im_a\Gamma_a}\big(m_\mu^2 F_2^{5}(s_{13}) +(s_{12}+s_{13})F_3^{5}(s_{13}) \big)\notag \\
&+ \frac{[k_e]_{21}-[k_E]_{21} }{s_{23}-m_a^2-im_a\Gamma_a}\big(m_\mu^2 F_2(s_{13}) +(s_{12}+s_{13})F_3(s_{13}) \big)  \bigg]+(1\leftrightarrow 2)\,.
\end{align}
where we have suppressed the $\mu\to e$ superscript which should appear on all the form factors.

%%%%%%%%%%%%%%%%%%%%%%%%%%%%%
\subsection[$\mu \to e \gamma$]{\boldmath{$\mu \to e \gamma$}}
\label{sec:muegamma}
%%%%%%%%%%%%%%%%%%%%%%%%%%%%%
The partial decay width for $\mu \to e \gamma$ is given by 
\begin{align}
\Gamma(\mu\to e\gamma)=\frac{m_\mu^3}{8\pi}\left(1-\frac{m_e^2}{m_\mu^2}\right)\Big[ |F_2^{\mu\to e}(0)|^2+|F_2^{5,\mu\to e}(0)|^2\Big]
\end{align}
with 
\begin{align}\label{eq:FandF5at0}
F_2^{\mu\to e}(0)&= -\frac{m_\mu e Q_\mu}{16 \pi^2 f^2}\big([k_E]_{12}-[k_e]_{12}\big)\left(\frac{1}{4}c_{\mu \mu} \, g_1(0,m_\mu,m_a) + \frac{\alpha}{4\pi} c_{\gamma \gamma} \, g_2(0,m_\mu,m_a)\right),\\
F_2^{5,\mu\to e}(0)&=-\frac{m_\mu e Q_\mu}{16 \pi^2 f^2}\big([k_E]_{12}+[k_e]_{12}\big)\left(\frac{1}{4} c_{\mu \mu} \, g_1(0,m_\mu,m_a) + \frac{\alpha}{4\pi} c_{\gamma \gamma} \, g_2(0,m_\mu,m_a)\right),
\end{align}
and where the loop functions $g_1(0,m_i,m_a)$ and $g_2(0,m_i,m_a)$ are given in Eqns.~\eqref{eq:gzeroq2} and \eqref{eq:ggamgamzeroq2}. Similar equations hold for radiative tau decays with obvious replacements.

%%%%%%%%%%%%%%%%%%%%%%%%%%%%%%
\subsection[$\mu \to e$ Conversion]{\boldmath{$\mu \to e$} Conversion}
\label{sec:mueconvo}
%%%%%%%%%%%%%%%%%%%%%%%%%%%%%%
Experiments searching for $\mu\to e$ conversion in the presence of an atomic nucleus have put strong limits on the branching ratio $\text{Br}(\mu\, \text{Au}\to e\,\text{Au})<7.0\times 10^{-13}$, which was measured by the SINDRUM-II collaboration \cite{Bertl:2006up} and looked for conversion in the presence of a gold target. Future experiments aim for increased sensitivity by multiple orders of magnitude, for example Mu2e \cite{Bartoszek:2014mya} and COMET \cite{Adamov:2018vin} which will use aluminum as a stopping target material and hope to reach limits as low as $\text{Br}\sim\mathcal{O}(10^{-17})$. We limit ourselves here to the case that only ALP-lepton and/or ALP-photon couplings are present. Then only the Feynman diagrams that are also responsible for $\mu\to e\gamma$ will contribute. Using results from Ref.~\cite{Kuno:1999jp}, we may write
\begin{equation}\label{eq:muNeN0}
\text{Br}(\mu N\to e N)=\frac{8\alpha_{\text{\tiny{QED}}}^5m_\mu Z_{\text{eff}}^4 Z F_p^2}{\Gamma_\text{capt}}\left(\left|F_2(-m_\mu^2)+F_3(-m_\mu^2)\right|+\left|F_2^{5}(-m_\mu^2)+F_3^{5}(-m_\mu^2)\right| \right)\,,
\end{equation}
where $Z_{\text{eff}}$ is the effective atomic charge, $F_p^2$ is the nuclear matrix element squared, $\Gamma_\text{capt}$ is the total muon capture rate, and we suppress the $\mu\to e$ superscript on the form factors. The numerical values for these quantities for the cases of gold and aluminum can be found in~\cite{Cornella:2019uxs, Cirigliano:2009bz, Kitano:2002mt}. %The factor $\xi^2$ reads
For heavy ALPs, i.e., $m_a>m_\mu$, the evaluation at $q^2=0$ is a good approximation and simplifies the calculation.

%%%%%%%%%%%%%%%%%%%%%%%%%%%%%%
\subsection{Muonium-antimuonium oscillations}
\label{sec:muonium}
%%%%%%%%%%%%%%%%%%%%%%%%%%%%%%
Muonium is a bound state of an antimuon and an electron ($\mu^+ e^-$) which can oscillate with antimuonium ($\mu^- e^+$) in the presence of $e$-$\mu$ flavor-violating interactions. The LFV ALP can mediate these transitions via both $s$- and $t$-channel tree-level diagrams~\cite{Hou:1995dg,Endo:2020mev}. In both cases, we have $s\approx t \approx m_\mu^2$; where the equality becomes exact in the limit that both the electron mass and the binding energy of muonium are taken to be zero (both are very small relative to $m_\mu$). This means that there are two limits in which the ALP propagators tend to a constant, and so the effects of the ALP can be mapped onto effective four-fermion operators; either $m_a \ll m_\mu$ or $m_a \gg m_\mu$. In the limit $m_a \ll m_\mu$,
\begin{equation}
\mathcal{H}_{\text{eff}}^{m_a \ll m_\mu} =-\frac{1}{4 f^2}\left([k_e]_{12}+[k_E]_{12} \right)^2\, (\bar{\mu} e) (\bar{\mu} e)-\frac{1}{4 f^2}\left([k_E]_{12}-[k_E]_{12} \right)^2 \,(\bar{\mu} \gamma^5 e)(\bar{\mu} \gamma^5 e),
\end{equation}
while in the limit $m_a \gg m_\mu$:
\begin{equation}
\mathcal{H}_{\text{eff}}^{m_a \gg m_\mu} =\frac{m_\mu^2}{4 m_a^2 f^2}\left([k_e]_{12}+[k_E]_{12} \right)^2\, (\bar{\mu} e) (\bar{\mu} e)+\frac{m_\mu^2}{4 m_a^2 f^2}\left([k_E]_{12}-[k_E]_{12} \right)^2 \,(\bar{\mu} \gamma^5 e)(\bar{\mu} \gamma^5 e).
\end{equation}
The muonium-antimuonium transition probability is then given in the $m_a \ll m_\mu$ limit by~\cite{Hou:1995dg,Endo:2020mev}
\begin{align}
P^{\,m_a \ll m_\mu}=\frac{\tau_\mu^2}{2\pi^2a_B^6}\frac{1}{f^4}\bigg[&|c_{0,0}|^2\Big| 4[k_E]_{12}[k_e]_{12}-\delta_B\big([k_e]_{12}-[k_E]_{12}\big)^2\Big|^2\notag\\
&+|c_{1,0}|^2 \Big| 4[k_E]_{12}[k_e]_{12}+\delta_B\big([k_e]_{12}-[k_E]_{12}\big)^2\Big|^2 \bigg]\,,
\end{align}
and in the $m_a \gg m_\mu$ limit by
\begin{align}\label{eq:muonium}
P^{\,m_a \gg m_\mu}=\frac{\tau_\mu^2}{2\pi^2a_B^6m_a^4}\frac{m_\mu^4}{f^4}\bigg[&|c_{0,0}|^2\Big| 4[k_E]_{12}[k_e]_{12}-\delta_B\big([k_e]_{12}-[k_E]_{12}\big)^2\Big|^2\notag\\
&+|c_{1,0}|^2 \Big| 4[k_E]_{12}[k_e]_{12}+\delta_B\big([k_e]_{12}-[k_E]_{12}\big)^2\Big|^2 \bigg]\,,
\end{align}
where the muon lifetime $\tau_\mu=3.34\times 10^{18}$\,GeV$^{-1}$ and the muonium Bohr radius $a_B=2.69\times 10^5$ GeV$^{-1}$. The population probabilities of the muonium angular momentum states $c_{J, m_J}$ and the value of $\delta_B$ depend on the experimental setup. Specifically, we define $\delta_B$ in terms of the magnetic field $B$ as $\delta_B \equiv\left(1+X^2\right)^{-1/2}$, with $X$ the dimensionless parameter
\begin{equation}
X=\frac{\mu_B B}{a}\left(g_e + \frac{m_e}{m_\mu}g_\mu \right) \approx 6.24\, \frac{B}{\text{Tesla}},
\end{equation}
where $\mu_B=e/(2m_e)$ is the Bohr magneton, $g_e\approx g_\mu \approx 2$ are the magnetic moments of the electron and muon, and $a\approx 1.864 \times 10^{-5}$\,eV is the muonium 1S hyperfine splitting.  

The strongest constraint on the transition probability has been reported by the MACS collaboration which obtained $P< 8.3\times 10^{-11}$ at $90\%$ CL~\cite{Willmann:1998gd}. For the MACS experiment, the population probabilities have been estimated as $|c_{0,0}|^2=0.32$ and $|c_{1,0}|^2=0.18$ and the magnetic field is $B=0.1$\,Tesla, giving $\delta_B=0.85$~\cite{Hou:1995dg,Horikawa:1995ae}.

%%%%%%%%%%%%%%%%%%%%%%%%%%%%%
\subsection{\boldmath The anomalous magnetic moment of the muon and the electron}
\label{sec:gminus2}
%%%%%%%%%%%%%%%%%%%%%%%%%%%%%

Precise SM predictions for the anomalous magnetic moment $a_\mu=(g-2)_\mu/2$ have been calculated  using experimental input from LEP measurements of the R-ratio to determine the hadronic vacuum polarization contributions terms by the $g-2$ theory initiative paper (TI)~\cite{Aoyama:2020ynm} and, alternatively, by using only input from lattice calculations by the Budapest, Marseille and Wuppertal (BMW) collaboration (BMW)~\cite{Borsanyi:2020mff}.   
These theory predictions disagree at the level of $2$ standard deviations. The comparison to the 
 combination of the measurements of $a_\mu$ from the Brookhaven~\cite{Bennett:2006fi} and Fermilab~\cite{Abi:2021gix} experiments 
\begin{align}
\Delta a_\mu^{\rm TI}&=a_\mu^{\rm exp}-a_\mu^{\rm TI}=(25.1\pm 5.9)\times 10^{-10}\,,\\
\Delta a_\mu^{\rm BMW}&=a_\mu^{\rm exp}-a_\mu^{\rm BMW}=(10.7\pm 6.9)\times 10^{-10}\,,
\end{align}
leads to a tension with the TI prediction with a statistical significance of $4.2 \sigma$, whereas the BMW determination is in better agreement with the measured value. In the following we will use the TI value and suppress the superscript $\Delta a_\mu= \Delta a_\mu^\text{TI}$,
and discuss the coupling structure of a potential ALP explanation for this tension.\footnote{ If the BMW calculation is correct it would imply a tension between the experimental value of the R-ratio and the BMW prediction with different, potentially interesting implications~\cite{Crivellin:2020zul}.}

Furthermore, a slight deviation of the electron anomalous magnetic moment $a_e=(g-2)_e/2$ has been observed. The central value of $a_e^\text{exp}$ \cite{Hanneke:2008tm, Hanneke:2010au} deviated from the SM prediction \cite{Aoyama:2017uqe} previously, but is now statistically more significant due to an improved measurement of the fine-structure constant \cite{Parker:2018vye} in Caesium atoms, which contributes to the error budget of the SM prediction. There exists a competing measurement of the finestructure constant in Rubidium~\cite{Morel:2020dww}, which would result in a deviation in the opposite direction,
\begin{align}
\Delta a_e^\text{Cs}&=(-88\pm 36)\times 10^{-14}\,,\\
\Delta a_e^\text{Rb}&=(48\pm 30)\times 10^{-14}\,.
\end{align}
The statistical significance of the deviation is $2.4\,\sigma$ and $1.6\,\sigma$, respectively. Interestingly, if rescaled by the lepton masses, one finds that the relative size of the effects in the anomalous magnetic moments are
\begin{align}
\frac{\Delta a_e^\text{Cs}}{\Delta a_\mu}\approx -15.0 \frac{m_e^2}{m_\mu^2}\,,\\
\frac{\Delta a_e^\text{Rb}}{\Delta a_\mu}\approx 8.1 \frac{m_e^2}{m_\mu^2}\,.
\end{align}
There are several ALP contributions to the lepton anomalous magnetic moments. At the one-loop level, there are penguin diagrams with the ALP attached only to fermion lines as well as Barr-Zee diagrams with the ALP connected to fermions and the photon. At two-loop level there is also a contribution from the ALP-photon coupling only \cite{Marciano:2016yhf}, and from the ALP-fermion coupling only, shown in Figure~\ref{fig:formfactor2loop}. For the case of flavor-conserving ALP couplings the different contributions have been discussed in \cite{Leveille:1977rc, Haber:1978jt, Chang:2000ii,Marciano:2016yhf, Bauer:2017ris,Buen-Abad:2021fwq}, and one finds with \eqref{eq:F2iii} at one loop, 
\begin{align}\label{eq:alpamudiag}
\Delta a_\mu =-\frac{m_\mu^2 c_{\mu\mu}^2}{16 \pi^2 f^2}\left[h_1(x_\mu) +\frac{2\alpha}{\pi}  \frac{c_{\gamma\gamma}}{c_{\mu\mu}} \Big(\log\frac{\mu^2}{m_\mu^2}-h_2(x_\mu)  \Big)\right]\,,
\end{align}
where $x_\mu = m_a^2/m_\mu^2$ and we have neglected the contribution from Barr-Zee diagrams with internal $Z$ bosons, which are suppressed by the $Z$ vector coupling $(1-4s_w^2)\approx \alpha$.  The loop functions \eqref{eq:h1h2} for vanishing and large $x_\mu$ are given by $h_{1,2}(0)=1$ and $h_1(x_\mu\gg 1)\approx (2/x_\mu)(\log x_\mu -11/6)$, $h_2(x_\mu \gg 1)\approx (\log x_\mu+3/2 )$. Note that the contribution proportional to $c_{\mu\mu}^2$ has the wrong sign to explain the deviation of $a_\mu^\text{exp}$ with respect to the SM value, but the analogous expression for $\Delta a_e^\text{Cs}$ would have the correct sign. 

For ALPs with lepton flavor-violating couplings, significant additional contributions to the anomalous magnetic moment of the muon can arise from the tau or the electron in the loop. For the case of a tau in the loop, we find from the rightmost diagram in Figure~\ref{fig:formfactors}
\begin{align}\label{eq:alpamutau}
\Delta a_\mu = \frac{m_\mu m_\tau}{16\pi^2\, f^2}\, \text{Re}\big[[k_e]_{23} [k_E]_{23}^* \big]\, h(x_{\tau})+\mathcal{O}\Big(\frac{m_\mu}{m_\tau}\Big) \,,
\end{align}
where the function $h(x)$ is given in \eqref{eq:hloop}, and $h(x)=1$ for $x\to 0$ and $h(x)=0$ for $x\gg 1$.
The right diagram in Figure~\ref{fig:formfactors} with an electron in the loop also contributes to $a_\mu$ as
\begin{align}\label{eq:alpamue}
\Delta a_\mu =  -\frac{ m_\mu^2}{32\pi^2 f^2} \left(\left| [k_E]_{12} \right|^2 +\left|[k_e]_{12} \right|^2 \right) j(x_\mu)+\mathcal{O}\Big(\frac{m_e}{m_\mu}\Big) \,,
\end{align}
where the function $j(x)$ is given in \eqref{eq:jloop} and $j(x)=1$ for $x\to 0$ and  $j(x)=0$ for $x\gg 1$. 
The anomalous magnetic moment of the electron receives a contribution from the tau running in the loop (analogous to \eqref{eq:alpamutau} with the replacement $\mu \to e$) as well as from the muon in the loop with the replacements $\mu \to e$ and $\tau \to \mu$ in \eqref{eq:alpamutau}.

Notably, equation \eqref{eq:alpamutau} can have either sign, whereas the sign of \eqref{eq:alpamue} is fixed and positive for $m_a>m_\mu$. The contributions from \eqref{eq:alpamutau} and \eqref{eq:alpamue} can have the right sign to explain the anomalous magnetic moment of the muon.
In the following we discuss three different scenarios to address the tension in $a_\mu$ and $a_e$ with an ALP coupling to leptons and photons. We assume a different set of two of these couplings to be dominant, but a combination of these different mechanisms might be feasible as well. In the following we will explore the possibility to explain both $\Delta a_\mu$ and $\Delta a_e^\text{Cs}$ with

%%%
%%%%%%%%%%%%%
\begin{figure}[t]
\begin{center}
\includegraphics[width=0.95\textwidth]{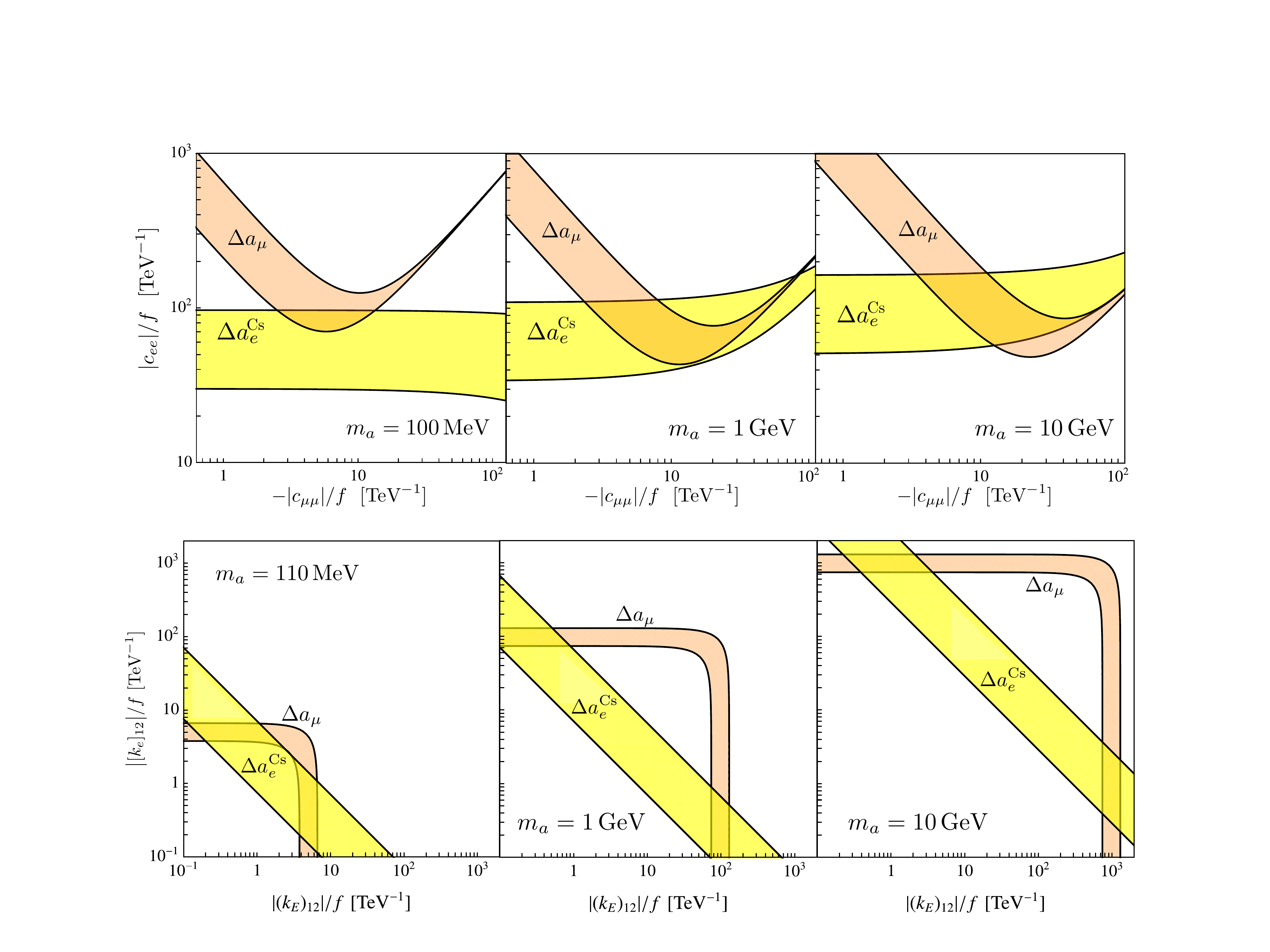}
\caption{ \label{fig:amuoptions} Parameter space for which $\Delta a_\mu$ (orange) and $\Delta a_e^\text{Cs}$ (yellow) can be explained at 95\% CL by flavor-conserving ALP couplings to muons and electrons for $m_a=100\,$MeV, $1\,$GeV and $10\,$GeV. The corresponding parameter space for $\Delta a_e^\text{Rb}$ is not shown as the deviation is $< 2 \sigma$.}
\end{center}
\end{figure}
%%%%%%%%%%%%
%%%

\paragraph{I. Large effective photon couplings.}\label{sec:largegamma}

A contribution from \eqref{eq:alpamudiag} can explain $\Delta a_\mu$ for large ALP couplings to photons and a relative sign between the photon and muon coupling, $-c_{\gamma\gamma}^\text{eff}/c_{\mu\mu}\sim 10\!-\!30$~\cite{Leveille:1977rc, Haber:1978jt, Chang:2000ii,Marciano:2016yhf, Bauer:2017ris}. The effective ALP photon coupling is given by \eqref{eq:Cgaga} and for $m_a \gg  m_{\ell} $ simplifies to $c_{\gamma\gamma}^\text{eff}\approx c_{\gamma\gamma} + \sum_{\substack{\ell }} c_{{\ell}{\ell}}$ where $\ell=e,\mu,\tau$,  when only the contribution of lepton loops is taken into account. For an ALP heavier than the electron, a large ALP-photon coupling can be induced even if $c_{\gamma\gamma}$ is small or vanishing at tree-level. An explanation of $\Delta a_\mu$ therefore requires non-universal ALP-lepton couplings $-c_{ee}/c_{\mu\mu}\approx 10\!-\!30$ and $m_a>2m_e$. The relative sign is important since $c_{\gamma\gamma}^\text{eff}c_{\mu\mu}<0$ is necessary to explain $\Delta a_\mu$. Interestingly, an ALP-electron coupling of this magnitude and sign can simultaneously explain $\Delta a_e^\text{Cs}$, since both terms in \eqref{eq:alpamudiag} are negative. A similar combined explanation with $\Delta a_e^\text{Rb}$ would not be possible. In Figure~\ref{fig:amuoptions} we show the parameter space for which $\Delta a_\mu$ and $\Delta a_e^\text{Cs}$  can be explained at 95\% C.L. by an ALP with $m_a=0.1,1, 10$ GeV  in orange and yellow, respectively. The ALP contribution to $\Delta a_e$ is almost constant in $c_{\mu\mu}$. For large $|c_{\mu\mu}|$ and $m_a > 2m_\mu$, the effective photon coupling is reduced and $c_{ee}$ needs to be larger to compensate. For small $|c_{\mu\mu}|$, the only sizeable contribution to $\Delta a_\mu$  comes from the $c_{ee}$ induced photon coupling and $c_{ee}$ needs to be large to explain the muon anomalous magnetic moment. Sizeable values of $|c_{\mu\mu}|$ increase the $c_{\gamma\gamma}^\text{eff}$-independent term and $c_{ee}$ needs to be large again to overcome this contribution. The combination of these two effects leads to the parabolic shape of the orange region. For $c_{\mu\mu}/f \approx -\mathcal{O}(10)/$TeV and $c_{ee}/f \approx \mathcal{O}(100)/$TeV both anomalies can be explained without additional ALP couplings. Even though such an explanation is possible for ALPs with masses $m_a< 1$ GeV, it is in tension with constraints from beam dump searches for ALPs coupled to leptons as shown on the left panel of Figure~\ref{fig:flavorconstraintsandleptons}. Any ALP model aiming to explain the anomalous magnetic moment of the muon \emph{or} the electron would need to be heavy enough or have additional couplings to evade these bounds.

\paragraph{II. Flavor-violating ALP couplings to $\mu$ and e.} 

For small or vanishing ALP couplings to leptons, $c_{\gamma\gamma}$, $c_{ee}\ll 1$, an explanation for $\Delta a_\mu$ and $\Delta a_e$ can in principle be provided by flavor-changing ALP couplings to muons and electrons. The diagram on the right of Figure~\ref{fig:formfactors}  with external muons and internal electrons gives rise to expression \eqref{eq:alpamue} and the same diagram with external electrons and internal muons gives the leading order contribution to $\Delta a_e$ given by \eqref{eq:alpamutau} with the replacements $\mu \to e$ and $\tau \to \mu$. These two contributions can have opposite signs and the right magnitude to explain both $\Delta a_e^\text{Cs}$ and $\Delta a_\mu$. As was pointed out by the authors of~\cite{Endo:2020mev}, the constraint imposed by muonium-antimuonium oscillations excludes such an explanation for $\Delta a_\mu$ for all values of $m_a$. The parameter space for an explanation of $\Delta a_e$ is strongly constrained as well, but for ALPs with masses $m_a\gtrsim 2$ GeV, $\Delta a_e^\text{Cs}$ can be accommodated. 

\paragraph{III. Flavor-violating ALP couplings to $\tau$, $\mu$ and e.} 
A contribution to both $a_\mu$ and $a_e$ arises from tau leptons in the loop of the rightmost diagram of Figure~\ref{fig:formfactors} if the external fermions are muons or electrons, respectively. This contribution is given by \eqref{eq:alpamutau} (for the electron with the replacement $\mu \to e$) and chirally enhanced by the tau mass. Even though \eqref{eq:alpamutau} can have either sign, it requires both $[k_E]_{32} $ and $ [k_e]_{31}$ to be non-zero (and $[k_E]_{31}, [k_e]_{31}\neq 0$ in the case of $\Delta a_e$)
 and a simultaneous explanation of both anomalies is ruled out by the ALP contribution to $\mu \to e\gamma$
\begin{align}
\Gamma(\mu\to e\gamma)=\frac{m_\mu^3m_\tau^2\alpha}{1024\pi^4 f^4}\left(1-\frac{m_e^2}{m_\mu^2}\right) \Big[ |[k_e]_{23}[k_E]_{31}|^2 +|[k_E]_{23}[k_e]_{31}|^2\Big] g_3(0,m_\tau, m_a)^2
\end{align}
with $g_3(q^2,m_\tau,m_a)$ given in \eqref{eq:g3loop}. 
From the constraint on $\mu\to e\gamma$ follows that for $m_a=1$ GeV
\begin{align}
\Big( \big|[k_e]_{23}[k_E]_{31}|^2 +|[k_E]_{23}[k_e]_{31}\big|^2\Big)^{1/2} \leq 2\times 10^{-5} \frac{f^2}{\text{TeV}^2}\,.
\end{align}
In order to explain $\Delta a_\mu$ or $\Delta a_e^\text{Cs}$ one needs coefficients $\text{Re}[[k_E]_{32}^*[k_e]_{32}]\approx4$ and $\text{Re}[[k_E]_{31}^*[k_e]_{31}]\approx 0.32$ for $f =1$ TeV. This conclusion does not change for different ALP masses. An explanation of either $\Delta a_\mu$ or $\Delta a_e^\text{Cs}$ can be obtained from tau-flavor-violating ALP couplings if $c_{\tau\tau}<1$. In Figure~\ref{fig:leptonconstraints2} and~\ref{fig:leptonconstraints3} we show the parameter space in the $m_a-c_{\tau\mu}$ and $m_a-c_{\tau e}$ plane for which the measured values can be reproduced in orange and yellow assuming $[k_e]_{ij}=[k_E]_{ij}=c_{ij}/\sqrt{2}$. A sizeable value of either $c_{\mu\mu}$ or $c_{ee}$ could provide a contribution large enough to explain the tension in the respective other magnetic moment, as shown in Figure~\ref{fig:amuoptions}, however this would imply a large $C^\text{eff}_{\gamma\gamma}$ coupling, and is ruled out by $\tau \to \mu \gamma$ or $\tau \to e \gamma$ bounds. This constraint can only be avoided if other contributions to $C^\text{eff}_{\gamma\gamma}$ cancel the contribution induced by $c_{\ell\ell}$. Therefore, a hybrid explanation of this sort in which one anomaly is explained by off-diagonal couplings and the other is explained by diagonal couplings is under tension. 
A better option is to explain $\Delta a_\mu$ via $\tau-\mu$ couplings, and $\Delta a_e$ via $\mu-e$ couplings (viable for ALP masses above around a GeV), while keeping the diagonal couplings rather small.

%%%%%%%%%%%%%%%%%%%%%%%%%%%%%
\subsection{\boldmath The electric dipole moment of the muon and the electron}
\label{sec:EDM}
%%%%%%%%%%%%%%%%%%%%%%%%%%%%%

\begin{figure}[t]
	\begin{center}
		\includegraphics[width=0.5\textwidth]{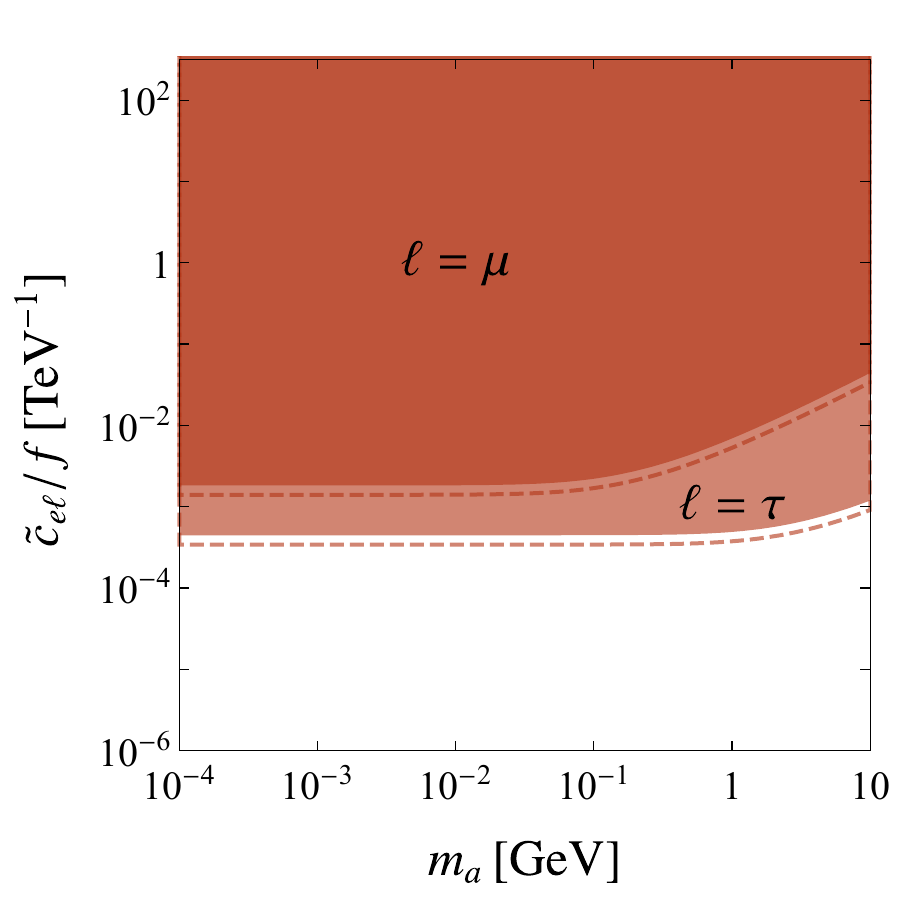}
		\caption{Bounds on ALP induced eEDM with $\tilde{c}_{e\ell}\equiv\sqrt{\text{Im}\left([k_E]^*_{j 1}[k_e]_{j 1}\right)}$,  assuming universal ALP couplings to leptons $c_{ee}/f=c_{\mu\mu}/f=c_{\tau\tau}/f=1 \,$TeV$^{-1}$ and all other Wilson coefficients zero at tree-level.\label{fig:EDMconstraints}}
	\end{center}
\end{figure}

The SM prediction for the electric dipole moment of the electron (eEDM) is $|d_e|\lesssim10^{-37}\rm ecm $ ~\cite{Bernreuther:1991,Booth:1993} and for the electric dipole moment of the muon ($\mu$EDM) $|d_\mu|\lesssim10^{-25}\rm ecm $ ~\cite{Bennett:2008}. By measurements, they are excluded by $|d_e|<1.1\times10^{-29}\rm ecm $ ~\cite{ACME:2018} and $|d_\mu|<1.9\times10^{-19}\rm ecm $ ~\cite{Bennett:2008}. An ALP with non-vanishing off-diagonal lepton couplings can generate an EDM at 1-loop level through the rightmost diagram in Figure~\ref{fig:formfactors} where both external leptons are of the same flavor and the internal ones belong to either of the other\footnote{Note that for a general spin-0 field with CP-even and -odd couplings additional contributions arise \cite{Kirpichnikov:2020lws,DiLuzio:2020oah}.}. The eEDM is then defined in terms of the form factor $F_2^{5, i\rightarrow i}$ as
\begin{equation}
|d_e|=\frac{|F_2^{5, i\rightarrow i}(q^2=0)|}{2}.
\end{equation}
with $F_2^{5,i\to i}$ given in eq.~\eqref{eq:fullF25same}. When using the limit $m_e\ll m_k$, where $k=\mu,\tau$, this simplifies to
\begin{equation}
F_2^{5, i\rightarrow i}(q^2=0)=-\frac{m_keQ_k}{32\pi^2f^2}i \,\text{Im} \left([k_E]^*_{j 1}[k_e]_{j 1}\right)h(x_k),
\end{equation}
where $h(x)$ is given in eq.~\eqref{eq:hloop}. The $\mu$EDM can be taken as a constraint on the $\tau-\mu$ couplings, since the $\mu-e$ couplings are already much more strongly constrained by the measurement of the eEDM.
The constraints on the imaginary parts of the off-diagonal ALP couplings to leptons are shown in Figure~\ref{fig:EDMconstraints}.

%%%
%%%%%%%%%%%%%
%%%%%%%%%%%%
%%%

%%%
%%%%%%%%%%%%%
\begin{figure}[t]
\begin{center}
\includegraphics[width=1\textwidth]{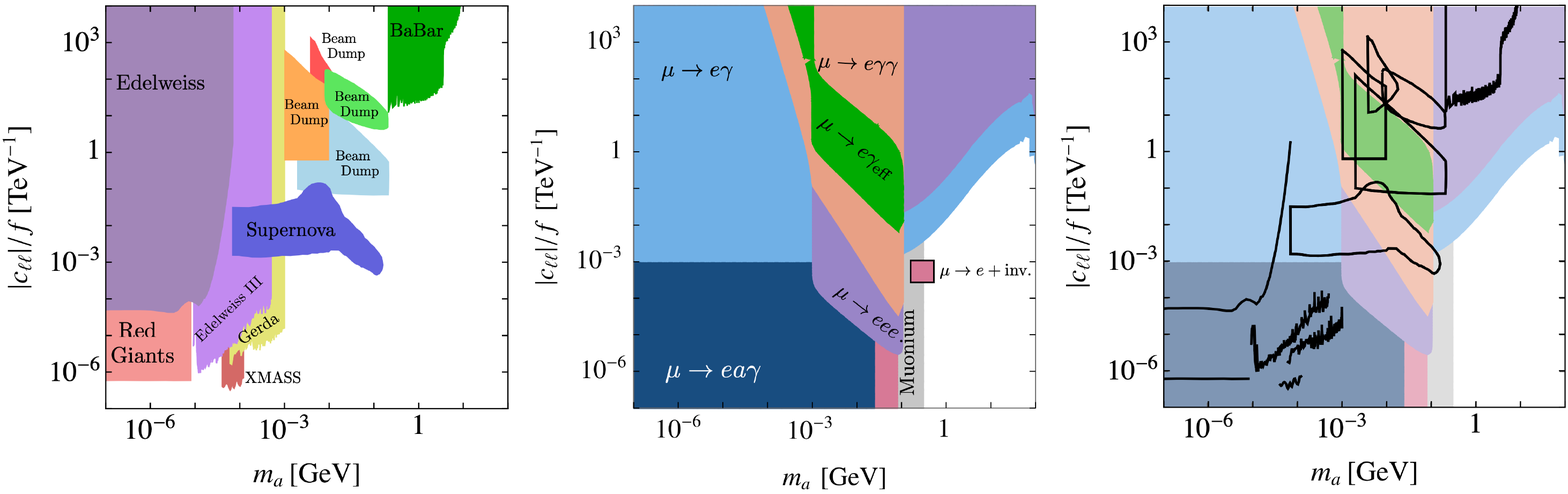}
\caption{Left: Astrophysical, beam dump, and collider constraints on ALP couplings to leptons $c_{\ell\ell}=c_{e}-c_L$ (as in Figure~\ref{fig:flavorconstraintsandleptons}). Center: Bounds on ALP mediated flavor off-diagonal $\mu \to e$ transition with $c_{\mu e}/f\equiv\sqrt{|[k_E]_{21}|^2+|[k_e]_{21}|^2}/f=1\,$TeV$^{-1}$ as a function of the universal ALP coupling to leptons, $c_{ee}/f=c_{\mu\mu}/f=c_{\tau\tau}/f$, and the ALP mass, $m_a$. All other Wilson coefficients are set to zero at tree-level. Right: Overlay of the flavor constraints (shown in the central panel) in color and the astrophysical, beam dump and collider limits (shown in the left panel) depicted by black contour lines.
\label{fig:leptonconstraintsnew}}
\end{center}
\end{figure}
%%%%%%%%%%%%
%%%

%%%%%%%%%%%%%%%%%%%%%%%%%%%%%
\subsection{\boldmath Interplay of flavor-violating and flavor-conserving ALP couplings to leptons}
\label{sec:Interplay}
%%%%%%%%%%%%%%%%%%%%%%%%%%%%%
In contrast to the quark sector, charged lepton flavor symmetry is conserved in the SM and any charged lepton flavor-violating effects vanish in the limit of zero neutrino masses. ALP couplings that conserve lepton flavor in the UV, e.g.,~a flavor-universal coupling to lepton doublets $c_L$, will therefore only induce flavor-conserving interactions at low energy scales. If, however, the ALP couplings are not flavor-universal, flavor off-diagonal ALP couplings are induced by the rotation into the charged lepton mass eigenbasis.  
For example, in an ALP model in which contributions to both $a_\mu$ and $a_e$ arise from ALP couplings $|c_{ee}|\gg c_{\mu\mu}$ as discussed in Section~\ref{sec:largegamma}, the flavor off-diagonal ALP couplings $[k_e]_{ij}, [k_E]_{ij}$ do not automatically vanish as in the case of flavor-universal ALP couplings. Similarly, for a UV theory which contains any of the ALP couplings $[k_e]_{ij}, [k_E]_{ij}$, there is no reason to expect universal flavor-conserving ALP couplings in the charged lepton mass basis. 

As discussed in Section~\ref{sec:ALPtolep}, flavor-conserving ALP couplings to leptons are already severely constrained by astrophysical, beam dump and collider experiments. We show the excluded parameter space again in the left panel of Figure~\ref{fig:leptonconstraintsnew}. 
Some of the flavor observables discussed earlier in this section, such as e.g., muonium oscillations, depend solely on flavor off-diagonal ALP couplings and are independent of the flavor-diagonal ALP couplings $c_{\ell\ell}$. Other observables, such as the decay process $\mu \to 3 e$, depend on both flavor-violating and -conserving ALP couplings.

In order to compare the experimental sensitivities to the various ALP couplings, we show the constraints from flavor observables for a single flavor off-diagonal ALP coupling $c_{\mu e}/f\equiv \big(|[k_E]_{21}|^2+|[k_e]_{21}|^2\big)^{1/2}/f=1/\text{TeV}$ as a function of the ALP mass $m_a$ and the universal flavor-conserving ALP coupling $|c_{\ell\ell}|/f$ in the centre panel of Figure~\ref{fig:leptonconstraintsnew}. While the constraints from muonium oscillations, $\mu \to ea(\text{invisible})$ and $\mu\to e a \gamma$ are independent of $c_{\ell\ell}$, limits from $\mu \to eee$ and $\mu \to e \gamma\gamma$ require the ALP to decay and become irrelevant below $|c_{\ell \ell}|\lesssim 10^{-6}/$ TeV.\footnote{A detailed description of the experimental limits shown here can be found in Section \ref{sec:discussion2} below.} The latter constraints are also only relevant for ALP masses $2m_e< m_a < m_\mu$.

In the right panel of Figure~\ref{fig:leptonconstraintsnew} we compare constraints from charged lepton flavor observables (central panel) with limits on flavor-conserving ALP couplings from astrophysical, beam dump and collider experiments (left panel). As we can see, the flavor bounds are highly competitive and outperform astrophysical, beam dump and collider experiments throughout the entire mass range under consideration. Note that this comparison neglects the fact that a non-zero value of $c_{\mu e}/f$ might alter the parameter space excluded by some of the astrophysical, beam dump and collider experiments, possibly reducing their importance further.

We note that most of the constraints in the left panel of Figure~\ref{fig:leptonconstraintsnew} only apply to the ALP-electron coupling. In fact, for ALP masses $m_a< 2m_e$ a coupling structure with $c_{ee}/f = 10^{-6}/$ TeV, $c_{\mu\mu}/f = c_{\tau\tau}/f= 1/\text{TeV}$ is still allowed while ALP masses in the range $ 2m_e< m_a < 2m_\mu $ only require ALP couplings $c_{ee}/f = 10^{-3}/$ TeV, $c_{\mu\mu}/f = c_{\tau\tau}/f= 1/\text{TeV}$ and for $m_a> 2m_\mu$, $c_{ee}/f =c_{\mu\mu}/f = c_{\tau\tau}/f= 1/\text{TeV}$ is still unconstrained. We will thus choose a mass dependent ALP coupling structure to show the maximal reach of different experimental observables in the following discussion of lepton flavor observables.

%%%%%%%%%%%%%%%%%%%%%%%%%%%%%
\subsection{\boldmath Discussion of constraints from lepton flavor-violating observables}
\label{sec:discussion2}
%%%%%%%%%%%%%%%%%%%%%%%%%%%%%

\begin{figure}[t]
\begin{center}
\includegraphics[width=0.85\textwidth]{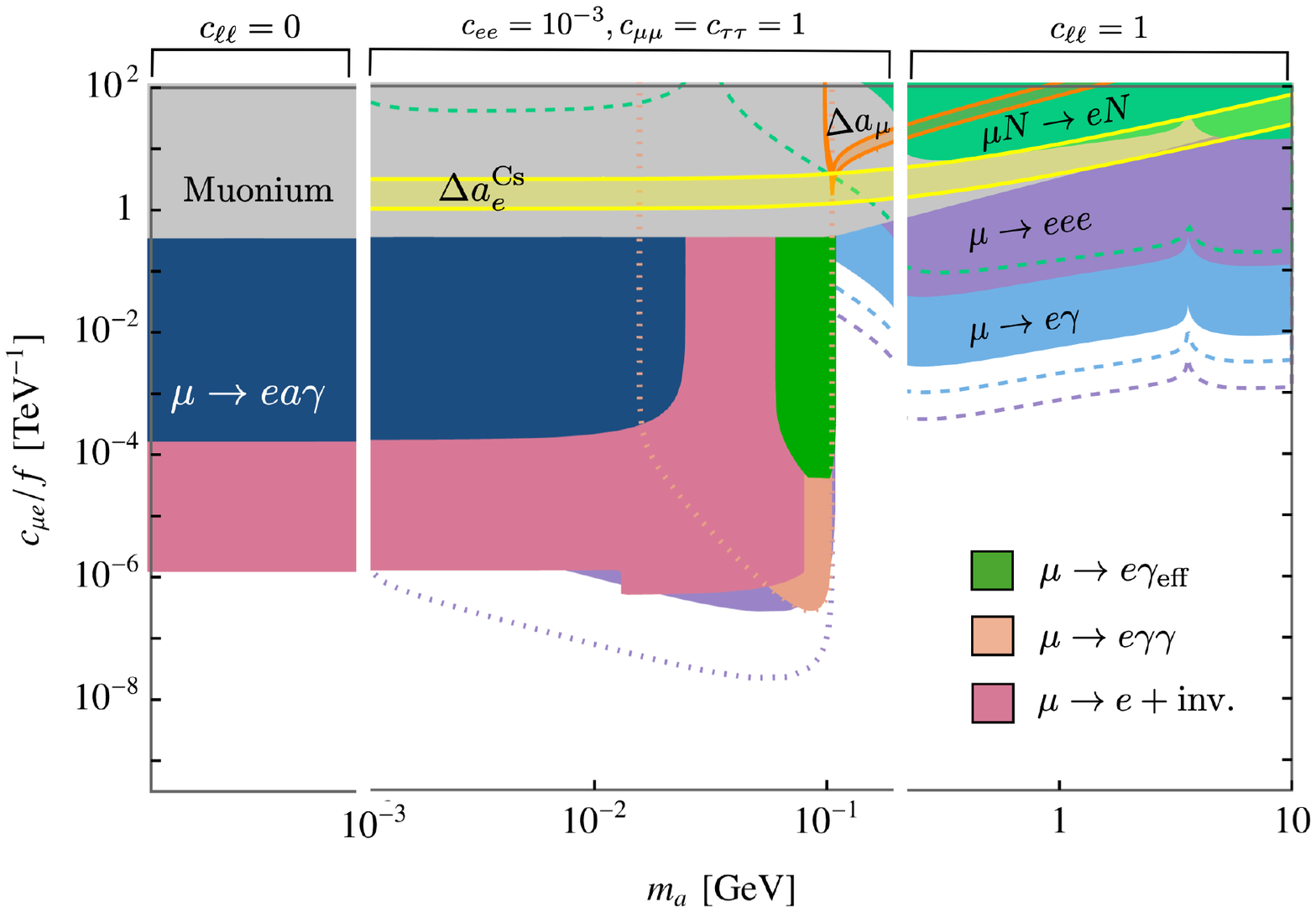}
\caption{Bounds on ALP mediated flavor off-diagonal transitions between muons and electrons with $c_{\mu e}\equiv\sqrt{|[k_E]_{21}|^2+|[k_e]_{21}|^2}$,  assuming universal ALP couplings to leptons as indicated above the plot for the different ALP mass regions.\label{fig:leptonconstraints1}}
\end{center}
\end{figure}

\begin{figure}[t]
\begin{center}
\includegraphics[width=0.85\textwidth]{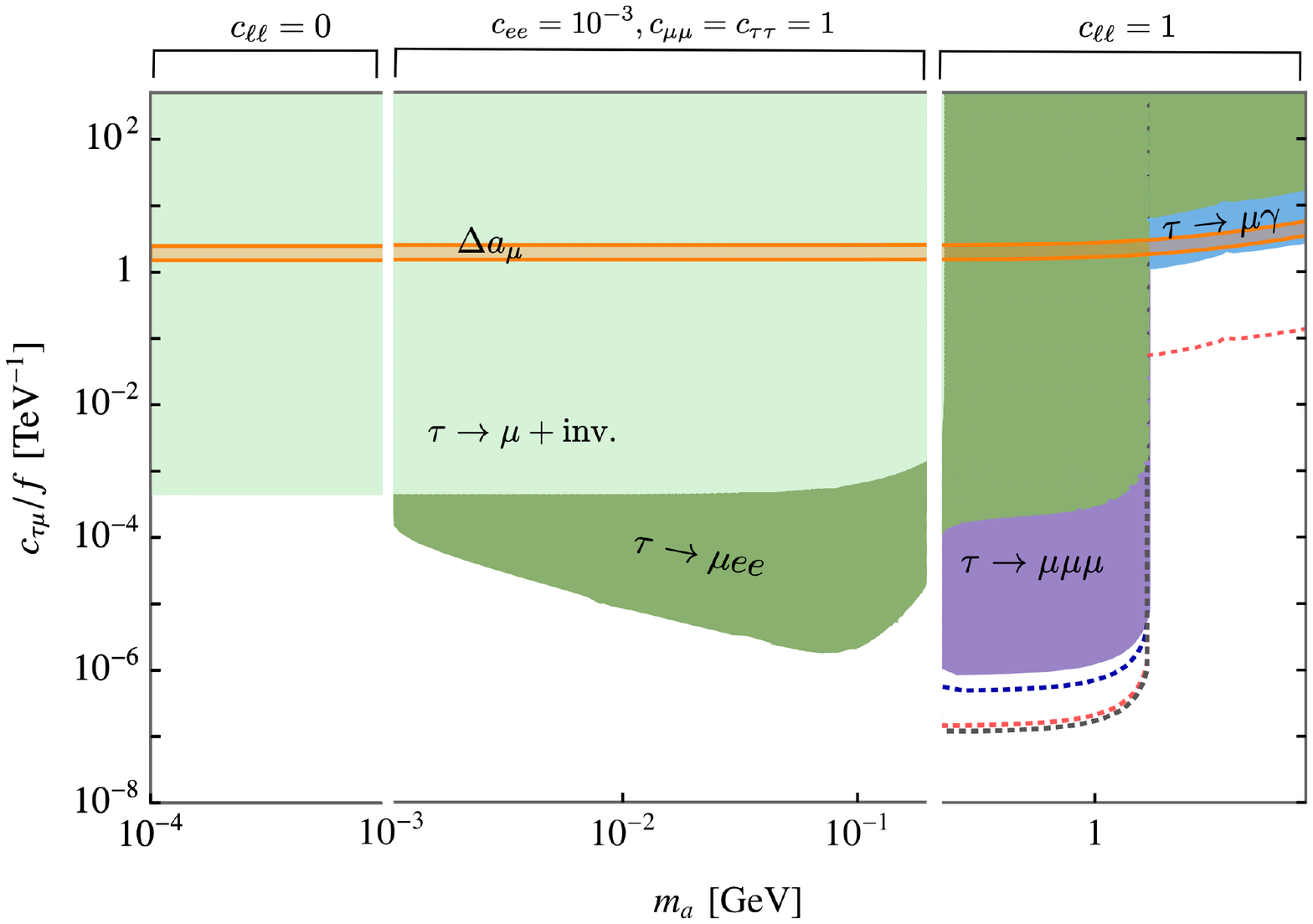}
\caption{Bounds on ALP mediated flavor off-diagonal transitions between taus and muons with $c_{\tau \mu}\equiv\sqrt{|[k_E]_{32}|^2+|[k_e]_{32}|^2}$, assuming universal ALP couplings to leptons as indicated above the plot for the different ALP mass regions.\label{fig:leptonconstraints2}}
\end{center}
\end{figure}

\begin{figure}[t]
\begin{center}
\includegraphics[width=0.85\textwidth]{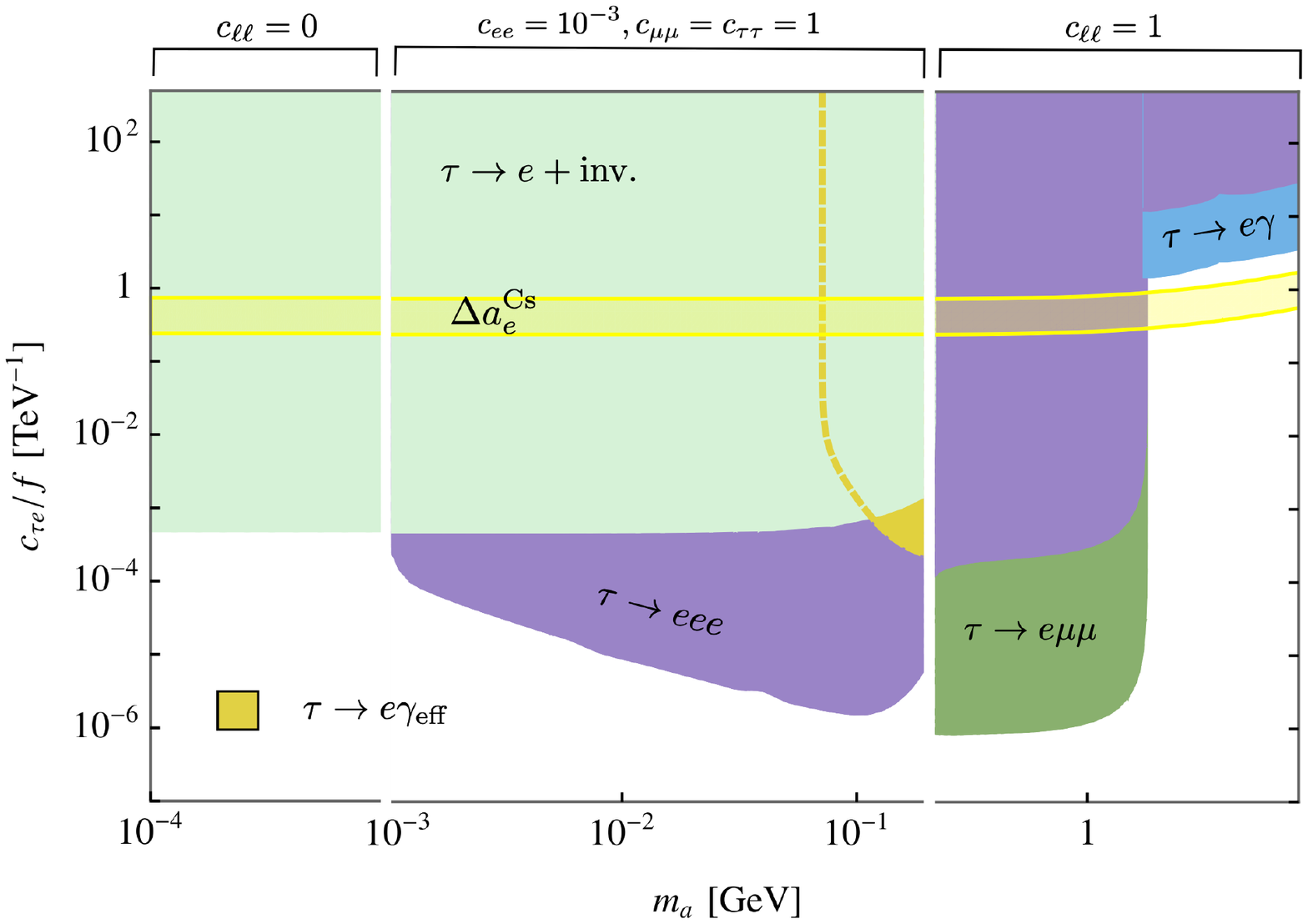}
\caption{Bounds on ALP mediated flavor off-diagonal transitions between taus and electrons with $c_{\tau e}\equiv\sqrt{|[k_E]_{31}|^2+|[k_e]_{31}|^2}$, assuming universal ALP couplings to leptons as indicated above the plot for the different ALP mass regions. \label{fig:leptonconstraints3}}
\end{center}
\end{figure}

We present constraints on the ALP-induced $\mu \to e $, $\tau \to \mu $ or $\tau \to e$ transitions in Figures~\ref{fig:leptonconstraints1},~\ref{fig:leptonconstraints2} and~\ref{fig:leptonconstraints3}, respectively. In each case we assume that only a single flavor-changing lepton coupling is present and that the flavor-diagonal ALP couplings to leptons are chosen such that they are not excluded by any of the constraints shown in Figure~\ref{fig:leptonconstraintsnew},  

\begin{equation}
\begin{aligned}\label{eq:cllvalues}
\frac{|c_{ee}|}{f}=\frac{|c_{\mu\mu}|}{f}=\frac{|c_{\tau\tau}|}{f}\equiv\frac{|c_{\ell\ell}|}{f}=0\,,\qquad &\text{for} \qquad m_a<2m_e\,,\\
\frac{|c_{ee}|}{f}=\frac{10^{-3}}{\text{TeV}}\,,\quad  \frac{|c_{\mu\mu}|}{f}=\frac{|c_{\tau\tau}|}{f}=\frac{1}{\text{TeV}}\,,\qquad &\text{for } \qquad 2m_e<m_a<2m_\mu\,,\\
\frac{|c_{ee}|}{f}=\frac{|c_{\mu\mu}|}{f}=\frac{|c_{\tau\tau}|}{f}\equiv\frac{|c_{\ell\ell|}}{f}=\frac{1}{\text{TeV}}\,,\qquad &\text{for } \qquad m_a>2m_\mu\,
\end{aligned}
\end{equation}
The flavor-conserving ALP couplings are relevant for the branching ratios and decay lengths of the ALP, which can decay into leptons, or photons through the loop-induced coupling given in Section \eqref{sec:decaysleptonsphotons}. 

The results presented in this section are useful to constrain UV models in which one coupling dominates over the others. However, in the absence of additional assumptions, a UV completion, in which a horizontal global symmetry group is broken to produce a pseudo-Nambu Goldstone boson, could induce all possible flavor off-diagonal couplings to leptons. A discussion of lepton flavor-violating ALP decays in the context of such explicit UV models can be found in~\cite{Bjorkeroth:2018dzu}.\\

%%%
%%%%%%%%%%%%%
%%%%%%%%%%%%
%%%

We show the excluded parameter space by experimental searches sensitive to 
$c_{\mu e}\equiv\big(|[k_E]_{21}|^2+|[k_e]_{21}|\big)^{1/2}$ in Figure~\ref{fig:leptonconstraints1}. The different values of $c_{ee}, c_{\mu\mu}$ and $c_{\tau\tau}$ given in \eqref{eq:cllvalues} are indicated above the plot which is split at $m_a=2m_e$ and $m_a=2m_\mu$.

For masses $m_a > m_\mu$, the lepton flavor-changing transitions $\mu \to e\gamma$ and $\mu\to 3 e$ are induced by the form factors defined in~\ref{sec:formfactors} 
and the four-fermion operators obtained by integrating out the ALP.  
The excluded parameter space is shown in light blue and purple in Figure~\ref{fig:leptonconstraints1} and their relative strength (for $m_a > m_\mu$) reflects the expected hierarchy between the muon decay widths from the additional factor of $\alpha$ and the phase space factor in $\Gamma(\mu \to 3e)$ compared to $\Gamma(\mu \to e\gamma)$ for this mass region \cite{Blondel:2013ia,Bauer:2016rxs}. The situation changes for masses $m_a< m_\mu$, for which the ALP can be produced on-shell in muon decays.
Constraints from $\mu \to e a$ with subsequent decays $a\to \gamma\gamma$, $a\to e^+e^-$ and $a\to \text{invisible}$ are shown in orange, purple and red, respectively, and provide stronger constraints than $\mu \to e\gamma$ in a mass range of a few MeV\,$ <m_a<m_\mu $. The limits are obtained by SINDRUM for $\mu \to 3e$~\cite{Bellgardt:1987du} and LAMPF for $\mu \to \gamma\gamma e$~\cite{Bolton:1988af}. 

If the ALP decay is delayed, this parameter space cannot be excluded even if the decay still happens within the detector. A search for resonances in the dataset without the strong cut on the time of detection of the decay products would be sensitive to much smaller ALP masses. This limit on $\mu\to e\gamma\gamma$ has been improved recently by \cite{Baldini:2020okg} and expands the excluded region of our model in the range of $20$ MeV$<m_a<35$ MeV. The Collaboration states limits for muon branching ratios for different lifetimes in bins of $1$ MeV and we have adapted the appropriate limit by calculating the ALP lifetime in the respective mass region.
We further show constraints from $\mu \to e a \to e \gamma \gamma$ transitions where the ALP is boosted such that the opening angle between the two collimated photons from the ALP decay is below the angular resolution of the experiment. The excluded parameter space is obtained from the limit set by the MEG collaboration~\cite{TheMEG:2016wtm} and is shown in dark green in Figure~\ref{fig:leptonconstraints1}. 

The decay $a\to \text{invisible}$ is defined as an ALP leaving the detector before decaying. Details of the calculation of lifetimes effects can be found in Appendix~\ref{appendix:decaylengths}. The corresponding constraint on the ALP-lepton coupling is derived from the limits on the branching ratio of $\mu\to ea(\text{inv.})$ obtained by the TWIST collaboration~\cite{Bayes:2014lxz}, and is sensitive to the ALP decay length which is set by the ALP coupling to electrons in this mass range. For masses $13\,\text{MeV}<m_a<80$ MeV, the bound is largely independent of the angular distribution of the electrons, whereas for masses $m_a>80$ MeV, the bound depends on whether the decay is (an)isotropic. The angular distribution depends on the relative values of $[k_E]_{21}$ and $[k_e]_{21}$ and we use the most conservative bound from~\cite{Bayes:2014lxz} in this mass region. 

%%%
%%%%%%%%%%%%%
\begin{figure}[t]
\begin{center}
\includegraphics[width=1\textwidth]{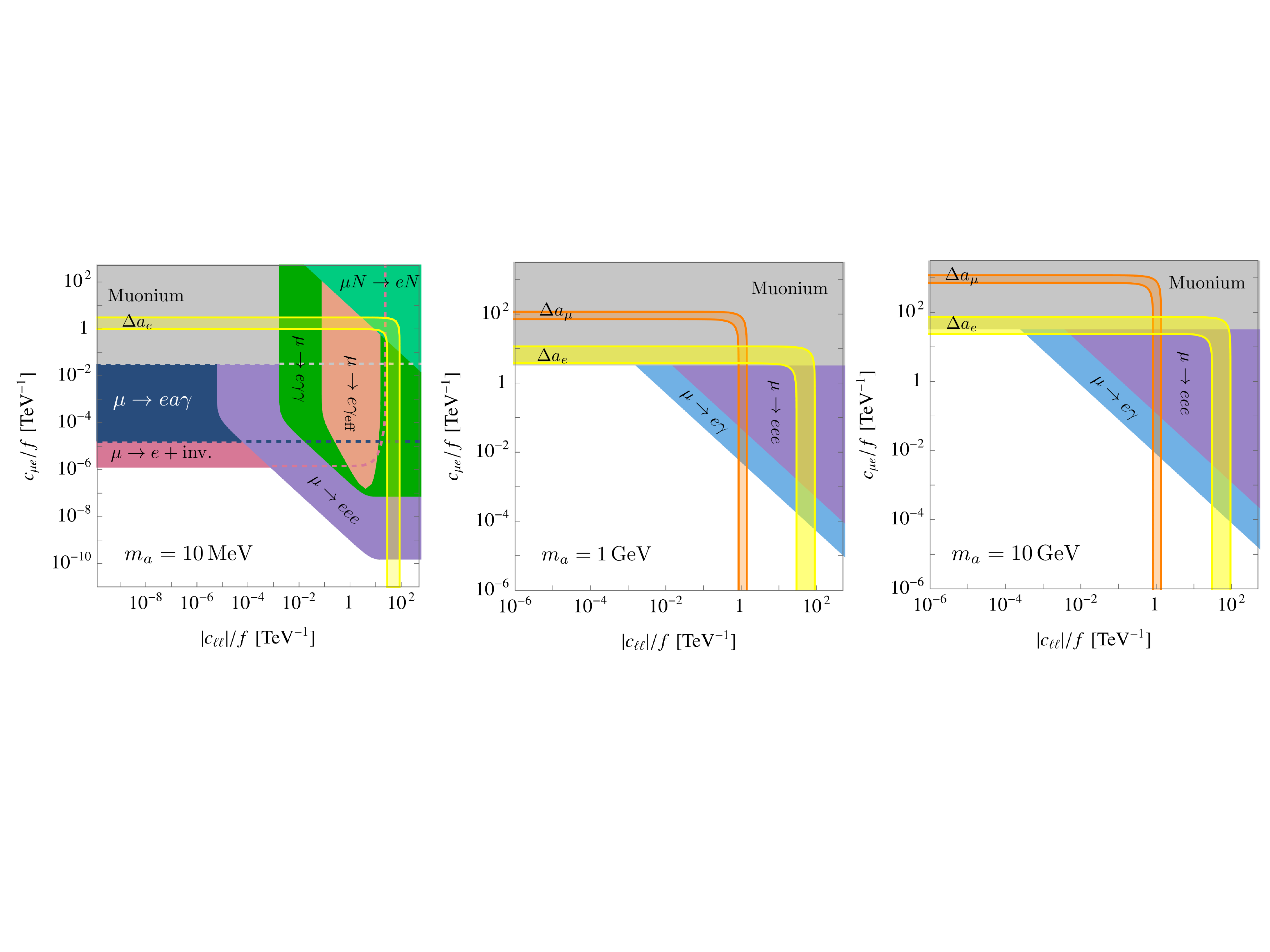}
\caption{Bounds on ALP mediated flavor off-diagonal transitions between muons and electrons with $c_{\mu e}\equiv\sqrt{|[k_E]_{21}|^2+|[k_e]_{21}|^2}$, assuming universal ALP couplings to leptons $c_{\ell\ell}\equiv c_{ee}/f=c_{\mu\mu}/f=c_{\tau\tau}/f$ and $m_a=10$ MeV,  $1$ GeV  and $10$ GeV, respectively. All other Wilson coefficients are set to zero at tree-level.\label{fig:leptonconstraints1b}}
\end{center}
\end{figure}
%%%%%%%%%%%%%%%

A slightly weaker constraint is derived from searches for the decay $\mu\to ea\gamma$ shown in dark blue in Figure~\ref{fig:leptonconstraints1}. The decay $\mu\to ea\gamma$ can be regarded as a $\mu\to ea$ decay with additional initial or final state radiation, where the ALP leaves the detector before decaying. Past searches for this type of decay have been performed with the Crystal Box detector \cite{Bolton:1988af}. The experiment required large photon and electron energies of $E_e>38-43$ MeV and $E_\gamma>38$ MeV, respectively. Here, we take the most conservative limits on the energy cuts for our plots. Though theoretically sub-dominant when compared with $\mu\to ea$ due to the additional radiation, the angular distribution is less dependent on the chiral structure of the ALP couplings and therefore can be almost competitive in constraining parameter space of ALP couplings and masses. Future searches at the upcoming MEG II experiment could exceed current bounds from TWIST by a factor of $5$, assuming optimal conditions and relaxed energy and angular cuts~\cite{Calibbi:2020jvd}.

The lifetime of the ALP strongly affects the reach of the different experiments. The constraint from the measurement of muonium-antimuonium oscillations from the MACS experiment~\cite{Willmann:1998gd} shown in gray is weaker than other constraints throughout the ALP mass range, but relevant for masses $m_a>m_\mu$, because it is independent of $c_{\ell\ell}$, whereas both the constraints from $\mu\to eee$ and $\mu\to e\gamma$ vanish for $c_{ee}\to 0$~\cite{Endo:2020mev}. The form factors~\eqref{app:formfactors} entering the $\mu \to 3e$ and $\mu \to e \gamma$ amplitudes also contribute to $\mu \to e$ conversion in $\mu N\to e N$ transitions, because under the assumption that only ALP couplings to leptons are present at tree-level only diagrams with internal photons contribute.  The constraint from the SINDRUM-II collaboration~\cite{SINDRUMII:2006dvw} shown in green in Figure~\ref{fig:leptonconstraints1} is therefore weaker throughout the parameter space and not enhanced by on-shell ALP exchange. Since the form factors vanish if the ALP coupling to photons is zero, muon conversion is not sensitive if $c_{\ell\ell}=0$ as considered here for $m_a<2m_e$. For the ALP couplings considered here, even the significant improvement in sensitivity expected at Mu2e~\cite{Mu2e:2014fns} and COMET~\cite{COMET:2018auw} shown by the green dashed contour cannot compete with the constraints from $\mu \to 3e$ and $\mu \to e \gamma$.

In Figure~\ref{fig:leptonconstraints1} we also 
show projections for future lepton flavor experiments indicated by the dashed lines. The dashed blue  contours show the sensitivity reach of MEGII~\cite{TheMEG:2016wtm} and the dashed purple contours indicate the future sensitivity of Mu3E~\cite{PerrevoortPHD, Perrevoort:2018ttp}.

The parameter space for which the anomalous magnetic moment of the electron  $\Delta a_e^\text{Cs}$ can be explained is shown in yellow in Figure~\ref{fig:leptonconstraints1}. The dominant contribution from ALP flavor-violating couplings is independent of $c_{ee}$, but requires both $[k_E]_{21}$ and $[k_e]_{21}$ to be non-zero and we choose $c_{\mu e} =|[k_E]_{21}|=|[k_e]_{21}|$ here. A successful explanation requires couplings $\text{Re}\big[[k_E]_{21}^* [k_e]_{21}\big] \lesssim -1$ which is excluded for all values of $m_a$ for $f=1$ TeV.  An explanation of the anomalous magnetic moment of the muon is only possible for $m_a> m_\mu$ and ruled out by $\mu \to e \gamma$ and $\mu \to eee$ for all values of $m_a$ as indicated by the orange contour. 
In order to understand how the parameter space preferred by the anomalous magnetic moment of the muon and electron changes as a function of the flavor-diagonal ALP couplings $c_{\ell\ell}$, we show the
exclusion contours and sensitivity reach of the various experimental searches in the $c_{\ell\ell}-c_{\mu e}$ plane for fixed ALP masses $m_a=10$ MeV, $m_a=1$ GeV and $m_a=10$ GeV in the left, centre and right panel of Figure~\ref{fig:leptonconstraints1b}, respectively.

Any explanation of $\Delta a_e^\text{Cs}$ or $\Delta a_\mu$ is only possible for $m_a> m_\mu$ and requires very small values of $c_{\mu e}< 10^{-4}$. For $m_a\gtrsim 2$ GeV, $\Delta a_e^\text{Cs}$ can also be explained if $c_{\mu e}\approx 50$ TeV$^{-1}$. Otherwise it is ruled out by the constraint from muonium-antimuonium oscillations~\cite{Endo:2020mev}.\\

%%%
%%%%%%%%%%%%%
%%%%%%%%%%%%
%%%

Constraints on the lepton flavor-changing ALP coupling $c_{\tau\mu }\equiv\big(|[k_E]_{32}|^2+|[k_e]_{32}|\big)^{1/2}$ are shown in Figure~\ref{fig:leptonconstraints2}. Flavor off-diagonal couplings of the ALP to muons and electrons, or to taus and electrons are assumed to be zero here and we assume flavor-diagonal couplings in the three different ALP mass regions as given in \eqref{eq:cllvalues} and indicated above the plot. 
The decay $\tau \to \mu \gamma$ is induced by the form factors $F_2(0)$ and $F_2^5(0)$ given in \eqref{eq:FandF5at0} and the parameter space excluded by the limit from BaBar~\cite{Aubert:2009ag} is shown in light blue in Figure~\ref{fig:leptonconstraints2}. The decay $\tau \to \mu e e$ is excluded for off-shell ALPs for $c_{\tau\mu}/f \gtrsim 10\,$TeV$^{-1}$ and depicted in dark green. For on-shell ALPs, the constraints are significantly stronger and searches for the decays $\tau \to \mu e e$ and $\tau \to \mu \mu\mu$, shown in dark green and purple, are excluded for values down to $c_{\tau\mu}/f \gtrsim 10^{-6}-10^{-4}$\,TeV$^{-1}$ for $m_a>2m_\mu$ and $m_a>2m_e$, respectively. For both decays the most stringent measurements come from Belle~\cite{Hayasaka:2010np}. 
ALP decays into photons for collimated photons that cannot be distinguished from a single photon leads to a $\tau \to \mu \gamma_\text{eff}$ final state. The constraint on this branching ratio is currently too weak to lead to any relevant constraint in the scenario considered here.

We further show the constraint from invisible ALP decays obtained by the ARGUS collaboration \cite{Albrecht:1988vy, Albrecht:1995ht} in light green in Figure~\ref{fig:leptonconstraints2}. Here, invisible decays are defined again as the ALP leaving the detector before decaying and details of the calculation of the ALP lifetime are given in Appendix~\ref{appendix:decaylengths}. For masses $2 m_\mu < m_a<  m_\tau$, the constraint is irrelevant, because the decay width of the ALP is determined by the partial decay width into muons. Below the muon pair threshold, the constraint is constant in $m_a$. While the ALP lifetime changes significantly for $m_a< 2 m_e$ the bound on $c_{\tau\mu}$ is unaffected because almost 100\% of the ALPs produced decay outside the detector into photons $a\to \gamma\gamma$. 
The ALP contribution to the anomalous magnetic moment of the muon is dominated by the diagram with a tau in the loop. In contrast to flavor-conserving ALP couplings, which are purely axial, this diagram can contribute with the right sign to address the tension between the measurement and the SM prediction if $\text{Re}\big[[k_e]_{23} [k_E]_{23}^* \big]>0$. We show the corresponding parameter space assuming 
$c_{\tau\mu}=|[k_e]_{23}|=| [k_E]_{23}|$ in orange in Figure~\ref{fig:leptonconstraints2}. 
However, the parameter space for which the ALP contribution is large enough to explain the tension is excluded by searches for $\tau\to \mu \gamma$ decays. Finally, we show projections for the sensitivity of future ALP searches by dashed contours. The dashed red line corresponds to the reach of a future high energy $e^+e^-$ collider FCC-ee for $\tau \to \mu \gamma$ and $\tau \to 3 \mu$ decays~\cite{Dam:2018rfz}. The blue and black dashed contours are projections for the sensitivity for $\tau \to 3 \mu$ at LHCb and Belle II~\cite{Kou:2018nap},  respectively. \\

%%%
%%%%%%%%%%%%%
%%%%%%%%%%%%
%%%

Figure~\ref{fig:leptonconstraints3} shows the bounds on a dominant LFV coupling $c_{\tau e }\equiv \big(|[k_E]_{31}|^2+|[k_e]_{31}|^2\big)^{1/2}$. The constraints look similar to those on $c_{\tau \mu}$ shown in Figure~\ref{fig:leptonconstraints2}. We again assume all other flavor-violating ALP couplings to vanish and assume flavor-diagonal couplings in the three different ALP mass regions as given in \eqref{eq:cllvalues} and indicated above the plot. 

BaBar searches for $\tau \to e \gamma$~\cite{Aubert:2009ag} exclude the parameter space in light blue. Searches for the three-body decays $\tau \to e\mu\mu$ and $\tau \to 3e$ from Belle~\cite{Hayasaka:2010np} only yield meaningful limits for on-shell ALPs and are shown in green and purple, respectively. ALPs with macroscopic decay lengths are excluded by the search for $\tau \to e\, +$ invisible by ARGUS~\cite{Albrecht:1988vy, Albrecht:1995ht} shown in light green. For boosted ALPs with subsequent decays $\tau \to e a \to e \gamma\gamma$ the photon pair cannot be reconstructed by the detector, 
and the yellow parameter space is excluded by the limit on the $\tau \to e\gamma$ branching ratio obtained by BaBar \cite{Aubert:2009ag}. 

The large contribution from the tau loop to the anomalous magnetic moment of the electron $\Delta a_e^\text{Cs}$ can explain the observed deviation from the SM prediction in the yellow band region assuming $c_{\tau e}=|[k_E]_{31}|=|[k_e]_{31}|$ here. The limit on $\tau \to e \gamma$ does not exclude this possible explanation, but the limits from $\tau \to e\mu\mu$ and $\tau \to e+X$ searches rule it out for almost all masses with $m_a<m_\tau$.\\

The results presented in this section may be compared with constraints obtained in the recent work of Ref.~\cite{Cornella:2019uxs}. Many of our bounds are very similar to theirs (accounting for differences in coupling normalisation), however some of our constraints, for example from $\mu \to 3e$, $\mu \to e \gamma \gamma$ and $\tau \to \mu ee$, extend to lower masses. The difference arises from how we account for long-lived decays of the ALP. The authors of Ref.~\cite{Cornella:2019uxs} assume that if the lab-frame decay length of the ALP is larger than $1\,$m, the ALP will escape the detector. They moreover take the decaying lepton to be at rest in the lab frame, which is true for $\mu \to e$ decays, but not for $\tau$ decays which have been measured at $B$ factories. We instead estimate the fraction of ALPs which will decay sufficiently promptly, taking into account the boost of the ALP and the geometry of the detector. The formulae we use, and the assumptions and approximations made, are described in Appendix~\ref{appendix:decaylengths}. We find that even for rather long decay lengths, a significant number of ALPs still decay within the detector due to the exponential nature of decay, resulting in bounds even for rather low ALP masses. 
Moreover, our analysis of the effects of boosted ALPs allow us to draw the constraints for $\ell_j\to \ell_i \gamma\gamma$ decays in which the lab frame opening angle is narrow enough that it is mistaken for $\ell_j \to \ell_i \gamma$. 

However we do not consider LFV couplings in combination with hadronic or quark flavor-violating processes, as the authors of Ref.~\cite{Cornella:2019uxs} do. Their analysis therefore takes into account measurements that ours does not.

\clearpage
\section{Conclusions} 
\label{sec:conclusions}

Axions or axion-like particles are pseudo Nambu-Goldstone bosons that originate from a spontaneously broken global symmetry in UV extensions of the SM. The ALP coupling structure is determined by the details of this UV theory and the coupling strengths to SM particles sensitively depend on it, not least in being inversely proportional to the new physics scale. 

In this paper, we have explored the sensitivity of quark and lepton flavor-changing processes within a general ALP effective field theory focussing on the MeV-GeV mass range. To do this, we have defined the effective ALP Lagrangian at the UV scale, and at and below the electroweak scale. We discussed RG running and matching effects and highlight the unavoidable contributions to quark flavor-changing couplings they induce. Below the QCD scale, the ALP couplings to QCD resonances are described by the chiral Lagrangian which we used to calculate the charged and neutral $K \to \pi a$ amplitudes, and to derive the ALP couplings to nucleons taking into account the finite ALP mass. We discussed possible ALP decay modes and exotic Higgs and Z-boson decays into ALPs.

To study flavor-changing processes in the quark sector, we have calculated a variety of processes, including rare meson decays, flavor oscillations of neutral mesons and the chromomagnetic dipole moment of the top-quark, in terms of ALP flavor-changing and flavor-conserving couplings at the scale of the measurement. The most sensitive observables are rare meson decays to an on-shell ALP. The sensitivity of the experimental measurements depends strongly on the branching ratios of the ALP and its lifetime. We have presented individual plots for each such measurement, showing the dependence of the resulting exclusion regions on the ALP decay length. 
Given that quark flavor-changing couplings are unavoidably induced by renormalisation group evolution and matching effects from the new physics scale down to the scale of the measurements, we further derived bounds on benchmark ALP models, in which only a single ALP coupling to gauge bosons or a single flavor-universal ALP coupling to a fermion species is present in the UV. This lets us compare the constraints from flavor observables on flavorless or flavor-diagonal ALP couplings with limits from collider searches, e.g., Higgs, $Z$-boson and $\Upsilon$ decays into ALPs, beam dumps, and astrophysical experiments within the same parameter spaces. We highlight the complementarity and competitiveness of flavor bounds which set some of the most stringent constraints in the MeV-GeV mass range, even for the most flavorless of ALP models. In particular, the measurements of $K^+\to \pi^+ a$ and $K_L\to \pi^0 a$, where $a$ escapes the detector, provide the strongest flavor constraints for ALPs lighter than a few hundred MeV. Searches for $B^0\to K^* a(\mu^+\mu^-)$ often provide the dominant constraints for $m_a> 2m_\mu$. We emphasize that future searches for $B\to K^{(*)}a$ with further ALP decay modes, in particular $a\to \gamma \gamma$ and $a\to e^+ e^-$, could probe currently unconstrained regions of parameter space.

We discussed current experimental anomalies and critically examined whether they could be the first sign of an ALP. Regarding the neutral $B$-physics anomalies, we find that a heavy ALP cannot account for the discrepancies in $R_K$ or $R_K^*$. 
The deviation in the low $q^2$-bin of $R_{K^*}$ can be accounted for by a light ALP with $200\,\text{MeV } < m_a < 210$\,MeV and $\text{Br}(a \to e^+ e^-)=1$. However, such a light ALP is not sufficient to address the observed discrepancies in $R_K$, or the high $q^2$-bin of $R_K^*$. 
The ATOMKI Beryllium and Helium anomalies can in principle be simultaneously explained by an ALP with a mass of $17$\,MeV and couplings to electrons and nucleons. However, a combined explanation is already ruled out by the measurement of rare kaon decays. We found that a small region of parameter space accounting for the Helium transition remains unconstrained by kaon decays but is in strong tension with beam dump constraints on the ALP-electron coupling.
The KTeV anomaly can be explained by an ALP but requires large ALP couplings to electrons as well as a large ALP-pion mixing, which can either be achieved by sizeable ALP-gluon or -quark couplings or by an ALP with a mass close to that of the pion.   

We have further studied lepton flavor-violating ALP couplings. In contrast to the quark sector, lepton flavor-violating couplings are not induced at loop-level if they are zero in the UV theory and LFV observables are therefore only sensitive to ALPs which have explicitly LFV couplings in the UV. 
We have derived general expressions for lepton form factors from ALP loops including the full mass and $q^2$ dependence and the most important two-loop diagrams with ALP-fermion couplings. 
We discussed the effects of the ALP lifetime in detail and carefully took them into account in our calculation of exclusion contours and projections for experiments looking for lepton flavor-violating decays. Our results were shown in three benchmark models in which we allowed for a single flavor-violating coupling to be present in addition to ALP-mass dependent flavor-diagonal ALP-lepton couplings which agree with current bounds from astrophysics, beam dumps, collider searches and quark flavor bounds induced by RG running.
This leads to a range of constraints from rare muon and tau decays, $\mu \to e$ conversion and muonium-antimuonium oscillations.
In agreement with the expectation for heavy lepton flavor-violating new physics, we find that for $m_a > m_\mu$, searches for $\mu \to e \gamma$ provide the strongest constraints on flavor-violating $\mu$-$e$-$a$ couplings, and similarly, searches for tau decays $\tau \to \ell\gamma$ yield the strongest constraints on flavor-violating $\tau$-$\ell$-$a$ couplings for $m_a>m_\tau$. For lighter ALPs, $2m_e  < m_a < m_\mu$, the muon decay $\mu \to 3 e $ is enhanced by the ALP going on-shell. Current limits are up to five orders of magnitude stronger than the limit from $\mu \to e\gamma$ in this region of parameter space, similar to constraints from the other on-shell observables $\mu \to e \gamma\gamma$ and $\mu \to e a$(invisible), when taking into account all constraints on flavor-diagonal ALP-lepton couplings. 
 A similar hierarchy is present for the analogous tau decays. We have further explored the parameter space for which MEGI and MEGII are sensitive to the decay $\mu \to e a\to e \gamma\gamma$ where the two photons are so collimated that they mimic a single photon $\gamma_\text{eff}$ in the detector, as well as for the tau decays $\tau \to e\gamma_\text{eff}$ and $\tau \to \mu\gamma_\text{eff}$. We find that upcoming data from the Mu3E experiment is projected to provide the best sensitivity on $c_{\mu e}$ for ALPs with masses $m_a>2m_e$. 

The anomalous magnetic moments of the electron $a_e$ and the muon $a_\mu$ receive contributions from flavor-violating and flavor-diagonal ALP-lepton couplings. We present a comprehensive analysis of all possible ALP contributions and show the parameter space for which an ALP could explain the observed tension between the experimental measurements and the SM predictions for the anomalous magnetic moments. A simultaneous explanation of both $\Delta a_\mu$ and $\Delta a_e^\text{Cs}$ is possible for an ALP with non-universal couplings of opposite signs to electrons and muons, and ALP masses of $m_a\geq$\,GeV, but is ruled out for purely flavor off-diagonal ALP couplings to leptons. However, ALPs with masses $m_a\geq$\,GeV and flavor-changing couplings $c_{\mu e}$ can address either the anomalous magnetic moment of the muon or of the electron if a sufficiently large flavor-diagonal ALP coupling to leptons is present. Similarly, ALPs with masses $m_a> m_\tau$ and either a $c_{\tau \mu}$ or $c_{\tau e}$ coupling could explain $\Delta a_\mu$ or $\Delta a_e^\text{Cs}$, respectively.

Experiments sensitive to flavor-changing transitions involving quarks and leptons provide an important avenue to search for both flavorful and flavorless axions and axion-like particles. These searches are highly competitive and complementary to astrophysical, beam dump and collider observables and can set the most stringent constraints for ALP masses between a few hundred MeV and tens of GeV, thereby closing important gaps in parameter space and offering promising future opportunities to discover ALPs. %

\section*{Acknowledgements}

The research of M.N.\ and M.S.\ was supported by the Cluster of Excellence {\em Precision Physics, Fundamental Interactions and Structure of Matter\/} (PRISMA${}^+$ -- EXC~2118/1) within the German Excellence Strategy (project ID 39083149). S.R.\ acknowledges support from the INFN grant~SESAMO. MB is supported by a UKRI Future Leadership Fellowship.

\appendix
\section[{\boldmath Contributions to the $K\to\pi a$ decay amplitudes from $SU(3)$ 27-plet operators}] {\boldmath Contributions to the $K\to\pi a$ decay amplitudes from $SU(3)$ 27-plet operators}
\label{app:A}

Here we report the contributions from the 27-plet operators to the $K^+\to \pi^+ a$ and  
$K_L\to \pi^0 a$ amplitudes discussed in Section~\ref{sec:chiral}. In analogy to \eqref{N8def} we define 
where 
\begin{equation}\label{N8def}
   N^i_{27} = - \frac{G_F}{\sqrt2}\,V_{ud}^* V_{us}\,g^i_{27}\,f_\pi^2  
\end{equation}
with $i=1/2, 3/2$
and find for the charged kaon decay
\begin{align}
i\mathcal{A}_{27}^{1/2}(K^-\to \pi^- a)&=\frac{N_{27}^{1/2}}{4f}\bigg[
\frac{m_a^2-4m_K^2+3m_\pi^2}{m_\eta^2-m_a^2}\big(2c_{GG}(2m_a^2-3m_\eta^2+m_\pi^2)+m_a^2(2c_{ss}-c_{uu}-c_{dd})\big)\notag\\
&+m_K^2(-16c_{GG}+3c_{dd}+4c_{uu}-7c_{ss})+m_a^2(4c_{GG}+c_{ss}-c_{uu})\notag\\
&+m_\pi^2(12c_{GG}+7c_{ss}-4c_{uu}-3c_{dd})+(m_K^2-m_a^2+m_\pi^2)([k_D+k_d]_{11}-[k_D+k_d]_{22})
\bigg]\,,\\
i\mathcal{A}^{3/2}_{27}(K^-\to \pi^- a)&=
\frac{N_{27}^{3/2}}{4f}\bigg[\frac{m_K^2-m_a^2}{m_\eta^2-m_a^2}\big(2c_{GG} (3m_\eta^2-2m_a^2-m_\pi^2)-m_a^2(2c_{ss}-c_{uu}-c_{dd})\big) \notag \\
&+\frac{m_a^2+3m_K^2-4m_\pi^2}{m_a^2-m_\pi^2}m_a^2 (c_{dd}-c_{uu})\notag\\
&+m_a^2(4c_{GG}-c_{dd}+c_{ss}) +m_\pi^2(3c_{dd}-4c_{uu}+c_{ss})\notag\\
&+m_K^2(-4c_{GG}+4c_{uu}-3c_{dd}-c_{ss})+(m_K^2+m_\pi^2-m_a^2)([k_D+k_d]_{11}-[k_D+k_d]_{22})\bigg]\,,
\end{align}
and for the neutral kaon decay we find
\begin{align}
i\mathcal{A}_{27}^{1/2}(K^0\to \pi^0 a)&=\frac{N_{27}^{1/2}}{4\sqrt2 f}\bigg[\frac{m_a^2-4m_K^2+3m_\pi^2}{m_a^2-m_\eta^2}
(2c_{GG}(3m_\eta^2-2m_a^2-m_\pi^2)+m_a^2(c_{uu}+c_{dd}-2c_{ss}) )\notag\\
&+\frac{m_a^2-2m_K^2+m_\pi^2}{m_a^2-m_\pi^2}m_a^2(c_{dd}-c_{uu})\notag\\
&+ m_a^2(4c_{GG}-c_{dd}+c_{ss})+m_\pi^2(12c_{GG}^2+7c_{ss}-5c_{dd}-2c_{uu})\notag\\
&+m_K^2(-16c_{GG}-7c_{ss}+5c_{dd}+2c_{uu})+(m_\pi^2+m_K^2-m_a^2)([k_D+k_d]_{11}-[k_D+k_d]_{22}) \bigg]\,,\\
i\mathcal{A}^{3/2}_{27}(K^0\to \pi^0 a)&=\frac{N_{27}^{1/2}}{2\sqrt2 f}\bigg[\frac{m_K^2-m_a^2}{m_\eta^2-m_a^2}
(2c_{GG}(2m_a^2-3m_\eta^2+m_\pi^2)-m_a^2(c_{uu}+c_{dd}-2c_{ss}) )\notag\\
&+\frac{m_a^2-2m_K^2+m_\pi^2}{m_a^2-m_\pi^2}m_a^2(c_{uu}-c_{dd})\notag\\
&+ m_a^2(-4c_{GG}+c_{dd}-c_{ss})+m_\pi^2(2c_{dd}-c_{ss}-c_{uu})\notag \\
&+m_K^2(4c_{GG}+c_{ss}-2c_{dd}+c_{uu} )+(3m_\pi^2-m_K^2+m_a^2)([k_D+k_d]_{11}-[k_D+k_d]_{22}) \bigg]\,.
\end{align}

\section{Form factors}
\label{app:formfactors}
The loop functions for the form factors in Section~\ref{sec:FormFactors} are given by the Feynman integrals
\begin{align}
g_1(q^2,m_i,m_a)&=2\,\int_0^1 dx\,\int_0^{1-x} dy\, \frac{1-x-yx}{\Delta^{\prime\prime}_{i \to j} } \,, \\
g_2(q^2,m_i,m_a)&=-\int dx\,\int_0^{1-x} dy
\bigg[4\log \frac{m_i^2}{\mu^2} +2\log \Delta_{i \to j}+2\log \Delta^\prime_{i \to j}-4 \delta_2-
 \left(\frac{x(1-y)}{\Delta^\prime_{i \to j}}- \frac{y x}{\Delta_{i \to j}}\right)\bigg]\,,\\
 l_1(q^2,m_i,m_a)&=2\,\int_0^1 dx\,\int_0^{1-x} dy\, \frac{1-x-yx-2y^2}{\Delta^{\prime\prime}_{i \to j} },\\ 
l_2(q^2,m_i,m_a)&=\int dx\,\int_0^{1-x} dy
 \left(\frac{x(1-y)}{\Delta^\prime_{i \to j}}- \frac{y x}{\Delta_{i \to j}}\right)\,,
\end{align}
and
\begin{align}
\Delta_{i \to j}&= y\frac{m_a^2}{m_i^2}-\frac{q^2}{m_i^2}y(1-x-y)-xy \,, \\
\Delta^\prime_{i \to j}&=  (1-x-y)\frac{m_a^2}{m_i^2}-\frac{q^2}{m_i^2}y(1-x-y)+x(1-y) \,, \\
\Delta^{\prime\prime}_{i \to j} &= x \frac{m_a^2}{m_i^2}+ (1-x-yx)-\frac{q^2}{m_i^2} y(1-x-y) \,.
\end{align}
The scheme dependent constant $\delta_2$ arises from the treatment of the Levi Civita symbol in $d$ dimensions, and for us $\delta_2=-3$.

%%%%%%%%%%%%%%%%%%%%%%%%%%%%%
\section{Measurements and SM predictions for flavor observables}
\label{app:measurements}
%%%%%%%%%%%%%%%%%%%%%%%%%%%%%
The measured values and SM predictions for observables used to derive constraints are given in Tables~\ref{tab:longlivedobs} to \ref{tab:LFVobs}.

\begin{table}[h]
\begin{center}
\scalebox{0.8}{
\begin{tabular}{l c c c}
\toprule
Observable & Mass Range [MeV] & Measurement & SM prediction\\
\midrule
$\text{Br} (K^+ \to \pi^+ X)$ & $0<m_{X}<261~(*)$ & \cite{CortinaGil:2021nts} (search) & - \\
$\text{Br} (K^+ \to \pi^+ X)$ & $110<m_{X}<155$ & \cite{CortinaGil:2020zwa} (search) & -\\
$\text{Br} (K_L \to \pi^0X)$ & $0<m_{X}<261$ & \cite{Ahn:2018mvc} (search)& - \\
$\text{Br} (B^+ \to K^+ \bar{\nu} \nu)$ & $0<m_{\nu \nu}<4785$ & $<1.6 \times 10^{-5}$ \cite{Lees:2013kla}& $(4.0 \pm 0.5)\times 10^{-6}$ \cite{Buras:2014fpa} \\
$\text{Br} (B^0 \to K^{*0} \bar{\nu} \nu)$ & $0<m_{\nu \nu}<4387$ & $<1.8 \times 10^{-5}$ \cite{Grygier:2017tzo}& $(9.2 \pm 1.0)\times 10^{-6}$ \cite{Buras:2014fpa} \\
$\text{Br} (\Upsilon \to \gamma a(\text{invisible}))$ & $m_a< 9200$ &  \cite{delAmoSanchez:2010ac} (search) & -\\
\bottomrule
\end{tabular}}
\end{center}
\caption{Observables relevant for a long lived ALP. Bounds are at 90\% CL. 
$(*)$: cuts are applied to exclude the region around $m_{\pi}$ ($100<m_{X}<161$ MeV).}
\label{tab:longlivedobs}
\end{table}

\begin{table}[h]
\begin{center}
\scalebox{0.8}{
\begin{tabular}{l c c c}
\toprule
Observable & Mass Range [MeV] & Measurement & SM prediction\\
\midrule
$\text{Br} (K^+ \to \pi^+ \gamma \gamma)$ & $m_{\gamma\gamma}< 108$& $<  8.3\times 10^{-9}$ \cite{Artamonov:2005ru}  & $6.1 \times 10^{-9}$ \cite{Gerard:2005yk} \\
$\text{Br} (K^+ \to \pi^+ \gamma \gamma)$ & $220<m_{\gamma\gamma}< 354$& $(9.65\pm 0.63) \times 10^{-7}$ \cite{Ceccucci:2014oza}& $(10.8\pm 1.7) \times 10^{-7}$ \cite{Gerard:2005yk}$\dagger$\\
$\text{Br} (K_L \to \pi^0 \gamma \gamma)$ & $m_{\gamma\gamma}< 110$& $<  0.6\times 10^{-8}$ \cite{Lai:2002kf}  & $(8^{+7}_{-5})\times 10^{-8}$ \cite{DAmbrosio:1996kjn}$*$\\
$\text{Br} (K_L \to \pi^0 \gamma \gamma)$ & $m_{\gamma\gamma}< 363~(\ddag)$& $(1.29\pm 0.03\pm 0.05)\times 10^{-6}$ \cite{Abouzaid:2008xm}  & $1.12 \times 10^{-6}$ \cite{DAmbrosio:1996kjn}\\
\bottomrule
\end{tabular}}
\end{center}
\caption{Observables with a photon pair in the final state. Bounds are at 90\% CL. $(\ddag)$: cuts are applied to exclude the region around the pion pole ($100< m_{\gamma\gamma}<160$ MeV). ($\dagger$: calculated from results in the given reference. Error bars estimated from varying parameter $\hat{c}$ between its quoted errors.) ($*$: calculated from results in the given reference. Error bars estimated from varying parameter $a_V$ between its quoted errors.) 
\label{tab:gamgamobs}}
\end{table}

\begin{table}[h]
\begin{center}
\scalebox{0.8}{
\begin{tabular}{l c c c}
\toprule
Observable & Mass Range [MeV] & Measurement & SM prediction\\
\midrule
$\text{Br}(K^+ \to \pi^+ a (e^+ e^-))$ & $m_{a} < 100$ & $< 8 \times 10^{-7}$ \cite{Baker:1987gp} & - \\
$\text{Br}(K_L \to \pi^0 e^+ e^-)$ & $140 < m_{ee} <362$ & $<2.8 \times 10^{-10}$ \cite{AlaviHarati:2003mr} & $\left(3.1^{+1.2}_{-0.8}\right)\times 10^{-11}$ \cite{Buchalla:2003sj}\\
$\text{Br}(B^+ \to \pi^+ e^+ e^-)$ & $140<m_{ee} <5140$ & $<8.0\times 10^{-8}$ \cite{Wei:2008nv} &$(2.26^{+0.23}_{-0.19})\times 10^{-8}$ \cite{Li:2014uha}\\
$d\text{Br}/dq^2(B^0 \to K^{*0} e^+ e^-)_{[0.0004,0.05]}$ & $20 < m_{ee} <224$ & $(4.2 \pm 0.5)\times 10^{-6}$ GeV$^{-2}$ \cite{Aaij:2015dea} & $(3.3\pm 0.7)\times 10^{-6}$ GeV$^{-2}$\\
$d\text{Br}/dq^2(B^0 \to K^{*0} e^+ e^-)_{[0.05,0.15]}$ & $224 < m_{ee} <387$ & $(2.6 \pm 1.0)\times 10^{-7}$ GeV$^{-2}$ \cite{Aaij:2015dea} & $(3.9\pm 0.8)\times 10^{-7}$ GeV$^{-2}$\\
$R_{K^*} [0.045,1.1]$ & $212 < m_{ee} <1049 $ & $0.66^{+0.11}_{-0.07}\pm 0.03$ \cite{Aaij:2014ora}  & $0.906 \pm 0.028$ \cite{Bordone:2016gaq}\\
$\text{Br}(D^0\to \pi^0 e^+ e^-)$ & $m_{ee}< 1730~(\dagger)$ & $<4 \times 10^{-6}$ \cite{Ablikim:2018gro} & $1.9 \times 10^{-9}$~\cite{Fajfer:2001sa}\\
$\text{Br}(D^+\to \pi^+ e^+ e^-)$ & $200<m_{ee}< 1730~(*)$ & $<1.1 \times 10^{-6}$ \cite{Lees:2011hb} & $9.4 \times 10^{-9}$~\cite{Fajfer:2001sa}\\
$\text{Br}(D_s^+ \to K^+ e^+ e^-)$ & $200<m_{ee}< 1475~(*)$ & $<3.7 \times 10^{-6}$ \cite{Lees:2011hb} & $9.0 \times 10^{-10}$~\cite{Fajfer:2001sa}\\
\bottomrule
\end{tabular}
}\end{center}
\caption{Observables with an electron pair in the final state. Bounds are at 90\% CL. Here we only include observables for which the electron invariant mass can be below or near the dimuon threshold, on the grounds that above it muonic observables will generically provide stronger bounds. Predictions without accompanying citations have been calculated using \texttt{flavio}~\cite{david_straub_2018_1218732}. In the measurements of the $D_{(s)}$ branching ratios, cuts are applied to exclude the region around the $\phi$ resonance. For the Babar measurements with a $(*)$, the excluded region is $950<m_{ee}< 1050$ MeV, while the BESIII measurement with a $(\dagger)$ excludes the region $935<m_{ee}< 1053$ MeV. Since the long-distance contributions to these decays peak around this excluded resonance, we take the SM prediction to be only due to the short-distance contributions, as calculated in Ref.~\cite{Fajfer:2001sa}.}
\label{tab:electronicobs}
\end{table}

\begin{table}[h]
\begin{center}
\scalebox{0.8}{
\begin{tabular}{p{4.2cm} p{4.3cm} p{4.5cm} c}
\toprule
Observable & Mass Range [MeV] & Measurement & SM prediction\\
\midrule
$\text{Br}(K_L \to \pi^0 \mu^+ \mu^-)$ & $210 < m_{\mu\mu} < 350$ & $< 3.8 \times 10^{-10}$ \cite{AlaviHarati:2000hs} & $(1.5\pm 0.3) \times 10^{-11}$ \cite{Isidori:2004rb}\\
$\text{Br}(B^+ \to K^+ a(\mu^+ \mu^-))$ & $250<m_a<4700 ~(\dagger)$& \cite{Aaij:2016qsm} (search)& -\\
$\text{Br}(B^0 \to K^{*0} a(\mu^+ \mu^-))$ & $214<m_a<4350 ~(\dagger)$& \cite{Aaij:2015tna} (search)& -\\
$\text{Br}(J/\psi \to \gamma a(\mu^+ \mu^-))$ & $212 < m_{\mu\mu} < 3000$ & \cite{Ablikim:2015voa} (search) & -\\
$\text{Br}(\Upsilon \to \gamma a(\mu^+ \mu^-))$ & $212 < m_{\mu\mu} < 9200$ & \cite{Lees:2012iw} (search) & -\\
$\text{Br}(B^+ \to \pi^+ \mu^+ \mu^-)$ & $211<m_{\mu\mu} <5140~(\ddag)$ & $(1.83\pm 0.25)\times 10^{-8}$ \cite{Aaij:2015nea} &$(2.26^{+0.23}_{-0.19})\times 10^{-8}$ \cite{Li:2014uha}\\
$\text{Br}(B_s^0\to \mu^+ \mu^-)$ &  $ 5320< m_{\mu\mu} < 6000$ & $(2.69^{+0.37}_{-0.35}) \times 10^{-9}$ \cite{ATLAS-CONF-2020-049}&  $(3.66\pm 0.14) \times 10^{-9}$ \cite{Beneke:2019slt}\\
$\text{Br}(B^0\to \mu^+ \mu^-)$ &  $4900 < m_{\mu\mu} < 6000$ & $(0.6^{+0.7}_{-0.7}) \times 10^{-10}$ \cite{ATLAS-CONF-2020-049} &$(1.03\pm 0.05) \times 10^{-10}$ \cite{Beneke:2019slt} \\
$\text{Br}(D^+\to \pi^+ \mu^+ \mu^-)$ & $250 < m_{\mu\mu} < 1730~(*)$  & $< 7.3\times 10^{-8}$ \cite{Aaij:2013sua} &$9.4 \times 10^{-9}$~\cite{Fajfer:2001sa} \\
$\text{Br}(D_s^+ \to K^+ \mu^+ \mu^-)$ & $200 < m_{\mu\mu} < 1475~(**)$ & $< 21 \times 10^{-6}$ \cite{Lees:2011hb} & $9.0 \times 10^{-10}$~\cite{Fajfer:2001sa}\\
\bottomrule
\end{tabular}}
\end{center}
\caption{Observables with a muon pair in the final state. Bounds are at 90\% CL.
$(\dagger)$: cuts are applied to exclude regions around the $J/\psi$, $\psi(2S)$ and $\psi(3370)$ resonances. $(\ddag)$: cuts are applied to exclude charmonium resonance regions ($8.0<m_{\mu\mu}^2<11.0 \,\text{GeV}^2$ and $12.5<m_{\mu\mu}^2<15.0 \,\text{GeV}^2$ are excluded).$(*)$: a large region containing the $\eta$, $\rho/\omega$ and $\phi$ resonances is excluded ($525<m_{\mu\mu}< 1250$ MeV).$(**)$: cuts are applied to exclude the region around the $\phi$ resonance ($990<m_{\mu\mu}< 1050$ MeV). Since the long-distance contributions to the $D_{(s)}$ decays peak around the excluded resonance(s), we take the SM prediction to be only due to the short-distance contributions, as calculated in Ref.~\cite{Fajfer:2001sa}.}
\label{tab:muonicobs}
\end{table}

\begin{table}
\begin{center}
\scalebox{0.8}{
\begin{tabular}{l c c c}
\toprule
Observable & Mass Range [MeV] & Measurement & SM prediction\\
\midrule
$\text{Br}(B^+\to K^+ \tau^+ \tau^-)$ & $3552< m_{\tau\tau} < 4785$ & $<2.25 \times 10^{-3}$ \cite{TheBaBar:2016xwe}  & \\
$\text{Br}(\Upsilon \to \gamma a(\tau\tau))$ & $3500 < m_{\tau\tau} < 9200$ & \cite{Lees:2012te} (search) & -\\
\bottomrule
\end{tabular}}
\end{center}
\caption{Observables with a tau pair in the final state. Bounds are at 90\% CL.
}
\label{tab:tauonicobs}
\end{table}

\begin{table}
\begin{center}
\scalebox{0.8}{
\begin{tabular}{l c c c}
\toprule
Observable & Mass Range [MeV] & Measurement & SM prediction\\
\midrule
$\text{Br}(\Upsilon \to \gamma a(\text{hadrons}))$ & $290 < m_{\text{hadrons}} < 7100$ & \cite{Lees:2011wb} (search) & -\\
\bottomrule
\end{tabular}}
\end{center}
\caption{Observables relevant for hadronic decays of the ALP.}
\label{tab:hadronicobs}
\end{table}

\begin{table}[t]
\begin{center}
\scalebox{0.8}{
\begin{tabular}{l c c}
\toprule
Observable & Mass Range [MeV] & Measurement\\
\midrule
$\text{Br}(\mu \to 3 e)$ & - &$<1.0\times 10^{-12}$ \cite{Bellgardt:1987du}  \\
$\text{Br}(\mu \to e \gamma)$ & -  & $<4.2\times 10^{-13}$ \cite{TheMEG:2016wtm}  \\
$\text{Br}(\mu \to e \gamma \gamma)$ & $0 < m_a < 105$ & $<7.2\times 10^{-11}$ \cite{Bolton:1988af}  \\
$\text{Br}(\mu \to e a(\text{invisible}))$ &  $0 < m_a < 13$ & $<5.8 \times 10^{-5}$ \cite{Bayes:2014lxz}\\
$\text{Br}(\mu \to e a (\text{invisible}))$ & $13 < m_a < 80$ & $\lesssim 10^{-5}$ \cite{Bayes:2014lxz}\\
$\text{Br}(\mu \to e \gamma a (\text{invisible}))$ & $0 < m_a < 105$ & $ < 1.1\times 10^{-9}$ \cite{Bolton:1988af}\\
$\text{Br}(\mu N \to e N)$ & - & $ <7.0\times 10^{-13}$ \cite{SINDRUMII:2006dvw}\\
$\text{Br}(\tau \to e a(\mathrm{invisible}))$ &$0 < m_a < 1600$ &  $<2.7 \times 10^{-3}$ \cite{Albrecht:1995ht} \\
$\text{Br}(\tau \to \mu a(\mathrm{invisible}))$ & $0 < m_a < 1600$ &$<5 \times 10^{-3}$ \cite{Albrecht:1995ht} \\
$\text{Br}(\tau \to 3 \mu)$ &$211 < m_a < 1671$ &$<2.1\times 10^{-8}$ \cite{Hayasaka:2010np}  \\
$\text{Br}(\tau \to 3 e)$ & $200 < m_a < 1776$&$<2.7\times 10^{-8}$ \cite{Hayasaka:2010np}  \\
$\text{Br}(\tau^- \to \mu^- e^+ e^-)$ & $200 < m_a < 1776$ & $<1.8\times 10^{-8}$ \cite{Hayasaka:2010np}  \\
$\text{Br}(\tau^- \to e^- \mu^+ \mu^-)$ & $211 < m_a < 1776$ & $<2.7\times 10^{-8}$ \cite{Hayasaka:2010np}  \\
$\text{Br}(\tau \to \mu \gamma)$ & - &$<4.4\times 10^{-8}$ \cite{Aubert:2009ag}  \\
$\text{Br}(\tau \to e \gamma)$ & - &$<3.3\times 10^{-8}$ \cite{Aubert:2009ag}  \\
\bottomrule
\end{tabular}}
\caption{Lepton flavor-violating observables. Where a mass range for $m_a$ is given, the range refers to masses that are consistent with the experimental cuts and for which the decay can proceed via a resonant ALP. For some of the observables (for example $\mu\to 3 e$), an ALP lying outside of this mass range may still be constrained by the experiment, if it can mediate the decay off-shell.
Where the mass range is left blank, the measurement can never involve a resonant ALP.}
\label{tab:LFVobs}
\end{center}
\end{table}

\begin{table}
\begin{center}
\begin{tabular}{c c c c}
\toprule
Decay & Experiment & Initial state & Time cut (ns) \\
\midrule
$\mu \to 3 e$ & SINDRUM~\cite{Bellgardt:1987du} & at rest & 0.8 \\
$\mu \to e \gamma \gamma$ & Crystal Box~\cite{Bolton:1988af}& at rest & 2.5 \\
$\mu \to e \gamma $ & MEG \cite{TheMEG:2016wtm} & at rest & 0.7 \\
$\mu\to e \gamma a(\text{invisible}) $ & Crystal Box \cite{Bolton:1988af} & at rest & 1.5\\
\bottomrule
\end{tabular}
\end{center}
\caption{\label{tab:timecuts} Cuts on the decay time of the ALP that should be applied in various LFV experiments.}
\end{table}

\clearpage

\section{Details of experimental cuts} \label{app:E}
In this Appendix we describe how we approximate the effects of experimental cuts in order to obtain our bounds.

\subsection{Lab frame lifetimes}
A few of the measurements we use to set bounds require cuts on the time for the ALP to decay in the lab frame. 
The momentum of an ALP produced in a decay $M_1 \to M_2 \, a$ in the rest frame of the decaying $M_1$ is given by
\begin{align}
\label{eq:alpmommesonframe}
p_{a}(m_{M_1},m_{M_2},m_a) 
= \, &\frac{1}{2m_{M_1}}\sqrt{\left(m_{M_1}^2-(m_{M_2}+m_{a})^2 \right)\left(m_{M_1}^2-(m_{M_2}-m_{a})^2 \right)} \, ,
\end{align}
from which the Lorentz factor for the boost of the ALP in the rest frame of the decaying $M_1$ can be found as $\sqrt{\gamma_a^2 - 1} = p_{a}(m_{\mu},m_{e},m_a) / m_a$. The fraction of ALPs which decay within a time $t^\text{max}$ in the rest frame of the decaying $M_1$ is
\begin{equation}
f_{\tau<t^\text{max}} = \frac{1}{\tau_0 \gamma_a} \int_0^{t^\text{max}} dt \exp \left(- \frac{t}{\tau_0 \gamma_a} \right),
\end{equation}
where $\tau_0$ is the proper lifetime of the ALP. We summarise the time cuts that we use in Tab.~\ref{tab:timecuts}. Sometimes the actual cut as done by the experiment is on the time difference between the detection of various particles in the final state; but we approximate the effects of this by taking this time difference to be a cut on the maximum lab-frame lifetime of the ALP.

\subsection{Decay lengths}
\label{appendix:decaylengths}
Whether the ALP is long-lived enough to escape a detector -- and be constrained by measurements with final state missing energy -- will depend on its proper lifetime, its lab frame boost and the size of the detector. Conversely, for an ALP to be detectable via its visible decay modes, it must decay sufficiently promptly. This is an important consideration especially for decays to pairs of photons or electrons; in some regions of parameter space (in particular below the $\mu^+ \mu^-$ threshold and for small couplings) the ALP can be rather long-lived. Therefore the fraction of ALPs that would decay within the detector must be taken into account before bounds from, e.g.,~$K\to \pi \gamma \gamma$ measurements can be applied. 
A summary of the relevant measurements and parameters are given in Table~\ref{tab:decaylengths}, with the necessary formulae explained below.

\subsubsection{Initial state at rest in the lab frame}
In some experiments the decaying meson is at rest, or has zero transverse momentum, in the lab frame. Then the fraction of ALPs produced in the decay which escape a detector of transverse radius $R_{\text{max}}$ is
\begin{equation}
\label{eqn:Ftransverse}
F_T(m_{M_1},m_{M_2},m_a,R_{\text{max}})=\int_{0}^{\pi/2}\sin \theta \,d\theta \exp\left(-\frac{m_a R_{\text{max}}}{\tau_0\, |p^T_{LAB}|} \right)
\end{equation}
with (using Eqn.~\eqref{eq:alpmommesonframe} above)
\begin{equation}
p_{LAB}^T=p_{a}(m_{M_1},m_{M_2},m_a) \sin \theta,
\end{equation}
and where $\tau_0$ is the proper lifetime of the ALP.

\subsubsection{Initial state boosted in the lab frame}
\label{sec:Bfactaus}
In the case that the initial state is longitudinally boosted in the lab frame, with a boost defined by the Lorentz factor $\gamma_{M_1}$, then the longitudinal momentum of the ALP in the lab frame is given by
\begin{equation}
p^L_{LAB}(\beta_{M_1},\gamma_{M_1}) = \gamma_{M_1}\left(E_a+ \beta_{M_1} \, p_a \cos \theta \right) .
\end{equation}
where $p_{a}=p_{a}(m_{M_1},m_{M_2},m_a)$ as given in Eqn.~\eqref{eq:alpmommesonframe}, and $E_a^2=p_a^2+m_a^2$. Then the fraction of ALPs produced in this decay which escape a detector of length $L_{\text{max}}$ is
\begin{equation}
\label{eqn:Flongitudinal}
F_L(m_{M_1},m_{M_2},m_a,\beta_{M_1},\gamma_{M_1},L_{\text{max}})=\int_{0}^{\pi/2}\sin \theta \,d\theta \exp\left(-\frac{m_a L_{\text{max}}}{\tau_0\, |p^L_{LAB}(\beta_{M_1},\gamma_{M_1})|}\right).
\end{equation}

If the direction of the boost of the initial particle is unknown, things become more complicated. This is the situation, for example, of decaying $\tau$ leptons at $B$ factories such as BaBar and Belle, in which $\tau$ pairs are produced at an unknown angle $\theta$ from asymmetric beams.
The probability that the ALP will escape a cylindrical volume with transverse radius $x^T_{max}$ and longitudinal length $\pm z_{max}$ is
\begin{align}
\label{eq:pescape}
&F_{\theta} \left(m_{M_1},m_{M_2},m_a,\gamma,z_{max}, x^T_{max}\right) =\\& \frac{1}{2\pi} \int_0^{2\pi} d \phi_a \int_0^{\pi/2} \sin \theta \,d \theta \int_0^{\pi/2} \sin \theta_a \,d \theta_a \exp\left(-\frac{z_{max}}{z_0} \right)\exp\left(-\frac{x^T_{max}}{x^T_0} \right).\nonumber
\end{align}
Here, $\gamma$ is the boost of the centre-of-mass (CM) in the lab frame,\footnote{In the case of BaBar and Belle, the $e^-$ beam has 9 GeV energy in the lab frame (and defines the $+z$ direction), and the $e^+$ beam has 3.1GeV energy. This means the boost of the CM frame is $\gamma \beta$=0.56.} and
\begin{align}
z_0 &= \frac{c \tau_0}{m_a} \left( p_a \cos \phi_a \sin \theta_a \sin \theta + \gamma_{M_1} \left(\beta_{M_1} E_a + p_a \cos \theta_a\right)\cos \theta\right)\\
x^T_0 &=  \frac{c \tau_0}{m_a} \sqrt{\left(p_a \cos \theta_a \sin \theta_a \cos \theta- \gamma_{M_1} \left(\beta_{M_1} E_a + p_a \cos \theta_a\right) \sin \theta \right)^2 + \left(p_a \sin \phi_a \sin \theta_a \right)^2 } \label{eq:xT0taus}
\end{align}
where $p_a = p_a (m_{M_1},m_{M_2},m_a)$ is given in Eqn.~\eqref{eq:alpmommesonframe}. The boost of the decaying $M_1$ along the direction of its momentum in the lab frame is given by $\gamma_{M_1}$, with
\begin{equation}
\gamma_{M_1} \beta_{M_1} = \frac{1}{m_{M_1}} \sqrt{\left( p_{M_1} \sin \theta \right)^2 + \gamma^2\left(\beta E_{M_1} +  p_{M_1} \cos \theta\right)^2 },
\end{equation}
where $E_{M_1} = \sqrt{s}/{2}$, $p_{M_1} = \sqrt{s/4-m_{M_1}^2}$.

\begin{table}
\scalebox{0.8}{
\begin{tabular}{c c c l l}
\toprule
Decay & Experiment & Initial state & Dimension (m) & Fraction\\
\midrule
$K^+ \to \pi^+$+inv. & NA62~\cite{CortinaGil:2021nts} & boosted & $L_{\text{max}} =140$ & $F_L(m_{K},m_{\pi},m_a,\beta_{K}^\text{NA62},\gamma_{K}^\text{NA62},L_{\text{max}})$  \\
$K_L \to \pi^0$+inv. & KOTO~\cite{Ahn:2018mvc} & boosted & $L_{\text{max}} =4.148$ & $F_L(m_{K},m_{\pi},m_a,\beta_{K}^\text{KOTO},\gamma_{K}^\text{KOTO},L_{\text{max}})$ \\
$B \to K$+inv. & BaBar~\cite{Lees:2013kla} & at rest (T) & $R_{\text{max}} =3.0$ &  $F_T(m_B, m_K, m_a, R_{\text{max}})$\\
$B \to K^*$+inv. & Belle~\cite{Grygier:2017tzo} & at rest (T) & $R_{\text{max}} =3.0$ &  $F_T(m_B, m_K^*, m_a, R_{\text{max}})$\\
$K^+ \to \pi^+ \gamma \gamma$ & E949~\cite{Artamonov:2005ru} & at rest & $R_{\text{max}} =1.45$ & $1-F_T(m_K, m_\pi, m_a, R_{\text{max}})$ \\
$K^+ \to \pi^+ \gamma \gamma$ & NA62~\cite{Ceccucci:2014oza} & boosted & $L_{\text{max}} =140$ & $1-F_L(m_{K},m_{\pi},m_a,\beta_{K}^\text{NA62},\gamma_{K}^\text{NA62},L_{\text{max}})$ \\
$K_L^0 \to \pi^0 \gamma \gamma$ & NA48~\cite{Lai:2002kf} & boosted & $L_{\text{max}} =140$ & $1-F_L(m_{K},m_{\pi},m_a,\beta_{K}^\text{NA48},\gamma_{K}^\text{NA48},L_{\text{max}})$ \\
$K_L^0 \to \pi^0 \gamma \gamma$ & KTeV~\cite{Abouzaid:2008xm} & boosted & $L_{\text{max}} =105$ & $1-F_L(m_{K},m_{\pi},m_a,\beta_{K}^\text{KTeV},\gamma_{K}^\text{KTeV},L_{\text{max}})$ \\
$B \to \pi \,l^+ l^-$& Belle~\cite{Wei:2008nv} & at rest (T) & $R_{\text{max}} =0.005$ &  $1-F_T(m_B, m_\pi, m_a, R_{\text{max}})$\\
\midrule
$\mu \to e a$ (inv.) & TWIST~\cite{Bayes:2014lxz} & at rest & $R_{\text{max}} =0.165$& $F_T(m_\mu, m_e, m_a, R_{\text{max}})$\\
$\tau \to 3 \mu$ & BaBar~\cite{Hayasaka:2010np} & boosted & $\lbrace z_{\text{max}}, x^T_{\text{max}}\rbrace=\lbrace 0.03, 0.005\rbrace$ & $1-F_{\theta} \left(m_{\tau},m_{\mu},m_a,\gamma_{\text{CM}}^\text{BaBar},z_{max}, x^T_{max}\right)$\\
$\tau \to \mu e e$ & BaBar~\cite{Hayasaka:2010np} & boosted & $\lbrace z_{\text{max}}, x^T_{\text{max}}\rbrace=\lbrace 0.03, 0.005\rbrace$ & $1-F_{\theta} \left(m_{\tau},m_{\mu},m_a,\gamma_{\text{CM}}^\text{BaBar},z_{max}, x^T_{max}\right)$\\
$\tau \to 3 e$ & BaBar~\cite{Hayasaka:2010np} & boosted & $\lbrace z_{\text{max}}, x^T_{\text{max}}\rbrace=\lbrace 0.03, 0.005\rbrace$ & $1-F_{\theta} \left(m_{\tau},m_{e},m_a,\gamma_{\text{CM}}^\text{BaBar},z_{max}, x^T_{max}\right)$\\
$\tau \to e \mu \mu$ & BaBar~\cite{Hayasaka:2010np} & boosted & $\lbrace z_{\text{max}}, x^T_{\text{max}}\rbrace=\lbrace 0.03, 0.005\rbrace$ & $1-F_{\theta} \left(m_{\tau},m_{e},m_a,\gamma_{\text{CM}}^\text{BaBar},z_{max}, x^T_{max}\right)$\\
\bottomrule
\end{tabular}}
\caption{\label{tab:decaylengths} Summary of maximum/minimum ALP decay lengths for relevant experiments, and the fraction of ALPs that pass the decay length cuts, in terms of the $F$ functions given in the text. ``At rest (T)'' means that the initial state is at rest in the transverse plane of the experiment. The various Lorentz factors involved are taken to be $\beta_{K}^{\text{NA48}}\gamma_{K}^{\text{NA48}}=\beta_{K}^{\text{NA62}}\gamma_{K}^{\text{NA62}}=75~\text{GeV}/m_K$~\cite{NA62:2017rwk}, $\beta_{K}^{\text{KTeV}}\gamma_{K}^{\text{KTeV}}=70~\text{GeV}/m_K$~\cite{Arisaka:1992ch}, $\beta_{\text{CM}}^{\text{BaBar}}\gamma_{\text{CM}}^{\text{BaBar}}=0.56$, $\beta_{K}^{\text{KOTO}}\gamma_{K}^{\text{KOTO}}=1.5~\text{GeV}/m_K$~\cite{Masuda:2015eta}.}
\end{table}

\subsection{Two photons mimicking one}
\subsubsection[$\mu \to e \gamma_{\text{eff}}$]{\boldmath{$\mu \to e \gamma_{\text{eff}}$}}
The measurement of $\mu \to e \gamma$~\cite{TheMEG:2016wtm} can also set bounds on the $\mu \to e a$ process with subsequent $a\to \gamma\gamma$ decay, if the two photons land within a distance smaller than the resolution of the detector. This can happen if the ALP is sufficiently boosted, and/or it decays sufficiently close to the photon detector. The distance between the muon decay point and the LXe photon detectors is approximately 1m, and the spatial resolution of the LXe detector is 5mm. 

If we approximate the lab frame opening angle of the photons by\footnote{It turns out that the bound does not change noticeably if this is calculated more carefully}
\begin{equation}
\cos \theta \approx \beta_a,
\end{equation}
 then the ALP's decay will mimic a single photon event if
\begin{equation}
2 \tan \theta \left(L_{\text{detector}}- L_{\text{decay}}\right)<\sigma
\end{equation}
where $\sigma$ is the resolution of the photon detector. In MEG, $L_{\text{detector}}=1$m, $\sigma=5$mm, and so a $\mu \to e \gamma\gamma$ event will look like $\mu \to e \gamma$ if $L_{\text{min}} <L_{\text{decay}}< 1$m, where
\begin{equation}
L_{\text{min}}=L_\text{detector} -\frac{\sigma}{2\tan \theta}
\end{equation}
The fraction of ALPs which will decay within this range is calculated as
\begin{equation}\label{eq:fLeff}
f_{L_{\text{min}}<L<1.0}=\frac{1}{L_a}\int_{L_{\text{min}}}^{1.0}dx \exp \left(-\frac{x}{L_a} \right)
\end{equation}
where $L_a = \sqrt{\gamma_a^2 - 1} \, \tau_0$ is the lab-frame decay length of the ALP.

\subsubsection[$\tau \to \mu \gamma_{\text{eff}}$ and $\tau \to e \gamma_{\text{eff}}$]{\boldmath{$\tau \to \mu \gamma_{\text{eff}}$ and $\tau \to e \gamma_{\text{eff}}$}}
The bounds on $\tau \to \mu \gamma$ and $\tau \to e \gamma$ were measured at the Babar experiment~\cite{Aubert:2009ag}. The initial $\tau$ is boosted in the lab frame, as described in Sec.~\ref{sec:Bfactaus}.
The radius of the Babar electromagnetic calorimeter is 1.375m, and it is segmented into square crystals of dimension 47$\times$47mm~\cite{Aubert:2001tu}. Then, similarly to the case for $\mu \to e \gamma_\text{eff}$, defining $R_\text{EMC}=1.375$m and $\sigma=4.7 \times 10^{-3}$m, the ALP's decay will mimic a single photon event if $R_{\text{min}} <R_{\text{decay}}< R_{\text{EMC}}$, where
\begin{equation}
R_{\text{min}}=R_{\text{EMC}} -\frac{\sigma}{2\tan \theta}
\end{equation}
where $R_\text{decay}$ is the radial distance at which the ALP decays. Here, the boosts of the ALPs are distributed according to the initial momentum of the $\tau$, and the angle of the ALP's momentum relative to the lab frame, so the fraction of decays which will mimic a single photon event is given by 
\begin{align}
\label{eq:pgammaeff}
f_{R_{\text{min}}<R<R_{\text{EMC}}} = \frac{1}{2\pi} \int_0^{2\pi} d \phi_a \int_0^{\pi/2} \sin \theta \,d \theta \int_0^{\pi/2} \sin \theta_a \,d \theta_a \, \frac{1}{x^T_0}\int_{R_\text{min}}^{R_{\text{EMC}}} dr \exp\left(-\frac{r}{x^T_0} \right),
\end{align}
where $x^T_0$ is given in Eqn.~\eqref{eq:xT0taus}, with $M_1=m_\tau$, and $M_2=m_\mu$ for $\tau \to \mu \gamma_\text{eff}$, or $M_2=m_e$ for $\tau \to e \gamma_\text{eff}$.

Instead of performing the full angular integration in \eqref{eq:pgammaeff}, we adapt \eqref{eq:fLeff} to set the limits shown in Figure~\ref{fig:leptonconstraints2}, Figure~\ref{fig:leptonconstraints3}, Figure~\ref{fig:ctaueplots} and Figure~\ref{fig:ctaumuplots}. We define the maximal decay length for an ALP produced in $\tau$ decays as the maximal transverse distance it travels from the interaction point $L_a^{\text{max},\tau}=\text{Max}(x_0^T)$ for any angle $\theta_a, \theta$ and $\phi_a$ and define the fraction of ALPs that decay before the corresponding ECAL component of the respective experiment as
\begin{align}\label{eq:ftaulim}
f^\tau_{L_{\text{min}}<L<L_\text{max}}=\frac{1}{L_a^{\text{max},\tau}}\int_{L_{\text{min}}}^{L_\text{max}}dx \exp \left(-\frac{x}{L_a^{\text{max},\tau}} \right)\,.
\end{align}
We checked that \eqref{eq:ftaulim} is a good approximation to \eqref{eq:pgammaeff} for the parameter space shown in Figure~\ref{fig:leptonconstraints2}, Figure~\ref{fig:leptonconstraints3}, Figure~\ref{fig:ctaueplots} and Figure~\ref{fig:ctaumuplots}.

\subsection[Binned $B\to K^{(*)} e^+ e^-$]{\boldmath{Binned $B\to K^{(*)} e^+ e^-$}}

The bounds from the differential distribution of $\text{Br}(B\to K^{*} e^+ e^-)$ at LHCb \cite{Aaij:2015dea} were calculated as follows. The longitudinal momentum distribution of $B^0$ mesons produced at $7/8$ TeV collision energy was approximated by taking an average $B$ transverse momentum of $\langle p_T \rangle=\,$5.5 GeV and using the measured $B^0$ pseudorapidity distribution given in \cite{Aaij:2013noa}. The longitudinal momentum at a given pseudorapidity $y$ is given by
\begin{equation}
p_L =\frac{1}{2}e^{-y} \left(e^{2y}-1 \right)\sqrt{m_B^2+\langle p_T\rangle ^2}.
\end{equation}
From this, the distribution of longitudinal boosts of the CM frame ($\beta_{B}^{\text{LHCb}}$and  $\gamma_{B}^{\text{LHCb}}$) can be derived. We then assume that an ALP will be detected in this measurement if it decays to a pair of electrons within $L_{\text{max}}^{\text{LHCb}}$=0.74m longitudinal distance. The fraction of ALPs which decay within this distance is $1-F^{\text{LHCb}}$, where $F^{\text{LHCb}}=F_L(m_{B},m_{K^*},m_a,\beta_{B}^{\text{LHCb}},\gamma_{B}^{\text{LHCb}},L_{\text{max}}^{\text{LHCb}})$ is defined in Eqn.~\eqref{eqn:Flongitudinal} above.
We approximate the effects of finite experimental resolution of the electron pair invariant mass by using a Gaussian smearing function, following the method in Ref.~\cite{Altmannshofer:2017bsz}. The smearing function is defined
\begin{equation}
\mathcal{G}(q_{\text{min}}, q_{\text{max}})=\frac{1}{\sqrt{2 \pi}r_e}\int_{q_{\text{min}}}^{q_{\text{max}}}d|q| \exp\left(- \frac{(|q|-m_a)^2}{2 r_e^2} \right)
\end{equation}
where the resolution is taken to be $r_e=10$ MeV~\cite{Altmannshofer:2017bsz,Ilten:2015hya}. Then the total NP contribution to a bin is given by
\begin{equation}
\langle \text{Br}\left(B \to K^* e e \right) \rangle \Big|_{q_{\text{min}}}^{q_{\text{max}}}=\left(1-F^{\text{LHCb}} \right)\times  \mathcal{G}(q_{\text{min}}, q_{\text{max}}) \times  \text{Br}\left(B \to K^* a \right)\times  \text{Br}\left(a \to e^+ e^- \right).
\end{equation}

\newpage
\addcontentsline{toc}{section}{References}
\bibliographystyle{JHEP}
\bibliography{references}

\end{document}